\documentclass[11pt,a4paper]{article}
\usepackage{graphicx} 
\usepackage{jheppubmod}
\usepackage{amsmath}
\usepackage{amssymb}
\usepackage{amsthm}
\usepackage{setspace}
\usepackage{xcolor}

\usepackage{array,hhline,multirow}
\usepackage{genyoungtabtikz}
\usepackage{simpler-wick} 
\usepackage{pifont}
\usepackage{tikz-cd}
\usepackage{mathtools}
\usepackage{pdflscape} 
\usepackage{faktor}

\usepackage{seqsplit}

\usepackage{xstring}

\usetikzlibrary{trees}
\usetikzlibrary{decorations.pathmorphing}
\usetikzlibrary{decorations.markings}
\usetikzlibrary{shapes.misc}
\usetikzlibrary{decorations.pathreplacing}
\usetikzlibrary{matrix}
\usetikzlibrary{arrows.meta, positioning}

\tikzset{
    arr/.tip={Triangle[fill=black, scale=1]},
  threept/.style={
    circle,
    draw,
    inner sep=2pt,
  },
  diff/.style={
    rectangle,
    draw,
    inner sep=3pt,
    minimum size=10pt
  },
    smalldiff/.style={
    rectangle,
    draw,
    inner sep=3pt,
    minimum size=3pt
  },
  twopt/.style={
    circle,
    draw,
    fill=black,
    inner sep=1.5pt,
    minimum size=1pt
  },
    bowtie/.style={
    draw=none,
    inner sep=0pt,
    minimum size=1pt,
    font=\normalsize
  },
  cross/.style={
    cross out,
    draw=black, 
    line width=1pt,
    minimum size=7pt, 
    inner sep=0pt,
    outer sep=0pt
  },
  scalar/.style={
    thick,
    densely dashed,
    postaction={
      decorate,
      decoration={
        markings,
        mark=at position 0.7 with {\arrow{latex}}
      }
    }
  },
    scalar2/.style={
    postaction={
      decorate,
      decoration={
        markings,
        mark=at position 0.7 with {\arrow{latex}}
      }
    }
  },
  spinning/.style={
    thick,
    postaction={
      decorate,
      decoration={
        markings,
        mark=at position 0.7 with {\arrow{latex}}
      }
    }
  },
  spinning no arrow/.style={
    thick,
  },
  finite with arrow/.style={
    decoration={
      snake,
      amplitude=1pt,
      segment length=6pt,
      post length=6pt
    },
    decorate,
    thick,
  -{Latex[length=5pt, width=4pt]}
  },
   doublespinning/.style={
    thick, double distance=1.5pt,
    postaction={
      decorate, very thick,
      decoration={
        markings,
        mark=at position 0.5 with {\arrow{To}}
      }
    }
  },
  finite/.style={
    decoration={
      snake,
      amplitude=1pt,
      segment length=6pt,
    },
    decorate,
    thick
  }
}

\DeclareMathOperator{\str}{str}
\DeclareMathOperator{\sdet}{sdet}

\newcommand\cN{\mathcal{N}}
\newcommand\nn\nonumber

\newcommand{\dota}{\dot{\alpha}}

\newcommand{\hX}{\hat{X}}
\newcommand{\dfb}{{\dot{\mathfrak{b}}}}
\newcommand{\fb}{{\mathfrak{b}}}
\newcommand{\fc}{{\mathfrak{c}}}

\newcommand{\fD}{{\mathfrak{D}}} 
\newcommand{\dfc}{{\dot{\mathfrak{c}}}}

\newcommand\bPi{\bar{\Pi}}

\def \rep{\Psi}
\def\frep{\Phi}
\def\finite{\Psi}

\def\ula{{\underline \lambda}}
\def\umu{{\underline \mu}}

\def \da{\dot \alpha}
\def \db{\dot \beta}

\def \bD{{\bar D}}
\def \udfa{{\underline {\dot{\mathfrak{a}}}}}
\def \udfb{{\underline {\dot{\mathfrak{b}}}}}
\def \ufa{{\underline{\mathfrak{a}}}}
\def \ufb{{\underline{\mathfrak{b}}}}

\def \ufc{{\underline{\mathfrak{c}}}}
\def \da{\dot \alpha}
\newcommand{\diagramEnvelope}[1]{#1}
\newcommand{\cO}{\mathcal{O}}
\newcommand{\cD}{\mathcal{D}}

\newcommand{\cA}{\mathcal{A}}
\newcommand{\cB}{\mathcal{B}}
\newcommand{\cC}{\mathcal{C}}

\newcommand{\cF}{\mathcal{F}}
\newcommand{\fa}{{\mathfrak{a}}}
\newcommand{\dfa}{{\dot{\mathfrak{a}}}}

\newcommand{\bX}{{\bar{X}}}

\newcommand{\bW}{\bar{W}}

\newcommand{\GL}{\mathrm{GL}}
\newcommand{\SL}{\mathrm{SL}}
\newcommand{\Gr}{\mathrm{Gr}}
\newcommand{\SU}{\mathrm{SU}}

\newcommand{\bea}{\begin{equation}\begin{aligned}}
\newcommand{\eea}[1]{\label{#1}\end{aligned}\end{equation}}
\newcommand{\beq}{\begin{equation}}
\newcommand{\eeq}{\end{equation}}


\title{Superconformal Weight Shifting Operators
}
\author{Tobias Hansen,}
\author{Paul Heslop,}
\author{Hector Puerta-Ramisa}
\affiliation{Department of Mathematical Sciences, Durham University, \\
Durham, DH1 3LE, United Kingdom}
\emailAdd{tobias.p.hansen@durham.ac.uk}
\emailAdd{paul.heslop@durham.ac.uk}
\emailAdd{hector.puerta-ramisa@durham.ac.uk}
\abstract{
We develop a framework for constructing superconformal blocks for correlators of general supermultiplets  in theories with $\mathrm{SU}(m,m|2n)$ symmetry, such as four-dimensional $\mathcal{N}\!=\!2$ and $\mathcal{N} \!=\! 4$ conformal theories. 
We use  analytic superspace, viewed as the super-Grassmannian $\mathrm{Gr}(m|n,2m|2n)$, which includes
4D Minkowski space ($m\!=\!2,n\!=\!0$).  
In this formalism, superblocks for non-half-BPS correlators are analogous to non-supersymmetric conformal blocks for correlators of fields with spin.
We construct $\mathrm{SU}(m,m|2n)$-covariant differential operators which generalise the existing conformal weight-shifting operators, and thus allow us to derive all superconformal blocks from the known half-BPS blocks.
Our results provide a framework from which to advance the conformal bootstrap in 4D supersymmetric settings, with potential extensions to lower and higher-dimensional SCFTs. 
The Grassmannian formalism is also seen to offer a natural and often simpler alternative to the embedding space formalism of non-supersymmetric CFTs.

}

\numberwithin{equation}{section}
\begin{document}

\maketitle
\section{Introduction}

Conformal blocks provide a natural basis for four-point correlation functions in conformal field theories (CFTs), with coefficients given by products of operator product expansion (OPE) coefficients,  the key data defining a CFT. 
This structure lies at the heart of the conformal bootstrap programme, which seeks to determine CFT data by enforcing consistency conditions such as crossing symmetry and unitarity. Perturbatively, known four-point functions can be decomposed into blocks to extract analytic data, which in turn constrains higher-order correlators. Non-perturbatively, even when explicit expressions for correlators are unavailable, the interplay of crossing symmetry, the conformal block expansion, and unitarity imposes strong constraints that can significantly restrict, and in some cases fully determine, the space of consistent CFTs~\cite{Mack:1969rr,Polyakov:1974gs,FERRARA1973161,Ferrara:1974nf,Rattazzi:2008pe,Caracciolo:2009bx,Poland:2011ey,El-Showk:2012cjh,Poland:2018epd}.

Blocks for correlators of four scalar fields (scalar blocks) are well understood~\cite{Dolan:2000ut,Dolan:2003hv,Dolan:2011dv} and, although there are well developed techniques for computing blocks for correlators of operators with spin  (spinning blocks)~\cite{Costa:2011mg,Costa:2011dw,Penedones:2015aga,Costa:2016hju,Costa:2016xah,Cuomo:2017wme,Karateev:2017jgd,Erramilli:2019njx}, the bulk of bootstrap work has so far been restricted to using scalar blocks. This is partly due to the technical complication of working with fields with spin. 
However, the full bootstrap programme of general CFTs requires the input of blocks for all operators,
and indeed using spinning blocks has been shown to give better bounds~\cite{,Dymarsky:2017xzb,Reehorst:2019pzi,He:2023ewx,Dymarsky:2017yzx,Chang:2024whx}.

Conformal field theories with extended supersymmetry (SCFTs) provide the best understood examples of 4D quantum field theories, with $\cN=4$ super Yang-Mills (SYM) the ultimate example. 
Therefore, they provide the perfect arena to explore bootstrap techniques. This is especially interesting in the strong coupling regime of $\cN=4$ SYM as it gives access to quantum gravity results via AdS/CFT~\cite{Maldacena:1997re}.

In SCFTs,
the blocks of conformal multiplets in the same supermultiplet assemble together into superconformal blocks (superblocks). 
So far, in 4 dimensions, only the superblocks of half-BPS supermultiplets are known explicitly. These were originally derived in~\cite{Dolan:2001tt,Dolan:2004mu,Nirschl:2004pa}
by decomposing the correlator into various functions by solving Ward identities and giving the  contributions of each superblock to these functions. In~\cite{Doobary:2015gia} all superblocks were written down explicitly in a manifestly superconformal way, making use of analytic superspace. Other papers which have considered and made use of half BPS blocks in a variety of different methods/formalisms include~\cite{Beem:2014zpa,Bobev:2015jxa,Bissi:2015qoa,Li:2016chh,Bobev:2017jhk}. Bootstrap results in these SCFTs have made primary use of half BPS superblocks (see~\cite{Poland:2018epd,Bissi:2022mrs,Heslop:2022xgp} for reviews). 
Examples are numerical bootstrap results constraining $\cN=2$ and $\cN=4$ theories non-perturbatively~\cite{Beem:2013qxa,Beem:2014zpa,Beem:2016wfs,Chester:2021aun,Chester:2024bij} as well as direct analytic  bootstrapping of strong coupling $\cN=4$ correlators in order to obtain perturbative quantum gravity amplitudes~\cite{Aprile:2017bgs,Aprile:2017qoy,Caron-Huot:2018kta,Alday:2018pdi,Aprile:2019rep,Abl:2020dbx,Aprile:2020mus,Huang:2021xws,Drummond:2022dxw,Alday:2022uxp,Alday:2022xwz,Alday:2023jdk,Alday:2023mvu}.

The non-half-BPS  superblocks on the other hand are not explicitly known (although  there has been some work involving  superblocks beyond half-BPS in various theories~\cite{Fitzpatrick:2014oza,Khandker:2014mpa,Lemos:2016xke,Li:2017ddj,Cornagliotto:2017dup,Ramirez:2018lpd,Kos:2018glc,Buric:2019rms,Buric:2020buk,Buric:2020qzp,Li:2018mdl,Rakshit:2023baq}).
In this paper, we develop a formalism to compute  non-half-BPS superblocks in 4D $\cN=2,4$ SCFTs. While the focus is on the aforementioned cases, the formalism can also be applied to non-supersymmetric CFT in 1 and 4 dimensions, as well as certain 1D supersymmetric theories.

To achieve this,  we will be using a natural and unified formulation of theories on various flat (super)spaces exhibiting superconformal symmetry, which we review in section \ref{sec:cosetspace}. In particular, we consider theories with superconformal  symmetry $\SU (m,m|2n)$ using  $(m,n)$ analytic superspace~\cite{Galperin:1984av,Howe:1995md,Hartwell:1994rp,Howe:2001je,Heslop:2001dr,Heslop:2001gp,Heslop:2001zm,Heslop:2002hp,Heslop:2003xu,Doobary:2015gia,Heslop:2022xgp}. 
This superspace is realised as a coset space of $\SU (m,m|2n)$:
a flag supermanifold isomorphic to the (super)Grassmannian space $\Gr(m|n, 2m|2n)$ \cite{Howe:1995md,Heslop:2003xu}. 
We develop the formulation of analytic superspace directly on both the super-coset and super-Grassmannian, rather than using the more standard coordinate approach. 
This has the advantage of completely manifesting the full superconformal symmetry and, as we will show, is especially well suited to non-half-BPS superblocks.
Moreover, this allows us to work with 1D and 4D $\cN=0,2,4$ CFTs in a completely unified manner, with each case corresponding to a different choice of $m$ and $n$. 

Our methods can be viewed as a simple generalisation of 4D CFT ($m=2$, $n=0$) where we simply promote the usual 2-component Weyl spinor indices $\alpha,\dot \alpha$ to superindices $\fa,\dfa$ running over an $(m|n)$ vector space. 
Thus, in this context, half-BPS operators are direct generalisations of 4D scalar operators, and non-half-BPS operators are then analogous to 4D spinning operators.

A major advancement in the construction of spinning blocks was the introduction of weight-shifting operators~\cite{Karateev:2017jgd}.
These are covariant differential operators that map between conformal representations, enabling a universal and algebraic method for generating all spinning blocks from scalar seeds.
We generalise these weight-shifting operators in 4D to the supersymmetric $(m,n)$ case, in the process yielding a reformulation of the  4D weight-shifting operators on the  Grassmannian $\Gr(2,4)$. 
The complication then largely reduces to dealing with the finite-dimensional representations carried by the $\SL(m|n)$ superindices.  We do this with the aid of some rather beautiful but not so well known results involving Hermitian Young projectors and their transition operators~\cite{alcock2017compact,Alcock-Zeilinger:2016cva}, equivalent to Young's semi-normal units~\cite{garsia2020lectures}. 
Strictly speaking, not all finite-dimensional $\SL(m|n)$ representations can  be directly obtained with   superindices, but rather only those leading to (half-)integer scaling dimension. Representations with general scaling dimensions can however be simply deduced from these with the help of quasi-tensors~\cite{Heslop:2001zm,Heslop:2003xu}.

We thus obtain supersymmetric weight-shifting operators which map from half-BPS multiplets to general multiplets, and can be used to give  superblocks of general operators.
As a by-product, the Grassmannian formalism also often gives  simpler expressions for 4D non-supersymmetric weight-shifting operators than the embedding space expressions. In particular, when acting  on chiral fields the weight-shifting operators can be viewed as  single  partial derivatives.

The paper proceeds as follows.

In section \ref{sec:cosetspace}, after a review of the analytic superspace construction, we provide an extensive treatment of (super)conformal representations on this Grassmannian space in terms of Young diagrams, and describe the most general solution to the $\SU (m,m|2n)$ Ward identities for functions of up to four points. Note that the formalism manifests the superconformal symmetry and thus the SUSY Ward identities are straightforwardly solved.

In section \ref{sec:weightshifting1}, we construct superconformal weight-shifting operators~\eqref{eq:collectedWSO} which are able to map between representations of $\SU (m,m|2n)$ generalising  the embedding space $\mathrm{SO}(d+1,1)$ weight-shifting operators of \cite{Karateev:2017jgd}.
We show how they map different  4D $\cN=2,4$ superconformal multiplets in section \ref{sec:genWSO}. 
In particular, the simplest (fundamental) weight-shifting operator acts by adding a box to a (supersymmetric) Young diagram describing the transformation of a field under the isotropy subgroup of the superspace. The position of the new box contains the information on how the different quantum numbers (dimension, spin, $R$-symmetry) and $Q,\bar{Q}$ shortening conditions are changed.\footnote{A full translation between 4D $\cN=2,4$ superconformal representations and our coset space description in terms of Young diagrams is given in appendix \ref{sec:repexamples}.}

The weight-shifting operators map to $\SU (m,m|2n)$-covariant operators rather than invariant ones, meaning that one needs to construct invariant combinations of them in order to obtain physically relevant (superconformally invariant) quantities. 
In section \ref{sec:WSOinvariants}, to understand how to construct such quantities, we begin by reviewing the helpful diagrammatic notation introduced in \cite{Karateev:2017jgd}. We then formulate the algebra of superconformal weight-shifting operators in this language, which works analogously to the conformal case described in \cite{Karateev:2017jgd}.
In particular, this algebra includes Schur's lemma (known as the bubble diagram): any $\SU (m,m|2n)$-invariant combination of weight-shifting operators must either vanish or be an isomorphism, as we explicitly verify. 
Thus, superconformally invariant compositions must always involve differential operators acting at different points of an $N$-point function.
The algebra of weight-shifting operators can then be used to understand how such invariant compositions act, leading to the construction of the differential basis of three-point structures, which was first described in the non-supersymmetric  case in~\cite{Costa:2011dw}. 
Indeed, one can form a basis of tensor structures for a three-point function by acting with different invariant compositions of weight-shifting operators on a simpler three-point structure, e.g.\ three-point function of half-BPS supermultiplets. 
We show how to construct this basis and that it is complete in a number of cases.

Then, in section \ref{sec:blocks}, we use the differential basis created by our superconformal weight-shifting invariants to construct superblocks.
In the non-supersymmetric setting, all spinning conformal blocks can be constructed from the scalar conformal blocks via weight-shifting techniques~\cite{Karateev:2017jgd}.  In our superconformal setting,  we are similarly able to construct superconformal blocks in terms of the known half-BPS superblocks. After a review of general superblocks on the coset space, we illustrate the method by providing some explicit examples of new superconformal blocks using weight-shifting operators, involving shifting both external and exchanged representations. We do a number of checks that they agree with the known 1D and 4D results. We discuss accessing non-integer dimensions in section~\ref{sec:nonintegerdim}, which largely works equivalently to the existing conformal weight shifting operators, with some important caveats when mapping from protected to long multiplets. Finally, we summarise the weight-shifting methodology in section~\ref{sec:blocksummary}.

We leave many technical details to the appendices.

\section{Superconformal representations on analytic superspace}\label{sec:cosetspace}
General flat superspaces can be viewed as coset spaces of the superconformal group~\cite{Howe:1995md}, just as Minkowski space can be viewed as a  coset of the conformal group 
$\mathbb{M} = \mathrm{SO}(d+2) / \{ L_{\mu \nu}, D, K_{\mu}\}.$
We will focus on a special coset space  of the generalised superconformal symmetry group $\SU (m,m|2n)$ which we complexify to~$\SL(2m|2n;\mathbb{C})$. 
The coset space is equivalent to 
 the maximal super-Grassmannian space $\Gr(m|n, 2m|2n)$, and is a particularly simple  example of  analytic superspace~\cite{Galperin:1984av,Howe:1995md,Hartwell:1994rp,Howe:2001je,Heslop:2001dr,Heslop:2001gp,Heslop:2001zm,Heslop:2002hp,Heslop:2003xu,Doobary:2015gia}.
 Theories with $\SU (m,m|2n)$ superconformal symmetry, and hence  which fit into this framework, include many cases of interest such as $4$D $\cN=2,4$ analytic superspaces ($m=2$, $n=1,2$), $4$D Minkowski space ($m=2$, $n=0$) and theories with compact $\SU (2n)$ symmetry ($m=0$). But there are also advantages in considering the generalised $m,n$ case which has a universal structure (see for example~\cite{Doobary:2015gia,Aprile:2017xsp,Aprile:2025nta} where bosonised $(m,0)$ superblocks are used as a tool for exploring superconformal cases). Note also that while the formalism we introduce and develop here is particularly natural for studying theories in 1 and 4 dimensions, there are also extensions involving generalised Grassmannian spaces which can be used for other dimensions~\cite{Howe:1994ms,Howe:2000nq,Heslop:2004du,Aprile:2021pwd}.

\subsection{Analytic superspace}\label{sec:setup}

The analytic superspace we consider here can be viewed as a natural generalisation of ordinary 4D Minkowski space in two-component Weyl spinor form. The latter has coordinates given by the $2\times 2$ matrix $x^{\alpha \dot \alpha}$ related to standard Lorentz notation via Pauli matrices 
$x^{\alpha \dot \alpha}= v^\mu \sigma_\mu^{\alpha \dot \alpha}$.  
Then, $(m,n)$ analytic superspace is viewed in exactly the same way by uplifting all $2\times 2$ matrices to $(m|n)\times (m|n)$ supermatrices. 
For example, position coordinates uplift as 
\begin{equation}
x^{\alpha \da}   \qquad \rightarrow \qquad x^{ \fa \dfa}= \left(\begin{array}{c|c}
        x^{\alpha \da} & \rho^{\alpha a'} \\\hline
         \bar{\rho}^{a \da} & y^{aa'} \end{array}\right)
\end{equation}
Weyl spinor indices $\alpha, \dot \alpha$ uplift to $m|n$ superindices $\fa=(\alpha|a),  \dfa=(\dot \alpha|a')$ where now $\alpha,\dot \alpha $ take on $m$ values and $a,a'$ $n$ values. 
These are superindices, meaning that the $a,a'$ parts are `fermionic' in the sense that symmetrising the superindex corresponds to antisymmetrising the internal index part $a$ and vice versa. In the $m=2$ case, the coordinates $x^{\alpha \dot \alpha}$ are precisely Minkowski coordinates whereas $y^{aa'}$ are internal coordinates used to deal with the 
internal $SU(2n)$ group of the $2n$-extended superconformal group. The coordinates $\rho,\bar \rho$ are then Grassmann odd superspace coordinates.

One can view 4D Minkowski space as a coset of the conformal group and relatedly  as the Grassmannian of 2-planes in 4D. Both of these concepts uplift to the full superconformal group. Both the supercoset and super-Grassmannian formulations of analytic superspace then manifest the full superconformal symmetry directly, and, as we will show, are particularly suited for the construction of weight-shifting operators and blocks.

\subsubsection{Detailed setup}
We define $(m,n)$ analytic superspace as a coset of $\SL(2m|2n)$, whose columns we permute to $\SL(m|2n|m)$, with a parabolic isotropy group given by $(m|2n|m)\times(m|2n|m)$ matrices with an upper block triangular form with $(m|n)\times(m|n)$ blocks
\begin{equation}\label{eq:H}
    H= \left\{ h\in \SL(m|2n|m) \ : \  h^{\cA}{}_{\cB}=\begin{pmatrix}
        A^{\fa}_{\ \fb} & 0 \\ C_{\dfa \fb} & D_{\dfa}^{\ \dfb}
    \end{pmatrix}\right\}.
\end{equation}
Here $\cA,\cB$ run over all $(m|2n|m)$ and $\fa, \dfa$ over $(m|n)$, analogously to $\alpha, \da$ indices in four dimensional Minkowski space.\footnote{For the case $m=2,n=0$ this is more than an analogy and the indices really are the Weyl spinor indices $\alpha, \da$.}
Thus elements in the coset $g \sim h g$ are given by
\begin{equation}\label{eq:g}
    g \coloneq \begin{pmatrix}
        X^\mathfrak{a}_{\ \mathcal{A}} \\ \bW_{\dfa \cA}
    \end{pmatrix}, \quad X^\mathfrak{a}_{\ \mathcal{A}}  \sim A^{\fa}_{\ \fb} X^{\fb}_{\ \cA} ,\quad \bW_{\dfa \cA} \sim C_{\dfa \fb} X^{\fb}_{\ \cA} + D_{\dfa}^{\ \dfb} \bW_{\dfb \cA}.
\end{equation}
Note that the top half of the coset, $X^\mathfrak{a}_{\ \mathcal{A}} $, can be identified with the Grassmannian $\Gr(m|n, 2m|2n)$ of $(m|n)$-planes in $\mathbb{C}^{2m|2n}$. 
Since we can always completely gauge away the lower half using~\eqref{eq:g} e.g.\ by setting $\bW \sim ( 0 \ \ \delta^{\dfb}_{\dfa})$, both the Grassmannian and coset descriptions are equivalent. The equivalence relation~\eqref{eq:g} can also be used to fix coordinates on the coset (and Grassmannian) as
\begin{equation}\label{eq:sconfgauge}
    g\sim \begin{pmatrix}
        \delta^{\fa}_{\  \fb} & x^{\fa \dfb} \\ 0 & \delta^{\ \dfb}_{\dfa}
    \end{pmatrix}, \quad  x^{ \fa \dfa}= \left(\begin{array}{cc}
        x^{\alpha \da} & \rho^{\alpha a'}\\
         \bar{\rho}^{a \da} & y^{aa'} 
    \end{array} \right),
\end{equation}
where $x^{ \fa \dfa}$ is the analytic superspace coordinate including the coordinates of Minkowski space $x^{\alpha \da}$, coordinates describing a coset space of the internal $\SU (2n)$ group, $y^{aa'}$ as well as the Grassmann odd variables $\rho^{\alpha a'}$ and $\bar{\rho}^{a \da}$. We call the fixing in \eqref{eq:sconfgauge} the (superspace) \emph{coordinate frame}. Note that most past work using analytic superspace has used this coordinate frame, but we instead here use the Grassmannian, $X^\fa_{\ \cA}$,  which makes weight-shifting operators particularly natural.

For a given element $X$ on the Grassmannian, the orthogonal element $\bX$ can be obtained by taking the inverse of $g$. We define it from \eqref{eq:g} as
\begin{equation}\label{inv}
    g^{-1}\coloneq \begin{pmatrix}
        W^{\cA}_{\ \ \fa} & \bX^{\cA \dfa}
    \end{pmatrix}.
\end{equation}
It follows straightforwardly (from $g^{-1}g=gg^{-1}=I$) that 
\begin{gather} \label{eq:ginvg}
        X^{\mathfrak{b}}_{\ \mathcal{A}}W^{\cA}_{\ \ \fa} = \delta^{\mathfrak{b}}_{\fa}, \qquad 
        X^{\mathfrak{b}}_{\ \mathcal{A}}\bX^{\cA \dfa}=0, \qquad
       \bW_{ \dfa \cA} W^{\cA}_{\ \ \fa} = 0, \qquad 
       \bW_{ \dot{\mathfrak{b}} \cA} \bX^{\cA \dfa}  = \delta_{\mathfrak{b}}^{\dfa}, \nn \\
    W^{\cB}_{\ \fa}X^{\fa}_{\ \cA}+ \bX^{\cB \dfa} \bW_{\dfa \cA}=\delta^{\cB}_{\cA}.    
\end{gather}
In the coordinate frame, the orthogonal Grassmannian becomes
\begin{equation}
    \bX \sim \begin{pmatrix}
        -x^{\fa \dfb} \\ \delta^{\ \dfb}_{\dfa}
    \end{pmatrix}.
\end{equation}

Recall that the case $(m,n)=(2,0)$ corresponds to (complexified) Minkowski space in 4 dimensions. 
Indeed, the conformal group of this space is $\mathrm{SO}(4,2)\sim \SU (2,2)$, whose complexification is $\SL(4, \mathbb{C})$. Then, Minkowski space can be viewed as the coset of the conformal group by the stabiliser of a point, i.e.\ the subgroup of $2\times 2$ block lower triangular matrices, consisting of dilatation, rotations and special conformal transformations. This is equivalent to the Grassmannian $\Gr(2,4)$, with coordinates $X^\alpha_{\ A}$, $\bX^{A \da}$ ($A=1,\dots,4$; $\alpha, \da =1,2$). These are related to the embedding space coordinate $X^M$ ($M=-1,0,\dots, 4$), as follows
\begin{equation}
    \Sigma_{AB}^M X_M = X^{\alpha}_{\ A} X^\beta_{\ B} \epsilon_{\alpha \beta}, \qquad X_M \bar{\Sigma}^{M,AB} = \epsilon_{\da\db}\bX^{A\da} \bX^{B \db} 
\end{equation}
where $\Sigma_{AB}^M$ and $ \bar{\Sigma}^{M,AB}$ are 6D sigma matrices. The embedding space method \cite{Dirac:1936fq,Costa:2011mg} is the most used method for studying spinning conformal blocks in non-supersymmetric CFTs. The Grassmannian/flag supermanifold picture provides a generalisation to spaces with supersymmetry, but also provides some advantages even in the non-SUSY case.

\subsection{Fields and representations}\label{sec:fieldsandreps}
An important feature of $(m,n)$ analytic superspace is that all unitary irreducible representations of the superconformal group are given by \emph{unconstrained} analytic superfields~\cite{Heslop:2001zm}. This means that shortening conditions are automatically satisfied by a multiplet, whereas in other superspaces they have to be imposed by hand via differential constraints. In this subsection, we show the explicit construction of such fields both on the supercoset space $H\backslash \SL(2m|2n)$ and on the Grassmannian $\Gr(m|n, 2m|2n)$. Recall from section~\ref{sec:setup} that the Grassmannian picture is that which `gauges' away all $W, \bW$ dependence. So, functions on the Grassmannian depend on the orthogonal planes $X$ and $\bX$ only.
In practice, the coset picture is most useful for carefully deriving weight-shifting operators. However, correlators will be shown to depend explicitly on $X,\bX$ and thus it is the Grassmannian picture which will be favoured. 

\subsubsection{Overview of fields on analytic superspace}
The analogy of analytic superspace with 4D Minkowski space means that  techniques for understanding fields and correlators in non-supersymmetric 4D CFT directly uplift to the full supersymmetric case. 
Representations of the 4D conformal group, realised as  fields on Minkowski space, are specified by the dilatation weight $\Delta$ together with left and right spins $s,\bar{s}$ specifying the representations of the left and right $\SL(2)$ spin groups. 
It is standard to denote  a 4D conformal field as 
\begin{equation}
    \cO^\tau_{\alpha_1..\alpha_{s},\dot \alpha_1..\dot \alpha_{\bar{s}}}(x),
\end{equation}
where $\tau$ is the twist $\tau= \Delta{-}(s{+}\bar{s})/2$ and the Weyl spinor indices $\alpha$ are completely symmetrised (dotted and undottes indices separately).
We use twist and not dimension as it will be shown to uplift to analytic superspace more directly.

Note the left and right $\SL(2)$ groups are subsets of $\GL(2)\subset \SU(2,2)$ which contains the dilatation operator on its diagonal. 
Viewed this way, we could also allow more general two-row $\GL(2)$ representations, with Young diagrams $\ula_L=[\lambda_L^1,\lambda_L^2], \ \ula_R=[\lambda_R^1,\lambda_R^2]$, giving
$$\cO^\tau_{\alpha_1..\alpha_{s},\dot \alpha_1..\dot \alpha_{\bar{s}}}(x^{\alpha \dot \alpha}) \ \rightarrow \  \cO^\gamma_{\ula_L,\ula_R}(x^{\alpha \dot \alpha}),$$ 
with
$$ \gamma= \tau - \lambda_L^2-\lambda_R^2,\qquad \quad s=\lambda_L^1-\lambda_L^2,\qquad \quad \bar{s}=\lambda_R^1-\lambda_R^2, $$ 
where the case  $\lambda_L^2=\lambda_R^2=0$ gives back the standard writing.
Now there is additional freedom in choosing the value of the second row of the Young diagrams.
In the free scalar theory, $\gamma$ is interpreted as the number of fundamental scalars used to build the operator, and the value of $\lambda_{L/R}^2$ as the number of contracted Weyl spinor indices. For example, the following two scalar operators of dimension 4 would be written differently as:  $\phi^4=\cO^\gamma_{\bullet,\bullet}$ whereas $\phi \Box\phi+\dots=\cO^2_{[1,1],[1,1]}$. 
In the interacting theory, these two mix and are no longer distinguishable.

This notation in terms of $\gamma$ and $\GL(2)$ Young diagrams now directly uplifts to superconformal representations in $(m,n)$ analytic superspace: 
\begin{equation}
    \cO^\gamma_{\ula_L,\ula_R}(x^{\alpha \dot \alpha}) \ \rightarrow\      \cO^\gamma_{\ula_L,\ula_R}(x^{\fa\dfa})
\end{equation}
where the right hand side encodes a full superconformal representation via $\GL(m|n)$ Young diagrams. Since we have lifted Weyl spinor indices to $(m|n)$ indices carrying $\GL(m|n)$ representations, the main new technical feature arises from understanding and classifying finite-dimensional irreducible representations of $\GL(m|n)$ via the symmetries of these indices. 
This can nicely be done via Young diagrams which are no longer restricted to two-row only. 

In the rest of the section, we present the details of this formulation. We have also included a full translation analytic superspace representations and 4D $\cN=2$ and $\cN=4$ superconformal representations in appendix \ref{sec:repexamples}.

\subsubsection{Coset representations}
A field $\cO$ on a  coset space $H\backslash G$ is a representation of $G$, given as an equivariant map $\cO:G\rightarrow V$, \begin{align}\label{cosetfield}
\cO(hg)= R(h)\cO(g),\end{align} where $V$ is a representation space of $H$. In our case
$G=\SL(2m|2n)$ and $H$ is the subgroup~\eqref{eq:H}.
Since this subgroup is block lower triangular, this splits into a representation of its block diagonal subgroup 
\begin{equation}\label{eq:levisubgroup}
    \mathrm{S}(\GL(m|n)\times \GL(m|n)) \cong \SL(m|n)\times \SL(m|n) \times \mathbb{C}^*,
\end{equation}
known as the Levi subgroup. 
The Levi subgroup of $(2,0)$ analytic superspace is $\SL(2)\times \SL(2) \times \mathbb{C}^*$ and we recover the well known 4D conformal representations (with the left and right $\SL(2)$ acting on the Weyl spinor $\alpha, \dot \alpha$ indices). 
More generally, a representation will be labelled by a charge $\gamma$, as well as by two finite-dimensional~\cite{Heslop:2001zm} $\SL(m|n)$ irreducible representations which we can label by 
 Young diagrams $\ula_L$ and $\ula_R$ for the  left- and right-handed super-rotations respectively.
In practice, this means that fields will carry $(m|n)$  (super)indices: $\fa_1,\fa_2\dots$ for left, $\dfa_1,\dfa_2\dots$ for right, and these will be symmetrised by a  (Hermitian) Young symmetriser $\Pi^{\ula_L}_1,\bar \Pi^{\ula_R}_1$ of shape given by the respective Young diagram $\ula_L$ and according to the Young tableau $\ula_L(1)$\footnote{Note that a Young tableau is a Young diagram where each box is labelled by elements from an ordered set which we will take to be the ordered superindices $\fa_1, \fa_2, \dots$ or often simply $1,2,..$.} obtained by filling the indices in order in the Young diagram, e.g.,
\begin{align}\label{ytindices}
    \ula_L(1) = \begin{tikzpicture}[x=13pt, y=13pt, baseline={(current bounding box.center)}]
    \scalebox{1.1}{ 
        \tyng(0,0,5,4,2,1);
    }
    \node at (0.55, 0.55) {\small $\fa_{\scriptscriptstyle 1}$};
    \node at (1.65, 0.55) {\small $\fa_{\scriptscriptstyle 2}$};
    \node at (2.75, 0.55) {\small $\fa_{\scriptscriptstyle 3}$};
    \node at (3.85, 0.55) {\small $\fa_{\scriptscriptstyle 4}$};
    \node at (4.95, 0.55) {\small $\fa_{\scriptscriptstyle 5}$};
    \node at (0.55, -0.55) {\small $\fa_{\scriptscriptstyle 6}$};
    \node at (1.65, -0.55) {\small $\fa_{\scriptscriptstyle 7}$};
    \node at (2.75, -0.55) {\small $\fa_{\scriptscriptstyle 8}$};
    \node at (3.85, -0.55) {\small $\fa_{\scriptscriptstyle 9}$};
    \node at (0.55, -1.65) {\small $\fa_{\scriptscriptstyle 10}$};
    \node at (1.65, -1.65) {\small $\fa_{\scriptscriptstyle 11}$};
    \node at (0.55, -2.75) {\small $\fa_{\scriptscriptstyle 12}$};
\end{tikzpicture}\, , \qquad \ula_R(1) = \begin{tikzpicture}[x=13pt, y=13pt, baseline={(current bounding box.center)}]
    \scalebox{1.1}{ 
        \tyng(0,0,6,3,1);
    }
    \node at (0.55, 0.55) {\small $\dfa_{\scriptscriptstyle 1}$};
    \node at (1.65, 0.55) {\small $\dfa_{\scriptscriptstyle 2}$};
    \node at (2.75, 0.55) {\small $\dfa_{\scriptscriptstyle 3}$};
    \node at (3.85, 0.55) {\small $\dfa_{\scriptscriptstyle 4}$};
    \node at (4.95, 0.55) {\small $\dfa_{\scriptscriptstyle 5}$};
    \node at (6.05, 0.55) {\small $\dfa_{\scriptscriptstyle 6}$};
    \node at (0.55, -0.55) {\small $\dfa_{\scriptscriptstyle 7}$};
    \node at (1.65, -0.55) {\small $\dfa_{\scriptscriptstyle 8}$};
    \node at (2.75, -0.55) {\small $\dfa_{\scriptscriptstyle 9}$};
    \node at (0.55, -1.65) {\small $\dfa_{\scriptscriptstyle 10}$};
\end{tikzpicture}\, . 
\end{align}
 The subscript `1' on the projector $\Pi^{\ula}_1$ indicates that the indices are symmetrised according to the Young tableau $\ula(1)$, the first standard Young tableau in a lexicographical ordering. We will use different standard Young tableaux later, see appendix~\ref{app:yt} for more details.
We will thus write our operators as 
\begin{align}
\cO^{\gamma}_{\ula_L, \ula_R}\coloneq\cO^{\gamma}_{\underline{\fa},\underline{\dfa}}(X,\bar W),  \qquad \text{where}\qquad \Pi^{\ula_L}_1\cO^{\gamma}_{\underline{\fa},\underline{\dfa}}=\cO^{\gamma}_{\underline{\fa},\underline{\dfa}},\quad \bar\Pi^{\ula_R}_1\cO^{\gamma}_{\underline{\fa},\underline{\dfa}}=\cO^{\gamma}_{\underline{\fa},\underline{\dfa}}
\end{align}
so the explicit $\ula_L,\ula_R$ dependence will often not be displayed to avoid cluttered notation, and underlined indices represent collective indices symmetrised according to the respective Young tableau.

All $\GL(m|n)$ Young diagrams fit into a hook of vertical thickness $m$ and horizontal thickness $n$ 
as shown below
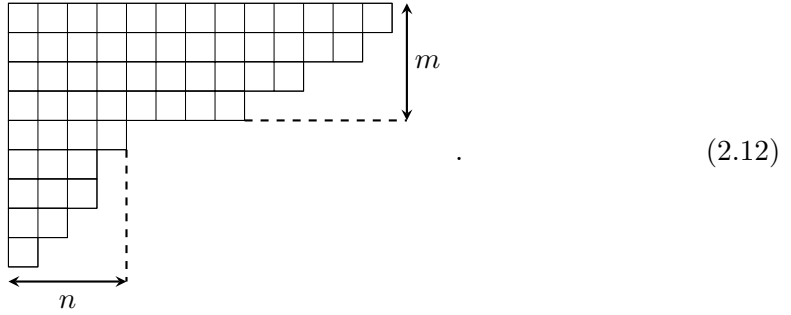
\begin{equation}\label{hook}
    \qquad \begin{tikzpicture}[x=13pt,y=13pt, baseline={(current bounding box.center)},scale=.85]
    \tyng(0,0,13,12,10,8,4,3,3,2,1);
      \draw[thick, stealth-stealth](0,-8.5) to (4,-8.5);
      \node at (2,-9.2) {$n$};
     \node at (14.2,-1){$m$};
      \draw[thick, stealth-stealth](13.5,1) to (13.5,-3);
      \draw [thick, dashed] (4,-4) to (4,-8.5);
      \draw [thick, dashed] (8,-3) to (13.5,-3);
\end{tikzpicture}\, .
\end{equation}
For example, for the $(2,0)$ case (4D Minkowski space) this is a two-row Young diagram, and for $(2,2)$ (4D $\cN=4$) it is two arbitrarily long rows and columns.

The infinitesimal action of the isotropy group, \eqref{eq:H} and \eqref{eq:g}, 
is $\delta g = \delta h \, g$ and thus the generators of this are $\str (\delta h \, g \partial_g) $
where
\begin{align}\label{eq:generators}
    (g \partial_g)^{\cA}_{\ \mathcal{B}}& = \left(\begin{array}{cc}
        (X \partial_X)^{\fa}_{\ \fb} & (X\partial_{\bW})^{\fa \dfb} \\ 
        (\bW\partial_X)_{\dfa \fb} & (\bW\partial_{\bW})^{\ \dfa}_{ \dot{\mathfrak{b}}}
    \end{array}\right) \coloneq \left( \begin{array}{cc}
        (m_L)^{\fa}_{\ \fb} &  k^{\fa \dfb} \\ 
        p_{\dfa \fb}  & -(m_R)^{\ \dfa}_{ \dot{\mathfrak{b}}}
    \end{array}\right)\,,
    \end{align}
    and $\str$ denotes the supertrace (that is trace over external $\alpha,\dot \alpha$ indices minus trace over  internal $a,a'$ indices).
So the infinitesimal isotropy group transformation is
$\str (\delta h \, g \partial_g)= \str(a m_{\!L}{-}d m_{\!R}{+}ck)$ where $a,c,d$ are the infinitesimal versions of $A,C,D$ in $h$~\eqref{eq:H}.

Now, a {\em function} $f(X,\bar W)$ on the coset space satisfies $f(hg)=f(g)$ and thus is annihilated by $k,m_L,m_R$. 
However, fields $\cO^{\gamma}_{\ula_L, \ula_R}\coloneq\cO^{\gamma}_{\underline{\fa},\underline{\dfa}}(X,\bar W)$ transform non trivially as~\eqref{cosetfield} with the precise transformation under the two $\GL(m|n)$ groups as dictated by the relevant indices, resulting in
\begin{align}\label{eq:primary}
     k^{\mathfrak{c} \dfb} \cO^{\gamma}_{\underline{\fa},\underline{\dfa}}&
     =0, \nn
     \\
        (m_L)^{\mathfrak{c}}{}_{\fb} \cO^{\gamma}_{\underline{\fa},\underline{\dfa}}&
        = -\gamma_L
        \delta^{\mathfrak{c}}_{\fb} \cO^{\gamma}_{\underline{\fa}, \underline{\dfa}}-\sum_{i=1}^s\delta_{ \fa_i}^{\mathfrak{c} }\cO^{\gamma}_{\fa_1 \dots \fa_{i-1}\fb\fa_{i+1}\dots \fa_{s}, \underline{\dfa}} ,
        \\
          (m_R)_{\dfb}^{\ \dot{\mathfrak{c}} }\cO^{\gamma}_{\underline{\fa},\underline{\dfa}} &=
          -\gamma_R
\delta^{\dot{\mathfrak{c}}}_{\dfb}\cO^{\gamma}_{\underline{\fa}, \underline{\dfa}} -\sum_{i=1}^{\bar{s}}\delta_{ \dfa_i}^{\dot{\mathfrak{c}} }\cO^{\gamma}_{\underline{\fa}, \dfa_1 \dots \dfa_{i-1}\dfb\dfa_{i+1}\dots \dfa_{\bar{s}}}  \nn,
\end{align}
where we have defined $\gamma_L= (\gamma + \tilde \gamma)/2$ and $\gamma_R= (\gamma-\tilde\gamma)/2$ as left-handed and right-handed charges and we discuss $\tilde \gamma$ shortly. 

Note that we have only considered operators with lower indices so far. In the non-supersymmetric cases ($m=0$ or $n=0$) lower indices can be exchanged for upper indices, or vice versa using an epsilon tensor. 
However, in the supersymmetric case ($m,n\neq 0$), this is not the case and the upstairs and downstairs indices correspond to different representations. 
Furthermore, unitary representations of $\SU (m,m|2n)$ are given by operators with lower indices only~\cite{Heslop:2001zm,Heslop:2003xu}. 
We thus assume our physical operators have lower indices. We will however also make use of non-unitary fields in finite-dimensional  $\SL(2m|2n)$ representations in order to construct weight-shifting operators. These will have upper $\SL(m|n)$ indices as we will discuss shortly.

There is an important subtlety in~\eqref{eq:primary} due to the fact that the coset space is a coset of the symmetry group $\SL(2m|2n)$ rather than $\GL(2m|2n)$.
The fact that the symmetry group has unit determinant means that only the (super)traceless part of $\delta h$ acts, thus $\str(a m_L{-}d m_R{+}ck)$ is a symmetry only for $a,c,d$ such that $\str(a)+\str(d)=0$. But  $\tilde \gamma$ in~\eqref{eq:primary} is precisely multiplied by $\str(a)+\str(d)$ and so does not act under this symmetry. 
On the other hand, the unit determinant constraint 
$    \sdet g=1$
means that the coordinates $g=(X,\bar W)$ are not completely independent.
This identity can then be used to assign any arbitrary value we like for $\tilde \gamma$, since
\begin{align}
    (m_L)^\fa_{\ \fb} \sdet g = \delta^\fa_{\fb} \sdet g, \qquad \qquad 
    (m_R)^{ \ \dfa}_{\dfb} \sdet g = -\delta^\dfa_{\dfb} \sdet g\ .
\end{align}
So, our operators (or more precisely $n$-point correlation functions of operators) can be viewed as functions of $X,\bar W$,  subject to $\sdet g=1$, satisfying~\eqref{eq:primary} for arbitrary $\tilde \gamma$. In practice, we will find it useful to fix a particular value of $\tilde \gamma$ and then  we will write $\cO^{\gamma_L,\gamma_R}_{\ula_L,\ula_R} \sim \cO^{\gamma_L+\gamma_R}_{\ula_L,\ula_R}$.

In fact, for {\em any} representation, if
a further symmetrisation of indices is imposed {\em after}  acting with $m_L,m_R$, then the entire action can be written as  a single term. 
More precisely, we project onto any irreducible representation $\umu_L$ in the tensor product $\Box \otimes \ula_L$ by applying the appropriate Young projector $\Pi^{\umu_L}_i$ on the indices of $\ula_L$ and $\fa$ (in that order):
\begin{align}\label{eq:mLconstant}
        \Pi^{\umu_L}_i
    (m_L)^{\fb}_{\ \fa} \cO_{\ula_L, \ula_R}^{\gamma_L, \gamma_R} &= B_{\umu_L\!(i)}^{\gamma_L} \Pi^{\umu_L}_i 
\delta^{\fb}_{\fa}\cO_{\ula_L, \ula_R}^{\gamma_L, \gamma_R},
& \ \ \
B_{\umu_L\!(i)}^{\gamma_L} &\coloneq 
-\gamma_L - C_{\umu_L\!(i)},
\end{align}
where $C_{\umu_L\!(i)}$ is  the length minus the height of the position of the final box in the Young tableau $\umu_L(i)$ (i.e.\ the one in $\umu_L$ but not in $\ula_L$, which is associated with the extra index $\fa$). 
This is known as the \emph{content} of the last box (see~\cite{garsia2020lectures}):
\begin{equation}\label{eq:content}
    C_{\umu_L\!(i)}= L_{\umu_L(i)}-H_{\umu_L(i)},
\end{equation}
where
\begin{equation}\label{Youngprojector}
    \Pi^{\umu_L}_i:   \ula_L \otimes  \young(\fa)\longmapsto \umu_L(i)= \ \raisebox{-0.3cm}{\begin{tikzpicture}[x=13pt, y=13pt, baseline={(current bounding box.center)}, scale=1]
    \draw [thick] (0,1) -- (6,1) -- (6,0) -- (5,0) -- (5,-1) --
                                (3,-1) -- (3,-2) -- (2,-2) -- (1,-2) -- (1,-3) --
                                (0,-3) -- cycle;
    \node at (1.7,-0.3) {$\ula_L(1)$};
    \draw [ultra thick, dashed] (4,-2)to (4,-1) to (3,-1) to (3,-2) to (4,-2);
    \draw [<->] (0, -3.5) -- (3, -3.5);
    \draw [<->] (6.5, 1) -- (6.5, -1);
    \node at (1.5,-4.4) {$L_{\umu_L(i)}$};
    \node at (8.1, -0.1) {$H_{\umu_L(i)}$};
     \node at (3.5, -1.5) {\small $\fa$};
\end{tikzpicture}}\, .
\end{equation}
We write $\overline{\umu_L\!(i)}=\ula_L\!(1)$ where the bar on the RHS denotes the tableau with the last box (i.e.\ added box in this case) removed.  The label $i$ simply specifies this particular standard Young tableau.
The analogous constant for $m_R$ can be obtained by swapping $\gamma_L$ by $\gamma_R$ and $\ula_L$ by $\ula_R$:
\begin{align}\label{eq:mRconstant}
        \Pi^{\umu_R}_i
    (m_R)^{\ \dfb}_{\dfa} \cO_{\ula_L, \ula_R}^{\gamma_L, \gamma_R} &=  B_{\umu_R(i)}^{\gamma_R} (\dfa) \Pi^{\umu_R}_i 
\delta^{\dfb}_{\dfa}\cO_{\ula_L, \ula_R}^{\gamma_L, \gamma_R}, 
& \ \ \
B_{\umu_R\!(i)}^{\gamma_R}&\coloneq-\gamma_R - C_{\umu_R\!(i)}. 
\end{align}

The identities~\eqref{eq:mLconstant} and~\eqref{eq:mRconstant} in fact follow, for Hermitian Young projectors,  from the original work of Jucys, Murphy and Young where it is part of a bigger story developed by Murphy, nicely summarised in the lecture notes~\cite{garsia2020lectures} (see Theorem 2.13 with k=n).\footnote{The particular case $k=n$ should also work for the usual Young projectors. We checked this for Young tableaux with up to 7 boxes.} We explain this in more detail in appendix~\ref{app:yt}. In section~\ref{sec:WSOinvariants}, this relation will also be shown to be related to $\SL(2m|2n)$ (and hence superconformal) $6j$ symbols and the so-called bubble coefficient of weight-shifting operators \cite{Karateev:2017jgd}.

\subsubsection{Fields on the Grassmannian}\label{sec:Grassmannianfields}

It is useful to consider operators as functions of the Grassmannian  $X,\bar X $ rather than  the coset space  $X, \bar W$ \eqref{eq:g}. To do this, we first consider them as functions of the coset $g$ and its inverse $g^{-1}$, so of all variables $X,\bar X, W, \bar W$  subject to the identities~\eqref{eq:ginvg}.  
Then, the isotropy group  operators $k,m_L,m_R$ defined in~\eqref{eq:primary} extend to the differential operators $g\partial_g-\partial_{g^{-1}}g^{-1}$ and thus become\footnote{One can relate derivatives of inverse  variables using $(\partial_{g^{-1}})^{\cA}_{\ \cB}  =-g^{\cA}_{\ \cD} g^{\cC}_{\ \cB} (\partial_g)^\cD_{\ \cC}$.}
\begin{equation}\label{kinXXb}
    \begin{aligned}
            k^{\fa \dfa}&=(X\partial_{\bar W} -\partial_{W}\bar X)^{\fa \dfa}, \qquad &p_{\dfa \fa}&=(\bar W \partial_X-\partial_{\bar X}W)_{\dfa \fa}, \\(m_L)^\fa_{\ \fb}&=(X\partial_X -\partial_W W)^\fa_{\ \fb},
&(m_R)_{\dfa}^{\ \dfb}&=(-\bar W \partial_{\bar W}+\partial_{\bar X}\bar X)_{\dfa}^{\ \dfb}\,,
    \end{aligned}
\end{equation}
where the derivatives are all understood to act first, the ordering indicating the index contractions (see appendix~\ref{app:superalgebra}).

From here, one can show that it is always possible to write operators (or correlators thereof) satisfying~\eqref{eq:primary} (in particular $k\cO=0$)
to  depend on $X,\bar X$ only (and not on $W,\bar W$). This is precisely the Grassmannian perspective and we show with an explicit change of variables in appendix~\ref{app:xxbproof} that this is possible. We can immediately see from~\eqref{kinXXb} that in these variables, $k\cO=0$ is automatically satisfied, but also $m_L=X\partial_X$ and $m_R=\partial_\bX\bX$.  The  constraints~\eqref{eq:primary} thus become
\begin{equation}\label{eq:primary2}
    \begin{split}
        (X\partial_X)^{\mathfrak{c}}{}_{\fb} \cO^{\gamma}_{\underline{\fa},\underline{\dfa}}\hspace{0.1cm}
        = -\gamma_L \delta^{\mathfrak{c}}_{\fb} \cO^{\gamma}_{\underline{\fa},\underline{\dfa}}-\sum_{i=1}^s\delta_{ \fa_i}^{\mathfrak{c} }\cO^{\gamma}_{\fa_1 \dots \fa_{i-1}\fb\fa_{i+1}\dots \fa_{s}, \underline{\dfa}} \, ,
        \\
         (\partial_{\bX}\bX)^{\ \dfc}_{ \dot{\mathfrak{b}}}\cO^{\gamma}_{\underline{\fa},\underline{\dfa}} =
          -\gamma_R\delta^{\dot{\mathfrak{c}}}_{\dfb}\cO^{\gamma}_{\underline{\fa},\underline{\dfa}} -\sum_{i=1}^{\bar{s}}\delta_{ \dfa_i}^{\dot{\mathfrak{c}} }\cO^{\gamma}_{ \underline{\fa},\dfa_1 \dots \dfa_{i-1}\dfb\dfa_{i+1}\dots \dfa_{\bar{s}}} \, .
    \end{split}
\end{equation}

\subsubsection{Finite-dimensional representations}

The operators we have been considering correspond to unitary irreducible representations of the superconformal group. These are infinite-dimensional and have lower indices in analytic superspace.
It is instructive to also consider finite-dimensional representations of $\SL(2m|2n)$ which can also be written explicitly on the coset space and/or the Grassmannian. These will not correspond to unitary representations of the real form $\SU (m,m|2n)$ when $m\neq0$.
These finite-dimensional reps are written with upstairs rather than downstairs $\fa,\dfa$ indices, 
and satisfy:
\begin{equation}\label{eq:primary3}
    \begin{split}
        (m_L)^{\mathfrak{b}}{}_{\fc} \finite^{ \underline{\fa},\underline{\dfa}}
 	&=- \gamma_L \delta^{\fb}_{\fc} \finite^{ \underline{\fa},\underline{\dfa}} +\sum_{i=1}^s\delta^{ \fa_i}_{\mathfrak{c} }\finite^{\fa_1 \dots \fa_{i-1}\fb\fa_{i+1}\dots \fa_{s},\underline{\dfa}} \,,
 	\\
 	(m_R)_{\dfc}^{\ \dot{\mathfrak{b}}}\finite^{ \underline{\fa},\underline{\dfa}} &=- \gamma_R \delta^{\dfb}_{\dfc} \finite^{ \underline{\fa},\underline{\dfa}} +
 \sum_{i=1}^{\bar{s}}\delta^{ \dfa_i}_{\dot{\mathfrak{c}} }\finite^{ \underline{\fa}, \dfa_1 \dots \dfa_{i-1}\dfb\dfa_{i+1}\dots \dfa_{\bar{s}}} \,.
    \end{split}
\end{equation}
The finite-dimensional representations we will use for deriving weight-shifting operators in section \ref{sec:weightshifting1} have $\gamma=\tilde \gamma=0$, in which case the above has the following solution on the Grassmannian,
\begin{equation}\label{eq:finiterep}
 \finite^{\underline \fa, \underline \dfa}(X, \bX)= \finite^{\fa_1 \dots \fa_p \dfa_1 \dots \dfa_q}(X, \bX)= \finite^{\cA_1 \dots \cA_p}_{\cB_1 \dots \cB_q} X^{\fa_1}_{\ \cA_1} \dots X^{\fa_p}_{\ \cA_p} \bX^{\cB_1 \dfa_1} \dots \bX^{\cB_q \dfa_q},
\end{equation}
where $\finite^{\underline \cA}_{\underline \cB}$ is the representation expressed in terms of tensor indices, whereas $\finite^{\underline \fa, \underline \dfa}(X,\bar X)$ is the corresponding representation as an equivariant map \eqref{cosetfield} on the coset space. 
The charges $\gamma_L, \gamma_R$ in \eqref{eq:primary3} can only take on very specific values for the representation to be finite-dimensional. They depend on $m,n$ since, for instance, in $(m,0)$, they would correspond to terms of the form $\epsilon_{\fa_1 \dots \fa_m} X^{\fa_1}_{\ \cA_1} \dots X^{\fa_m}_{\ \cA_m}$ in \eqref{eq:finiterep}.

Note that \eqref{eq:primary3} also describes the transformation properties of a more general infinite-dimensional $\SL(2m|2n)$ representation with upper indices (which would be a non-unitary rep of the real form $\SU (m,m|2n)$). These can have arbitrary values of the charges $\gamma_L$, $\gamma_R$.
 
Unlike the finite-dimensional case discussed above, for unitary representations, the general solution to \eqref{eq:primary2} cannot be written down explicitly on the Grassmannian as functions of $X,\bar X$. However, the general solutions of $N$-point  correlation functions can be written down, as we explore in depth in section \ref{sec:Wardids}. 

\subsubsection{Types of representations}\label{sec:classification}

We now classify some different types of fields/representations we will come across,
i.e.\ the tensor representations of the full Levi subgroup $S\left(\GL(m|n) \times \GL(m|n)\right)$, 
and introduce some  vocabulary that will be used throughout the article.

\paragraph{Scalar fields:}
The simplest type of field transforms under the trivial representation of $S\left(\GL(m|n) \times \GL(m|n)\right)$ and we will write these as $\cO^\gamma\coloneq\cO^\gamma_{\bullet,\bullet}$ (where throughout the paper we denote the trivial representation by $\bullet$). These correspond to half-BPS representations in the supersymmetric case and scalars in the non supersymmetric case. We will simply refer to them as scalar fields in all cases however.

\paragraph{Symmetric representations:} An important class of representations are those for which $\ula_L=\ula_R\coloneq\ula$ which we will call  \emph{symmetric} representations. To simplify the notation we will write them  as $\cO^\gamma_{\ula}\coloneq \cO^\gamma_{\ula,\ula}$. These are the representations appearing in the OPE of two scalars. Thus, they are important for the study of half-BPS superconformal blocks \cite{Doobary:2015gia,Aprile:2021pwd,Aprile:2025nta}, which are the seed functions for the weight-shifting operators we derive in this work. They correspond to the symmetric traceless representations in the non-supersymmetric $(m,n)=(2,0)$ CFT case. Many superconformal multiplets of interest are described by symmetric representations in analytic superspace. 
Examples include 4D $\cN=4$ quarter-BPS (and other semi-short multiplets) and the $\cN=2$ stress-tensor multiplet or any multiplet associated with the Higgs branch, for example (see appendix~\ref{sec:repexamples}).

\paragraph{Chiral representations:}
Another simple but important class of  representation is the chiral representation for which $\ula_R$ is the trivial representation, $\ula_R=\bullet$, and anti-chiral, those with $\ula_L=\bullet$. These are generalisations of the usual 4D chiral fermions $\psi_{(\alpha \dots \beta)}$ and $\bar{\psi}_{(\da \dots \db)}$ and describe many relevant 4D $\cN=2$ representations, as shown in detail in appendix \ref{sec:repexamples}. As will be seen they have the simplest possible weight-shifting operators. 
Note that these are not present as physical operators in $\cN=4$ SYM, which requires $|\ula_L|=|\ula_R|$. 

\paragraph{General representations:}

For general representations, $\cO^\gamma_{\ula_L,\ula_R}$, it is useful to consider the intersection of the left and right representations, $\ula= \ula_L \cap \ula_R$ so that $\cO^\gamma_{\ula_L,\ula_R}=\cO^\gamma_{\ula + \umu_{L},\ula + \umu_{R}}$, where $\umu_{L,R}$ is the skew Young diagram corresponding to the remaining boxes added on. This is denoted as
$    \umu_{L,R} = \ula_{L,R}/\ula$,
e.g.
\begin{align}
\ula_L=\begin{tikzpicture}[x=13pt, y=13pt, baseline={(current bounding box.center)},scale=0.9]
    \draw (1,-2) rectangle ++(1,-1); 
    \draw (6,1) rectangle ++(1,-1); 
    \draw (0,-3) rectangle ++(1,-1); 
    \draw [ultra thick, dashed] (0,1) -- (6,1) -- (6,0) -- (5,0) -- (5,-1) --
                                (3,-1) -- (3,-2) -- (2,-2) -- (1,-2) -- (1,-3) --
                                (0,-3) -- cycle;
    \node at (2,-0.5) {$\ula$};
    \node (pL) at (4.5,  -3.5) {$\umu_L$};
    \draw[->, thick] (pL) -- (2.2,-2.5); 
    \draw[->, thick] (pL) -- (6.5,-0.2);  
    \draw[->, thick] (pL) -- (1.2,-3.5); 
\end{tikzpicture}, \qquad \ula_R=
\begin{tikzpicture}[x=13pt,y=13pt,baseline=(align),scale=0.9]
    \draw (4,-1) rectangle ++(1,-1); 
    \draw (5,0) rectangle ++(1,-1); 
    \draw (3,-1) rectangle ++(1,-1); 
    \draw [ultra thick, dashed] (0,1) -- (6,1) -- (6,0) -- (5,0) -- (5,-1) --
                                (3,-1) -- (3,-2) -- (2,-2) -- (1,-2) -- (1,-3) --
                                (0,-3) -- cycle;
     \node at (2,-0.5) {$\ula$};
        \node (pR) at (4.5, -4.5) {$\umu_R$};
    \draw[->, thick] (pR) -- (4.5,-2.2); 
    \draw[->, thick] (pR) -- (5.5,-1.2); 
    \draw[->, thick] (pR) -- (3.5,-2.2); 
    \node (align) at (0,-1.2) {};
\end{tikzpicture}\,.
\label{asymmetricreps}
\end{align}
When considering 4D $\cN=4$ supermultiplets (more generally, any $m=n$ superspace), the representations must satisfy $|\ula_L|=|\ula_R|$ and thus are described by their {\em level of asymmetry}.\footnote{The constraint $|\ula_L|=|\ula_R|$ is due to $\SL(2m|2n)$ not being semi-simple for $m=n$. See appendix \ref{sec:repexamples} for details.} In particular, we define \emph{$N$-asymmetric} states as representations that can be made symmetric by removing, minimally,  $N$ boxes from both left and right Young diagrams. The diagram \eqref{asymmetricreps} is an example of $3$-asymmetric representation.

\paragraph{Short representations:} 
When a (super) primary field has certain quantum numbers which either saturate the continuous unitarity bound, or occur in the isolated bounds,  certain (super) descendants are required to vanish by unitarity. This takes the form of a differential constraint on the primary (super) field. The simplest example is that of fundamental scalars and Dirac spinor, which satisfy the wave and Dirac equation as their shortening condition, respectively. These generalise straightforwardly to analytic superspace via the $(2,0)\mapsto (m,n)$ lift presented throughout this section, thus yielding the corresponding fundamental superfields for different $m,n$. In superconformal field theory, there are further types of shortening conditions describing the vanishing of one or more whole conformal multiplets inside the supermultiplet. These are known as $Q$-shortening conditions and are present in an analytic superspace field $\cO^\gamma_{\ula_L,\ula_R}$ if and only if $\ula_L$ and $\ula_R$ do not fill the entire $m \times n$ rectangle. For example, in 4D $\cN=4$, a supermultiplet with 
\begin{equation}
    \ula_L=\ula_R = [1^\mu ]\ =\  \begin{tikzpicture}[scale=0.65, baseline=(current bounding box.center)]
        \draw[thick] (0,0) rectangle (0.5,-2.25);
        \draw[<->] (0.75,0) -- (0.75,-2.25) node[midway,right] {$\mu$};
    \end{tikzpicture},
\end{equation}
is short as is does not fill the second column of the box $[n^m]=[2,2]$. In particular, this describes isolated representations: those with no nearby unitary representations as one varies the dimension. These are half-BPS for $\mu=0$ (charge $\gamma$) and quarter-BPS otherwise. See appendix \ref{sec:repexamples} for a full translation between (super) conformal representations and analytic superspace fields in various relevant choices of $m,n$. 

The key property of short representations in analytic superspace is that they are unconstrained, i.e. satisfy the differential constraints automatically by  virtue of their analyticity on the compact internal space.
This was proven in \cite{Heslop:2001zm} and makes this formalism ideal for weight-shifting operators as they too will map shortening conditions correctly. We will also show that this applies to non-supersymmetric $(n\neq 0)$ representations too due to the unified nature of our construction.

\paragraph{Long representations:}
\label{longops}
Finally we mention here special types of representations that only appear in the supersymmetric case $n,m>0$, known as long representations. Long representations with integer dilatation weight can be described as operators $\cO^\gamma_{\ula_L,\ula_R}$ where the $\SL(2|2)$ irreps $\ula_L,\ula_R$ 
are described by Young diagrams containing a full $m\times n$ rectangle. 
Non-perturbatively, the conformal dimensions of long operators can become anomalous and so we also wish to describe cases with non-integer dimensions. 
These are still given by $\cO^\gamma_{\ula_L,\ula_R}$, where $\ula_{L,R}$ are finite irreps of $\SL(2|2)$, but in this case these finite-dimensional irreducible representations can not be described directly as tensors. 
Indeed, finite-dimensional irreducible representations of supergroups can have non-integer quantum numbers. 
This can be formally understood in terms of Young diagrams by allowing rows and/or columns to become non-integer. 
Such generalised Young diagrams were discussed in this context in~\cite{Heslop:2001zm,Heslop:2003xu} where they were called `quasi-tensors'. 
Specifically, section 5.2 of~\cite{Heslop:2001zm} gives an explicit $\cN=2$ example of this analytic continuation.
In practice, one simply uses tensors with integer row lengths, and since the dimensions of related irreducible representations are all the same due to the Grassmann odd nature of the supergroup, one can then analytically continue the row length parameter. We discuss this in the context of superconformal blocks and our weight shifting operators in section \ref{sec:nonintegerdim}.

Another interesting feature of certain long $\SL(m|n)$ representations is the equivalence of  different Young diagrams under $\SL(m|n)$ (but not under $\GL(m|n)$). In the non-supersymmetric case this is simply the equivalence of the completely antisymmetric representation and the trivial representation, shown by the existence of the constant $\epsilon$-tensor, which is invariant under $\SL$ but not $\GL$.  One can also describe the supersymmetric equivalence of long multiplets via a supersymmetric  generalisation of the $\epsilon$ tensor, the difference being it has both  upstairs and downstairs indices, $\mathcal{E}^{\ufa}_{\ufb}$, where $\ufa,\ufb$ are collective indices in two representations which differ for $\GL$ but not for $\SL$.  These points are all discussed in more detail in~\cite{Heslop:2001zm,Heslop:2003xu}. We don't consider such cases in this paper explicitly.

\subsection{Ward identities and correlation functions}\label{sec:Wardids}
In $(m,n)$ analytic superspace, superconformal Ward identities are straightforward to set up and solve~\cite{
Heslop:2003xu}. In this work, we work directly on the Grassmannian and/or coset rather than with the superspace coordinates \eqref{eq:sconfgauge} as done in~\cite{
Heslop:2003xu}. Crucially, the output is completely equivalent.  In particular, note that in the Grassmannian approach $\SL(2m|2n)$ acts linearly on the $\cA$ indices. Thus, constructing superconformal invariants amounts to contracting all free $\cA$ indices of the coordinates $X^\fa_{\ \cA},\bar X^{\cA \dfa}$. $N$-point correlation functions are then  functions of $X_i$ and $\bX_i$ where $i=1,\dots,N$, which are $\SL(2m|2n)$ invariant (all free $\cA$ indices are contracted) and which satisfy the equations~\eqref{eq:primary2} at every point. 
In practice, this means they should have open $\fa\dots,\dfa \dots$ indices at point $i$ as dictated by the corresponding operator $\cO_{\fa\dots,\dfa \dots}(X_i,\bar X_i)$.
In practise we focus on the Grassmannian formulation rather than the coset one, taking advantage of the fact that the constraint $k\cO=0$ is then automatically solved (see discussion below~\eqref{kinXXb}.)

The key object for writing solutions to the Ward identities is  
\begin{equation}\label{eq:Xij}
    X_{ij}^{\fa_i \dfa_j} \coloneq (X_{i})^{\fa_i}_{\ \cA} (\bX_j)^{\cA \dfa_j} \sim (x_i - x_j)^{\fa \dfa},
\end{equation}
where $x^{\fa \dfa}$ is the superspace coordinate \eqref{eq:sconfgauge} and the subscript on the indices denotes the local (pointwise) nature of the isotropy group. 
This combination is conformally invariant, but transforms under the isotropy group as 
dictated by its indices (i.e.\ like~\eqref{eq:primary2}).
This is a $(m|n)$-dimensional square matrix so it admits an inverse and a determinant, given respectively by
\begin{align}\label{eq:propagator}
    (\hX_{ji})_{\dfa_j\fa_i } &\coloneq (X_{ij}^{\fa_i \dfa_j})^{-1},&
        g_{ij}&\coloneq \text{sdet} (\hX_{ij}).
\end{align}
The inverse $\hX_{ji}$ then transforms according to its (lower) indices under the isotropy group. The determinant $g_{ij}$ on the other hand  transforms as a scalar with $\gamma_i=\gamma_j=1$, $\tilde \gamma_i=-\tilde \gamma_j=-1$ (see~\eqref{eq:primary2}).
In fact, we have that
\begin{align}\label{g=g}
    g_{ij}=g_{ji}
\end{align}
due to the fact that $\sdet g=1$.
\footnote{To see this note that we can always write $X_i=a_i(1,x_i)$, $\bar X_i=(-x_i,1)^Td_i$, then $g_{ij}= 1/(\sdet d_i \sdet a_j)\sdet(x_{i}{-}x_{j})= g_{ji}$ since $\sdet h_i=1 \Rightarrow \sdet a_i =\sdet d_i$. For a more direct derivation see appendix~\ref{app:superalgebra} }

In summary, correlators are built from $X_{ij},\hX_{ij},g_{ij}$ connecting the local $\SL(m|n)$ indices as dictated by the correlators. Let us see this in practice for low numbers of points.

A subtlety we will not explore further in this paper which is particularly important for half BPS  correlators at 5-points and higher,
 but will also be relevant at lower points for correlators with external long operators, is the existence of the $\epsilon$ discussed at the end of section~\ref{sec:classification}. This constant invariant tensor can be used to obtain more general solutions of the Ward identities than those described here. We leave an investigation of this for future work.

\subsubsection{Two points}

In this case, there is a unique object which transforms correctly under the local isotropy group, as dictated by the indices and $\gamma$-charge. The most general non-zero two-point structure is
\begin{equation}\label{eq:twopoint}
    \langle \cO_{\underline{\fa} ,\underline{\dfa}}^{\gamma}(X_1) \bar{\cO}_{\underline{\fb} ,\underline{\dfb}}^{\gamma}(X_2) \rangle = A \, (g_{12})^{\gamma} (\hX_{12} \dots \hX_{12})_{\underline{\dfa} ,\underline{\fb}} (\hX_{21} \dots \hX_{21})_{\underline{\dfb} ,\underline{\fa}},
\end{equation}
where $A$ is an arbitrary constant (which can be set to 1 by normalising the operator appropriately). Then the symmetrisation of indices imposed on the indices of the operators on the LHS must be imposed on the RHS. Crucially this means  $\underline{\dfa}$ and $\underline{\fb}$ must be symmetrised in the same Young tableau, similarly for $\underline{\dfb}$ and $\underline{\fa}$. Note that this is simply the condition that the two operators must be conjugates of each other, as in CFT. By identifying $X_{ij}\sim x_i-x_j$, the structure maps to the two-point structure in analytic superspace coordinates given in~\cite{Heslop:2003xu}. 

\subsubsection{Three points}
Here we want combinations of $X_{ij}$ and $\hX_{ij}$ for $i,j=1,2,3$ which will solve the Ward identities. Because of the additional insertion, it is possible to  contract local $\fa, \dfa $ indices. For example, we can construct the following combinations
\begin{equation}\label{eq:Yijk}
    (Y_{i,jk})_{\dfa_i \fa_j} \coloneq (\hX_{ij})_{\dfa_i \fa_j} X_{jk}^{\fa_j \dfa_k} (\hX_{ki})_{\dfa_k \fa_i} \sim (x_i{-}x_k)^{-1}-(x_i{-}x_j)^{-1}.
\end{equation}
Note that as a consequence of~\eqref{eq:ginvg}, one can show that
\begin{equation}
    Y_{i,jk}=-Y_{i,kj} \,,
\end{equation}
and thus, at three points, we can omit the second and third labels and identify $Y_i$ with the positive permutation of $(i,j,k)$. See appendix \ref{app:superalgebra} for a derivation of the above. Just as with two-point structures, three-point structures require the right transformation under the local determinant as dictated by the local $\gamma_i$ charges. This can be ensured by including appropriate powers of $g_{ij}$, $i,j=1,2,3$.

A general three-point structure for three generic unitary fields $\cO_{\ula^i_{L}, \ula^i_R}^{\gamma_i}$ is given by~\cite{Heslop:2003xu}
\begin{equation}\label{threepointstructure}
 \langle \cO^{\gamma_1}_{\ula^1_{L}, \ula^1_R}(X_1) \cO^{\gamma_2}_{\ula^2_{L}, \ula^2_R}(X_2)\cO^{\gamma_3}_{\ula^3_{L}, \ula^3_R}(X_3)\rangle \supseteq  K\hspace{-.5mm}\prod_{i=1,2,3}\hspace{-.5mm}\Pi^{\ula_L^i}(X_i) \bPi^{\ula_R^i}(X_i) \hspace{-.5mm}\bigotimes_{(ij) \in S_3}\hspace{-.5mm}\hX_{ij}^{\otimes a_{ij}} \hspace{-.5mm}\bigotimes_{i=1,2,3}\hspace{-.5mm} (Y_{i})^{\otimes b_i} ,
\end{equation}
where $K=g_{12}^{\gamma_{12,3}}g_{23}^{\gamma_{23,1}}g_{31}^{\gamma_{31,2}}$ with $\gamma_{ij,k}=(\gamma_i{+}\gamma_j{-}\gamma_k)/2$.
The powers $a_{ij}$ and $b_i$ depend on the number of boxes in the relevant Young diagram and satisfy
\begin{equation}
     a_{ji}+a_{ki}+b_i= |\ula_L^i|, \qquad \qquad a_{ij}+a_{ik}+b_i= |\ula_R^i|, \qquad \qquad (i,j,k)\in S_3.
\end{equation}
The structure \eqref{threepointstructure} is the same for all $m$ and $n$, the only possible difference being in the shapes of the Young diagrams which contribute: for given $m,n$ any term with any Young diagram $\ula_L^i,\ula_R^i$ not fitting within the hook shape~\eqref{hook} will vanish.

We note that the above structure does not vanish when Grassmann odd coordinates are turned off. However, such nilpotent structures can exist in supersymmetric theories even at three points, when long operators are involved. To include such structures in our formalism, it suffices to include the $\SL(m|n)$ constant invariant epsilon tensor $\mathcal{E}^{\ufa}_{\ufb}$ mentioned at the end of section~\ref{sec:classification} in a way that is consistent with the index structure as discussed in~\cite{Heslop:2003xu}. These would only occur when at least one of the $SL(m|n)$ irreps is long and we don't consider them explicitly in this paper.

In general, there are different $\SL(2m|2n)$ invariant three-point tensor structures \eqref{threepointstructure} transforming in the same tensor product of representations at the different points. This is expected and indeed also happens in non-supersymmetric CFT \cite{Costa:2011mg}.
Thus, a physical three-point function is a linear combination of tensor structures given by
\begin{equation}\label{eq:threepointfunction}
    \langle \cO_1 \cO_2 \cO_3 \rangle = \sum_{k} \lambda^{(k)}_{123} \langle \cO_1 \cO_2 \cO_3 \rangle^{(k)},
\end{equation}
where the $\lambda^{(k)}_{123}$ are the OPE coefficients and we denoted the different tensor structures by superscripts. 

The counting of three-point structures is very straightforward in the $(m,n)$ Grassmannian space and follows from the fact that tensor products of $Y_i$ only contain tensors that have the same representation on dotted and undotted indices.
For example, consider the following correlator 
\begin{equation}
    \langle \cO^{\gamma_1}_{\ula_L, \ula_R} \cO^{\gamma_2} \cO^{\gamma_3}_{\umu_L, \umu_R}\rangle\,.
\end{equation}
Then, if $|\ula_L|=|\ula_R|=\lambda$ and $|\umu_L|=|\umu_R|=\mu\geq \lambda$ (recall this is a condition on 4D $\cN=4$ reps), there are up to $\lambda+1$ different tensor structures, which are schematically given by
\begin{equation}\label{eq:threepointstructures}
 \ K\, \underbrace{(Y_1 \dots Y_1)}_{\lambda-k} \underbrace{(Y_3 \dots Y_3)}_{\mu-k} \underbrace{(\hX_{13} \dots \hX_{13})}_{k} \underbrace{(\hX_{31} \dots \hX_{31})}_{k}, \qquad k=0,1,..,\lambda\,,
\end{equation}
where we omitted the indices to avoid clutter but their appropriate symmetrisation according to $\ula_L,\ula_R,\umu_L,\umu_R$ should be implicitly understood. Some of the above structures may vanish due to the asymmetry between the pointwise left and right $\SL(m|n)$ representations. In particular, if both representations are symmetric, none of the above structures vanishes and thus the bound of $\lambda+1$ different structures is saturated. 
On the contrary, if $\umu_L$ and $\umu_R$ are $\lambda$-asymmetric (see  section \ref{sec:classification} for the definition), all but the last three-point structure above vanish.

Correlators involving short supermultiplets must satisfy the appropriate shortening conditions at each point. This requirement imposes additional constraints on the space of tensor structures. The standard approach is to construct the larger space of tensor structures following the process described above, and then apply each differential constraint separately, thus producing constraints on the OPE coefficients $\lambda_{123}^{(k)}$ in~\eqref{eq:threepointfunction}. However, because fields are automatically unconstrained in analytic superspace, such differential constraints need not be applied. Instead, it suffices to ensure that the three-point structures \eqref{threepointstructure} are analytic in the internal coordinates $y_{ij}$ \eqref{eq:sconfgauge} ($i,j=1,2,3$) \cite{Heslop:2003xu}.

The counting and construction of three-point structures is very relevant for our discussion of $\SL(2m|2n)$ weight-shifting operators and blocks in sections \ref{sec:WSOinvariants} and \ref{sec:blocks}, since weight-shifting operators use the fact that blocks are given by a gluing of three-point structures \cite{Simmons-Duffin:2012juh}.

\subsubsection{Four points}

As in a non-supersymmetric CFT, higher point functions are not as constrained. Indeed, while 2 and 3 point functions are fixed up to constants, any solution to the $4$ point Ward identities can be multiplied by a  class function of the `cross-ratio matrix'~\cite{Heslop:2003xu}
\begin{equation}\label{eq:Z}
    (Z)^{\fa_1}_{\ \fb_1}\coloneq(X_{12}\hX_{24}X_{43}\hX_{31})^{\fa_1}_{\ \fb_1}.
\end{equation} 
A class function is a function on the group which is invariant under matrix conjugation, so 
\begin{equation}\label{class}
    f(R^{-1} Z R)=f(Z), \quad \forall R \in \SL(m|n).
\end{equation}
All  such functions satisfy $kf=m_lf=m_Rf =0$ at all points as all indices have been locally contracted,  and so such functions can multiply any solution to the Ward identities to produce another solution.

The class functions~\eqref{class} are functions of the eigenvalues of the matrix $Z$ (since there is an $R$ that diagonalises $Z$). 
These eigenvalues are precisely the $\SL(2m|2n)$-invariant cross-ratios.
Moreover, these functions have additional symmetry properties due to the existence of a subset of the matrices $R$ which preserves a diagonal $Z$.
In particular, in the non supersymmetric case ($m$ or $n=0$) the functions  will be symmetric under arbitrary permutations of the eigenvalues, i.e.\ they are symmetric functions of the cross ratios. For example, in the $(2,0)$ space, the two eigenvalues are the Dolan-Osborn coordinates $z,\bar{z}$, which relate to the usual 4D cross-ratios $u,v$ as follows
\begin{equation}\label{eq:4Dcrossratios}
    u=z \bar{z}=\det Z, \qquad v=(1-z)(1-\bar{z}) = \det(1-Z).
\end{equation}
For general $m$ there are $m$ eigenvalues which we denote
 $z_1,\dots,z_m$.

In the supersymmetric case, the eigenvalues also include the cross-ratios of the internal compact manifold $w_1,\dots,w_n$, together with a $\mathbb{Z}_2$ grading, such that
\begin{equation}\label{eq:strsdet}
    \text{str} Z = \sum_{i=1}^{m} z_i - \sum_{i=1}^{n} w_i, \qquad \text{sdet} Z = \frac{\prod_{i=1}^{m}z_i}{\prod_{i=1}^{n}w_i}.
\end{equation}
In this case the functions of cross ratios resulting from~\eqref{class} are known as supersymmetric functions~\cite{stembridge1985characterization}.
A supersymmetric function of $z_1,\dots,z_m|w_1,\dots,w_n$ is one which is doubly symmetric in the two sets $z_i$ and $w_j$ and satisfies the additional constraint~\cite{stembridge1985characterization}
\begin{equation}\label{susyfunction}
    \partial_t f(z_1,\dots,t,\dots,z_m|w_1,\dots,t,\dots,w_n)=0.
\end{equation}
Such functions can be expanded in an appropriate basis of (super)symmetric functions e.g.\ super Schur polynomials \cite{Dolan:2003hv, Doobary:2015gia, Aprile:2021pwd}. 

Scalar  (e.g.\ half-BPS) correlators depend on only a single such class function but more general correlators will have a number of independent tensor structures, each of which can be multiplied by a scalar class function.
To enumerate the different tensor structures, we first transfer indices at points $i=2,3,4$ to point 1, using $\hX_{1i}$ and $\hX_{i1}$, and then 
transfer all dotted indices at point 1 to undotted indices 
using $Y:=Y_{1,42}$~\eqref{eq:Yijk}.

In this way, the problem is reframed as that of finding a tensor invariant at a single point, with only undotted indices. This can be a function of $Z^\fa_{\ \fb}$~\eqref{eq:Z} only. So we have 
\begin{align}
    \langle \cO_{\ufa_1 \udfa_1}\cO_{\ufa_2 \udfa_2}\cO_{\ufa_3 \udfa_3}\cO_{\ufa_4 \udfa_4}\rangle = P \,Y_{\udfa_1 \ufc_1}\prod_{i=2}^4 \left((Y^{-1}\hX_{1i})^{\ufc_i}_{\ \ufa_i}(\hX_{i1})_{\udfa_i\ufb_i}  \right)f_{\ufa_1 \ufc_2\ufc_3\ufc_4}^{\ufc_1\ufb_2\ufb_3\ufb_4}(Z)\ ,\label{4ptsol}
\end{align}
where $P$ is a scalar prefactor absorbing the $\gamma$-charges using products of $g_{ij}$, $i,j=1,2,3,4$. 
Since $Y_{1,43}=Y_{1,42}Z$ and 
$Y_{1,23}= Y_{1,43}-Y_{1,42}$, different choices of $Y$ in~\eqref{4ptsol} do not effect this structure. 
Note that in writing down a basis for the tensor $f$ one can string together $Z$s to make $(Z^p)^\fa_{\ \fb}$, including $p=0$ (the Kronecker delta) and negative $p$ (powers of $Z^{-1}$). 
Furthermore, each independent term can then come with its own class function. 
The general solution at higher points can be phrased in exactly the same way, the difference being that the final tensor $f$ will be a function of $n-3$ different $Z$'s.\footnote{The three-point case can be phrased similarly, but the final tensor can only be given in terms of Kronecker deltas, $\delta_\fa^\fb$, or epsilon tensors, giving nilpotent terms in the $m,n\neq0$ case \cite{Heslop:2003xu}. }

For example, consider the following correlator with one non-singlet operator at point 1 
and three singlets
\begin{align}
    \langle  \cO_{\fa \dfa}^{\gamma_1} \cO^{\gamma_2} \cO^{\gamma_3} \cO^{\gamma_4}\rangle \, ,
\end{align}
corresponding to one quarter-BPS operator and three half-BPS ones in 4D $\cN=4$.
Then, the general solution to the superconformal Ward identities~\eqref{4ptsol} is
\begin{align}\label{WI1}
    \langle  \cO_{\fa \dfa}^{\gamma_1} \cO^{\gamma_2} \cO^{\gamma_3} \cO^{\gamma_4}\rangle = P \times Y_{1,42\dfa \fb} f^\fb{}_\fa(Z)\,,
\end{align}
where $P$ is a scalar prefactor: a monomial in $g_{ij}$ which absorbs all dependence on the local $\gamma_i$ weights.
$f^\fb{}_\fa(Z)$ is a matrix function of $Z$, so generated by $(Z^n)^\fb{}_\fa$, with  coefficients $g_n(Z)$ which are themselves scalar class functions:
\begin{align}
f^{\fb}_{\ \fa}(Z)=\sum_k (Z^k)^\fb_{\ \fa}\,g_k(Z),\qquad 
g_k(GZG^{-1})=g_k(Z)\ . \label{F}   
\end{align}
When $n\neq0$  there will then be additional restrictions on the allowed generators for a given correlator of fixed charges which arise from analyticity in the internal variables.
Furthermore, the function $f$ \eqref{F} can be diagonalised by diagonalising $Z$, giving
\begin{equation}
    f^{\fa}_{\ \fb} (z_1, \dots, z_m | w_1, \dots, w_n) \sim \text{diag}\left(f_1(z_i|w_j), \dots, f_{m+n}(z_i|w_j)\right).
\end{equation}
Then the consequence of $f$ being a matrix function~\eqref{F} is firstly that 
the component functions $f_i$ are symmetric separately in the $z$-variables and the $w$-variables, but with the symmetry excluding $z_i$ (or excluding $w_{i-m}$ if $i>m$). So for example, $f_1$ is a symmetric function of all $z_i$ where $i\neq 1$, and separately a symmetric function of all $w$-variables. Furthermore, in the supersymmetric case, the functions $f_i$ also obey 
\begin{equation}\label{F2}
    \partial_t f_k(z_i|w_j)|_{z_p=w_q=t} = 0,\qquad \quad   \forall \  p,q \quad \text{such that} \quad \left\{ \begin{array}{ll} p \neq k \qquad \qquad  &k\leq m\\
     q \neq k{-}m \qquad &k> m
\end{array}
\right. \ \ .
\end{equation}
This generalises the SUSY Ward identities of half-BPS correlators~\eqref{susyfunction}.

Similarly to \eqref{WI1}, a correlator with two non-singlet operators with one index at points 1 and 2 
and two singlets at 3 and 4 has a very similar solution
\begin{align}\label{WI2}
    \langle  \cO_{\fa_1}^{\gamma_1} \cO^{\gamma_2}_{\dfa_2} \cO^{\gamma_3} \cO^{\gamma_4}\rangle  = P \times (\hX_{21})_{\dfa_2 \fb_1} f^{\fb_1}{}_{\fa_1}(Z)\ ,
\end{align}
with $f$ again satisfying the properties~\eqref{F},~\eqref{F2}.
We will come back to both of these cases when discussing superconformal blocks in section \ref{sec:blocks}.

\section{$\SL(2m|2n)$ weight-shifting operators}\label{sec:weightshifting1}

Weight-shifting operators of a group $G$ are differential operators which relate two different irreducible representations of $G$. They are based on earlier mathematical work where they are known as translation functors \cite{ZuckermanFunctors,jantzen_highest_weight_modules,knapp_vogan_cohomological_induction}, and were more recently derived for the conformal group $\mathrm{SO}(d+1,1)$ \cite{Karateev:2017jgd}, in which they provide a tool for deriving expressions for spinning conformal blocks, both seeds and non-seeds. 
In this section, we briefly review the relevant theory and then derive weight-shifting operators \eqref{eq:collectedWSO} for $\SL(2m|2n)$ on the supercoset space (and the super-Grassmannian) which we constructed in section \ref{sec:cosetspace}. 
We show how they interact with $\cN=2$ and $\cN=4$ superconformal multiplets, i.e.\ their spin, dimension and $R$-symmetry representation, but also with their $Q, \bar{Q}$ shortening conditions. 
As a key application, these differential operators offer a systematic derivation of all 4D $\cN=2,4$ superconformal blocks involving external superconformal multiplets beyond half-BPS, which we will explore in subsequent sections.

\subsection{Covariant differential operators}
\label{sec:cov}
Weight-shifting operators are associated with finite-dimensional representations $\rep$ of the symmetry group $G$, and are maps $\cO\mapsto \cO'$, where $\cO'$ is contained in the tensor product of $\rep$ and $\cO$~\cite{Karateev:2017jgd}. 
In particular, for any
\begin{equation}\label{tensor}
 \cO' \in   \rep \otimes \cO \,,
\end{equation}
there is a differential operator $\cD_{\overline{\rep}}: \cO \mapsto \cO'$
where $\overline{\rep}$ is the dual representation of $\rep$. 
For sufficiently generic representations
$\cO$,
the representations $\cO'$ appearing in the tensor product~\eqref{tensor} are precisely those whose highest weight states are obtained by adding all the weights of the representation $\rep$ to the highest weight state of 
$\cO$.
\footnote{In particular, the number of operators on the RHS of~\eqref{tensor}, and thus the number of weight-shifting operators associated with $\rep$,  should be equal (in the generic case) to the dimension of the finite representation in question.}
This can be seen very clearly in analytic superspace, since finite-dimensional representations can be explicitly written  as fields~ $\finite^{\underline \fa \underline \dfa}(X,\bar X)=\finite^{\underline{\cA}}_{\ \underline{\cB}} X^{\underline \fa}_{\ \underline \cA}\bar X^{\underline \cB \underline \dfa}$~\eqref{eq:finiterep}, and so  can $\cO$ and $\cO'$ (or at least their correlators). Then we can explicitly construct the field $\cO'$ out of $\finite$ and $\cO$, and write the result  as follows
\begin{equation}\label{eq:weightshiftingdef}
    \cO'= \finite^{\underline{\cA}}_{\ \underline{\cB}} \cD^{\underline{\cB}}_{\ \underline{\cA}}\cO\,,
\end{equation}
with $\cD_{\bar \rep}=\cD^{\underline{\cB}}_{\ \underline{\cA}}$ the corresponding weight-shifting operator. 

To construct the operator $\cO'$ from $\cO$ and $\finite$, one employs the standard procedure for constructing primary fields from other primary fields in a CFT, namely take combinations of the fields $\finite, \cO$ with the momentum operator $p$ applied and then  fix the coefficients of each term by insisting that $kO'=0$.
The same procedure can be employed here, taking $p$ to be the generator of translations in analytic superspace, and $k$ the special conformal transformations of the superspace, both given on the coset by~\eqref{eq:generators}. 
We will construct $\cO'$~\eqref{eq:weightshiftingdef}, hence $\SL(2m|2n)$ weight-shifting operators, both in a general analytic superspace (without specifying the coordinates) in section \ref{sec:cosetWSO} and in the Grassmannian space (\ref{sec:grassmannianWSO} and \ref{sec:genWSO}). We will use the latter in calculations involving correlation functions, since these are most easily defined in terms of Grassmannian planes $X, \bX$.
\subsection{Fundamental weight-shifting operators}
We first focus on the case where $\finite=\frep$, the fundamental of $\SL(2m|2n)$, and so the field \eqref{eq:finiterep} is explicitly given by
\begin{equation}\label{fund}
    \Phi^\fa(X)=X^\fa_{\ \cA}\Phi^\cA.
\end{equation}
By symmetry, one can then deduce the weight-shifting operators arising from tensor products involving the antifundamental $\overline{\frep}$. General weight-shifting operators corresponding to more complicated $\SL(2m|2n)$ representations can be constructed from products of fundamental and antifundamental operators. We will show this in section \ref{sec:genWSO}.

\subsubsection{Construction in analytic superspace}\label{sec:cosetWSO}

There are two ways to construct primaries from the fundamental $\Phi^\fa(X)$ and a unitary irreducible representation on analytic superspace, $\cO^\gamma_{\underline \lambda_L,\underline \lambda_R}=\cO^\gamma_{\underline \fa,\underline \dfa}$. The first is to simply multiply them together and contract one of the indices:
\begin{align}\label{Xwsop}
    \Phi^\fa\cO^{ \gamma}_{\fa_1..\fa_{i-1}\fa \fa_{i+1}\dots, \underline \dfa}\,.
    \end{align}

The number of independent operators one obtains this way is equal to the number of ways of obtaining $\ula_L$ in the tensor product of the fundamental with another representation $\ula'_L$.

We can realise the independent operators nicely using Clebsch-Gordan coefficients 
for the tensor product of representations $\ula'_L\otimes\Box \rightarrow \ula_L$.
A concrete way to realise the Clebsch-Gordan operators  is via (Hermitian) Young projectors and their corresponding transition operators
\begin{align}\label{eq:Xwso}
    \cO^{\prime \gamma}_{\ula'_L\ula_R}=\cO^{\prime \gamma}_{\underline \fa, \underline \dfa}     &= 
    \frep^\fb  \Pi^{\ula_L}_{i,1}\cO^{ \gamma}_{\underline\fa\fb, \underline \dfa} 
    \,.
\end{align}
Here, the new Young diagram $\ula'_L$ is obtained from that of the original $\ula_L$  by deleting a single box. 
The $\Pi$ is a Young transition operator on the indices $\underline \fa,\fb$, taking the lexicographically ordered standard Young tableau $\ula_L(1)$ to the Young tableau $\ula_L(i)$ with the final index in the position we wish to delete to obtain $\ula'_L$. 
So, for example 
\begin{equation}\label{eq:1youngsym}
    \ula_L(1) = \begin{tikzpicture}[x=13pt, y=13pt, baseline={(current bounding box.center)}]
    \scalebox{1.1}{ 
        \tyng(0,0,5,4,3);
    }
    \draw [ultra thick, dashed] (1.1,-3.3) to (1.1,-2.2) to (0,-2.2) to (0,-3.3) to (1.1,-3.3);
    \node at (0.55, 0.55) {\small $\fa_{\scriptscriptstyle 1}$};
    \node at (1.65, 0.55) {\small $\fa_{\scriptscriptstyle 2}$};
    \node at (2.75, 0.55) {\small $\fa_{\scriptscriptstyle 3}$};
    \node at (3.85, 0.55) {\small $\fa_{\scriptscriptstyle 4}$};
    \node at (4.95, 0.55) {\small $\fa_{\scriptscriptstyle 5}$};
    \node at (0.55, -0.55) {\small $\fa_{\scriptscriptstyle 6}$};
    \node at (1.65, -0.55) {\small $\fa_{\scriptscriptstyle 7}$};
    \node at (2.75, -0.55) {\small $\fa_{\scriptscriptstyle 8}$};
    \node at (3.85, -0.55) {\small $\fa_{\scriptscriptstyle 9}$};
    \node at (0.55, -1.65) {\small $\fa_{\scriptscriptstyle 10}$};
    \node at (1.65, -1.65) {\small $\fa_{\scriptscriptstyle 11}$};
    \node at (0.55, -2.75) {\small $\fb$};
    \node at (2.75, -1.65) {\small $\fa_{\scriptscriptstyle 12}$}; 
\end{tikzpicture} \,  \quad  \xrightarrow{\displaystyle \ \Pi^{\ula_L}_{i,1}\ }\quad
\ula_L(i) = \begin{tikzpicture}[x=13pt, y=13pt, baseline={(current bounding box.center)}]
    \scalebox{1.1}{ 
        \tyng(0,0,5,4,2,1);
    }
    \draw [ultra thick, dashed] (3.3,-2.2) to (3.3,-1.1) to (2.2,-1.1) to (2.2,-2.2) to (3.2,-2.2);
    \node at (0.55, 0.55) {\small $\fa_{\scriptscriptstyle 1}$};
    \node at (1.65, 0.55) {\small $\fa_{\scriptscriptstyle 2}$};
    \node at (2.75, 0.55) {\small $\fa_{\scriptscriptstyle 3}$};
    \node at (3.85, 0.55) {\small $\fa_{\scriptscriptstyle 4}$};
    \node at (4.95, 0.55) {\small $\fa_{\scriptscriptstyle 5}$};
    \node at (0.55, -0.55) {\small $\fa_{\scriptscriptstyle 6}$};
    \node at (1.65, -0.55) {\small $\fa_{\scriptscriptstyle 7}$};
    \node at (2.75, -0.55) {\small $\fa_{\scriptscriptstyle 8}$};
    \node at (3.85, -0.55) {\small $\fa_{\scriptscriptstyle 9}$};
    \node at (0.55, -1.65) {\small $\fa_{\scriptscriptstyle 10}$};
    \node at (1.65, -1.65) {\small $\fa_{\scriptscriptstyle 11}$};
    \node at (0.55, -2.75) {\small $\fa_{\scriptscriptstyle 12}$};
    \node at (2.75, -1.65) {\small $\fb$}; 
\end{tikzpicture} \, , 
\end{equation} 
where 
\begin{equation}
    \ula_L'(1) \equiv \overline{\ula_L(i)} = \begin{tikzpicture}[x=13pt, y=13pt, baseline={(current bounding box.center)}]
    \scalebox{1.1}{ 
        \tyng(0,0,5,4,2,1);
    }
    \node at (0.55, 0.55) {\small $\fa_{\scriptscriptstyle 1}$};
    \node at (1.65, 0.55) {\small $\fa_{\scriptscriptstyle 2}$};
    \node at (2.75, 0.55) {\small $\fa_{\scriptscriptstyle 3}$};
    \node at (3.85, 0.55) {\small $\fa_{\scriptscriptstyle 4}$};
    \node at (4.95, 0.55) {\small $\fa_{\scriptscriptstyle 5}$};
    \node at (0.55, -0.55) {\small $\fa_{\scriptscriptstyle 6}$};
    \node at (1.65, -0.55) {\small $\fa_{\scriptscriptstyle 7}$};
    \node at (2.75, -0.55) {\small $\fa_{\scriptscriptstyle 8}$};
    \node at (3.85, -0.55) {\small $\fa_{\scriptscriptstyle 9}$};
    \node at (0.55, -1.65) {\small $\fa_{\scriptscriptstyle 10}$};
    \node at (1.65, -1.65) {\small $\fa_{\scriptscriptstyle 11}$};
    \node at (0.55, -2.75) {\small $\fa_{\scriptscriptstyle 12}$};
\end{tikzpicture}\, .
\end{equation}
The transition operator should be interpreted via Kronecker deltas $\delta^{\fb}_{\fa}$ (see appendix~\ref{app:yt} for full details of the Young transition operators etc.\ and in particular~\eqref{permdel} for the interpretation in terms of $\delta^{\fb}_{\fa}$).
Then using~\eqref{fund} we  write out~\eqref{eq:Xwso} as
\begin{align}\label{eq:Xwso2}
   \cO^{\prime \gamma}_{\underline \fa', \underline \dfa}     &= 
    \frep^\cA X^{\fb'}_{\ \cA} \left(\Pi^{\ula_L}_{i,1}\right)_{\underline\fa'\fb'}^{\underline\fa\fb}\cO^{ \gamma}_{\underline\fa\fb, \underline \dfa} 
    \,.
\end{align}
and hence read off the independent weight-shifting operators $\cD:\ula_L \mapsto \ula_L'$ with $|\ula_L'|= |\ula_L|-1$ from \eqref{eq:weightshiftingdef} and \eqref{eq:Xwso} as
\begin{align}\label{eq:Xwso3}
   ( \cD_\cA)^{\ula_L}_{\ula'_L} \coloneq ( \cD_\cA)^{\underline \fa \fb}_{\underline \fa'} = X^{\fb'}_{\ \cA} \left(\Pi^{\ula_L}_{i,1}\right)^{\underline \fa \fb}_{\underline \fa' \fb'}\, ,
\end{align}
consisting of a multiplicative operator $X$ together with a Young transition operator $\Pi^{\ula_L}_{1,i}$ which ensures that $\overline{\ula_L(i)}=\ula_L'(1)$.
The counting of independent cases is  the same as the number of ways of subtracting  a box from the Young diagram $\ula_L$, which for a generic representation in the $\SL(2m)$ case would be $m$: one at the end of each row.\footnote{For a non generic case, we can delete a box at the bottom right of each rectangle consisting of a maximal number of rows of equal length in the Young diagram.}
This counting accounts for precisely half the total number of weight-shifting operators which should be $2m$, the dimension of the fundamental representation $\frep$. 

The other, less trivial way to construct  operators from $\frep$ and $\cO$ is using the momentum operator $p_{\dfa \fa}$.
For this we write the ansatz
\begin{align}\label{oprime}
    &\cO_{\ula_L \ula'_R}^{\prime\gamma} =\cO_{\underline\fa, \underline\dfa\dfb}^{\prime\gamma}\notag \\
   & = \bPi^{\ula'_R}_{1,i}
    \left( a_0 p_{\dfb \fb} \frep^\fb \cO_{\underline\fa, \underline\dfa}^\gamma + a_1
  \frep^\fb  p_{\dfb \fb}\cO_{\underline\fa, \underline\dfa}^\gamma+ a_2
  \frep^\fb  p_{\dfb \fa_1} \cO_{\fb \fa_2\dots,\underline\dfa}^\gamma+ a_3
  \frep^\fb  p_{\dfb \fa_2} \cO_{\fa_1\fb \fa_3\dots \underline\dfa}^\gamma +\dots\right)\notag\\&= 
  \bPi^{\ula'_R}_{1,i}
    \left(a_0 p_{\dfb \fb} \frep^\fb \cO_{\underline\fa, \underline\dfa}^\gamma + \frep^\fb\sum_{\ula'_L\in \ula_L\otimes \Box} a_{\ula'_L} 
\Pi^{\ula'_L}_j\  p_{\dfb \fb} \cO_{\underline\fa, \underline\dfa}^\gamma  \right)\,.
\end{align}
where $\ula'_R$ is the Young tableau obtained from $\ula_R$ by adding one box corresponding to the new index $\dfb$. 
The transition operator  $\bPi^{\ula'_R}_{1,i}$ first projects the indices $\underline \dfa$ of $\ula_R$ 
together with the new index $\dfb$ 
according to the corresponding Young tableau $\ula_R'(i)$, with $\dfb$ in the new box, and then transforms this into the lexicographically ordered form (see~\eqref{eq:1youngsym} but this time from $\ula_R'(i)$ to $\ula_R'(1)$). 
The  Young symmetriser $\Pi^{\ula'_L}_j$ projects the indices $\ufa$ and $ \fb$, according to Young tableau $\ula'_L(j)$ which has the last index $\fb$ in the extra box.
 Here, it is not necessary to further transition to the fully lexicographically ordered case as the index is being contracted.
So completely explicitly, the term in the sum written out with indices is given by
\begin{align}
 \frep^\fb
\Pi^{\ula'_L}_j\  p_{\dfb \fb} \cO_{\underline\fa, \underline\dfa}^\gamma= \frep^\fb
\left(\Pi^{\ula'_L}_j\right)^{\underline\fa' \fb'}_{\underline\fa \fb}\  p_{\dfb \fb'} \cO_{\underline\fa', \underline\dfa}^\gamma   \ .
\end{align}
 Note $\Pi$ and $\bPi$ are denoted differently to emphasise that they act independently of each other: on $\GL(m|n)_L$ and $\GL(m|n)_R$ representation spaces, respectively.

The number of terms inside the sum in \eqref{oprime} corresponds to the number of different ways that a box can be added to $\ula_L$. 
Crucially, the complexity of the weight-shifting operator does not depend on the Young diagram being acted upon (i.e.\ $\ula_R \mapsto \ula'_R$ in the example above) but rather on the representation on the opposite side ($\ula_L$). Since $\GL(m|n)$ Young diagrams fit into a thick hook of height $m$ and length $n$~\eqref{hook}, there are at most $m+n$ positions for a new box. Thus, the number of terms in the weight-shifting operator is limited by the dimension of the superspace. For example, the sum in \eqref{oprime} will have at most 2 terms in 4D ($m=2, n=0$), at most 3 in 4D $\cN=2$ ($m=2, n=1$), and at most 4 in 4D $\cN=4$  ($m=2, n=2$).
 
The coefficients $a_{\ula'_L}$ in \eqref{oprime} and \eqref{wsgen} are fixed by demanding $k\cO'=0$,  using the commutator relations~\eqref{commutators}, schematically  $[k,p]=m_L+m_R$, $[k,m_L]=k$. 
Explicitly, applying $k^{\fc \dfc}$ to~\eqref{oprime} we obtain 
\begin{align}
    k^{\fc \dfc} \cO_{\underline\fa, \underline\dfa\dfb}^{\prime\gamma} &=  \bPi^{\ula'_R}_{1,i}
    \left( a_0(m{-}n)\delta_{\dfb}^{\dfc} \frep^\fc  \cO_{\underline\fa, \underline\dfa}^\gamma + \frep^\fb\sum_{\ula'_L\in \ula_L\otimes \Box} a_{\ula'_L} 
 \Pi^{\ula'_L}_{j}\  \left(\delta_\dfb^{\dfc} (m_L)^{\fc}_{\ \fb}+\delta_\fb^{\fc} (m_R)_\dfb^{\ \dfc}\right) \cO_{\underline\fa, \underline\dfa}^\gamma  \right)\notag\\
&=  \bPi^{\ula'_R}_{1,i}
    \left( a_0(m{-}n)\delta_{\dfb}^{\dfc} \frep^\fc  \cO_{\underline\fa, \underline\dfa}^\gamma + \frep^\fb\sum_{\ula'_L\in \ula_L\otimes \Box} a_{\ula'_L} 
 \Pi^{\ula'_L}_{j}\  \delta_\dfb^{\dfc}\delta_\fb^{\fc}
\left(B_{\ula'_L\!(i)}+B_{\ula'_R\!(j)}\right) \cO_{\underline\fa, \underline\dfa}^\gamma  \right),\label{kop2}
\end{align}
where to get the second line we used~\eqref{eq:mLconstant} and~\eqref{eq:mRconstant}.

Now we use the following key property of Hermitian Young Projectors (HYPs)
\begin{align}\label{addbox0}
\sum_{\ula'_L\in\Box\otimes \ula_L}  \Pi^{\ula'_L}_{j}= \Pi^{\ula_L}_1\,,
\end{align}
which is explained in detail in appendix~\ref{app:yt}. Here on the RHS is a HYP which only permutes the indices $\underline \dfa$ of $\ula_L$, leaving the single index $\fb$ untouched. Thus  if we set the coefficients 
\begin{align}
  a_{\ula'_L}=c_0c_{\ula'_L\!(j), \ula'_R(i)}, \qquad  c_{\ula'_L\!(j), \ula'_R(i)}=-\frac{1}{B_{\ula'_L\!(j)}+B_{\ula'_R(i)}}=\frac{1}{\gamma+C_{\ula'_L\!(j)}+C_{\ula'_R(i)}}\, ,\label{coeffs1}
\end{align}
where $c_0=(m{-}n)a_0$, then
the sum over Young tableaux in~\eqref{kop2} reduces to~\eqref{addbox0} and the entire sum in~\eqref{kop2} reduces to (minus) the first term (the resulting projector $ \Pi^{\ula_L}_1$ can be pulled through to hit $\cO_{\underline\fa, \underline\dfa}^\gamma$ giving the identity). Thus, with the coefficients~\eqref{coeffs1} we indeed obtain a good primary operator $ \cO_{\underline\fa, \underline\dfa\dfb}^{\prime\gamma}$.

The corresponding weight-shifting operator is then derived by writing \eqref{oprime} in the form of \eqref{eq:weightshiftingdef}. This is straightforwardly done by extracting $\Phi^\cA$ from  $\Phi^\fc(X)=X^{\fc}_{\ \cA} \Phi^\cA$ to directly obtain
\begin{align}\label{genwsop}
\cO_{\underline\fa,\underline\dfa\dfb }^{\prime\gamma}  = c_0\Phi^\cA \bPi^{\ula'_R}_{1,i}\left(  \left(p_{\dfb \fb} X^{\fb}_{\ \cA}\right)+X^\fb_{\ \cA}\!\!\!\!\sum_{\ula'_L\in \ula_L\otimes \Box} \!\!\!\!c_{\ula'_L\!(j), \ula'_R(i)} 
    \Pi^{\ula'_L}_{j} p_{\dfb \fb}\right)\cO_{\underline \fa,\underline\dfa}^\gamma,
\end{align}
from which one obtains the weight-shifting operator $\bar \cD:\ula_R \mapsto \ula_R'$ such that $|\ula_R'|= |\ula_R|+1$, given by
\begin{align}\label{wsgen2}
 (\bar \cD_{\cA})^{\ula_R}_{\ula'_R}\coloneq  (\bar \cD_{\cA})^{\underline \dfa}_{\underline \dfa' }\coloneq c_0\left(\bPi^{\ula'_R}_{1,i}\right)^{\underline \dfa \dfb}_{\underline \dfa'}\left( \left(p_{\dfb \fb} X^{\fb}_{\ \cA}\right)+X^\fb_{\ \cA}\!\!\!\!\sum_{\ula'_L\in \ula_L\otimes \Box} \!\!\!\!c_{\ula'_L\!(j), \ula'_R(i)}
    \Pi^{\ula'_L}_{j} p_{\dfb \fb}\right)\,,
\end{align}
with the coefficients~$c_{\ula'_L\!(j), \ula'_R(i)}$ given in~\eqref{coeffs1}. 
As with the previous weight-shifting operator, in the $(m,0)$ case and  for generic $\ula_L$ we see that there will be precisely $m$ of these, thus we have accounted for all $2m$ weight-shifting operators arising from the fundamental representation $\Phi$.

The above differential operator can be written in any chosen set of coordinates, e.g. coset $X,\bar{W}$, Grassmannian $X,\bX$ or directly in analytic superspace, by inserting the corresponding form of the momentum generator $p_{\dfa \fa}$ into \eqref{wsgen2}. For example, on the coset coordinates we have $p=\bW \partial_X$ and thus the differential operator in \eqref{wsgen2} becomes
\begin{equation}\label{wsgen}
    (m-n)\bW_{\dfb \cA}+X^\fb_{\ \cA}\!\!\!\!\sum_{\ula'_L\in \ula_L\otimes \Box} \!\!\!\!c_{\ula'_L\!(j), \ula'_R(i)}
    \Pi^{\ula'_L}_{j} (\bW \partial_X)_{\dfb \fb}.
\end{equation}

\subsubsection{Differential operators on the Grassmannian}\label{sec:grassmannianWSO}
Recall from section \ref{sec:Wardids}, that correlation functions are given explicitly in terms of Grassmannian planes and thus we wish to express our weight-shifting operators in terms of $X,\bX$ only. To do so, one way is to insert $p$ from \eqref{kinXXb} and remove all $W,\bW$ dependence using identities. However, this is cumbersome in the general case (see appendix \ref{app:cosettoGR} for an example). Thus, in this section we present a much more elegant derivation directly on the Grassmannian, where we only used consistency conditions of differential operators on such super-Manifolds.

Firstly, note that $(\partial_{ X})^\cA_{\ \fa}\cO$ and  $(\partial_{\bar X})_{\dfa \cA}\cO$ already transform correctly as operators~\eqref{eq:primary2} with one additional $\fa$ or $\dfa$ index. Thus, after appropriate young projection, the final result transforms in a new representation of the Levi subgroup. Moreover, with the perspective outlined in appendix \ref{app:xxbproof}, where $k$ is a measure of $W,\bW$ dependence, the derivatives $\partial_X$, $\partial_{\bX}$ commute with $k$. Thus they naively look like good weight-shifting operators by themselves. However, they are not well-defined operators on the Grassmannian because they do not preserve the hypersurface $X \bX = 0$. Indeed,
$$\partial_X{}^{\cA}_{\ \fa} (X \bX)^{\fb \dfb}= \delta^{\fb}_{\fa} \bX^{\cA \dfb} \neq 0\, , \qquad \partial_{\bX}{}_{\dfa \cA} (X \bX)^{\fb \dfb}= \delta^{\dfb}_{\dfa} X^{\fb}_{\ \cA} \neq 0\, .$$

Thus, to derive Grassmannian weight-shifting operators, we need to find a differential operator that preserves the orthogonality constraint $X \bX =0$.\footnote{Note the similarities with the ideal constraint of \cite{Karateev:2017jgd}.}
The fundamental weight-shifting operator which preserves the orthogonality constraint is:
\begin{align}\label{eq:wsoansatz1}\left(
\bar \cD_{ \cA}\right)^{\underline \dfa}_{\underline \dfa'}=\left(\bPi^{\ula'_R}_{1,i}\right)^{\underline \dfa\fb}_{\underline \dfa'}\bar D_{\dfb \cA}, \qquad   \bar D_{\dfb \cA}\coloneq(\partial_{\bar X})_{\dfb \cA} + \!\!\sum_{\ula'_L\in\Box\otimes \ula_L}c_{\ula'_L\!(j),\ula'_R(i)}(X)^{\fb}_{\ \cA} \Pi^{\ula'_L}_{1,j} (\partial_{\bar X}\partial_{X})_{\dfb \fb}.
\end{align}
where the coefficients are in~\eqref{coeffs1}.

To obtain this we wrote down~\eqref{eq:wsoansatz1} as an ansatz with arbitrary coefficients $c$, inspired by intuition gained from both the singlet and chiral results derived in the previous subsection as well as the form of the ansatz \eqref{wsgen} in the coset space. We then imposed the  constraint $X\bar X=0$ to fix the coefficients $c$. 

The constraint $X\bar X=0$ is implemented as follows
\begin{equation}\label{eq:idealconst}
    \bar \cD_{ \cA} \left(f(X,\bX) \cO^\gamma_{\ula_L\ula_R}\right)|_{X\bX=0}=0,
\end{equation}
where $f$ is a function in the ideal generated by $X\bar X$.
It is enough to take a linear term in $X\bX$, so $f=(X\bX)^{\fc \dfc} \mathcal{T}_{\fc \dfc}(X,\bX)$, where ${\mathcal T}$ is non-singular at $X\bar X=0$.

Then, inserting \eqref{eq:wsoansatz1} into~\eqref{eq:idealconst} and setting $X\bar X=0$ we get
\begin{align}
   0
   ={\mathcal T}_{\fc\dfc} \bPi^{\ula_R'}_{1,i}\left(\delta_{\dfb}^\dfc  X^\fc_{\ \cA}+
\sum_{\ula'_L\in\Box\otimes \ula_L} c_{\ula'_L\!(i),\ula'_R(j)}(X)^{\fb}_{\ \cA} \Pi^{\ula'_L}_{1,j}\left((m_L)^\fc_{\ \fb}\delta^\dfc_\dfb+(m_R)^\dfc_{\ \dfb}\delta^\fc_\fb\right)\right)\cO_{\ula_L\ula_R}^\gamma
   \, ,\label{idealcalcredux}
\end{align}
where we have simply commuted $(\partial_{\bar X}\partial_{X})_{\dfb \fb}$ past $(X\bar X)^{\fc \dfc}$ which gives
\begin{align}\label{commutator}
    [(\partial_{\bar X}\partial_{X})_{\dfb \fb},(X\bar X)^{\fc \dfc}]= 2(m-n) \delta^\fc_\fb\delta^\dfc_\dfb+(m_L)^\fc_{\ \fb}\delta^\dfc_\dfb+(m_R)^\dfc_{\ \dfb}\delta^\fc_\fb\ ,
\end{align}
and since $\mathcal T$ transforms as dictated by its indices (i.e.~\eqref{eq:primary2} with $\gamma=0$) we have that 
\begin{align}
\left((m_L)^\fc_{\ \fb}\delta^\dfc_\dfb+(m_R)^\dfc_{\ \dfb}\delta^\fc_\fb\right) \mathcal T_{\fc \dfc} = -2(m-n) \mathcal T_{\fb \dfb},
\end{align}
precisely cancelling the constant term in~\eqref{commutator}.
From~\eqref{idealcalcredux}, the process is identical as in the coset calculation starting with~\eqref{kop2}
and hence yielding the same  coefficients~\eqref{coeffs1}. 
Note from~\eqref{commutator} (and similar commutators), that we obtain the algebra $\mathfrak{sl}(2m|2n)$ if we identify $p=\partial_{\bar X}\partial_{X}$ and $k=X\bar X$ 
and add a constant to the $m_L,m_R$. This is reminiscent of the representation of the conformal group in spinor helicity space~\cite{Witten:2003nn} (with $k$ and $p$ swapping roles). It is interesting to note that, with this new perspective, the primary constraint $[k,\cD]=0$ is equivalent to the constraint \eqref{eq:idealconst}.

As a final note,  the above result can be derived by acting on $N$-point functions. Weight-shifting operators are in practice used on such functions and hence this is a natural approach.
We simply have to examine how $(X_1 \bX_1)^{\fa \dfa}$ (with the upper indices contracted locally) can appear in a correlation function involving an operator $\cO^{\gamma}_{\underline \dfa \underline \fa}$ at point 1. The only objects that the local indices in $X_1 \bX_1$ can be contracted against are $\hat X_{i1},\hat X_{1i}$ for some other points $i$. Such a term can only ever appear in the form
\begin{align}\label{idealcon}
  (\hat X_{j1}X_1 \bar X_1 X_{1i})_{\dfb_j \fb_i}f^{ \fb_i\dfb_j}_{\underline \dfa \underline \fa}=0 \, ,
\end{align}
where $f$ is an $N$-point structure (as constructed in section \ref{sec:Wardids}). 
As we show in appendix~\ref{app:superalgebra}, this implies all other identities that can arise from the orthogonality constraint, such as the three-point identity $Y_{123}+Y_{132}=0$.
Applying our ansatz~\eqref{eq:wsoansatz1} to~\eqref{idealcon} yields the expected result for the coefficients \eqref{coeffs1}.

\subsubsection{Examples} 
In the previous section, we derived the most general fundamental weight-shifting operator \eqref{eq:wsoansatz1}, capable of acting on any $(m,n)$ analytic superspace field $\cO^\gamma_{\ula_L,\ula_R}$. In practice, we will be using simpler versions of this differential operator obtained by acting on simpler representations. We consider here the most relevant examples for our applications to blocks in section \ref{sec:blocks}.

\paragraph{Scalar:}
Suppose we act on a scalar primary $\cO^{\gamma}$ with no indices. Then the sum in~\eqref{genwsop} has only one term, and the coefficient~\eqref{coeffs1} $c_{\ula'_L,\ula_R'}=c_{\Box, \Box}$ for a single box is just $1/\gamma$, so~\eqref{genwsop} reduces to
\begin{equation}\label{scalarwsop1}
    (\bar{\cD}_{\cA})_{\dfa} \cO^{\gamma} =  (\partial_{\bar X}+\frac1\gamma \partial_{\bX} \partial_X X)_{\dfa \cA} \cO^{\gamma}.
\end{equation}

There is a subtle point here. Recall from section \ref{sec:fieldsandreps} that, as a consequence of the symmetry group having unit determinant, the charges $\gamma_L=\tfrac12(\gamma+\tilde \gamma)$ and $\gamma_R=\tfrac12(\gamma-\tilde \gamma)$, which count the homogeneity in $X$ and $\bX$, respectively, are not good quantum numbers. This is why the operators are labelled by $\gamma$ alone. In particular, we have that two different $\GL(2m|2n)$ representations $(\gamma_L, \gamma_R)$ and   $(\gamma_L', \gamma_R')$ are equivalent representations of $\SL(2m|2n)$ if $\gamma_L+\gamma_R=\gamma_L'+\gamma_R'$. Related to this was the fact that the coset variables $X, \bX, W, \bW$ were constrained by the unit determinant constraint. 
Thus, in principle, we have to be careful when we differentiate as it can be inconsistent with this constraint. 
However, by comparing normalisations of the analytic superspace weight-shifting operator \eqref{wsgen2} and the Grassmannian version \eqref{eq:wsoansatz1}, one can see that the only result of making different for $\gamma_L,\gamma_R$ arises in the overall normalisation (see appendix \ref{app:cosettoGR}). We conclude that we can  make any choice of $\gamma_R$ with $\gamma_L=\gamma-\gamma_R$ (apart from $\gamma_R=0$) to obtain the same weight-shifting operator, as long as we are consistent throughout the calculation. 
So the weight-shifting operator acting on a scalar can be simplified further by choosing $\gamma_L$, equivalently writing $\cO^\gamma$ (or its correlation function) in terms of $\bX$ only. We then have that \eqref{scalarwsop1} simplifies to
\begin{equation}
    (\bar{\cD}_{\cA})_{\dfa} = \partial_{\bX}{}_{\dfa \cA}\ 
\end{equation}
when acting on a scalar operator.

\paragraph{General right-handed chiral:}
This is the case where the left-hand representation is still a singlet, but the right-hand one is non-trivial, that is, a chiral field $\cO^{\gamma}_{\bullet,\ula_R}\equiv\cO^{\gamma}_{\underline \dfa}$. This proceeds very similarly to the singlet case above~\eqref{scalarwsop1} but with the addition of the Young symmetriser $\Pi_{\ula'_R}$. 
All in all, the Grassmannian weight-shifting operator \eqref{eq:wsoansatz1} which acts on a chiral field $\cO^{\gamma}_{\bullet,\ula_R}$ by adding a box to $\ula_R$ is as follows
\begin{align}\label{wschiral}
      (\bar \cD_\cA)_{\underline \dfa'}^{\underline \dfa}=\left(\bPi^{\ula'_R}_{1,i}\right)^{\underline \dfa \dfb}_{\underline \dfa'} \bD_{\dfb \cA}, \qquad \qquad \bD_{\dfb \cA}\coloneq (\partial_{\bar X}+\frac1{\gamma+C_{\ula'_R(i)}} \partial_{\bar X}\partial_{ X} X )_{\dfa \cA}.
\end{align} 
As for the scalar case,  we can ensure that $\cO^\gamma_{\bullet,\ula_R}$ is independent of $X$ since all open indices only depend on $\bX$, in which case the second term vanishes and we are left with the remarkably simple differential operator 
\begin{align}\label{wschiral2}
	 \bD_{\dfb \cA} \cO^{\gamma}_{\underline \dfa}= (\partial_{\bar X})_{\dfb \cA} \cO^{\gamma}_{\underline \dfa}\,.
\end{align}

We note here that a nice by-product of this simple weight shifting operator is in giving a remarkably simple construction of (super)conformal primary operators from fundamental scalar fields ($\gamma=1$ singlets) by applying the above differential operator, $(\bar \partial_X)_{\dfa \cA}$ to $\cO^1|_{\gamma_L=0}:=\bar \Phi$ (which we 
assume to be chiral and depend only on $\bX$ not $X$) 
and $(\partial_X)^\cA{}_\fa$ derivatives to $\cO^1|_{\gamma_R=0}:=\Phi$ and contracting all the $\cA$ indices.
We will come back to this point, giving the explicit formula in the conclusions
(see~\eqref{ops}).

\paragraph{Action on non-trivial $\ula_L$:}
The next simple case to consider corresponds to a rectangular $\ula_L$.  This will be
of particular relevance to our computations of superconformal blocks of external quarter-BPS multiplets of section \ref{sec:blocks}. We will see that extra terms of higher order in derivatives are now inevitable, but these are necessary to map shortening conditions appropriately. 

Consider acting on a chiral field $\cO_{\Box, \bullet}^\gamma \equiv \cO^\gamma_{\fa}$ as an illustrative example, we have that
\Yboxdim{5pt}
\begin{equation}\Yvcentermath1
    \begin{split}
    \bD_{\dfa \cA} \cO^{\gamma}_{\fa} &= \partial_{\bX}{}_{\dfa \cA} \cO^{\gamma}_{\fa}+ \sum_{\ula_L' \in \left\{\yng(2),\, \yng(1,1) \right\}} c_{\ula_L'
    ,\yng(1)
    }X^{\fb}_{\ \cA} \Pi^{\ula_L'}_j (\partial_{\bX} \partial_X )_{\dfa \fb} \cO^{\gamma}_{\fa} \\
    & = \partial_{\bX}{}_{\dfa \cA} \cO^{\gamma}_{\fa}+ c_{\,\yng(2)
    ,\yng(1)
    } X^{\fb}_{\ \cA} (\partial_{\bX} \partial_X)_{\dfa ( \fb} \cO^{\gamma}_{\fa)} + c_{\,\yng(1,1)
    ,\yng(1)
    } X^{\fb}_{\ \cA} (\partial_{\bX} \partial_X)_{\dfa [\fb} \cO^{\gamma}_{\fa]},
    \end{split}
\end{equation}
where the coefficients can be read from \eqref{coeffs1} to be
\begin{align}\Yvcentermath1
    c_{\,\yng(2)
    ,\yng(1)
    } = \frac{1}{\gamma+1}, \qquad c_{\,\yng(1,1)
    ,\yng(1)
    } = \frac{1}{\gamma-1}.
\end{align}
Explicitly (anti)symmetrising the indices in the above equation can be written
\begin{equation}\label{eq:chiralWSO1}
    \bD_{\dfa \cA} \cO^{\gamma}_{\fa} =  \partial_{\bX}{}_{\dfa \cA} \cO^{\gamma}_{\fa} + \frac{\gamma}{\gamma^2-1} X^{\fb}_{\ \cA} (\partial_{\bX} \partial_X)_{\dfa \fb} \cO^{\gamma}_{\fa} - \frac{1}{\gamma^2-1} X^{\fb}_{\ \cA} (\partial_{\bX} \partial_X)_{\dfa \fa} \cO^{\gamma}_{\fb}.
\end{equation}
This weight-shifting operator will be used for computing quarter-BPS 4D $\cN=4$ superconformal blocks in section \ref{sec:blocks}. 
The above computation can be straightforwardly generalised to the general rectangular case $\ula_L= [\lambda^{\mu}]$ ($\mu$ rows of length $\lambda$) using
\begin{equation}
    c_{\,\begin{tikzpicture}[baseline={(current bounding box.center)},scale=0.17]
    \draw[thin] (0,0) rectangle (2,-2);
    \draw[thin] (2,0) rectangle (3,-1);
\end{tikzpicture}
, \begin{tikzpicture}[baseline={(current bounding box.center)},scale=0.17]   \draw[thin] (0,0) rectangle (1,-1);\end{tikzpicture}
} = \frac{1}{\gamma+ \lambda}, \qquad c_{\,\begin{tikzpicture}[baseline={(current bounding box.center)},scale=0.17]
    \draw[thin] (0,0) rectangle (2,-2);
    \draw[thin] (0,-2) rectangle (1,-3);
\end{tikzpicture}
, \begin{tikzpicture}[baseline={(current bounding box.center)},scale=0.17]   \draw[thin] (0,0) rectangle (1,-1);\end{tikzpicture}
} = \frac{1}{\gamma- \mu},
\end{equation}
where the bigger box denotes the rectangle $[\lambda^\mu]$. All in all, the explicit weight-shifting operator acting on a general rectangular antichiral field $\cO_{[\lambda^\mu],\bullet}^\gamma\equiv\cO^\gamma_{\ufa}$, where $\ufa=\Pi^{[\lambda^\mu]}_1 \fa_1 \dots \fa_{\lambda^\mu}$, is
\Yboxdim{13pt}
\begin{equation}\label{eq:chiralWSO2}
    \begin{split}
            \bD_{\dfa \cA} \cO^{\gamma}_{\ufa} &= \partial_{\bX}{}_{\dfa \cA} \cO^{\gamma}_{\ufa} + \frac{2\gamma + \lambda-\mu}{2(\gamma+ \lambda)( \gamma-\mu)}X^{\fb}_{\ \cA} (\partial_{\bX} \partial_X)_{\dfa \fb} \cO^{\gamma}_{\ufa}  \\ 
    & \quad \ - \frac{\lambda+\mu}{2(\gamma+ \lambda)( \gamma-\mu)}\Pi^{[\lambda^{\mu}]}_1 X^{\fb}_{\ \cA}  (\partial_{\bX} \partial_X)_{\dfa \fa_1} \cO^{\gamma}_{\fa_2 \dots \fa_{\lambda^{\scriptscriptstyle\mu}} \fb},
    \end{split}
\end{equation}
where in the last line, $\Pi^{[\lambda^{\mu}]}_1$ is a Young symmetriser of the indices $\fa_1, \dots, \fa_{\lambda^{\scriptscriptstyle \mu}}$ only (hence why we wrote $X^{\fb}_{\ \cA}$ contracting the $\fb$ index to the right of $\Pi$ and not to its left).

\subsubsection{Acting on non-unitary representations}\label{sec:WSOnonunitary}

So far, we have only considered weight-shifting operators arising from the tensor product of $\frep$ with unitary irreps $\cO^{\gamma}_{\ula_L, \ula_R}$. We found that this led to $X^{\fa}_{\ \cA}$ which acts by contracting an index $\fa$, i.e.\ removing a box from $\ula_L$, and to $\bD_{\dfa \cA}$ which acts by adding an index $\dfa$, i.e adding a box to $\ula_R$. 
However, there is an entire class of operators associated with $\frep$ that we will use but have not considered yet: those acting on representations with upper indices. Recall that this difference is not present in non-supersymmetric theories as representations with upper and lower indices are related by the totally antisymmetric tensor $\epsilon_{\fa_1 \dots \fa_m}$ of $\SL(m)$.
While we will not directly act on this kind of representations as they are non-unitary reps of $\SU (m,m|2n)$ when $m,n\neq0$ \cite{Heslop:2001zm}, they will still appear in intermediate steps of calculations even when we initially act on a unitary irrep. 

Let us denote such a representation by swapping the position of the representation labels $\tilde{\cO}_{\gamma}^{\ula_L, \ula_R}\equiv \tilde{\cO}_{\gamma}^{\ufa, \udfa}$. Then, as before, we want to find operators 
\begin{equation}
   \tilde{\cO}^{\ula_L', \ula_R'}_{\gamma} \in\frep \otimes \tilde{\cO}^{\ula_L, \ula_R}_{\gamma},
\end{equation}
which lead to weight-shifting operators transforming in $\overline{\frep}$.
By following the same process as before, we arrive at a very similar result
\begin{equation}
    \begin{aligned}
            \tilde{\cO}^{\ula_L', \ula_R} &= \Phi^{\cA}  \Pi^{\ula_L'}_{1,i} \left(X^{\fa}_{\ \cA} \tilde{\cO}^{\ula_L, \ula_R} \right), & \overline{\ula'_L(i)}&=\ula_L(1)\,, \\
        \tilde{\cO}^{\ula_L, \ula_R'} &= \Phi^{\cA} \bD_{\dfa \cA} \bPi^{\ula_R}_{i,1}\left(\tilde{\cO}^{\ula_L, \ula_R}\right), & \qquad \overline{\ula_R(i)}&=\ula'_R(1) \label{eq:upperWSO}\,.
    \end{aligned}
\end{equation}
Unlike in the lower index case, $X^{\fa}_{\ \cA}$ now adds an index $\fa$ to $\ula_L$ and symmetrises it accordingly, and $\bD_{\dfa \cA}$ \eqref{eq:wsoansatz1} now contracts an index $\dfa$ from $\ula_R$ as picked by $\bPi^{\ula_R}_{1,i}$. 

The derivation of the differential weight-shifting operator \eqref{eq:upperWSO} follows the same process outlined in the previous sections with the key difference of acting on upper indices by removing a box. In particular, the ansatz for $\bD_{\dfa \cA}$ above is as follows
\begin{align}\label{eq:wsoansatzfinite}
\bD_{\dfa \cA}=(\partial_{\bar X})_{\dfa \cA}  + \sum_{\ula_L'\in \ula_L \otimes \Box}\tilde{c}_{\ula_L'(j),\ula_R(i)}  (\partial_{\bar X}\partial_{X})_{\dfa \fb} \Pi^{\ula_L'}_{1,j}X^{\fb}_{\ \cA} ,
\end{align}
where $i,j$ label the specific standard Young tableau, giving the information of the position of the $\dfb,\fb$ indices being contracted, respectively.
The solution to the ansatz follows similarly to the unitary case \eqref{coeffs1} and is given by
\begin{equation}
    \tilde{c}_{\ula'_L\!(j), \ula_R(i)}= \frac{1}{\gamma - 2(m-n) -C_{\ula_L'(j)} - C_{\ula_R(i)}}.
\end{equation}

\subsection{General weight-shifting operators}\label{sec:genWSO}

In the previous subsection, we derived fundamental ($\frep$)  $\SL(2m|2n)$ weight-shifting operators~\eqref{eq:Xwso3}, \eqref{eq:wsoansatz1} and \eqref{coeffs1} from tensor products of the fundamental $\frep$ of $\SL(2m|2n)$ and a field $\cO^\gamma_{\ula_L, \ula_R}$. They are $\SL(2m|2n)$-covariant, transforming in the antifundamental $\overline{\frep}$, and are made of a multiplicative or differential operator with an accompanying permutation operator. The same step by step process can be followed to derive weight-shifting operators $\bar{\cD}^{\cA}$, $\cD^{\cA}$ arising from tensor products of the antifundamental $\overline{\frep}$ of $\SL(2m|2n)$, which will instead transform in $\frep$.\footnote{We will always denote a weight-shifting operator by the representation they transform in, which is the dual of the finite representation they arise from.}
All in all, the results for our (anti)fundamental $\SL(2m|2n)$ weight-shifting operators acting on unitary irreducible representations $\cO^{\gamma}_{\ula_L, \ula_R}$ of the superconformal group $\SU (m,m|2n)$ by mapping $\ula_{L,R
}\mapsto \ula_{L,R}'$  are
\begin{equation}\label{eq:collectedWSO}
   \begin{aligned}
       \frep: \qquad  \quad& \begin{aligned}
           ( \bar \cD^\cA)^{\ula_R}_{\ula'_R}\coloneq(\bar \cD^\cA)^{\underline \dfa \dfb}_{\underline \dfa'}&= \bX^{\cA \dfb'}\left(\bPi^{\ula_R}_{i,1}\right)^{\underline \dfa \dfb}_{\underline \dfa' \dfb'}, \qquad \quad \hspace{-1.6mm}& \overline{\ula_R(i)}&=\ula'_R(1), \\
           (\cD^\cA)^{\ula_L}_{\ula'_L}\coloneq(\cD^{\cA})^{\underline \fa}_{\underline \fa'} &= \left(\Pi^{\ula_L'}_{1,i}\right)^{\underline \fa \fb}_{\underline \fa' \fb'} D_{\ \fb}^\cA,\qquad \quad&  \overline{\ula'_L(i)}&=\ula_L(1),
       \end{aligned}  \\[0.2cm]
       \overline{\frep}: \qquad \quad & \begin{aligned}
           ( \cD_\cA)^{\ula_L}_{\ula'_L}\coloneq( \cD_\cA)^{\underline \fa \fb}_{\underline \fa'} &= X^{\fb'}_{\ \cA} \left(\Pi^{\ula_L}_{i,1}\right)^{\underline \fa \fb}_{\underline \fa' \fb'},\qquad \quad&  \overline{\ula_L(i)}&=\ula'_L(1), \\
           ( \bar \cD_\cA)^{\ula_R}_{\ula'_R}\coloneq\left(
\bar \cD_{ \cA}\right)^{\underline \dfa}_{\underline \dfa'}&=\left(\bPi^{\ula'_R}_{1,i}\right)^{\underline \dfa\fb}_{\underline \dfa'}\bar D_{\dfb \cA},\qquad \quad& \overline{\ula'_R(i)}&=\ula_R(1),
       \end{aligned}
   \end{aligned}
\end{equation}
 where~\eqref{eq:wsoansatz1}
 \begin{align}
     \bar D_{\dfa \cA}&\coloneq(\partial_{\bar X})_{\dfa \cA} + X^{\fa}_{\ \cA} \sum_{\ula'_L\in\Box\otimes \ula_L}c_{\ula'_L(j),\ula'_R(i)}\Pi^{\ula'_L}_{j} (\partial_{\bar X}\partial_{X})_{\dfa \fa}\,,\\
     D_{\ \fa}^\cA&\coloneq(\partial_X) {}^{\cA}_{\ \fa} + \bX^{\cA \dfa}\sum_{\ula_R' \in \ula_R \otimes \Box} c_{\ula_L'(i),\ula_R'(j)} \bPi^{\ula_R'}_i (\partial_{\bX} \partial_X)_{\dfa \fa}\,,
 \end{align}
 and~\eqref{coeffs1}
\begin{equation}\label{eq:coeffsss}
    c_{\ula'_L(j),\ula'_R(i)} = \frac{1}{\gamma+ C_{\ula'_L(j)}+C_{\ula'_R(i)}}.
\end{equation}
The indices $i,j$ on the Young projectors $\Pi$ simply specify the Young tableau with given shape $\ula_{L/R},\ula'_{L/R}$ and are uniquely specified by the conditions on removing the last box. In practice, the last box denotes the index being contracted/added. 
See appendix \ref{app:yt} for details on Young diagrams, tableaux and projectors.
Note that there is no explicit  $m,n$ dependence in the weight-shifting operators and its presence is only in the size of $\fa$ and $\cA$ indices and in the allowed non vanishing shape of the Young diagrams $\ula_L, \ula_R$ and $\ula_L',\ula_R'$. One can thus use \eqref{eq:collectedWSO} to compute quantities in theories with $\SL(2m|2n)$ symmetry in a unified manner, and fix $m,n$ at the end.

General weight-shifting operators $\cD_{\finite}$ transforming in a representation $\finite \in \frep^{\otimes p} \otimes \overline{\frep}^{\otimes q}$ can be constructed by successive application of the (anti)fundamental weight-shifting operators \eqref{eq:collectedWSO}.\footnote{This is automatic for the Grassmannian operators we considered here and follows more generally from the algebra of weight-shifting operators \cite{Karateev:2017jgd}, reviewed in section \ref{sec:WSOinvariants}.}
In this section, we will show how such differential operators shift the weights of 4D $\cN=0,2,4$ multiplets as well as their interaction with shortening conditions.
For this, we ignore the open $\SL(2m|2n)$ indices, which should be understood as being contracted by different weight-shifting operators acting on other points of an $N$-point structure. 

The diagram below depicts how the (first few) left Young diagrams are changed by successive application of $\cD^{\cA}$~\eqref{eq:collectedWSO} 
\Yboxdim{8pt}
\begin{equation}
\Yvcentermath1
    \begin{tikzpicture}[baseline={(current bounding box.center)},scale=0.85]
        \tikzstyle{node} = [draw, rectangle, minimum width=10cm, align=center]
        \node (a1) at (0.5,0) {$[\bullet, \bullet]$};
        \node (a2) at (2.5,0) {$[\,\yng(1)\,, \bullet]$};
        \node (a21) at (5,1) {$[\,\yng(1,1)\,, \bullet]$};
        \node (a22) at (5,-1) {$[\,\yng(2)\,, \bullet]$};
        \node (a211) at (7.5,2) {$[\,\yng(1,1,1)\,, \bullet]$};
        \node (a212) at (7.5,0) {$[\,\yng(2,1)\,, \bullet]$};
        \node (a222) at (7.5,-2) {$[\,\yng(3)\,, \bullet]$};
         \node (a2111) at (10,3) {$\cdots$};
         \node (a2112) at (10,1) {$\cdots$};
         \node (a2121) at (10,0) {$\cdots$};
         \node (a2221) at (10,-1) {$\cdots$};
         \node (a2222) at (10,-3) {$\cdots$};
        \draw[->] (a1)--(a2);
        \draw[->] (a2)--(a21);
        \draw[->] (a2)--(a22);
        \draw[->] (a21)--(a211);
        \draw[->] (a21)--(a212);
        \draw[->] (a22)--(a212);
        \draw[->] (a22)--(a222);
        \draw[->] (a211)--(a2111);
        \draw[->] (a211)--(a2112);
        \draw[->] (a212)--(a2112);
        \draw[->] (a212)--(a2121);
        \draw[->] (a212)--(a2221);
        \draw[->] (a222)--(a2221);
        \draw[->] (a222)--(a2222);
    \end{tikzpicture}\,,\label{Daction}
\end{equation}
where we can also reverse direction, going  from right to left, by applying  $\cD_A$.
At each node, one can also apply $\bar \cD_{\cA}$ \eqref{eq:collectedWSO} to construct representations with non-trivial right-handed $\SL(m|n)_R$ reps. 
 Using the translation tables of appendix \ref{sec:repexamples}, these diagrams show how our weight-shifting operators \eqref{eq:collectedWSO} relate 4D $\cN=2,4$ superconformal multiplets.
\Yboxdim{13pt}

\paragraph{4D conformal multiplets:} Let us first quickly consider how our Grassmannian differential operators shift the well-known 4D conformal representations labelled by $[\ell, \bar{\ell}]_{\Delta}$. $\GL(2)$ Young diagrams admit at most 2 rows and thus one can add an index in two different places. Consider $\cD^{\cA}$ \eqref{eq:collectedWSO} without loss of generality. Then, we have that $\ula_L$ can be shifted as follows
\begin{equation}\label{eq:4Drepshift}
\begin{tikzpicture}[x=14pt, y=14pt, baseline={(current bounding box.center)},scale=0.9]
    \draw [thick] (0,0) -- (10,0) -- (10,-1) -- (0,-1) -- (0,0);
    \draw [thick] (0,-1) -- (7,-1) -- (7,-2) -- (0,-2) -- (0,-1);
    \draw [ultra thick, dashed] (10,0)to (11,0) to (11,-1) to (10,-1) to (10,0);
     \draw [ultra thick, dashed] (7,-1)to (8,-1) to (8,-2) to (7,-2) to (7,-1);
     \node at (11.2,-0.2) [right] {$[\ell, \bar{\ell}]_{\Delta} \mapsto [\ell+1, \bar{\ell}]_{\Delta+1/2}$};
    \node at (8.2,-2) [right] {$[\ell, \bar{\ell}]_{\Delta} \mapsto [\ell-1, \bar{\ell}]_{\Delta+1/2}$};
\end{tikzpicture}\  ,
\end{equation}
where the dashed boxes correspond to the two possible locations for the new index added by $\cD^{\cA}$.
So in the notation of \cite{Karateev:2017jgd}, the above operations correspond to $\cD^{a}_{++0}$ and $\cD^a_{+-0}$, respectively. Similarly, $X^{\fa}_{\ \cA}$ is analogous to $\bar{\cD}_{a}^{-+0}$ and $\bar{\cD}_a^{--0}$.\footnote{Note that contracting and antisymmetrising a Weyl spinor index in 4D are equivalent operations.}  
An advantage of our formalism is that the same operator has two different actions, which only depend on Young symmetriser $\Pi^{\ula_L}$. Moreover, the differential operators $\cD^{\cA}$ and $\bar{\cD}_{\cA}$ often simplify substantially to simply $\partial_X$ and $\partial_{\bX}$ together with a subsequent symmetriser.\footnote{Indeed, since $X$ and $\bX$ are not independent, many $N$-point structures can be written independently of one variable, simplifying the differential weight-shifting operators.}

\paragraph{4D $\cN=2$ multiplets:}  4D $\cN=2$ superconformal representations are written in \cite{Cordova:2016emh} as 
\begin{equation}\label{eq:multiplet2}
    S \bar{S}[\ell, \bar{\ell}]^{(R;r)}_{\Delta} \,,
\end{equation}
where $S,\bar{S}$ are the shortening conditions for $Q$ and $\bar{Q}$, respectively. Then, translating~\eqref{Daction} to this formalism, the action of $\cD^{\cA}$ on (the first few) $\cN=2$ multiplets is as follows
\begin{equation}
\Yvcentermath1
    \begin{tikzpicture}[baseline={(current bounding box.center)},scale=0.85]
        \node (a1) at (0,0) {\small $ B \bar{B} [0,0]^{\scriptscriptstyle(\gamma;0)}_{\scriptscriptstyle\gamma}$};
        \node (a2) at (3,0) {\small $A_2 \bar{B} [0,0]^{\scriptscriptstyle(\gamma-1;2)}_{\scriptscriptstyle\gamma}$};
        \node (a21) at (6,1.25) {\small $L \bar{B}[0,0]^{\scriptscriptstyle(\gamma-2;3)}_{\scriptscriptstyle\gamma}$};
        \node (a22) at (6,-1.25) {\small $A_1 \bar{B}[1,0]^{\scriptscriptstyle(\gamma-1;3)}_{\scriptscriptstyle\gamma+\frac12}$};
        \node (a211) at (9,2.5) {\small $L \bar{B} [0,0]^{\scriptscriptstyle(\gamma-3;4)}_{\scriptscriptstyle\gamma}$};
        \node (a212) at (9,0) {\small $ L \bar{B}[1,0]^{\scriptscriptstyle(\gamma-2;4)}_{\scriptscriptstyle\gamma+\frac12}$};
        \node (a222) at (9,-2.5) {\small $A_1 \bar{B}[2,0]^{\scriptscriptstyle(\gamma-1;4)}_{\scriptscriptstyle\gamma+1}$};
         \node (a2111) at (11.5,3.75) {\small $\cdots$};
         \node (a2112) at (11.5,1.25) {\small $\cdots$};
         \node (a2121) at (11.5,0) {\small $\cdots$};
         \node (a2221) at (11.5,-1.25) {\small $\cdots$};
         \node (a2222) at (11.5,-3.75) {\small $\cdots$};
        \draw[->] (a1)--(a2);
        \draw[->] (a2)--(a21);
        \draw[->] (a2)--(a22);
        \draw[->] (a21)--(a211);
        \draw[->] (a21)--(a212);
        \draw[->] (a22)--(a212);
        \draw[->] (a22)--(a222);
        \draw[->] (a211)--(a2111);
        \draw[->] (a211)--(a2112);
        \draw[->] (a212)--(a2112);
        \draw[->] (a212)--(a2121);
        \draw[->] (a212)--(a2221);
        \draw[->] (a222)--(a2221);
        \draw[->] (a222)--(a2222);
    \end{tikzpicture}\,.
\end{equation}
See appendix~\ref{sec:repexamples} for the full correspondence between coset space representations and superconformal multiplets.
In particular, the diagram \eqref{eq:4Drepshift} can be completed by adding an arbitrary long column in which any added box contributes to the $\SU (2)$ part of the full $\mathrm{U} (2)$ $R$-symmetry representation. The remaining $\mathbb{C}^*$ charge corresponds to the difference between left and right Young diagrams and thus any shift of one side will shift it.
All other types of multiplets can be accessed by acting with the dual weight-shifting operator $\bar \cD_\cA$.

\paragraph{4D $\cN=4$ multiplets:} Because $\SL(4|4)$ is not semi-simple, there is an additional constraint that multiplets must satisfy, indicating that they are actually representations of $\mathrm{PSL}(4|4)$. In particular, 4D $\cN=4$ multiplets are described by $\SL(m|n)\times \SL(m|n)$ representations with the same number of boxes on each side. This is described in detail in appendix \ref{sec:repexamples}. Thus, to map between representations, it is necessary to apply differential operators in the adjoint of $\mathfrak{sl}(2m|2n)$ given by tensor products of \eqref{eq:collectedWSO}. We focus on   $\bar \cD_\cA \cD^\cB$ which gives physical  reps starting from the scalar rep.
The number of possible different operations grows rapidly so we only show the first two steps, together with their translation to the notation of \cite{Cordova:2016emh}, where  4D $\cN=4$ superconformal multiplets are written as
\begin{equation}\label{eq:N4multiplet}
    S \bar{S}[\ell, \bar{\ell}]^{(R_1,R_2,R_3)}_{\Delta}\, .
\end{equation}
The successive application of the adjoint weight-shifting operator $\bar \cD_\cA \cD^\cB$ on the first few $\cN=4$ multiplets is as follows 
\Yboxdim{8pt}
\begin{equation}\label{adjWSO}
\Yvcentermath1
    \begin{tikzpicture}[baseline=(a1),scale=0.85]
        \node (a1) at (0.5,0) {$[\bullet, \bullet]$};
        \node (a2) at (2.5,0) {$[\,\yng(1)\,,\,\yng(1)\,]$};
        \node (a21) at (5,1.5) {$[\,\yng(1,1)\,,\,\yng(1,1)\,]$};
        \node (a22) at (5.5,0.5) {$[\,\yng(2)\,,\,\yng(1,1)\,]$};
        \node (a23) at (5.5,-0.5) {$[\,\yng(1,1)\,,\,\yng(2)\,]$};
        \node (a24) at (5,-1.5) {$[\,\yng(2)\,, \,\yng(2)\,]$};
        \node (a211) at (7.5,3) {$[\,\yng(1,1,1)\,,\,\yng(1,1,1)\,]$};
        \node (a211d) at (9.2,3) {$\cdots$};
        \node (a213) at (7.5,1.7) {$\cdots$};
        \node (a223) at (8,1) {$\cdots$};
        \node (a243) at (7.5,-1.7) {$\cdots$};
        \node (a233) at (8,-1) {$\cdots$};
        \node (a241) at (7.5,-3) {$[\,\yng(3)\,,\,\yng(3)\,]$};
        \node (a241d) at (9.2,-3) {$\cdots$};
        \node (a221) at (8.5,0) {$[\,\yng(2,1)\,,\,\yng(2,1)\,]$};
        \node (a221d) at (10,0) {$\cdots$};
        \draw[->] (a1)--(a2);
        \draw[->] (a2)--(a21);
        \draw[->] (a2)--(a22);
        \draw[->] (a2)--(a23);
        \draw[->] (a2)--(a24);
        \draw[->] (a21)--(a211);
        \draw[dashed, ->] (a21)--(a213);
        \draw[dashed, ->] (a24)--(a243);
        \draw[dashed, ->] (a22)--(a223);
        \draw[dashed, ->] (a23)--(a233);
        \draw[->] (a24)--(a241);
        \draw[->] (a23)--(a221);
        \draw[->] (a22)--(a221);
    \end{tikzpicture}\,,
\end{equation}
where the dashed arrows indicate continuation. In general, for two Young diagrams $\ula_L$ and $\ula_R$ with $N$ and $M$ distinct rectangles, respectively, there will be $(N+1)\times(M+1)$ possible operations, that is, arrows coming out of their corresponding node.
Using table \ref{tab:4DN4reps1}, the above diagram translates to
\begin{equation}\label{eq:WSOmaps2}
\Yvcentermath1
    \begin{tikzpicture}[baseline=(a1),scale=0.9]
        \node (a1) at (-0.5,0) {\small$B \bar{B} [0,0]^{\scriptscriptstyle[0,\gamma,0]}_{\scriptscriptstyle\gamma}$};
        \node (a2) at (2.5,0) {\small$B \bar{B} [0,0]^{\scriptscriptstyle[1,\gamma-2,1]}_{\scriptscriptstyle\gamma}$};
        \node (a21) at (4.5,2) {\small$B \bar{B} [0,0]^{\scriptscriptstyle[2,\gamma-4,2]}_{\scriptscriptstyle\gamma}$};
        \node (a22) at (6,0.7) {\small$A_2 \bar{B} [0,0]^{\scriptscriptstyle[0,\gamma-3,1]}_{\scriptscriptstyle\gamma}$};
        \node (a23) at (6,-0.7) {\small$B \bar{A}_2 [0,0]^{\scriptscriptstyle[1,\gamma-3,0]}_{\scriptscriptstyle\gamma}$};
        \node (a24) at (4.5,-2) {\small$A_2 \bar{A}_2 [0,0]^{\scriptscriptstyle[0,\gamma-2,0]}_{\scriptscriptstyle\gamma}$};
        \node (a211) at (7.5,3) {\small$B \bar{B} [0,0]^{\scriptscriptstyle[3,\gamma-6,3]}_{\scriptscriptstyle\gamma}$};
        \node (a211d) at (9.4,3) {$\cdots$};
        \node (a213) at (7,1.7) {$\cdots$};
        \node (a223) at (8.5,1.1) {$\cdots$};
        \node (a243) at (7,-1.7) {$\cdots$};
        \node (a233) at (8.5,-1.1) {$\cdots$};
        \node (a241) at (7.5,-3) {\small$A_1 \bar{A}_1 [1,1]^{\scriptscriptstyle[0,\gamma-2,0]}_{\scriptscriptstyle\gamma+1}$};
        \node (a241d) at (9.4,-3) {$\cdots$};
        \node (a221) at (9.7,0) {\small$A_2 \bar{A}_2[0,0]^{\scriptscriptstyle[1,\gamma-4,1]}_{\scriptscriptstyle \gamma}$};
        \node (a221d) at (11.7,0) {$\cdots$};
        \draw[->] (a1)--(a2);
        \draw[->] (3,0.5)--(a21);
        \draw[->] (a2)--(a22);
        \draw[->] (a2)--(a23);
        \draw[->] (3,-0.5)--(a24);
        \draw[->] (a21)--(a211);
        \draw[dashed, ->] (a21)--(a213);
        \draw[dashed, ->] (a24)--(a243);
        \draw[dashed, ->] (a22)--(a223);
        \draw[dashed, ->] (a23)--(a233);
        \draw[->] (a24)--(a241);
        \draw[->] (a23)--(a221);
        \draw[->] (a22)--(a221);
    \end{tikzpicture}.
\end{equation}
\Yboxdim{5pt}
Observe that the upper branch corresponds to quarter-BPS multiplets. In particular,  $B\bar{B}[0,0]^{[2,0,2]}_{\Delta=4}$, the first non-trivial quarter-BPS field  in interacting $\SU (N)$ super Yang-Mills can be derived from the higher-KK mode $\cO^{4}$ (charge 4 half-BPS operator) as
\begin{equation}
    \Pi_{\yng(1,1)}(\bar{\cD}_{\cA_1} \bar{\cD}_{\cA_2})_{\dfa \dfb} \Pi_{\yng(1,1)}(\cD^{\cB_1} \cD^{\cB_2})_{\fa \fb} \cO^{4}.
\end{equation}
Note that as currently written, this is a superconformally covariant object. Indeed, in order to obtain the full physically relevant quantity, we need to  contract the  $\cA$ indices thus making the above expression invariant under superconformal transformations.
In section~\ref{sec:WSOinvariants}, we will consider such invariant compositions of weight-shifting operators.

\Yboxdim{13pt}
\section{Weight-shifting invariants}\label{sec:WSOinvariants}
Weight-shifting operators are $\SL(2m|2n)$-covariant, so they are not able to meaningfully act on invariant $N$-point functions and blocks.
The goal of this section is thus to construct $\SL(2m|2n)$-invariant compositions of the weight-shifting operators~\eqref{eq:collectedWSO} which map invariants to invariants while changing the representations (e.g. at the insertions of an $N$-point function). 
We recall that $\SL(2m|2n)$ acts linearly on the $\cA$ indices, so the construction of weight-shifting invariants is largely analogous to that tensor structures shown in section \ref{sec:Wardids}.
In particular, the naive guess of constructing one such invariant with a fundamental and antifundamental differential operator acting at the same point vanishes due to Schur's lemma.
The solution to this is to have each covariant operator act on a different insertion in an $N$-point function.
Then, the key result is that, when acting on three-point structures, invariant combinations of weight-shifting operators of the form
\begin{equation}
    \cD^{\cA}(X_i) \cD_{\cA} (X_j), \qquad i,j=1,2,3
\end{equation}
form a basis of $\SL(2m|2n)$ three-point tensor structures, generalising the conformal differential basis of \cite{Costa:2011dw,CastedoEcheverri:2015mkz,Karateev:2017jgd}. We provide several examples of this, including the construction of differential bases for correlators of short (super) multiplets --- a new feature unique to analytic superspace.
In section \ref{sec:blocks}, we will directly apply the differential basis of weight-shifting invariants to (super) conformal blocks by writing them as a gluing of three-point functions by means of a shadow-type integral \cite{Simmons-Duffin:2012juh} in $(m,n)$ analytic superspace.

\subsection{Diagrams and weight-shifting algebra}
We will be using a very useful diagrammatic notation for general  weight-shifting operators, 
$\cD_{\rep}$, introduced in \cite{Karateev:2017jgd}, which we will make extensive use of and which reflects the close relationship between the weight-shifting operators and the tensor product of representations~\eqref{tensor}, \eqref{eq:weightshiftingdef} (and in turn, three-point functions).
Edges with arrows denote representations,  a standard edge for a generic operator,  a dashed edge for a scalar (half-BPS) operator  and a curly edge for a finite dimensional representation
\begin{equation}
    \begin{tikzpicture}[baseline,scale=0.9]
        \draw[spinning] (0,0)--(2.25,0) node[midway, above] {$\cO^{\gamma}_{\ula_L, \ula_R}$};
        \draw[scalar] (3.5,0)--(5.75,0) node[midway, above] {$\cO^{\gamma}_{\bullet, \bullet}$};
        \draw[finite with arrow] (7,0)--(9.25,0);
        \node at (8.125, 0.1) [above] {$\Psi$};
        \node at (9.75,0) {$.$};
    \end{tikzpicture}
\end{equation}
Invariant products of representations i.e.\ $N$-point structures constructed in section \ref{sec:Wardids}, are denoted by a node with $N$ representations coming out of it. In particular, two and three point structures are given by
\begin{equation}
    \begin{split}
        \langle \cO^{\gamma}_{\ula_L, \ula_R} \cO^{\gamma}_{\ula_R, \ula_L}\rangle \ &=\ \begin{tikzpicture}[baseline=(baseline),scale=0.8]
    \node (a) at (0,0) [twopt] {};
        \draw[spinning] (a)--(1.8,0) node[right, right] {$\cO^{\gamma}_{\ula_R, \ula_L}$};
        \draw[spinning] (a)--(-1.8,0) node[left, left] {$\cO^{\gamma}_{\ula_L, \ula_R}$};
        \node (baseline) at (0,-0.1) {};
    \end{tikzpicture}, \\
    \langle \cO_1 \cO_2 \cO_3\rangle^{(i)} \ &= \ \begin{tikzpicture}[baseline={(current bounding box.center)},scale=.9]
\node (verttl) at (-0.7,0.9) [above] {$\cO_1$};
\node (vertbl) at (-0.7,-0.9) [below] {$\cO_2$};
\node (vertr) at (0.6,0) [threept] {$\scriptstyle i$};
\node (r) at (1.8,0) [right] {$\cO_3$};
\draw [spinning] (vertr) -- (r);
\draw [spinning] (vertr) -- (verttl);
\draw [spinning] (vertr) -- (vertbl);
\end{tikzpicture},
    \end{split}
\end{equation}
where the node can be empty if there is only one tensor structure.

Weight-shifting operators $\cD_{\finite}$ map the above structures to $\SL(2m|2n)$-covariant structures. Their action on a specific insertion is represented as

\vspace{-4mm}
\begin{equation}
        \cD_{\finite} : \cO^{\gamma}_{\ula_L \ula_R} \mapsto \cO'^{\gamma}_{\ula_L' \ula_R'} \quad \leftrightarrow \quad
        \diagramEnvelope{\begin{tikzpicture}[baseline=(baseline),scale=0.85]
            \node (dot) at (0,0) [threept] {$\scriptstyle \cD$};
            \node (O1) at (-2,0) [left] {$\cO^\gamma_{\ula_L,\ula_R}$};
            \node (O'1) at (2,0) [right] {$\cO^{\prime\gamma}_{\ula'_L,\ula'_R}$};
            \node (W) at (0.8,1.9) [above] {$\finite$};
            \draw[spinning] (-2,0)--(dot);
            \draw[spinning] (dot)--(2,0);
            \draw[finite with arrow] (dot)--(0.75,1.875);
            \node (baseline) at (0,-.1) {};
        \end{tikzpicture}}.
        \label{diagrams}
    \end{equation}
As explained in the previous section, a general weight-shifting operator $\cD_{\finite}$ can always be constructed from successive applications of the (anti)fundamental weight-shifting operators, $\cD_\frep,\cD_{\bar \frep}$ \eqref{eq:collectedWSO}. In this case, the label in each individual node is not necessary as it is uniquely determined by the initial and final states. For example, for $\cO_{\ula_L,\ula_R} \mapsto \cO_{\ula_L',\ula_R}$, $|\ula_L'|=1+|\ula_L|$, there is only one (anti)fundamental weight-shifting operator \eqref{eq:collectedWSO} that can achieve it: $\cD^{\cA}$.

The above statement can be made more precise by the algebra of weight-shifting operators, which relates consecutive applications of  weight-shifting operators,  $\cD_{\frep_1}\cD_{\frep_2}$, to linear combinations of single weight-shifting operators $\cD_{\finite}$, $\finite \in \frep_1 \otimes\frep_2$, and is represented diagrammatically as
\begin{equation}\label{eq:WSOalgebra}
        \diagramEnvelope{\begin{tikzpicture}[baseline={(current bounding box.center)}]
            \node (dot2) at (-0.375,-0.5) [threept] {\scriptsize $\cD$};
            \node (dot) at (1.5,-0.5) [threept] {\scriptsize $\cD$};
            \node (O1) at (0.6,-0.4) [above] { $\cO''$};
            \node (O''1) at (-1.875,-0.6) [below] {$\cO$};
            \node (O'1) at (3,-0.6) [below] {$\cO'$};
            \node (W1) at (-0.375,0.9) [above] {$\frep_1$};
            \node (W) at (1.5,0.9) [above] {$\frep_2$};
            \draw[spinning] (-1.875,-0.5)--(dot2);
            \draw[spinning] (dot2)--(dot);
            \draw[spinning] (dot)--(3,-0.5);
            \draw[finite with arrow] (dot)--(1.5,0.9);
            \draw[finite with arrow] (dot2)--(-0.375,0.9);
        \end{tikzpicture} 
         \displaystyle=\sum_{\finite\in  \frep_1\otimes \frep_2} \left\{ \begin{array}{ccc}
             \frep_1 & \cO & \finite\\
             \cO'& \frep_2 & \cO'' 
        \end{array}\right\}
        \begin{tikzpicture}[baseline={(current bounding box.center)}]
            \node (dot) at (1.5,-0.5) [threept] {\scriptsize $\cD$};
            \node (node) at (1.5,1) [threept] {};
            \node (O1) at (0,-0.6) [below] {$\cO$};
            \node (O'1) at (3,-0.6) [below] {$\cO'$};
            \node (W) at (1.6,0.25) [right] {$\finite$};
            \node (W1) at (0,1.75) [above] {$\frep_1$};
            \node (W2) at (3,1.75) [above] {$\frep_2$};
            \draw[spinning] (0,-0.5)--(dot);
            \draw[spinning] (dot)--(3,-0.5);
            \draw[finite with arrow] (dot)--(node);
            \draw[finite with arrow] (node)--(W1);
            \draw[finite with arrow] (node)--(W2);
        \end{tikzpicture}},
    \end{equation}
where the small vertex on the RHS is the relevant Clebsch-Gordan coefficient and the object in curly brackets are simply  the coefficients in this relation, known as 6$j$ symbols for the Lie group $G$. These coefficients are best known for $G=\SU (2)$ where they describe the crossing between tensor products of different spins in the way depicted in \eqref{eq:WSOalgebra}. 
Within the $(m,n)$ superspace, we are able to extend these ideas to the $\mathbb{Z}_2$-graded Lie group $\SL(2m|2n)$ and thus obtain superconformal $6j$ symbols.

We now consider the properties of weight-shifting operators that follow from the above algebra, as first shown in the conformal case in \cite{Karateev:2017jgd}, giving explicit realisations using our $\SL(2m|2n)$ weight-shifting operators \eqref{eq:collectedWSO}.
    
\subsection{Bubble}
If we set  $\frep_2=\overline{\frep}_1=\overline\frep$ in~\eqref{eq:WSOalgebra} and make the whole expression invariant by contracting indices, we get a bubble diagram. The sum on the right-hand side of \eqref{eq:WSOalgebra} will now only have one non-zero term, corresponding to $\finite=\mathbf{1}$, the trivial representation. This means that $\cO$ must be equal to $\cO'$.\footnote{Note that this is simply a realisation of Schur's lemma.}

We conclude that the conformally invariant composition of two weight-shifting operators
either vanishes or leaves the representation unchanged. This is known as the bubble and 
is summarised by the following equation
    \begin{align}\label{eq:bubble}
        \begin{tikzpicture}[baseline=(baseline),scale=0.85]
            \node (dot2) at (-1.5,0) [threept] {};
            \node (dot) at (0.75,0) [threept] {};
            \node (O1) at (-0.375,-0.1) [below] {$\cO''$};
            \node (O''1) at (-3.6,-0.1) [below] {$\cO$};
            \node (O'1) at (2.85,-0.1) [below] {$\cO'$};
            \node (W1) at (0.75,0.8) [above] {$\frep$};
            \draw[spinning] (-3.5,0)--(dot2);
            \draw[spinning] (dot2)--(dot);
            \draw[spinning] (dot)--(2.75,0);
            \draw[finite with arrow] (dot2) to[out=90, in=90](dot);
            \node (baseline) at (0,-0.1) {};
        \end{tikzpicture}
        = \dim {\frep} \left\{ \begin{array}{ccc}
             {\frep} & \cO & \mathbf{1}\\
             \cO'& \overline{\frep} & \cO'' 
        \end{array}\right\}\delta_{\cO, \cO'} \cO\ .
\end{align}
Upon first look, this property implies that we cannot perform non-trivial changes to representations. However, in practice, weight-shifting operators are used on correlation functions, i.e.\ there are more points to apply differential operators at, and thus more ways of constructing non-trivial conformally invariant expressions using weight-shifting techniques, which we will show in section \ref{sec:diffbasis}.

Let us show that the fundamental and anti-fundamental superconformal weight-shifting operators~\eqref{eq:collectedWSO} satisfy the bubble property~\eqref{eq:bubble}.
There are four different combinations to consider: $\cD_{\cA}\bar \cD^{\cA},\,\cD_{\cA} \cD^{\cA},\,\bar \cD_{\cA}\bar \cD^{\cA}$ and $
 \bar \cD_{\cA}\cD^{\cA}$. 
 
When neither weight-shifting operator is differential, the property $\cD_{\cA}\bar \cD^{\cA}=0$ is trivially satisfied on both the coset and Grassmannian following directly from $X^{\fa}_{\ \cA} \bX^{\cA \dfa} = 0$.
In this case the RHS vanishes, consistent with~\eqref{eq:bubble} since the final representation is always different from the initial one (we apply a right-hand shift followed by a left-hand shift).

The second and third cases are more subtle: if one operator adds an index and the other contracts, there are cases in which the final Young diagram is equal to the initial one and so the combination does not vanish.  
We only consider $\cD_{\cA} \cD^{\cA}$ as the conjugate case is very similar.
Now note that, unlike their multiplicative counterparts, differential weight-shifting operators have different forms on the coset~\eqref{wsgen} and Grassmannian~\eqref{eq:collectedWSO} and so we consider each case separately. The coset differential operators obey the bubble property more non-trivially: 
\begin{align}
    (\cD_{\cA})^{\ula'_L}_{\ula''_L} (\cD^{\cA})^{\ula_L}_{\ula'_L}  &= c_0 X^{\fb'}_{\ \cA} \Pi^{\ula_L'}_{i,1}\Pi^{\ula'_L}_{1,j}\bigg( W {}^{\cA}_{\ \fb} +\bX^{\cA \dfb}\sum_{\mu_R \in \Box \otimes \ula_R} c_{\ula_L'(j)\umu_R} \bPi^{\umu_R}_kp_{\dfb \fb}\bigg)  \nn \\
        & = c_0 X^{\fb}_{\ \cA} \Pi^{\ula_L'}_{i,j} W {}^{\cA}_{\ \fb}  = c_0 \delta_{\ula_L,\ula_L''}\cO_{\ula_L, \ula_R}^{\gamma}\, .\label{bubble2}
\end{align}
The $i,j,k$ are defined by $\overline{\ula'_L(j)}=\ula_L(1),\ \overline{\ula'_L(i)}=\ula''_L(1),\ \overline{\umu_R(k)}=\ula_R(1)$ where the barred tableau is the tableau with the final entry (and its box) removed. Thus 
$\delta_{i,j}= \delta_{\ula_L,\ula_L''}$.  In the second equality we used $X\bar X=0$ and the final equality uses $XW=I$~\eqref{eq:ginvg} as well as beautiful properties of Hermitian Young projectors. It is shown explicitly in appendix~\ref{app:yt}.

The above composition of weight-shifting operators using the Grassmannian form $\cD^\cA$~\eqref{eq:collectedWSO} is slightly trickier. The computation is similar to~\eqref{bubble2}, with $W \rightarrow \partial_X$, and thus the second line becomes
\begin{align}
(\cD_{\cA})^{\ula'_L}_{\ula''_L} (\cD^{\cA})^{\ula_L}_{\ula'_L} \cO_{\ula_L, \ula_R}^{\gamma} =\left(\Pi^{\ula_L'}_{i,j}\right)^{\underline \fa \fb}_{\underline \fa'' \fb'}(X\partial_X)^{\fb'}_{\ \fb}  \cO_{\underline \fa, \ula_R}^{\gamma} =  B^{\gamma_L}_{\ula'_L\!(i)} \delta_{\ula_L,\ula_L'}\cO_{\ula_L, \ula_R}^{\gamma}  \ .  
\end{align}
The second equality follows from the fact that $\Pi^{\ula_L'}_{i,j}=\Pi^{\ula_L'}_{i,j}\Pi^{\ula_L'}_{j}$(which follows from~\eqref{units}) after which we can use~\eqref{eq:mLconstant} to yield $B^{\gamma_L}_{\ula'_L\!(i)}$ multiplied by~\eqref{bubble2}.

Thus, the coefficient $B^{\gamma_L}_{\ula'_L\!(i)}$ defined by equation~\eqref{eq:mLconstant} is precisely the $6j$ symbol~\eqref{eq:bubble}: 
\begin{equation}\label{bubblecoeff}
      \dim \frep\left\{ \begin{array}{ccc}
             \frep & \cO^{\gamma}_{\ula_L, \ula_R} & \bullet\\
             \cO^{\gamma}_{\ula_L, \ula_R}& \overline{\frep} & \cO_{\ula'_L, \ula_R}^{\gamma}
        \end{array}\right\} \equiv B_{\ula'_L(i)}^{\gamma_L} = - \gamma_L - C_{\ula'_L(i)}.
\end{equation}
 The process is similar for the combination $\bX^{\cA \dfb} \bD_{\dfa \cA}$.
 Note that by comparing~\eqref{bubble2} and~\eqref{bubblecoeff} we can now fix the normalisation $c_0$ relating the coset and Grassmannian versions of the weight-shifting operators
\begin{equation}\label{norm}
    D_{\fa}^\cA: \quad c_0= - \gamma_L - C_{\ula'_L(i)}\qquad \qquad \bD_{\dfa \cA}: \quad c_0= - \gamma_R - C_{\ula'_R(i)} \,.
    \end{equation}
Note that this is  in complete agreement with the normalisation derived directly in~\eqref{wschiral}.

Finally, the case where both weight-shifting operators are differential is highly non-trivial.
Because each differential operator always adds a box to the corresponding Young diagram $\ula_L$, $\ula_R$ (or removes it in the case of non-unitary irreps \eqref{eq:wsoansatzfinite}), the combination should always vanish by the bubble property. 
We include it in its full generality in appendix \ref{app:hardbubble} as it is an instructive application of all the properties of Young projectors, the relations between coset space coordinates, and the coefficients $c_{\ula_L,\ula_R}$ of weight-shifting operators \eqref{eq:collectedWSO}. 
Here, we will consider an example. 

Suppose we act on $\cO_{\bullet,\bullet}^{\gamma}$ by adding a box to both left and right $\GL(m|n)$ representations. Without loss of generality, we add an undotted index first by acting with $\cD^{\cA}$ and then a dotted one with $\bar{\cD}_{\cA}$. Then, we claim that the following vanishes
\Yboxdim{5pt}
\begin{equation}
    (\bar{\cD}_{\cA})^{\bullet}_{\yng(1)}(\cD^{\cA})^{\bullet}_{\yng(1)}= c_0^2 \bigg( \bW_{\dfa \cA} +X^{\fb}_{\ \cA}\!\!\sum_{\Yvcentermath1 \umu_L \in \left\{ \yng(2),\, \yng(1,1) \right\}} c_{\umu_L,\yng(1)} \Pi^{\umu_L}_jp_{\dfa \fb}\bigg) \bigg( W {}^{\cA}_{\ \fa} -\bX^{\cA \dfb} c_{\,\yng(1)\,,\yng(1)} \,p_{\dfb \fa}\bigg).
\end{equation}
\Yboxdim{13pt}
where $\cD^{\cA}$ acts on a scalar hence its simplicity, whereas $\bar{\cD}_{\cA}$ acts on $\cD^{\cA} \cO_{\bullet} \sim \cO_{\Box, \bullet}$ so has a more complicated action.
We can use the commutators $[p,\bX]=-W$, $[p,W]=0$ and the identities between coset coordinates \eqref{eq:ginvg} to simplify the above to
    \Yboxdim{5pt}
\begin{equation}\label{eq:hardbubble2}
    \begin{split}
        \Yvcentermath1
        (\bar{\cD}_{\cA})^{\bullet}_{\Box} (\cD^{\cA})^{\bullet}_{\Box} \propto -  c_{\,\yng(1), \yng(1)}  p_{\dfa \fa} 
         +  \sum_{\umu_L} c_{\umu_L, \yng(1)} \delta^{\fb}_{\fc} \Pi^{\umu_L}_{j} p_{\dfa \fb}\delta^{\fc}_{\fa} + c_{\,\yng(1), \yng(1)} \sum_{\umu_L } c_{\umu_L, \yng(1)} \delta^{\fb}_{\fc}\Pi^{\umu_L}_{j} \delta^{\fc}_{\fa} p_{\dfa \fb},
    \end{split}
\end{equation}
where one must symmetrise the lower $\fa, \fb$ indices according to $\umu_L$ before contracting the $\fb$ index. Then, one can simplify further with the relation one can easily check is satisfied by the coefficients $c$
\begin{equation}\label{eq:ccdecomp}
    \Yvcentermath1
    c_{\,\yng(1), \yng(1)} c_{\umu_L, \yng(1)} = \pm (c_{\,\yng(1), \yng(1)} - c_{\umu_L, \yng(1)}),
\end{equation}
where the sign depends on whether $\umu_L$ is symmetric ($+$) or antisymmetric ($-$). Substituting this into \eqref{eq:hardbubble2} and recalling that $\fb$ is in position $j$ (i.e.\ last box), the equation vanishes as expected from the bubble property. See appendix \ref{app:hardbubble} for the full derivation in the general case.

In the Grassmannian form \eqref{eq:collectedWSO}, the bubble property calculation follows similar steps with $W \mapsto \partial_X$, $p\mapsto \partial_{\bX} \partial_X$.

\Yboxdim{13pt}

\subsection{Crossing and superconformal $6j$ symbols}\label{sec:6j}
By swapping one of the finite-dimensional representations $\frep_i$ in \eqref{eq:WSOalgebra} by an infinite-dimensional one $\cO$, the corresponding node becomes a three-point structure and thus the algebra of weight-shifting operators becomes an equation relating the action of weight-shifting operators at two different insertions. Diagrammatically, this looks like a \emph{crossing} equation for three-point functions given by
\begin{equation}\label{eq:crossing}
    \diagramEnvelope{\begin{tikzpicture}[baseline={(current bounding box.center)},scale=0.85]
	\node (vertL) at (0,0) [threept] {};
	\node (vertR) at (2,0) [threept] {\scriptsize$j $};
	\node (opW) at (-0.75,1.25) [left] {$\frep$};
	\node (opO1) at (-0.75,-1.25) [left] {$\cO'_1$};
	\node (opO2) at (2.75,1.25) [right] {$\cO_2$};
	\node (opO3) at (2.75,-1.25) [right] {$\cO_3$};	
	\node at (1,0.1) [above] {$\cO_1$};	
	\draw [finite with arrow] (vertL)-- (opW);
	\draw [spinning] (vertL)-- (opO1);
	\draw [spinning] (vertR)-- (vertL);
	\draw [spinning] (vertR)-- (opO2);
	\draw [spinning] (vertR)-- (opO3);
\end{tikzpicture}}
	\ =\
	\sum_{i,\cO'_2} \left\{ \begin{array}{ccc}
             \cO_2 & \cO_3 & \cO'_2\\
             \cO'_1& \frep & \cO_1 
        \end{array}\right\}_{i}^{j}
\diagramEnvelope{\begin{tikzpicture}[baseline={(current bounding box.center)},scale=0.85]
	\node (vertU) at (0,0.8) [threept] {};
	\node (vertD) at (0,-0.8) [threept] {\small$i$};
	\node (opO1) at (-1.25,-1.75) [left] {$\cO'_1$};
	\node (opW) at (-1.25,1.75) [left] {$\frep$};
	\node (opO2) at (1.25,1.75) [right] {$\cO_2$};
	\node (op) at (1.25,-1.75) [right] {$\cO_3$};	
	\node at (0.1,0) [right] {$\cO'_2$};	
	\draw [spinning] (vertD)-- (opO1);
	\draw [finite with arrow] (vertU)-- (opW);
	\draw [spinning] (vertD)-- (vertU);
	\draw [spinning] (vertU)-- (opO2);
	\draw [spinning] (vertD)-- (op);
\end{tikzpicture}},
\end{equation}
where the coefficients are again $6j$ symbols as in \eqref{eq:WSOalgebra}, but with the finite representations replaced by infinite-dimensional ones and thus the sum is now over $\cO'_2 \in \cO_{2} \otimes \frep$. We considered the fundamental representation $\frep$, which works similarly for the antifundamental $\overline{\frep}$. The crossing equation for general finite representation $\finite$ follows from successive application of the above equation and/or its antifundamental counterpart as we showed in section \ref{sec:genWSO}, which itself follows from \eqref{eq:WSOalgebra}.

The crossing diagram above corresponds to
\begin{equation}\label{eq:crossing2}
    \cD^{\cA}(X_1)\langle \cO_1 \cO_2 \cO_3\rangle^{(j)} = \sum_{i,\cO'_2} \left\{ \begin{array}{ccc}
             \cO_2 & \cO'_1 & \cO'_2\\
             \cO_3& \frep & \cO_1 
        \end{array}\right\}_{i}^{j}
        \cD^{\cA}(X_2) \langle \cO'_1 \cO'_2 \cO_3 \rangle^{(i)}.
\end{equation}
This equation provides a new and simple way of computing $\SL(2m|2n)$ $6j$ symbols using the weight-shifting operators \eqref{eq:collectedWSO}. 

Let us consider a simple example given by 
adding an $\dfa$ index at position 1 to a scalar three-point function
(we use dashed lines to indicate superspace scalars):
\begin{equation}\label{eq:scalar3pt}
    \langle \cO^{\gamma_1}(X_1) \cO^{\gamma_2}(X_2) \cO^{\gamma_3}(X_3)\rangle = g_{21}^{\gamma_{12,3}}g_{23}^{\gamma_{23,1}}g_{31}^{\gamma_{31,2}} = \begin{tikzpicture}[baseline={(current bounding box.center)},scale=.8]
            \node (dot2) at (0,0) [threept] {};
            \node (pt1) at (1.275,0.75) [above] {$\cO^{\gamma_3}$};
            \node (pt2) at (-1.275,0.75)[above] {$\cO^{\gamma_2}$};
            \node (pt3) at (0,-1.5){$\cO^{\gamma_1}$};
            \draw[scalar] (dot2) to (pt1);
            \draw[scalar] (dot2) to (pt2);
            \draw[scalar] (dot2) to (pt3);
        \end{tikzpicture},
\end{equation}
where $\gamma_{ij,k}=(\gamma_i+\gamma_j-\gamma_k)/2$.
We can add a box via the weight-shifting operator $\cD^{\cA}$~\eqref{eq:collectedWSO} which reduces simply to  $\partial_X$ if  we assume no $\bar X$ dependence. Then using~\eqref{a5} we get
\begin{equation}\label{eq:1}
    \begin{tikzpicture}[baseline={(current bounding box.center)},scale=0.8]
            \node (dot2) at (0,0) [threept] {};
            \node (pt1) at (1.275,0.75) [above] {$ \cO^{\gamma_3}$};
            \node (pt2) at (-1.275,0.75)[above] {$\cO^{\gamma_2}$};
            \node (pt3) at (0,-1.3) [threept]{};
            \node (O1) at (0.2,-0.6) [right]{$\cO^{\gamma_1}$};
            \node (O1p) at (1.275,-2) [below]{$\cO^{\gamma_1}_{\fa_1}$};
            \node (W) at (-1.275,-2) [below]{$\frep$};
            \draw[scalar] (dot2) to (pt1);
            \draw[scalar] (dot2) to (pt2);
            \draw[scalar] (dot2) to (pt3);
            \draw[spinning] (pt3) to (O1p);
            \draw[finite with arrow] (pt3) to (W);
        \end{tikzpicture} = g_{12}^{\gamma_{12,3}}g_{23}^{\gamma_{23,1}}g_{31}^{\gamma_{31,2}} \left( -\gamma_{12,3} (\bX_2 \hX_{21})^{\cA}_{\ \fa_1} -\gamma_{13,2} (\bX_3 \hX_{31})^{\cA}_{\ \fa_1}\right).
\end{equation}
Now we consider all the terms which can appear on the RHS of the crossing equation~\eqref{eq:crossing}. There are in fact only two contributing terms:
\begin{equation}\label{eq:2}
   \begin{tikzpicture}[baseline={(current bounding box.center)},scale=.8]
            \node (dot2) at (0,0) [threept] {};
            \node (pt1) at (0.75,1.275) [above] {$\cO^{\gamma_3}$};
            \node (O2) at (-0.73,0.15)[above] {$\cO^{\gamma_2}_{\dfa}$};
            \node (pt2) at (-1.5,0)[threept] {};
            \node (pt3) at (0.75,-1.275) [below] {$\cO^{\gamma_1}_{\fa_1}$};
            \node (O2p) at (-2.25,1.275) [above] {$\cO^{\gamma_2}$};
            \node (W) at (-2.25,-1.275) [below] {$\frep$};
            \draw[scalar] (dot2) to (pt1);
            \draw[spinning] (dot2) to (pt2);
            \draw[spinning] (dot2) to (pt3);
            \draw[scalar] (pt2) to (O2p);
            \draw[finite with arrow] (pt2) to (W);
        \end{tikzpicture}=
       g_{12}^{\gamma_{12,3}}g_{23}^{\gamma_{23,1}}g_{31}^{\gamma_{31,2}} (\bX_2 \hX_{21})^{\cA}_{\ \fa_1}\,, 
\end{equation}
and
\begin{equation}\label{eq:3}
\begin{split}
   \begin{tikzpicture}[baseline=(baseline),scale=.8]
            \node (dot2) at (0,0) [threept] {};
            \node (pt1) at (0.75,1.275) [above] {$\cO^{\gamma_3}$};
            \node (O2) at (-0.75,0.2)[above] {$\tilde{\cO}^{\fa}_{\gamma_2}$};
            \node (pt2) at (-1.5,0)[threept] {};
            \node (pt3) at (0.75,-1.275) [below] {$\cO^{\gamma_1}_{\fa_1}$};
            \node (O2p) at (-2.25,1.275) [above] {$\cO^{\gamma_2}$};
            \node (W) at (-2.25,-1.275) [below] {$\frep$};
            \draw[scalar] (dot2) to (pt1);
            \draw[spinning] (dot2) to (pt2);
            \draw[spinning] (dot2) to (pt3);
            \draw[scalar] (pt2) to (O2p);
            \draw[finite with arrow] (pt2) to (W);
            \node (baseline) at (0,-.15) {};
        \end{tikzpicture} 
&=g_{12}^{\gamma_{12,3}}g_{23}^{\gamma_{23,1}}g_{31}^{\gamma_{31,2}} \big(\gamma_{12,3}(\bX_2 \hX_{21})^{\cA}_{\ \fa_1}  + (m{-}n{-} \gamma_{2})(\bX_3 \hX_{31})^{\cA}_{\ \fa_1} \big),
        \end{split}
\end{equation}
where to obtain the second line one needs to use the key identity~\eqref{keyid}. 
Note that the second diagram includes a field with upper indices, i.e.\ a non-unitary representation of $\SU (m,m|2n)$ for $n\neq 0$. They are indeed needed as intermediate steps of crossing calculations involving physical fields, as we first discussed in section \ref{sec:WSOnonunitary}.

Now, by comparing \eqref{eq:1} with \eqref{eq:2} and \eqref{eq:3}, one can solve for the two coefficients, i.e.\ the $\SL(2m|2n)$ $6j$ symbols, in the crossing equation~\eqref{eq:crossing}
\begin{equation}\label{eq:6jsymbols1}
    \begin{split}
    \left\{ \begin{array}{ccc}
             \cO^{\gamma_2} &  
              \cO^{\gamma_3}& \cO^{\gamma_2}_{\dfa} \\\relax \cO^{\gamma_1}_{\dfa_1} &  {\frep} &  \cO^{\gamma_1} 
        \end{array}\right\}
        &= \frac{\gamma_{12,3}(2m-2n-\gamma_1 - \gamma_2 - \gamma_3)} { { 2}(\gamma_2 -m+n)}, \\
    \left\{ \begin{array}{ccc}
             \cO^{\gamma_2} & 
              \cO^{\gamma_3} & \tilde{\cO}^{\fa}_{\gamma_2} \\\relax \cO^{\gamma_1}_{\fa_1} &  {\frep} &  \cO^{\gamma_1} 
        \end{array}\right\}&= \frac{\gamma_{13,2}}{ (\gamma_2 -m+n)}. 
     \end{split}
\end{equation}
If we specialise to one-dimensional CFT by setting $m=1$ and $n=0$ and using the conversion in~\eqref{eq:1Dreps}, we recover the results of \cite{Karateev:2017jgd}.\footnote{Note that $\gamma \neq \Delta$ in general, see \eqref{eq:1Dreps} for details.}

There is a different way of adding an $\fa$ index at point $X_1$ in \eqref{eq:scalar3pt}, which yields the other two 1D conformal $6j$ symbols given in \cite{Karateev:2017jgd}, by using the weight-shifting operator $\cD^{\fa}_{\ \cA}=X^{\fa}_{\ \cA}$.
This adds an upper $\fa$ index which, as explained before, does not correspond to a unitary representation of the superconformal group $\SU (m,m|2n)$ for $n\neq 0$. Nevertheless, one can proceed similarly and work out the $6j$ coefficients for these cases too. 

The crossing operation is also one of the key ingredients for calculating seed $\SL(2m|2n)$ blocks, and thus we will come back to it in section \ref{sec:internalshift}.

\subsection{Differential basis for $\SL(2m|2n)$-invariant structures}\label{sec:diffbasis}
Recall that the bubble property \eqref{eq:bubble} states that $\SL(2m|2n)$ invariant combinations of weight-shifting operators acting at the same point are unable to change the representation. For this reason, one needs to act at different points in the $N$-point structure in order to perform non-trivial changes in the representations. That is, we want to construct invariant compositions of the weight-shifting operators \eqref{eq:collectedWSO} given by repeated action of
\begin{equation}\label{Dij}
   \fD_{ij}\coloneq \cD^{\cA}(X_i) \cD_{\cA}(X_j),
\end{equation}
which act at two different insertions $X_i,\, X_j$ of an $N$-point structure. 
The idea then is that  for a given third operator $\cO_3$ we can obtain any non-vanishing three-point function $\langle \cO_1 \cO_2 \cO_3 \rangle$ by acting with appropriate combinations of $\fD_{12}$ on any other non-vanishing three-point function $\langle \cO'_1 \cO'_2 \cO_3 \rangle$.
In particular we should be able to cover the full space  of tensor structures for a given three-point correlator by acting on a possibly unique 3-point function with different combinations of $\fD_{12}$'s.

Suppose we act on a three-point function $\langle\cO_1 \cO_2 \cO_3 \rangle$ with $\cD^{\cA}(X_1):\cO_1 \mapsto \cO_1'$, $\cD_{\cA}(X_2):\cO_2 \mapsto \cO_2'$. One can use crossing \eqref{eq:crossing} to better understand the action of $\fD_{12}$~\cite{Karateev:2017jgd},
\begin{equation}
        \cD^{\cA}(X_1) \cD_{\cA}(X_2) \langle \cO_{1} \cO_2  \cO_3 \rangle =\cD^{\cA}(X_1) \sum_{i, \cO_1''}\left\{ \begin{array}{ccc}
        \cO_1 & \cO_3 & \cO_1''\\
        \cO_2' & \overline{\frep} & \cO_2
    \end{array} \right\}_i \cD_{\cA}(X_1) \langle \cO_{1}''  \cO_2' \cO_3 \rangle^{(i)} ,
\end{equation}
where on the RHS $\cD_{\cA}:\cO_1'' \mapsto \cO_1$. Then, by the bubble property \eqref{eq:bubble}, only the term $\cO_1''=\cO_1'$ survives, leading to 
\begin{equation}
    \cD^{\cA}(X_1) \cD_{\cA}(X_2) \langle \cO_{1} \cO_2  \cO_3 \rangle =\dim {\frep} \left\{ \begin{array}{ccc}
             {\frep} & \cO'_1 & \mathbf{1}\\
             \cO'_1& \overline{\frep} & \cO_1
        \end{array}\right\} \sum_i \left\{ \begin{array}{ccc}
        \cO_1 & \cO_3 & \cO_1'\\
        \cO_2' & \overline{\frep} & \cO_2
    \end{array} \right\}_i \langle \cO_1' \cO_2' \cO_3\rangle^{(i)},
\end{equation}
where the coefficient before the sum is precisely the bubble coefficient $B^{\gamma_L}_{\ula_L(i)}$ \eqref{bubblecoeff}  
(with $\cO_1=\cO^{\gamma_1}_{\ula_L, \ula_R}$, $\cO'_1=\cO^{\gamma_1}_{\ula'_L, \ula_R}$ and $\overline{\ula_L(i)}=\ula'(1)$).
 The above results follow straightforwardly using the diagrammatic notation, in particular~\eqref{eq:bubble} and~\eqref{eq:crossing} (see  see~\eqref{diffbasis}).

In practice one can cover the space of different tensor structures on the RHS (labelled by $i$) by choosing different weight-shifting operators on the LHS (in practice by applying further weight-shifting operators). Indeed it is then useful to invert this and  define  a  basis of the three-point structures via the differential operators $\fD^{[i]}_{12}$. This is  known as the \emph{differential} basis, and we will denote it by square brackets $\langle \dots \rangle^{[i]}$ as follows
\begin{equation}
    \langle \cO_1' \cO_2' \cO_3\rangle^{[j]} \coloneq  \fD^{[j]}_{12}\langle \cO_{1} \cO_2  \cO_3 \rangle.
\end{equation}
We will show this inversion explicitly in terms of our $\SL(2m|2n)$ weight-shifting operators \eqref{eq:collectedWSO} shortly.
In terms of diagrams, we denote the differential basis by a square node
\begin{equation}
\begin{tikzpicture}[baseline=(baseline),scale=.7]
\node (verttl) at (-0.7,0.9) [above] {$\cO_1'$};
\node (vertbl) at (-0.7,-0.9) [below] {$\cO_2'$};
\node (vertr) at (0.6,0) [diff] {$\scriptstyle j$};
\node (r) at (1.6,0) [right] {$\cO_3$};
\draw [spinning] (vertr) -- (r);
\draw [spinning] (vertr) -- (verttl);
\draw [spinning] (vertr) -- (vertbl);
\node (baseline) at (0,-.15) {};
\end{tikzpicture}{ \coloneq} 
\begin{tikzpicture}[baseline=(baseline),scale=.7]
\node (tl) at (-1.6,1.6) [above] {$\cO_1'$};
\node (vertr) at (0.6,0) [threept] {};
\node (verttl) at (-0.7,0.9) [threept] {};
\node (vertbl) at (-0.7,-0.9) [threept] {};
\node (bl) at (-1.6,-1.6) [below] {$\cO_2'$};
\node (r) at (1.6,0) [right] {$\cO_3$};
\node (f1) at (0.3,0.4) [above] {$\cO_1$};
\node (f2) at (0.3,-0.4) [below] {$\cO_2$};
\node (phi) at (-0.7,0) [diff]{$\scriptstyle j$};
\draw [spinning] (verttl) -- (tl);
\draw [spinning] (vertbl) -- (bl);
\draw [spinning] (vertr) -- (r);
\draw [spinning] (vertr) -- (verttl);
\draw [spinning] (vertr) -- (vertbl);
\draw [finite] (verttl) -- (phi);
\draw [finite] (phi) -- (vertbl);
\node (baseline) at (0,-.15) {};
\end{tikzpicture}
{  =  B^{\gamma_L}_{\ula_L(i)} \sum_i \left\{ \begin{array}{ccc}
        \cO_1 & \cO_3 & \cO_1'\\
        \cO_2' & \frep_j & \cO_2
    \end{array} \right\}_i }\begin{tikzpicture}[baseline=(baseline),scale=.7]
\node (verttl) at (-0.7,0.9) [above] {$\cO_1'$};
\node (vertbl) at (-0.7,-0.9) [below] {$\cO_2'$};
\node (vertr) at (0.6,0) [threept] {$\scriptstyle i$};
\node (r) at (1.6,0) [right] {$\cO_3$};
\draw [spinning] (vertr) -- (r);
\draw [spinning] (vertr) -- (verttl);
\draw [spinning] (vertr) -- (vertbl);
\node (baseline) at (0,-.15) {};
\end{tikzpicture},
\label{diffbasis}
\end{equation}
where to simplify notation, we label the exchanged finite rep in the differential basis by a square node.

Note that in practice, we always wish to act on a three point function $\langle \cO_{1} \cO_2  \cO_3 \rangle$ which itself has a {\em unique} tensor structure. If the third operator is symmetric then we can act on the three-point function of that operator with  two scalars, which indeed has a unique tensor structure,  and in the next subsections we will examine this case explicitly in a number of examples.

For  a more general  third operator $\cO^{\gamma_3}_{\ula_L,\ula_R}$ consider $\ula= \ula_L \cap \ula_R$ and then pick any irreps $\umu_1,\umu_2$ such that $\ula_L \in \ula \otimes \mu_1$,  $\ula_R \in \ula \otimes \mu_2$. Then the three-point function
of (anti-)chiral operators $\cO_{\umu_1,\bullet}$ and $\cO_{\bullet, \umu_2}$ with $ \cO_{\ula_1, \ula_2}$ will have a unique tensor structure
\begin{align}\label{seed3pt}
    \langle \cO_{\umu_1,\bullet} \cO_{\bullet, \umu_2} \cO_{\ula_1, \ula_2}\rangle \sim \hX_{31}^{|\umu_1|}\hX_{23}^{|\umu_2|}Y_{3}^{|\ula|}\,. 
\end{align}
Thus we will build any three-point function $\langle \cO_1 \cO_2 \cO_{\ula_1, \ula_2}\rangle$ by acting with combinations of $\fD_{12}^{[i]}$ on this.

\subsubsection{An illustrative example}\label{sec:illustrative}
Before considering the more general construction, we show a simple explicit example of the differential basis using our $\SL(2m|2n)$ weight-shifting operators \eqref{eq:collectedWSO}.

Let us  consider $\langle \cO_\fa^{\gamma_1} \cO_\dfa^{\gamma_2} \cO_{\ula,\ula}^{\gamma_3} \rangle$. We know from section \ref{sec:Wardids} that the space of $\SL(2m|2n)$ invariant tensor structures is two-dimensional in this case. The two tensor structures in the standard basis are given by
\begin{equation}\label{eq:stdbasis1}
    \begin{split}
         \langle\cO^{\gamma_1}_{\fa} \cO^{\gamma_2}_{\dfa} \cO_{\ula}^\gamma\rangle^{(1)} &= K (\hX_{21})_{\dfa \fa} (Y_3^{|\ula|})_{\udfb \ufb}, \\
         \langle\cO^{\gamma_1}_{\fa} \cO^{\gamma_2}_{\dfa} \cO_{\ula}^\gamma\rangle^{(2)} &= K (\hX_{23})_{\dfa \fb} (\hX_{31})_{\dfb \fa} (Y_3^{|\ula|-1})_{\udfb \ufb}.
    \end{split}
\end{equation}
We thus expect to be able to construct two different weight-shifting operators which, when applied to the three point function $\langle \cO^{\gamma_1} \cO^{\gamma_2} \cO_{\ula}^{\gamma_3}\rangle$, yield linearly independent structures.
 The only possible such combinations are
\begin{equation}\label{WSOex}
    \begin{split}
        \fD^{[1]}_{\dfa_2 \fa_1} &\coloneq (\hX_{21})_{\dfa_2 \fa_1}, \\
        \fD^{[2]}_{\dfa_2 \fa_1} &\coloneq (\bar{D}_2)_{\dfa_2 \cA} (D_1)^{\cA}_{\ \fa_1}.
    \end{split}
\end{equation}
Note that in non-supersymmetric cases, $\hX_{ij}\coloneq(X_{ji})^{-1}$ can be constructed as a polynomial in $\bar X_i$ and $X_j$  (multiplied by $g_{ij}$, the inverse determinant of $X_{ji}$)
\begin{equation}\label{eq:Xhat}
    (\hX_{ji}) {}_{\dfa_j^1\fa_i^1} = \frac1{(m-1)! } \times{g_{ji}} \underbrace{(X_{ij} \dots X_{ij})}_{m-1} {}^{\fa_i^2 \dots \fa_i^m \dfa_j^2 \dots \dfa_j^m} {\epsilon_{\fa_i^1 \dots \fa_i^m} \bar{\epsilon}_{\dfa_j^1 \dots \dfa_j^m}}.
\end{equation}
In particular, in 4D the combination is simply $X_{12}g_{21}$ with spinor indices lowered by epsilon tensors. In a general supersymmetric setting where $m,n \neq 0$, the super determinant is instead non-polynomial.  The writing in diagrammatic form in terms of finite reps $\frep$ as in~\eqref{diffbasis} is then somewhat formal. 

We can then verify that the operators \eqref{WSOex} give rise to two linearly independent tensor structures
\begin{equation}\label{eq:diffbasis}
    \begin{split}
    \langle\cO^{\gamma_1}_{\fa} \cO^{\gamma_2}_{\dfa} \cO_{\ula}^{\gamma_3}\rangle^{[1]}&\equiv(\hX_{21})_{\dfa \fa} \langle \cO^{\gamma_1} \cO^{\gamma_2} \cO_{\ula}^{\gamma_3}\rangle\\ & =  \langle\cO^{\gamma_1}_{\fa} \cO^{\gamma_2}_{\dfa} \cO_{\ula}^{\gamma_3}\rangle^{(1)},  \\
    \langle\cO^{\gamma_1}_{\fa} \cO^{\gamma_2}_{\dfa} \cO_{\ula}^{\gamma_3}\rangle^{[2]}&\equiv (\partial_{\bX_2} \partial_{X_1})_{\dfa \fa} \langle \cO^{\gamma_1} \cO^{\gamma_2} \cO_{\ula}^{\gamma_3}\rangle \\ &= c_1 \langle\cO^{\gamma_1}_{\fa} \cO^{\gamma_2}_{\dfa} \cO_{\ula}^{\gamma_3}\rangle^{(1)} + c_2 \langle\cO^{\gamma_1}_{\fa} \cO^{\gamma_2}_{\dfa} \cO_{\ula}^{\gamma_3}\rangle^{(2)},
\end{split}
\end{equation}
where 
\begin{equation}\label{eq:c1c2coefs}
    c_1 = \gamma_{12,3} \left( \frac{1}{2} (\gamma_1 + \gamma_2 +\gamma_3)-m+n \right), \qquad c_2 = - (1-\delta_{\ula,\bullet}) (\gamma_3 + L_{\ula} -H_{\ula}),
\end{equation}
with $\gamma_{ij,k}=(\gamma_i+\gamma_j-\gamma_k)/2$. It is then evident that both tensor structures are linearly independent for all $m,n$ unless $\ula=\bullet$ or $\gamma_3=-L_{\ula} +H_{\ula}$. 
In terms of graphs we have that
\begin{equation}
    \begin{split}
        \begin{tikzpicture}[baseline={(current bounding box.center)},scale=.7]
\node (tl) at (-1.6,1.6) [left] {$\cO_{\fa}^{\gamma_1}$};
\node (vertr) at (0.6,0) [threept] {};
\node (verttl) at (-0.7,0.9) [threept] {};
\node (vertbl) at (-0.7,-0.9) [threept] {};
\node (bl) at (-1.6,-1.6) [left] {$\cO_{\dfa}^{\gamma_2}$};
\node (r) at (1.6,0) [right] {$\cO_{\ula}^\gamma$};
\node (f1) at (0.3,0.4) [above] {$\cO^{\gamma_1}$};
\node (f2) at (0.3,-0.4) [below] {$\cO^{\gamma_2}$};
\node (hX) at (-.7, 0) [diff] {$\scriptstyle 1$};
\draw [spinning] (verttl) -- (tl);
\draw [spinning] (vertbl) -- (bl);
\draw [spinning] (vertr) -- (r);
\draw [scalar] (vertr) -- (verttl);
\draw [scalar] (vertr) -- (vertbl);
\draw [finite] (verttl) -- (hX);
\draw [finite] (hX) -- (vertbl);
\end{tikzpicture} & = \begin{tikzpicture}[baseline={(current bounding box.center)},scale=.7]
\node (verttl) at (-0.7,0.9) [above] {$\cO_{\fa}^{\gamma_1}$};
\node (vertbl) at (-0.7,-0.9) [below] {$\cO_{\dfa}^{\gamma_2}$};
\node (vertr) at (0.6,0) [threept] {$\scriptstyle 1$};
\node (r) at (1.6,0) [right] {$\cO_{\ula}^\gamma$};
\draw [spinning] (vertr) -- (r);
\draw [spinning] (vertr) -- (verttl);
\draw [spinning] (vertr) -- (vertbl);
\end{tikzpicture},\\
\begin{tikzpicture}[baseline={(current bounding box.center)},scale=.7]
\node (tl) at (-1.6,1.6) [left] {$\cO_{\fa}^{\gamma_1}$};
\node (vertr) at (0.6,0) [threept] {};
\node (verttl) at (-0.7,0.9) [threept] {};
\node (vertbl) at (-0.7,-0.9) [threept] {};
\node (bl) at (-1.6,-1.6) [left] {$\cO_{\dfa}^{\gamma_2}$};
\node (r) at (1.6,0) [right] {$\cO_{\ula}^\gamma$};
\node (w) at (-0.7,0) [diff] {$\scriptstyle 2$};
\node (f1) at (0.3,0.4) [above] {$\cO^{\gamma_1}$};
\node (f2) at (0.3,-0.4) [below] {$\cO^{\gamma_2}$};
\draw [spinning] (verttl) -- (tl);
\draw [spinning] (vertbl) -- (bl);
\draw [spinning] (vertr) -- (r);
\draw [scalar] (vertr) -- (verttl);
\draw [scalar] (vertr) -- (vertbl);
\draw [finite] (verttl) -- (w);
\draw [finite] (w) -- (vertbl);
\end{tikzpicture} & = c_1 \begin{tikzpicture}[baseline={(current bounding box.center)},scale=.7]
\node (verttl) at (-0.7,0.9) [above] {$\cO_{\fa}^{\gamma_1}$};
\node (vertbl) at (-0.7,-0.9) [below] {$\cO_{\dfa}^{\gamma_2}$};
\node (vertr) at (0.6,0) [threept] {$\scriptstyle 1$};
\node (r) at (1.6,0) [right] {$\cO_{\ula}^\gamma$};
\draw [spinning] (vertr) -- (r);
\draw [spinning] (vertr) -- (verttl);
\draw [spinning] (vertr) -- (vertbl);
\end{tikzpicture} + c_2 \begin{tikzpicture}[baseline={(current bounding box.center)},scale=.7]
\node (verttl) at (-0.7,0.9) [above] {$\cO_{\fa}^{\gamma_1}$};
\node (vertbl) at (-0.7,-0.9) [below] {$\cO_{\dfa}^{\gamma_2}$};
\node (vertr) at (0.6,0) [threept] {$\scriptstyle 2$};
\node (r) at (1.6,0) [right] {$\cO_{\ula}^\gamma$};
\draw [spinning] (vertr) -- (r);
\draw [spinning] (vertr) -- (verttl);
\draw [spinning] (vertr) -- (vertbl);
\end{tikzpicture}.
    \end{split}
\end{equation}
The differential basis is essential for understanding how to shift external representations appearing in $\SL(2m|2n)$ blocks, and will be used for this purpose in section~\ref{sec:blocks}.

\subsubsection{General construction}\label{sec:diffbasisgeneral}

Having covered examples of the differential basis in both short and long cases,
we now give a construction of this basis for some more general families of correlators 
which we will make use of in section~\ref{sec:blocks} for constructing superconformal blocks. 
As usual, we work with unfixed $m,n$ so Young diagrams can be arbitrarily large. By fixing $m,n$ to cases of interest (e.g.\ 4D $\cN=0,2,4$), the methods we show here can be used to prove the existence of the differential basis in general.\footnote{It was shown for non-conserved symmetric traceless tensors in \cite{Costa:2011dw} using the embedding space formalism.} 
\paragraph{ $\langle \cO_{\umu,\bullet} \cO_{\bullet,\umu} \cO_{\ula}\rangle$:}
Such a correlator has $|\umu|+1$ independent tensor structures (for generic $\ula$, that is, $|\umu|\leq |\ula|$). In the standard basis $\langle \cO_{\umu,\bullet} \cO_{\bullet,\umu} \cO_{\ula}\rangle^{(k)}$ they are given by\footnote{Note the similarities with \eqref{eq:threepointstructures}.}
\begin{equation}\label{chiralstandardbasis}
    K\,  \underbrace{\hX_{21} \dots \hX_{21}}_{|\umu|-k}\  \underbrace{(\hX_{23} \hX_{31}) \dots(\hX_{23} \hX_{31})}_{k} \ \underbrace{Y_3 \dots Y_3}_{|\ula|-k}\ , \qquad k = 0, 1, \dots, |\umu|.
\end{equation}
As shown in the previous subsection, the way to construct the differential basis of tensor structures in the chiral case is by using the two invariant weight-shifting operators  $\fD^{[1]}_{\dfa_2 \fa_1},\fD^{[2]}_{\dfa_2 \fa_1}$    \eqref{WSOex} on a three-point correlator with one less index at points 1 and 2.

However if we recurse this process we can instead construct the differential basis by acting on $\langle \cO^{\gamma_1} \cO^{\gamma_2} \cO^{\gamma_3}_{\ula}  \rangle$,  which has a single structure. This is what we will be doing when constructing superconformal blocks in section \ref{sec:blocks}.
So we now need to apply one of the invariant differential operators   $\fD^{[1]}_{\dfa_2 \fa_1},\fD^{[2]}_{\dfa_2 \fa_1}$    \eqref{WSOex} at each step as follows
\begin{equation}\label{eq:wsoex10}
   \begin{tikzpicture}[baseline={(current bounding box.center)},scale=.8]
\node (verttl) at (-0.7,0.9) [above] {$\cO_{\umu,\bullet}^{\gamma_1}$};
\node (vertbl) at (-0.7,-0.9) [below] {$\cO_{\bullet,\umu}^{\gamma_2}$};
\node (vertr) at (0.6,0) [diff] {$\scriptstyle k$};
\node (r) at (1.6,0) [right] {$\cO_{\ula}^{\gamma_3}$};
\draw [spinning] (vertr) -- (r);
\draw [spinning] (vertr) -- (verttl);
\draw [spinning] (vertr) -- (vertbl);
\end{tikzpicture} \ \coloneq \ \begin{tikzpicture}[baseline={(current bounding box.center)},scale=0.8]
\node (tl2) at (-3.8,1.08) [threept] {};
\node (bl2) at (-3.8,-1.08) [threept] {};
\node (tl) at (-2.16,1.08) [threept] {};
\node (vertr) at (0.72,0) [threept] {};
\node (verttl) at (-0.84,1.08) [threept] {};
\node (vertbl) at (-0.84,-1.08) [threept] {};
\node (bl) at (-2.16,-1.08) [threept] {};
\node (r1) at (1.92,0) [right] {$\cO_{\ula}^{\gamma_3}$};
\node (f1) at (0.38,0.7) [above] {$\cO^{\gamma_1}$};
\node (f2) at (0.38,-0.7) [below] {$\cO^{\gamma_2}$};
\node at (-2.9,1.08) {$\cdots$};
\node at (-2.9,-1.08) {$\cdots$};
\node (D1) at (-0.84,0) [diff] {$\scriptstyle j_1$};
\node (D2) at (-2.16,0) [diff] {$\scriptstyle j_2$};
\node (D3) at (-3.8,0) [diff] {$\scriptstyle j_\mu$};
\node (O1) at (-4.6,1.09) [left] {$\cO^{\gamma_1}_{\umu,\bullet}$};
\node (O2) at (-4.6,-1.09) [left] {$\cO^{\gamma_2}_{\bullet,\umu}$};
\node at (-1.6,1.2) [above] {\small $\cO^{\gamma_1}_{\Box,\bullet}$};
\node at (-1.6,-1.2) [below] {\small $\cO^{\gamma_2}_{\bullet,\Box}$};
\draw [spinning] (verttl) -- (tl);
\draw [spinning] (vertbl) -- (bl);
\draw [spinning] (vertr) -- (r1);
\draw [scalar] (vertr) -- (verttl);
\draw [scalar] (vertr) -- (vertbl);
\draw [finite] (verttl) -- (D1);
\draw [finite] (D1) -- (vertbl);
\draw [finite] (tl) -- (D2);
\draw [finite] (D2) -- (bl);
\draw [finite] (tl2) -- (D3);
\draw [finite] (D3) -- (bl2);
\draw[thick] (tl)--(-2.5,1.08);
\draw[thick] (bl)--(-2.5,-1.08);
\draw[thick] (tl2)--(-3.4,1.08);
\draw[thick] (bl2)--(-3.4,-1.08);
\draw[spinning] (tl2)--(O1);
\draw[spinning] (bl2)--(O2);
\end{tikzpicture},
\end{equation}
applying an appropriate symmetriser at the end.\footnote{This is equivalent to symmetrising at each step by orthogonality of the Young projectors. See appendix \ref{app:yt} for details.} Na\"ively, there are $2^{|\mu|}$ ways of doing this. However since 
$
    [\partial_{\bX_2}\partial_{X_1}, \hX_{21}] \sim \hX_{21} \hX_{21},
$
we see that different orderings of the operators gives back linearly dependent expressions so only the following $|\umu|+1$ choices are independent
\begin{equation}\label{eq:mfrakD1}
    \begin{split}
        \mathfrak{D}^{[k]}_{[\umu,\bullet]_1, [\bullet, \umu]_2}&\coloneq \Pi^{\umu}_1(X_1) \bPi^{\umu}_1(X_2)(\fD^{[1]})^{\otimes ( |\umu|-k)} (\fD^{[2]})^{\otimes k} \\ & \equiv  \Pi^{\umu}_1(X_1) \bPi^{\umu}_1(X_2)(\underbrace{\hX_{21} \dots \hX_{21}}_{|\umu|-k}) \ \big(\underbrace{(\partial_{\bX_2} \partial_{X_1}) \dots (\partial_{\bX_2} \partial_{X_1})}_{k}\big),
    \end{split}
\end{equation}
for $k=0, \dots, |\umu|$. This matches the dimension of the space of tensor structures of $\langle \cO^{\gamma_1}_{\umu,\bullet} \cO^{\gamma_2}_{\bullet, \umu} \cO_{\ula}^{\gamma_2}\rangle$.

\paragraph{$\langle \cO_{\umu} \cO \cO_{\ula}\rangle$:}
This is another case that we will use for superblocks later. It can be reduced to the previous case by applying a suitable weight-shifting operator,  removing $|\umu|$ boxes from  point 2 and adding them to the right $\mathfrak{sl}(m|n)$ at point 1. This can be done by recursively applying an invariant combination of weight-shifting operators \eqref{eq:collectedWSO} given by
\begin{equation}\label{diffppex}
    \bD_1{}_{\dfa_1 \cA} \bX_2^{\cA \dfa_2} \equiv (\bD_1 \bX_2)^{\ \dfa_2}_{\dfa_1},
\end{equation}
where $\bX_2$ contracts an index at point 2 and $\bD_1$ adds a suitable symmetrised index at point 1.
Therefore, the differential basis for such structures is 
\begin{equation}\label{eq:diffopp11}
    \langle \cO_{\umu} \cO \cO_{\ula}\rangle^{[k]}\coloneq \underbrace{(\bD_1 \bX_2) \dots (\bD_1 \bX_2)}_{|\umu|}\langle \cO_{\umu,\bullet} \cO_{\bullet,\umu} \cO_{\ula}\rangle^{[k]}.
\end{equation}

\subsubsection{Correlators of short and conserved (super) multiplets}\label{subsec:shortdiffbasis}
Fields on analytic superspace were shown to be automatically irreducible for $n>0$ in \cite{Heslop:2001zm}. This means that given a choice of $\gamma,\ula_L,\ula_R$, no further differential constraints need to be imposed in order for the field to be irreducible. In other words, shortening conditions are automatically satisfied. It is thus expected that our weight-shifting operators are able to act on three-point functions of short supermultiplets and produce a correct basis that satisfies new shortening conditions. This is not possible in other approaches to (super) conformal field theory and thus we demonstrate it explicitly here. The main statement is that the differential basis for correlators of short supermultiplets is generated by combinations of weight-shifting operators that do not include the inverse $\hX_{ij}$, i.e. built out of the following weight-shifting invariants:
\begin{equation}\label{eq:invariantsforshort}
    X_{ij}^{\fa_i \dfa_j}, \qquad (X_i)^{\fa_i}_{\ \cA} (D_j)^{\cA}_{\ \fb_j}, \qquad (\bX_i)^{\cA\dfa_i}(\bar{D}_j)_{\ \dfb_j \cA}, \qquad (\bar{D}_i)_{\dfa_i \cA} (D_j)^{\cA}_{\ \fa_j}\,.
\end{equation}

Let us consider this with the illustrative example presented in section \ref{sec:illustrative}. Let $\gamma_1=\gamma_2=1$ and $\gamma=2,\, \ula=[\lambda]$ so that \eqref{eq:stdbasis1} is a correlator that in 4D ($m=2,n=0$) contains two fundamental spinors and a conserved symmetric traceless tensor of spin $\lambda$
\begin{equation}
    \langle \cO^1_{\fa} \cO^1_{\dfa} \cO_{[\lambda]}^2\rangle = g_{23}g_{31}\left( A (\hX_{21})_{\dfa \fa} (Y_{3}^{|\ula|})_{\udfb \ufb}+ B (\hX_{31})_{(\dfb_1 \fa} (\hX_{23})_{\dfa (\fb_1} (Y_{3}^{|\ula-1|})_{\udfb') \ufb')} \right),
\end{equation}
where $A$ and $B$ are OPE coefficients.
Such a correlator must satisfy the Dirac equation at points 1 and 2, and the continuity equation at point 3 (with $(m,n)$ generalisations obtained by simply replacing $x_{\alpha \dota}\mapsto \mathrm{x}_{\fa \dfa}$). A simple calculation shows that these differential constraints require $A=0$. Comparing with~\eqref{eq:c1c2coefs} for the chosen values of $\gamma_1,\gamma_2,\gamma,\ula$ we find $c_1=0$, $c_2=-1-\lambda$, i.e. producing the correct structure even when mapping from and to short representations:
\begin{equation}
    \langle \cO^1_{\fa} \cO^1_{\dfa} \cO_{[\lambda]}^2\rangle \propto (\partial_{\bX_2} \partial_{X_1})_{\dfa \fa} \langle \cO^{1} \cO^{1} \cO_{[\lambda]}^{2}\rangle.
\end{equation}

Now let us consider a more non-trivial and very relevant example given by the map
\begin{equation}
    \langle \cO^2 \cO^2 \cO^2_{\fa \dfa}\rangle \to  \langle \cO^2 \cO^2_{\fb \dfb}\cO^2_{\fa \dfa}\rangle.
\end{equation}
These involve dimension 2 scalars and conserved currents $J$ in 4D, and half-BPS and stress-tensor supermultiplets in 4D $\cN=2$. There is a single structure on the LHS given by
\begin{equation}\label{eq:seedshort}
    \langle \cO^2 \cO^2 \cO^2_{\fa \dfa}\rangle= g_{12} g_{23} g_{31}(Y_3)_{\dfa \fa},
\end{equation}
which also satisfies the conservation constraint at point 3. Now, before imposing conservation there are two structures given by
\begin{equation}
    \langle \cO^2 \cO^2_{\fb \dfb}\cO^2_{\fa \dfa}\rangle = g_{12} g_{23} g_{31} \left( A (\hX_{13})_{\dfb \fa} (\hX_{31} )_{\dfa \fb}+ B (Y_1)_{\dfb \fb} (Y_3)_{\dfa \fa}\right),
\end{equation}
where as it is well-known that conservation picks out the linear combination with $A=-B$ \cite{Schreier:1971um,Osborn:1993cr,Costa:2011mg}. The simplest differential operator (least number of derivatives) built out of the weight-shifting invariants~\eqref{eq:invariantsforshort} is 
\begin{equation}
    (\bX_2 \bar{D}_1)^{\dfc}_{\ \dfb} (\partial_{\bX_2} \partial_{X_1})_{\dfc \fb},
\end{equation}
where $\bar{D}$ is given by \eqref{eq:chiralWSO1} for $\gamma=2$. An explicit calculation of the above differential operator acting on the seed structure \eqref{eq:seedshort} yields the following linear combination
\begin{equation}
    (\bX_2 \bar{D}_1)^{\dfc}_{\ \dfb} (\partial_{\bX_2} \partial_{X_1})_{\dfc \fb} \langle \cO^2 \cO^2 \cO^2_{\fa \dfa}\rangle = \frac{4}{3}g_{12}g_{23} g_{31} \left( (\hX_{13})_{\dfb \fa} (\hX_{31} )_{\dfa \fb} - (Y_1)_{\dfb \fb} (Y_3)_{\dfa \fa} \right),
\end{equation}
as expected.

We note that while the automatic irreducibility property of analytic superspace needs $n\neq0$, the results solve shortening conditions for all $m,n$ and thus also apply to cases with $n=0$ such as 4D CFT, as demonstrated above

\section{Superconformal blocks from weight-shifting}\label{sec:blocks}

We are now ready to compute $\SL(2m|2n)$ blocks by using the weight-shifting operators \eqref{eq:collectedWSO} and their invariant combinations constructed in the previous section. After a review of (super)conformal blocks on the $\SL(2m|2n)$ coset space, we compute explicit non-half-BPS superblocks by acting with differential operators on the blocks of the correlator of four coset scalar (half-BPS) fields derived in \cite{Doobary:2015gia}. This includes shifts of both external and exchanged (internal) representations.

We have performed two non-trivial checks on our results. First, all computations can be applied to 1D CFT by setting $m=1,n=0$. In this simple scenario, all representations are scalar and thus adding a box to either Young diagram $\ula_L, \ula_R$ simply corresponds to shifting $\gamma_L, \gamma_R$, respectively. All such blocks were given in \cite{Doobary:2015gia, Aprile:2021pwd}. Second, by setting $m=2,n=0$ and matching tensor structures, we reproduce the 4D spinning conformal blocks of the CFTs4D package \cite{Cuomo:2017wme}. The details of the checks are given in appendix \ref{app:checks}. Finally, we discuss how to access non-integer dimensions in section \ref{sec:nonintegerdim}.

\subsection{$\SL(2m|2n)$ blocks}\label{subsec:blocks}
The superconformal blocks expansions of correlation functions of singlets of $\SL(m|n)\times \SL(m|n)$ were studied in depth in \cite{Doobary:2015gia,Aprile:2021pwd,Heslop:2022xgp,Aprile:2025nta}. These correspond to blocks for half-BPS correlators as outlined in section \ref{sec:fieldsandreps}. All exchanged operators in this case are symmetric and hence labelled by a single Young diagram $\ula\coloneq\ula_L=\ula_R$, a generalisation of the symmetric traceless operators in non-supersymmetric CFT, which are simply labelled by their spin $\ell$. The block expansion of a correlation function of singlets is
\begin{equation}
     \langle \cO^{\gamma_1}(X_1) \cO^{\gamma_2}(X_2) \cO^{\gamma_3}(X_3) \cO^{\gamma_4} (X_4)\rangle = \sum_{\gamma,\ula} A_{\gamma_1\gamma_2\gamma_3\gamma_4}^{\gamma \ula} \cF^{\gamma,\ula}_{\gamma_1\gamma_2\gamma_3\gamma_4} (X_i)
\end{equation}
with the block coefficients $A_{\gamma_1\gamma_2\gamma_3\gamma_4}^{\gamma \ula}$ giving data on OPE coefficients involving the exchanged operator. The $\SL(2m|2n)$ blocks $\cF^{\gamma,\ula}_{\gamma_1\gamma_2\gamma_3\gamma_4}$ are written diagrammatically as
\begin{equation}\label{eq:singletblock}
    \cF^{\gamma,\ula}_{\gamma_1\gamma_2\gamma_3\gamma_4} \coloneq\begin{tikzpicture}[baseline={(current bounding box.center)}, scale=0.8]
            \node (dot2) at (1.5,0) [threept] {};
            \node (pt1) at (2.25,1.275) [above] {$\cO^{\gamma_3}$};
            \node (Or) at (0,0.1)[above] {$\cO^{\gamma}_{\ula}$};
            \node (pt2) at (-1.5,0)[threept] {};
            \node (pt3) at (2.25,-1.275) [below] {$\cO^{\gamma_4}$};
            \node (O2p) at (-2.25,1.275) [above] {$\cO^{\gamma_2}$};
            \node (W) at (-2.25,-1.275) [below] {$\cO^{\gamma_1}$};
           \node (X) at (0,0) [bowtie] {$\bowtie$};
            \draw[scalar] (dot2) to (pt1);
            \draw[spinning] (dot2) to (X);
            \draw[spinning] (pt2) to (X);
            \draw[scalar] (dot2) to (pt3);
            \draw[scalar] (pt2) to (O2p);
            \draw[scalar] (pt2) to (W);
        \end{tikzpicture} \equiv P\times F^{\gamma,\ula} (Z),
\end{equation}
where the prefactor $P$ absorbs the dependence on the different $\mathbb{C}^*$ charges $\gamma_i$, e.g.\ 
\begin{equation}\label{eq:K4}
    P= g_{12}^{\frac{\gamma_1+\gamma_2}{2}} g_{34}^{\frac{\gamma_3+\gamma_4}{2}} \left(\frac{g_{24}}{g_{14}}\right)^{\frac{1}{2}\gamma_{21}}\left(\frac{g_{14}}{g_{13}}\right)^{\frac{1}{2}\gamma_{43}}\left(\frac{g_{13}g_{24}}{g_{12}g_{34}}\right)^{\frac{1}{2}\gamma},
\end{equation}
and hence $F^{\gamma \ula}$ is simply a function of the cross-ratio matrix $Z$, as explained in section \ref{sec:Wardids}.
The specific form of the function $F^{\gamma,\ula}$ in \eqref{eq:singletblock} is given in \cite{Doobary:2015gia} (see section 3.5). In this section, we will derive explicit forms of differential operators that act on the above singlet block to give new blocks for external non-trivial $\SL(m|n)\times \SL(m|n)$ representations, including those for non-symmetric exchanges.

Note that both singlet  and non-singlet  blocks are eigenfunctions of the quadratic Casimir, 
\begin{equation}\label{eq:Casimir}
    C_2(X_i, X_j) = \frac12(L_i + L_j)^{\cA}_{\ \cB}(L_i + L_j)^{\cB}_{\ \cA}\ .
\end{equation}
where the generators are
\begin{equation}
    L^{\cB}_{\ \cA}= X^{\fa}_{\ \cA} \frac{\partial}{\partial X^{\fa}_{\ \cB}} - \bX^{\cB\dfa}\frac{\partial}{\partial \bX^{\cA\dfa}}\ .
\end{equation}

Blocks for non singlet  correlators are more complicated to study. In particular due to the appearance of multiple different three-point tensor structures which we already discussed in section \ref{sec:Wardids}. 
Using the notation of \cite{Karateev:2017jgd}, we write a general superconformal block as the gluing of two three-point structures 
\begin{equation}
        [\cF_{\cO_r}^{(i,j)}]_{\bigotimes\rho_k} = \langle \cO_{\rho_1} \cO_{\rho_2} \cO_r \rangle^{(i)} \bowtie \langle \cO_r^\dagger \cO_{\rho_3}\cO_{\rho_4}\rangle ^{(j)}\label{ij},
\end{equation}
which in terms of the diagrammatic notation is given by
\begin{equation}
     [\cF_{\cO_r}^{(i,j)}]_{\bigotimes\rho_k}=\begin{tikzpicture}[baseline={(current bounding box.center)}, scale=0.8]
            \node (dot2) at (1.5,0) [threept] {$\scriptstyle j$};
            \node (pt1) at (2.25,1.275) [above] {$\cO_{\rho_3}$};
            \node (Or) at (-0.75,0.1)[above] {$\cO_r$};
            \node (Ord) at (0.75,0.1)[above] {$\cO_r^\dagger$};
            \node (pt2) at (-1.5,0)[threept] {$\scriptstyle i$};
            \node (pt3) at (2.25,-1.275) [below] {$\cO_{\rho_4}$};
            \node (O2p) at (-2.25,1.275) [above] {$\cO_{\rho_2}$};
            \node (W) at (-2.25,-1.275) [below] {$\cO_{\rho_1}$};
            \node (X) at (0,0) [bowtie] {$\bowtie$};
            
            \draw[spinning] (dot2) to (pt1);
            \draw[spinning] (dot2) to (X);
            \draw[spinning] (pt2) to (X);
            \draw[spinning] (dot2) to (pt3);
            \draw[spinning] (pt2) to (O2p);
            \draw[spinning] (pt2) to (W);
        \end{tikzpicture},
\end{equation}
where $\rho_k=[\gamma_k, \ula_L^{k}, \ula_R^{k}]$ is the representation of the operator inserted at point $X_k$.
The gluing operation generalises straightforwardly to the $(m,n)$ coset space as follows
\begin{equation}\label{eq:gluing}
    | \cO_{\ula_L, \ula_R}\rangle\bowtie \langle \cO_{\ula_R\ula_L}|= \int dg d\bar{g} | \cO_{\ula_L, \ula_R}(g) \rangle \langle \tilde{\cO}^{\ula_L, \ula_R}(g) \tilde{\cO}^{\ula_R, \ula_L}(\bar{g})\rangle \langle \cO_{\ula_R\ula_L} (\bar{g})|\,,
\end{equation}
where $\tilde{\cO}$ is the analogue to the shadow operator and has upper indices symmetrised according to $\ula_L, \ula_R$ which contract their corresponding lower indices. In particular, they transform as non-unitary reps according to \eqref{eq:primary3}.
For explicit constructions of the shadow integral in superconformal settings see \cite{Fitzpatrick:2014oza}.
However, note that as already pointed out in the conformal weight-shifting operator context \cite{Karateev:2017jgd}, the explicit form of $\bowtie$ is not necessary. Indeed, weight-shifting operators allow one to bypass having to perform such integral. In our superconformal setting, the only properties that will be used are bilinearity, $\SL(2m|2n)$ invariance and the following normalisation 
\begin{equation}\label{eq:gluingnormalisation}
    \langle \cO_{\ula_L, \ula_R} \cO^\dagger_{\ula_R, \ula_L} \rangle \bowtie \langle\cO _{\ula_L, \ula_R}| = \langle\cO_{\ula_L, \ula_R}|,
\end{equation}
which diagrammatically reads
\begin{equation}\label{eq:gluingnormalisationdiag}
  \begin{tikzpicture}[baseline=(align),scale= 0.85]
         \node (align) at (0,-.1) {};
        \node at (0.8,-0.1) [below] {$\cO$};
        \node at (-0.8,-0.1) [below] {$\cO^\dagger$};
        \node at (-2.4,-0.1) [below] {$\cO$};
        \node (vertR) at (1.5,0) {};
        \node (shad) at (-1.5,0) [twopt] {};
        \node (ex0) at (0, 0) [bowtie] {$\bowtie$};
        \node (vertL) at (-3,0){};
        \draw [spinning] (shad) -- (ex0);
        \draw [spinning] (vertR) --  (ex0) ;
        \draw [spinning] (shad)-- (vertL);
    \end{tikzpicture}\ =  \ \begin{tikzpicture}[baseline=(align),scale= 0.85]
         \node (align) at (0,-.1) {};
        \node at (-1,-0.1) [below] {$\cO$};
        \node (vertR) at (0,0) {};
        \node (vertL) at (-2,0){};
        \draw [spinning] (vertR) --  (vertL);
    \end{tikzpicture}\, .
\end{equation}

\subsubsection{Seed blocks}\label{sec:seedblockss}
There is an important class of blocks it is useful to focus on, known as \emph{seed blocks} for which  the degeneracy labelled by $i,j$ in~\eqref{ij} collapses. 
They thus  behave similarly to scalar blocks and superspace half-BPS blocks.
In CFT, certain four-dimensional spinning seed blocks were found in \cite{CastedoEcheverri:2016dfa} by directly solving the Casimir equation. Then, \cite{Karateev:2017jgd} derived new expressions for these blocks and more general exchanges in various dimensions in terms of weight-shifting operators.

In $(m,n)$ analytic superspace, a seed superblock for a general exchanged representation $\cO_{\ula_1, \ula_2}$ can be formed from taking two of the three point functions in~\eqref{seed3pt}
\begin{equation}\label{eq:seedblock}
    \langle \cO_{\umu_1,\bullet} \cO_{\bullet, \umu_2} \cO_{\ula_1, \ula_2}\rangle \bowtie \langle \cO_{\ula_2, \ula_1}  \cO_{\umu_2,\bullet} \cO_{ \bullet,\umu_1}\rangle, \qquad \ula_{i} \in \umu_{i} \otimes (\ula_1 \cap \ula_2),
\end{equation}
since  the constituent three point structures have a unique tensor structure.
The name seed is 
due to them being simple
blocks from which one can generate (or `seed')  other blocks for the same exchange by acting with differential operators. This operation is known as external shift and we will consider it in our superconformal setting in section \ref{sec:externalshift}. 
Note that as already pointed out in \cite{Karateev:2017jgd}, one can in fact also relate seed blocks to the  scalar blocks by using weight-shifting operators which also act on the exchanged operator, known as `internal shifts'. This also holds in the $(m,n)$ superspace we consider here and thus \emph{all} superconformal blocks can be related to the simplest superblock of half-BPS supermultiplets \eqref{eq:singletblock} by using the weight-shifting operators~\eqref{eq:collectedWSO}. This is done with a combination of external and internal shifts. We will perform internal shifts to derive seed blocks from scalar blocks in section \ref{sec:internalshift}.

\subsection{External shift}\label{sec:externalshift}
This corresponds to the shift of one or more external representations, leaving the exchanged representation fixed. Weight-shifting operators are able to perform this operation as they are able to form a basis of the relevant three-point structures, as explained in detail in section \ref{sec:diffbasis}.

We construct superblocks involving a symmetric exchanged operator $\cO_{\ula}^\gamma$ by applying external weight-shifting operators~\eqref{eq:collectedWSO} to the the known singlet (half-BPS) blocks in $(m,n)$ analytic superspace~\eqref{eq:singletblock} (which are the seed blocks for this operator).

We do so by applying the differential basis trick \eqref{eq:diffbasis} as follows
\begin{equation}\label{eq:externalshiftblock}
    \langle\cO_1 \cO_2 \cO_{\ula}^\gamma\rangle^{[i]} \bowtie \langle \cO_{\ula}^\gamma\cO_3 \cO_4 \rangle^{[j]}  = \fD_{12}^{[i]} \fD_{34}^{[j]} \langle\cO_1\cO_2\cO_{\ula}^\gamma\rangle \bowtie \langle \cO_{\ula}^\gamma\cO_3\cO_4 \rangle,
\end{equation}
where we recall that each $\fD_{pq}$~\eqref{eq:mfrakD1} denotes $\SL(2m|2n)$ invariant combinations of the superspace weight-shifting operators \eqref{eq:collectedWSO} acting at points $X_p$ and $X_q$. In the diagrammatic notation, this is given by
\begin{equation} 
    \diagramEnvelope{\begin{tikzpicture}[baseline=(shad),scale= 0.8]
        \node (vertL) at (-1.5,0) [diff] {\small $i$};
        \node (vertR) at (1.5,0) [diff] {\small $j$};
        \node (opO1) at (-2.25,-1.125) [below] {$\cO_1$};
        \node (opO2) at (-2.25,1.125) [above] {$\cO_2$};
        \node (opO3) at (2.25,1.125) [above] {$\cO_3$};
        \node (opW) at (2.25,-1.125) [below] {$\cO_4$};
        \node (shad) at (0,0) [bowtie] {$\bowtie$};
        \node at (0,0.15) [above] {$\cO^\gamma_{\ula}$};		
        \draw [spinning] (vertL)-- (opO1);
        \draw [spinning] (vertL)-- (opO2);
        \draw [spinning] (vertL)-- (shad);
        \draw [spinning] (vertR) -- (shad);
        \draw [spinning] (vertR)-- (opO3);
        \draw [spinning] (vertR)-- (opW);
    \end{tikzpicture}} 
    =
    \diagramEnvelope{\begin{tikzpicture}[baseline=(shad),scale= 0.8]
        \node (vertL) at (-1.5,0) [threept] {};
        \node (vertR) at (1.5,0) [threept] {};
        \node (opO1) at (-1.35,-0.45) [below] {$\cO^{\gamma_1}$};
        \node (opO2) at (-1.35,0.45) [above] {$\cO^{\gamma_2}$};
        \node (opO3) at (1.35,0.45) [above] {$\cO^{\gamma_3}$};
        \node (opW) at (1.35,-0.45) [below] {$\cO^{\gamma_4}$};
        \node (opO11) at (-3.6,-1.5) [below] {$\cO_1$};
        \node (opO22) at (-3.6,1.5) [above] {$\cO_2$};
        \node (opO33) at (3.6,1.5) [above] {$\cO_3$};
        \node (opO44) at (3.6,-1.5) [below] {$\cO_4$};
        \node (o1) at (2.25,-1.05) [threept] {};
        \node (o2) at (2.25,1.05) [threept] {};
        \node (o3) at (-2.25,-1.05) [threept] {};
        \node (o4) at (-2.25,1.05) [threept] {};
        \node (shad) at (0,0) [bowtie] {$\bowtie$};
        \node (D1) at (-2.625,0) [diff] {\small $i$};
        \node (D2) at (2.625,0) [diff] {\small $j$};
        \node at (0,0.15) [above] {$\cO^\gamma_{\ula}$};		
        \draw [scalar] (vertL)-- (o3);
        \draw [scalar] (vertL)-- (o4);
        \draw [spinning] (vertL)-- (shad);
        \draw [spinning] (vertR) -- (shad);
        \draw [scalar] (vertR)-- (o1);
        \draw [scalar] (vertR)-- (o2);
        \draw [finite] (o3) to (D1);
        \draw [finite] (o4) to (D1);
        \draw [finite] (o1) to (D2);
        \draw [finite] (o2) to (D2);
        \draw [spinning] (o3)-- (opO11);
        \draw [spinning] (o4)-- (opO22);
        \draw [spinning] (o1)-- (opO44);
        \draw [spinning] (o2)-- (opO33);
    \end{tikzpicture}}.
\end{equation}
By construction, these weight-shifting invariants commute with the Casimir element \eqref{eq:Casimir}, since
\begin{equation}
    \left[ (L_p+L_q)^{\cA}_{\ \cB}, \fD_{pq}^{[k]}\right]=0,
\end{equation}
and thus map solutions of the Casimir equation to other solutions:
\begin{equation}
    C_2 \fD_{pq} \cF^{\gamma \ula}_{\phi_1 \phi_2 \phi_3 \phi_4}  = \fD_{pq} C_2 \cF^{\gamma \ula}_{\phi_1 \phi_2 \phi_3 \phi_4} = c_{\gamma,\ula} \fD_{pq} \cF^{\gamma\ula}_{\phi_1 \phi_2 \phi_3 \phi_4} ,
\end{equation}
where $c_{\gamma \ula}$ is the Casimir eigenvalue.

We will now consider some simple examples involving non-trivial shifts in the $\SL(m|n)\times \SL(m|n)$ portion of the Levi subgroup. That is, adding boxes to the Young diagrams at the different points in \eqref{eq:singletblock}. 
\subsubsection{$\langle \cO_{\fa} \cO_{\dfa} \cO \cO \rangle$}
This correlator involves two 4D spin $1/2$ fermions in the $(m,n)=(2,0)$ theory, and two chiral  supermultiplets in $(2,1)$, i.e.\ 4D $\cN=2$ theories (such as the `extra-SUSY multiplet', which corresponds to $\gamma_1=\gamma_2=2$:  see section \ref{sec:repexamples} for a full outline of the correspondence). 
Because operators at points $X_3$ and $X_4$ are singlets whose OPE only includes symmetric representations, this is the only possible exchange in this case. All the blocks are thus given by the following gluing of three-point structures
\begin{equation}\label{eq:fermionicblock}
    \langle\cO^{\gamma_1}_{\fa} \cO^{\gamma_2}_{\dfa} \cO_{\ula}^\gamma\rangle \bowtie \langle \cO_{\ula}^\gamma\cO^{\gamma_3} \cO^{\gamma_4} \rangle.
\end{equation}
The left three-point structure admits two different tensor structures given by~\eqref{eq:stdbasis1}. We already constructed the differential basis for these tensor structures \eqref{eq:diffbasis}. We can then use this to write our first set of $\SL(m|n)\times \SL(m|n)$ covariant blocks as follows
\begin{equation}
    \begin{split}
        \langle\cO^{\gamma_1}_{\fa} \cO^{\gamma_2}_{\dfa}\cO_{\ula}^\gamma\rangle^{[1]} \bowtie \langle \cO_{\ula}^\gamma\cO^{\gamma_3} \cO^{\gamma_4} \rangle &= \fD^{[1]}_{\fa \dfa} \langle\cO^{\gamma_1} \cO^{\gamma_2} \cO_{\ula}^\gamma\rangle \bowtie \langle \cO_{\ula}^\gamma\cO^{\gamma_3} \cO^{\gamma_4} \rangle, \\
         \langle\cO^{\gamma_1}_{\fa} \cO^{\gamma_2}_{\dfa}\cO_{\ula}^\gamma\rangle^{[2]} \bowtie \langle \cO_{\ula}^\gamma\cO^{\gamma_3} \cO^{\gamma_4} \rangle &= \fD^{[2]}_{\fa \dfa}\langle\cO^{\gamma_1} \cO^{\gamma_2} \cO_{\ula}^\gamma\rangle \bowtie \langle \cO_{\ula}^\gamma\cO^{\gamma_3} \cO^{\gamma_4} \rangle,
    \end{split}
\end{equation}
with~\eqref{WSOex}
\beq
\fD^{[1]}_{\fa \dfa} = (\hX_{21})_{\dfa \fa}\,, \qquad
\fD^{[2]}_{\fa \dfa} = (\partial_{\bX_2}\partial_{X_1})_{\dfa \fa}\,.
\label{wso_ex1}
\eeq
We may now write the above blocks in a form that more closely resembles the solution to the Ward identities \eqref{WI2}. The first block is very simple, indeed
\begin{equation}\label{eq:xhatblock}
     \langle\cO^{\gamma_1}_{\fa} \cO^{\gamma_2}_{\dfa}\cO_{\ula}^\gamma\rangle^{[1]} \bowtie \langle \cO_{\ula}^\gamma\cO^{\gamma_3} \cO^{\gamma_4} \rangle = P \, \hX_{21}{}_{\dfa \fb} f^{[1]} {}^{\fb}_{\ \fa} (Z) ,\qquad f^{[1]} {}^{\fb}_{\ \fa} (Z)= \delta^{\fb}_{\fa} F^{\gamma \ula}(Z),
\end{equation}
where $F^{\gamma \ula}$ is just the scalar block \eqref{eq:singletblock} whose precise form is given in \cite{Doobary:2015gia} (section 3.5).
The second (differential) operator can be written in terms of a differential operator on the $\SL(2m|2n)$-invariant cross-ratio matrix $Z^{\fa}_{\ \fb}$~\eqref{eq:Z} as
\begin{equation}
     (\partial_{\bX_2} \partial_{X_1})_{\dfa \fb} P F^{\gamma \ula}(Z) =  P \hX_{21 \dfa \fb} D^{\fb}_{\ \fa} (Z, \partial_Z) F^{\gamma \ula}(Z)\equiv P \hX_{21 \dfa \fb}  f^{[2]} {}^{\fb}_{\ \fa}(Z)\,,
     \label{DZ10}
\end{equation}
where
\begin{align} \label{DZ1}
    D^{\fa}_{\ \fb} (Z, \partial_Z)={}&
\frac{1}{4} \left(\left(\gamma -\gamma _1-\gamma _2\right) \left(2 (m{-}n)-\gamma -\gamma _1-\gamma _2\right) \delta^{\fa}_{\fb}+\left(\gamma +\gamma _{21}\right) \left(\gamma -\gamma _{43}\right) Z^{\fa}_{\ \fb}\right) \nn \\
&+\frac{1}{2} \Big(\left(2 (m{-}n)-\gamma -\gamma _1-\gamma _2\right)   Z^\fa_{\ \fa_2}\delta^{\fa_1}_{\fb}
+2 Z^\fa_{\ \fb} Z^{\fa_1}_{\ \fa_2}-2 Z^\fa_{\ \fb}\delta ^{\fa_1}_{\fa_2} \nn\\
&+ \left(\gamma _1+\gamma _2 -\gamma\right) \delta^{\fa}_{\fa_2}Z^{\fa_1}_{\ \fb}+\left(2 \gamma +\gamma _{21}-\gamma _{43}\right) Z^\fa_{\ \fa_2}Z^{\fa_1}_{\ \fb}
\Big) (\partial_Z)_{\ \fa_1}^{\fa_2} \\
&+Z^\fa_{\ \fa_4} Z^{\fa_1}_{\ \fb} (Z^{\fa_3}_{\ \fa_2}-\delta^{\fa_3}_{\fa_2}) (\partial_Z)_{\ \fa_1}^{\fa_2} (\partial_Z)_{\ \fa_3}^{\fa_4}\,. \nn
\end{align}
In order to write this in terms of the usual $m+n$ cross-ratios, we can first write an operator acting on a function of the $m+n$ traces
\begin{equation}
F^{\gamma \ula}(Z) = F^{\gamma \ula}(t_k)\,, \quad
t_k \equiv \tfrac1k\text{str} (Z^k)\,, \quad k=1,\ldots,m+n\,.
\end{equation}
Using that
\begin{equation}
\frac{\partial t_k}{\partial Z^\fa_{\ \fb}} =   (Z^{k-1})^\fb_{\ \fa}\,,
\end{equation}
one finds that the operator in \eqref{DZ1} then reads
\begin{equation}
\begin{aligned}
{}& D^{\fa}_{\ \fb} (Z, \partial_Z)=
\frac{1}{4} \left(\left(\gamma -\gamma _1-\gamma _2\right) \left(2 (m-n)-\gamma -\gamma _1-\gamma _2\right) \delta^{\fa}_{\fb}+\left(\gamma +\gamma _{21}\right) \left(\gamma -\gamma _{43}\right) Z^{\fa}_{\ \fb}\right)\\
&+ \sum\limits_{k=1}^{m+n}  \left[
(m{-}n-\gamma)(Z^{k})^\fa_{\ \fb}+ (\text{str} (Z^k)-\text{str} (Z^{k-1})) Z^\fa_{\ \fb}
+\frac12 (2\gamma+\gamma_{21}-\gamma_{43}) (Z^{k+1})^\fa_{\ \fb}
\right] \partial_{t_k}\\
&+ \sum\limits_{k=1}^{m+n} \sum\limits_{l=1}^{k-1} (\text{str} (Z^l)-\text{str} (Z^{l-1}))(Z^{k+1-l})^\fa_{\ \fb}\partial_{t_k}
 \hspace{-1mm}+ \hspace{-1.5mm} \sum\limits_{k_1,k_2=1}^{m+n}\hspace{-1.5mm} 
 \left( (Z^{k_1 + k_2 + 1})^\fa_{\ \fb} - (Z^{k_1 + k_2})^\fa_{\ \fb} \right) \partial_{t_{k_1}}\partial_{t_{k_2}}.
\end{aligned}
\label{op_tk}
\end{equation}
We can further use the relation
\begin{equation}
t_k = \frac1k \left( \sum\limits_{i=1}^{m} z_i^k - \sum\limits_{j=1}^{n} w_j^k\right)\,,
\end{equation}
to write \eqref{op_tk} as a differential operator in the cross-ratios $z_i, w_j$. 
The Jacobian is (the inverse of) the (suitably supersymmetrised) Vandermonde matrix, 
\begin{equation}
    \partial_{z_i} = \sum_{k=1}^{m+n} z_i^{k-1} \partial_{t_k}\ , \qquad
    \partial_{w_j} = -\sum_{k=1}^{m+n} w_j^{k-1} \partial_{t_k}\ .
\end{equation}
So for example, the Jacobian matrix for the case $(m,n)=(2,1)$, i.e.\ 4D $\cN=2$, is given by
\begin{equation}
\left(
\begin{array}{c}
\partial_{t_1} \\
\partial_{t_2} \\
\partial_{t_3} \\
\end{array}
\right)
=
\left(
\begin{array}{ccc}
 \frac{w_1 z_2}{\left(z_2-z_1\right) \left(w_1-z_1\right)} & \frac{w_1 z_1}{\left(z_1-z_2\right) \left(w_1-z_2\right)} & \frac{z_1 z_2}{\left(z_1-w_1\right)
   \left(w_1-z_2\right)} \\
 \frac{w_1+z_2}{ \left(z_1-z_2\right) \left(w_1-z_1\right)} & \frac{w_1+z_1}{ \left(z_2-z_1\right) \left(w_1-z_2\right)} & \frac{z_1+z_2}{ \left(w_1-z_1\right)
   \left(w_1-z_2\right)} \\
 \frac{1}{ \left(z_2-z_1\right) \left(w_1-z_1\right)} & \frac{1}{ \left(z_1-z_2\right) \left(w_1-z_2\right)} & \frac{1}{ \left(z_1-w_1\right) \left(w_1-z_2\right)} \\
\end{array}
\right)
\left(
\begin{array}{c}
\partial_{z_1} \\
\partial_{z_2} \\
\partial_{w_1} \\
\end{array}
\right)\,.
\end{equation}
We checked that the two operators in \eqref{wso_ex1} produce the correctly shifted blocks in 1D CFTs in \eqref{1d_c1} and \eqref{1d_c2} and the known spinning conformal blocks in 4D CFTs in \eqref{4d_c1} and \eqref{check5}.

\subsubsection{Two general (anti-)chiral operators $\langle \cO_{\umu,\bullet} \cO_{\bullet,\umu} \cO \cO\rangle$}\label{sec:genchiral}
For $\umu=[\mu]$, this corresponds to a correlator of higher-spin current multiplets in 4D $\cN=2$, given by
\begin{equation}
    \cO_{[\mu],\bullet} \leftrightarrow A_1 \bar{B}[\mu-1,0]^{(\gamma-1;1+\mu)}_{\gamma + \frac12(\mu -1)} , \qquad \cO_{\bullet,[\mu]} \leftrightarrow B \bar{A}_1[0,\mu-1]^{(\gamma-1;-1-\mu)}_{\gamma + \frac12(\mu -1)}\,.
\end{equation}
See appendix \ref{sec:repexamples} for examples involving more general $\umu$.
Again, the only exchanged representation in the superconformal block expansion is symmetric $\cO^\gamma_{\ula}$, so the blocks are given by
\begin{equation}\label{eq:genchiralbl}
    \langle \cO_{\umu,\bullet}  \cO_{\bullet, \umu} \cO^{\gamma}_{\ula}\rangle^{(i)} \bowtie \langle\cO^{\gamma}_{\ula} \cO  \cO \rangle.
\end{equation}
If $|\ula|\geq |\umu|$, there are $|\umu|+1$ different tensor structures associated with the exchange, corresponding to the structures of the correlator on the LHS. In the standard basis they are given by \eqref{chiralstandardbasis} and, as explained in detail in section \ref{sec:diffbasis}, the differential basis is achieved by the differential operators $\mathfrak{D}^{[k]}$ \eqref{eq:mfrakD1}. One can then directly perform the operation \eqref{eq:wsoex10} to the LHS of the singlet block \eqref{eq:singletblock} to obtain the following form of the superconformal block \eqref{eq:genchiralbl} in the differential basis
\begin{equation}\label{eq:wsoex11}
    \begin{tikzpicture}[baseline={(current bounding box.center)}, scale=0.8]
\node (tl2) at (-3.8,1.08) [threept] {};
\node (bl2) at (-3.8,-1.08) [threept] {};
\node (tl) at (-2.16,1.08) [threept] {};
\node (vertr) at (0.72,0) [threept] {};
\node (verttl) at (-0.84,1.08) [threept] {};
\node (vertbl) at (-0.84,-1.08) [threept] {};
\node (vvt) at (4.68,1.03) [above] {$\cO^{\gamma_4}$};
\node (vvb) at (4.68,-1.03) [below] {$\cO^{\gamma_3}$};
\node (bl) at (-2.16,-1.08) [threept] {};
\node (r1) at (1.92,0.12) [above] {$\cO_{\ula}^\gamma$};
\node (r) at (1.92,0) [bowtie] {$\bowtie$};
\node (vv) at (3.12,0) [threept] {};
\node (f1) at (0.38,0.7) [above] {$\cO^{\gamma_1}$};
\node (f2) at (0.38,-0.7) [below] {$\cO^{\gamma_2}$};
\node at (-2.9,1.08) {$\cdots$};
\node at (-2.9,-1.08) {$\cdots$};
\node (D1) at (-0.84,0) [diff] {$\scriptstyle i_1$};
\node (D2) at (-2.16,0) [diff] {$\scriptstyle i_2$};
\node (D3) at (-3.8,0) [diff] {$\scriptstyle i_\mu$};
\node (O1) at (-4.6,1.09) [left] {$\cO^{\gamma_1}_{\umu,\bullet}$};
\node (O2) at (-4.6,-1.09) [left] {$\cO^{\gamma_2}_{\bullet,\umu}$};
\node at (-1.6,1.2) [above] {\small $\cO^{\gamma_1}_{\Box,\bullet}$};
\node at (-1.6,-1.2) [below] {\small $\cO^{\gamma_2}_{\bullet,\Box}$};
\draw [spinning] (verttl) -- (tl);
\draw [spinning] (vertbl) -- (bl);
\draw [spinning] (vertr) -- (r);
\draw [scalar] (vertr) -- (verttl);
\draw [scalar] (vertr) -- (vertbl);
\draw [finite] (verttl) -- (D1);
\draw [finite] (D1) -- (vertbl);
\draw [finite] (tl) -- (D2);
\draw [finite] (D2) -- (bl);
\draw [finite] (tl2) -- (D3);
\draw [finite] (D3) -- (bl2);
\draw[thick] (tl)--(-2.5,1.08);
\draw[thick] (bl)--(-2.5,-1.08);
\draw[thick] (tl2)--(-3.4,1.08);
\draw[thick] (bl2)--(-3.4,-1.08);
\draw[spinning] (tl2)--(O1);
\draw[spinning] (bl2)--(O2);
\draw[scalar] (vv)--(vvt);
\draw[scalar] (vv)--(vvb);
\draw[spinning] (vv)--(r);
\end{tikzpicture}.
\end{equation}

\subsubsection{$\langle \cO_{\dfa \fa} \cO \cO \cO\rangle$}\label{sec:extblockexample2}
The insertion at point $X_1$ has many relevant physical interpretations in $(m,n)$ analytic superspace. Firstly, in the $(2,0)$ theory, it  corresponds to a 4D vector field. Furthermore, note that for $\gamma=2$ it corresponds to the stress-tensor multiplet in 4D $\cN=2$. It is also the simplest example of a quarter-BPS multiplet in 4D $\cN=4$, given by $B \bar{B} [0,0]^{[1,\gamma-2,1]}_{\gamma}$. Note that, although it is absent in $\SU (N)$ SYM, it is always an intermediate step for mapping to other quarter-BPS multiplets from the corresponding half-BPS one, as shown by \eqref{eq:WSOmaps2}. See appendix \ref{sec:repexamples} for the full list of examples.

Again, the only possible exchange in this case is that of symmetric operators. Thus, all blocks are given by the following gluing of three-point structures
\begin{equation}\label{eq:block1}
    \langle\cO^{\gamma_1}_{\dfa \fa} \cO^{\gamma_2} \cO_{\ula}^\gamma\rangle \bowtie \langle \cO_{\ula}^\gamma\cO^{\gamma_3} \cO^{\gamma_4} \rangle.
\end{equation}
 The left three-point structure in \eqref{eq:block1} admits two different tensor structures given in the standard basis by \eqref{eq:threepointstructures}
\begin{equation}\label{eq:stdbasis}
    \begin{split}
         \langle\cO^{\gamma_1}_{\dfa \fa} \cO^{\gamma_2} \cO_{\ula}^\gamma\rangle^{(1)} &= K (Y_1)_{\dfa \fa} (Y_0^{|\ula|})_{\udfb \ufb}, \\
         \langle\cO^{\gamma_1}_{\dfa \fa} \cO^{\gamma_2} \cO_{\ula}^\gamma\rangle^{(2)} &= K (\hX_{10})_{\dfa \fb} (\hX_{01})_{\dfb \fa} (Y_0^{|\ula|-1})_{\udfb \ufb},
    \end{split}
\end{equation}
where we label the coordinate of the exchanged operator by $X_0$ and the $\dfb,\fb$ indices are to be symmetrised accordingly. This means that there are two different blocks for each exchange in the expansion of $\langle\cO_{\dfa \fa} \cO \cO \cO \rangle$. 

The differential basis for this case was constructed in section \ref{sec:diffbasisgeneral}. In particular, one simply has to use the invariant differential operator \eqref{diffppex} on the chiral structure of the block \eqref{eq:fermionicblock} which we previously derived. Thus one can immediately write a differential basis for \eqref{eq:block1} from the differential operators
\begin{equation}
    (\bar{\cD}_1 \bX_2)_{\dfa_1}^{\ \dfa_2} (\hX_{21})_{\dfa_2 \fa_1}, \qquad \qquad (\bar{\cD}_1 \bX_2)_{\dfa}^{\ \dfb} (\bar{\cD}_2 \cD_1)_{\dfb \fa}.
    \label{easybasis}
\end{equation}
However, the first term is not as simple as it can be as $\bar{\cD}_1$ acts last. It is more convenient to place the differential operators acting first on the scalar block as they have the simplest form (indeed they can often be reduced to $\partial_X$ or $\partial_{\bX}$ as explained in section \ref{sec:grassmannianWSO}). This will give the most straightforward form of the blocks. Thus, we define the differential basis for \eqref{eq:block1} as
\begin{equation}\label{eq:O01WSos1}
    \begin{split}
    \fD^{[1]}_{\dfa \fa} &= \hX_{12 \dfa \fc} (X_2)^{\fc}_{\ \cA} (\cD_1)^{\cA}{}_{\fa} \equiv (\hX_{12}X_2 \partial_{X_1})_{\dfa \fa}, \\
    \fD^{[2]}_{\dfa \fa} &=  (\bX_2)^{\cB \dfc} (\bar{\cD}_1)_{\dfa \cB} (\bar{\cD}_2)_{\dfc \cA}  (\cD_1)^{\cA}{}_{\fa} \equiv (\bar{\cD}_1 \bX_2)_{\dfa}^{\ \dfb} (\partial_{\bX_2} \partial_{X_1})_{\dfb \fa} ,
    \end{split}
\end{equation}
where $(\bar{\cD}_1)_{\dfa \cA}$ acts on $\partial_{X_1} \cO_\bullet \sim \cO_{\fa}$ and so we need the full weight-shifting operator, given in \eqref{eq:chiralWSO1}.

As in the chiral example, the action of these weight-shifting operators on the conformal blocks can be expressed in a form that explicitly solves the Ward identities \eqref{WI1}, given as a differential operator on the cross-ratio matrix $Z^{\fa}_{\fb}$~\eqref{eq:Z}. The first yields
\begin{equation}
\fD^{[1]}_{\dfa \fa} P F^{\gamma \ula}(Z)= P 
 Y_{1,24 \dfa \fa} D^{(1)\fb}_{\hspace{15pt} \fa} (Z, \partial_Z)
F^{\gamma \ula}(Z)\,,
\label{DZ20}
\end{equation}
with
\begin{equation}
\begin{aligned}
D^{(1)\fb}_{\hspace{15pt} \fa} (Z, \partial_Z) ={}& 
\frac{1}{2} \left(\gamma-\gamma _{43} \right) Z^{\fb}_{\ \fa}
+ \frac12 \left( \gamma_{21} - \gamma \right)
\delta^\fb_{\fa}
-(\delta^\fb_{\fa_2} - Z^{\fb}_{\ \fa_2}) Z^{\fa_1}_{\ \fa} (\partial_Z)^{\fa_2}_{\ \fa_1}\\
={}& \frac{1}{2} \left(\gamma-\gamma _{43} \right) Z^{\fb}_{\ \fa}
+ \frac12 \left( \gamma_{21} - \gamma \right)
\delta^\fb_{\fa}+ \sum\limits_{k=1}^{m+n}\left( (Z^{k+1})^\fb_{\ \fa} -  (Z^{k})^\fb_{\ \fa}\right) \partial_{t_k}\,.
\end{aligned}
\label{DZ2}
\end{equation}
The second operator in \eqref{eq:O01WSos1} can be expressed similarly, however we do not include it here as it is third order in weight-shifting operators, and up to fourth order in derivatives, and thus a rather long expression in terms of cross-ratios.
We did however check in \eqref{check3} that both operators in \eqref{easybasis} give the correct result in 1D and in \eqref{4d_c3} that \eqref{DZ20} gives one of the known spinning 4D blocks.
As we will discuss in section \ref{sec:blocksummary}, the method for using weight-shifting operators most effectively is to build blocks by recursion, i.e. adding boxes one-by-one and storing the intermediate results. This avoids complicated expressions.

\subsubsection{General symmetric insertion $\langle \cO_{\umu}\cO \cO \cO\rangle$}
For $\umu=[1^\mu]$, the first insertion in the above correlator is a general quarter-BPS supermultiplet $B \bar{B}[0,0]_{\gamma}^{\mu,\gamma-2\mu,\mu}$ in 4D $\cN=4$. For  $\mu=2$, $\gamma=4$ the first non-trivial quarter-BPS multiplet in interacting $\SU (N)$ super Yang-Mills, as studied in \cite{Bissi:2021hjk}.\footnote{It corresponds to $\cO_{02}$ in their notation.} 
See appendix~\ref{sec:repexamples} for the full list of examples. There are now $|\umu|+1$ structures associated with the symmetric exchange $\cO^\gamma_{\ula}$, as already shown in \eqref{eq:threepointstructures}. The process for constructing the corresponding $|\umu|+1$ weight-shifting operators was outlined in section \ref{sec:diffbasisgeneral}. In particular, we use \eqref{diffppex} as in \eqref{eq:diffopp11} to each of the corresponding general chiral blocks \eqref{eq:genchiralbl} generated by $\fD^{[i]}$ \eqref{eq:mfrakD1}, giving
\begin{equation}\label{eq:gensymex}
    \langle \cO_{\umu,\umu} \cO \cO^\gamma_{\ula}\rangle^{[i]} \bowtie \langle \cO^\gamma_{\ula}\cO \cO\rangle \equiv\bPi^{\umu}_1(X_1)(\bD_1 \bX_2)^{\otimes |\umu|} \langle \cO_{\umu,\bullet} \cO_{\bullet, \umu} \cO^\gamma_{\ula}\rangle^{[i]} \bowtie \langle \cO^\gamma_{\ula} \cO \cO\rangle.
\end{equation}
\Yboxdim{8pt}
As an application, we can write the three blocks for $\Yvcentermath1 \umu= \yng(1,1)$ as differential operators on the singlet block $\cF^{\gamma \ula}$ \eqref{eq:singletblock} as follows
\Yboxdim{5pt}
\begin{equation}
    \Yvcentermath1
    \langle \cO_{\yng(1,1),\yng(1,1)}^{\gamma_1} \cO^{\gamma_2} \cO^\gamma_{\ula}\rangle^{[i]} \bowtie \langle \cO^\gamma_{\ula}\cO^{\gamma_3} \cO^{\gamma_4}\rangle = (\bD_1 \bX_2)_{[\dfb_1}^{\ \dfb_2} (\bD_1 \bX_2)_{\dfa_1]}^{\ \dfa_2} \fD^{[i]}_{[\dfa_2 \dfb_2] [\fa_1 \fb_1]} \cF^{\gamma \ula}_{\gamma_1 \gamma_2 \gamma_3 \gamma_4},
\end{equation}
which for $\gamma_1=4$ is very relevant to $\cN=4$ SYM as explained above. For completeness, we write the relevant $\fD^{[i]}$ operators \eqref{eq:mfrakD1} below
\begin{equation}
    \begin{split}
        \fD^{[1]}_{[\dfa_2 \dfb_2] [\fa_1 \fb_1]}&= \hX_{21} {}_{[\dfb_2 [\fb_1} \hX_{21}{}_{\dfa_2] \fa_1]}, \\ 
        \fD^{[2]}_{[\dfa_2 \dfb_2] [\fa_1 \fb_1]}&= \hX_{21} {}_{[\dfb_2 [\fb_1}(\partial_{\bX_2}\partial_{X_1})_{\dfa_2] \fa_1]}, \\ 
        \fD^{[3]}_{[\dfa_2 \dfb_2] [\fa_1 \fb_1]}&= (\partial_{\bX_2}\partial_{X_1})_{[\dfb_2 [\fb_1} (\partial_{\bX_2}\partial_{X_1})_{\dfa_2] \fa_1]}.
    \end{split}
\end{equation}

\Yboxdim{13pt}

\subsection{Internal shift and seed blocks}
\label{sec:internalshift}
So far we have only discussed blocks with symmetric reps as exchanged operators, which can be generated by applying external weight-shifting operators on scalar blocks. As discussed in section \ref{subsec:blocks}, blocks with  more general exchanged operators can similarly be generated from the seed blocks~\eqref{eq:seedblock}. 
To derive the seed blocks themselves, we can use weight-shifting operators which  shift the internal representation just as for the non-supersymmetric case~\cite{Karateev:2017jgd}. Let us see how this works in practice.

Consider the seed block for an exchanged operator $\cO^\gamma_{\ula, \ula+1}$ where $\ula+1$ denotes a Young diagram obtained from adding a box to $\ula$ (so $\ula \cap (\ula+1) = \ula$).  Then the corresponding seed block is~\eqref{eq:seedblock}
 with $\umu_1=\Box$, $\umu_2=\bullet$, so an insertion at point 1 of a spin $1/2$ fermion $\psi_{\alpha}$ in 4D or a chiral vector multiplet $A_2 \bar{B}[0,0]_{\gamma}^{(\gamma-1,2)}$ in 4D $\cN=2$:
\begin{equation}\label{eq:seedblockex}
    [\cF^{\gamma, \ula,\ula+1}_{\text{seed}}]_{\fa \dfa} \coloneq\langle \cO^{\gamma_1}_{\fa} \cO^{\gamma_2} \cO^\gamma_{\ula, \ula+1}\rangle \bowtie \langle \cO_{\ula+1,\ula}^\gamma \cO^{\gamma_3}_{\dfa} \cO^{\gamma_4} \rangle = \diagramEnvelope{\begin{tikzpicture}[baseline=(align),scale= 0.8]
    \node (vertL) at (-1.5,0) [threept] {};
    \node (vertR) at (1.5,0) [threept] {};
    \node (opO1) at (-2.25,-1.125) [below] {$\cO^{\gamma_2}$};
    \node (opO2) at (-2.25,1.125) [above] {$\cO^{\gamma_1}_{\fa}$};
    \node (opO3) at (2.25,1.125) [above] {$\cO^{\gamma_3}_{\dfa}$};
    \node (opW) at (2.25,-1.125) [below] {$\cO^{\gamma_4}$};
    \node (shad) at (0,0) [bowtie] {$\bowtie$};
    \node at (-0.8,0.1) [above] {$\cO^\gamma_{{\scriptscriptstyle \ula, \ula+1}}$};
    \node at (0.8,-0.1) [below] {$\cO^\gamma_{{\scriptscriptstyle \ula+1, \ula}}$};
    \draw [scalar] (vertL)-- (opO1);
    \draw [spinning] (vertL)-- (opO2);
    \draw [spinning] (vertL)-- (shad);
    \draw [spinning] (vertR) -- (shad);
    \draw [spinning] (vertR)-- (opO3);
    \draw [scalar] (vertR)-- (opW);
    \node (align) at (0,-.1) {};
\end{tikzpicture}}.
\end{equation}
 Indeed, the single LHS three-point structure is
\begin{equation}
    \langle \cO^{\gamma_1}_{\fa} \cO^{\gamma_2} \cO^\gamma_{\ula, \ula+1}\rangle = K  \Pi^{\ula} \bPi^{\ula+1}\left((Y_0)^{\otimes |\ula|} \hX_{01} \right),
\end{equation}
where we denote the exchanged insertion by $X_0$ as before.

The aim of this section, like in the previous section, is to relate the above block to the singlet block for an exchanged symmetric representation $\cO^{\gamma}_{\ula, \ula}$ given in \eqref{eq:singletblock}.
The difference with the previous cases is that now we are also changing the exchanged representation, that is, mapping $\cO^\gamma_{\ula, \ula} \mapsto \cO^\gamma_{\ula, \ula+1}$.

Let us consider the case $\ula=\bullet$ in detail.
The first step is to rewrite the three-point structure on the right-hand side in the differential basis, but this time involving point~$X_0$ 
\begin{equation}
    \langle \cO_{\fb}^\gamma (X_0) \cO^{\gamma_3}_{\dfa} (X_3)\cO^{\gamma_4} (X_4)\rangle = \partial_{\bX_3}{}_{\dfa \cA} \partial_{X_0}{}^{\cA}_{\ \fb} \langle \cO^\gamma (X_0)\cO^{\gamma_3} (X_3)\cO^{\gamma_4}(X_4) \rangle.
\end{equation}
Now, in order to express \eqref{eq:seedblockex} as derivatives on a scalar block, we need to move the weight-shifting operator acting on the insertion of the exchanged operator $X_0$ through the shadow integral by integrating by parts and performing crossing on the two-point function kernel in the integral. In terms of diagrams, this corresponds to the following operation
\begin{equation}\label{eq:gluecrossing}
    \begin{tikzpicture}[baseline=(align),scale= 0.75]
        \node (align) at (0,-.1) {};
        \node (vertR) at (1.5,0) [threept] {};
         \node (vertL) at (-3,0) [left] {$\cdots$};
        \node (opO3) at (3,2) [above] {$\cO^{\gamma_3}_{\dfa}$};
        \node (opW) at (2.25,-1.125) [right] {$\cO^{\gamma_4}$};
        \node at (0.8,-0.1) [below] {$\cO^\gamma$};
        \node at (-0.8,-0.1) [below] {$\cO^\gamma_{\fb}$};
        \node at (-2.4,-0.1) [below] {$\cO^\gamma_{\dfb}$};
         \node (shad) at (-1.5, 0) [bowtie] {$\bowtie$};
        \node (3p) at (2.25, 1.125) [threept] {}; 
        \node (ex0) at (0, 0) [threept] {};
        \draw [spinning] (ex0) -- (shad);
        \draw [scalar] (vertR) -- (ex0);
        \draw [scalar] (vertR)-- (3p);
        \draw [spinning] (3p)-- (opO3);
        \draw [scalar] (vertR)-- (opW);
        \draw [finite with arrow] (ex0) to[out=90, in=130] (3p);
        \draw [spinning] (vertL)-- (shad);
    \end{tikzpicture} 
    \longrightarrow \
    \begin{tikzpicture}[baseline=(align),scale= 0.75]
        \node (align) at (0,-.1) {};
        \node (vertR) at (1.5,0) [threept] {};
         \node (vertL) at (-3,0) [left] {$\cdots$};
        \node (opO3) at (3,2) [above] {$\cO^{\gamma_3}_{\dfa}$};
        \node (opW) at (2.25,-1.125) [right] {$\cO^{\gamma_4}$};
        \node at (0,-0.2) [below] {$\cO^\gamma$};
        \node at (-2.4,-0.1) [below] {$\cO^\gamma_{\dfb}$};
        \node (shad) at (-1.5,0) [threept] {};
        \node (3p) at (2.25, 1.125) [threept] {}; 
          \node (ex0) at (0, 0) [bowtie] {$\bowtie$};
        \draw [scalar] (shad) -- (ex0);
        \draw [scalar] (vertR) -- (ex0);
        \draw [scalar] (vertR)-- (3p);
        \draw [spinning] (3p)-- (opO3);
        \draw [scalar] (vertR)-- (opW);
        \draw [finite with arrow] (shad) to[out=90, in=130] (3p);
        \draw [spinning] (vertL)-- (shad);
    \end{tikzpicture}.
\end{equation}
As shown in the non-supersymmetric case in \cite{Karateev:2017jgd}, one can relate differential operators acting on each side of the gluing by using the crossing equation \eqref{eq:crossing}. 
This works equivalently in our $(m,n)$ superconformal setting.
In particular, one can turn the three-point crossing equation \eqref{eq:crossing} into a crossing for two-point functions by setting one of the insertions of the three-point function to the trivial representation. In our present case, the relevant two-point crossing is
\begin{equation}\label{eq:twopointcrossing}
    \begin{tikzpicture}[baseline=(align),scale= 0.85]
         \node (align) at (0,-.1) {};
        \node at (0.8,-0.1) [below] {$\cO^\gamma_{\fb}$};
        \node at (-0.8,-0.1) [below] {$\cO^\gamma$};
        \node at (-2.4,-0.1) [below] {$\cO^\gamma$};
        \node (vertR) at (1.5,0) {};
        \node (shad) at (-1.5,0) [twopt] {};
        \node (ex0) at (0, 0) [threept] {};
        \node (vertL) at (-3,0){};
        \draw [scalar] (shad) -- (ex0);
        \draw [spinning] (ex0) -- (vertR);
        \draw [finite with arrow] (ex0) to (0,1.5);
        \draw [scalar] (shad)-- (vertL);
    \end{tikzpicture} = - \gamma \begin{tikzpicture}[baseline=(align),scale= 0.85]
         \node (align) at (0,-.1) {};
        \node at (-2.3,-0.1) [below] {$\cO^\gamma$};
        \node at (-0.7,-0.1) [below] {$\cO^\gamma_{\dfb}$};
        \node at (0.9,-0.1) [below] {$\cO^\gamma_{\fb}$};
        \node (vertR) at (-3.0,0) {};
        \node (shad) at (0.0,0) [twopt] {};
        \node (ex0) at (-1.5, 0) [threept] {};
        \node (vertL) at (1.5,0){};
        \draw [spinning] (shad) -- (ex0);
        \draw [scalar] (ex0) -- (vertR);
        \draw [finite with arrow] (ex0) to (-1.5,1.5);
        \draw [spinning] (shad)-- (vertL);
\end{tikzpicture},
\end{equation}
where the coefficient of proportionality is found by direct calculation. Indeed,
\begin{equation}
    \partial_{X_2}{}^{\cA}_{\ \fb} \langle \cO^\gamma(X_1) \cO^\gamma (X_2)\rangle = -\gamma \bX_{1}^{\cA \dfb} \langle \cO^\gamma_{\dfb}(X_1) \cO_{\fb}^\gamma(X_2) \rangle
\end{equation}
It is important to note that equivalently to the crossing equation \eqref{eq:crossing}, when the finite representation is more general, one should expect more terms on the RHS, one for each different weight-shifting operator in the corresponding representation.

Now, equation \eqref{eq:twopointcrossing} can then be turned into a crossing of the form of \eqref{eq:gluecrossing} by acting with a shadow integral on both sides and using the normalisation \eqref{eq:gluingnormalisation}. 
The result is that both sides of \eqref{eq:gluecrossing} are related by a coefficient of proportionality, which is precisely the one governing the crossing of two-point functions \eqref{eq:twopointcrossing}. Namely,
\begin{equation}\label{eq:gluecrossing2}
    \begin{tikzpicture}[baseline=(align),scale= 0.75]
        \node (align) at (0,-.1) {};
        \node (vertR) at (1.5,0) [right] {};
         \node (vertL) at (-3,0) [left] {};
        \node at (0.8,-0.1) [below] {$\cO^\gamma$};
        \node at (-0.8,-0.1) [below] {$\cO^\gamma_{\fb}$};
        \node at (-2.4,-0.1) [below] {$\cO^\gamma_{\dfb}$};
         \node (shad) at (-1.5, 0) [bowtie] {$\bowtie$};
        \node (3p) at (0, 1.5) [above] {}; 
        \node (ex0) at (0, 0) [threept] {};
        \draw [spinning] (ex0) -- (shad);
        \draw [scalar] (vertR) -- (ex0);
        \draw [finite with arrow] (ex0) to (3p);
        \draw [spinning] (vertL)-- (shad);
    \end{tikzpicture} 
    = -\gamma
    \begin{tikzpicture}[baseline=(align),scale= 0.75]
        \node (align) at (0,-.1) {};
        \node (vertR) at (1.5,0) [right] {};
         \node (vertL) at (-3,0) [left] {};
        \node at (0,-0.2) [below] {$\cO^\gamma$};
        \node at (-2.4,-0.1) [below] {$\cO^\gamma_{\dfb}$};
        \node (shad) at (-1.5,0) [threept] {};
        \node (3p) at (-1.5, 1.5) [above] {}; 
          \node (ex0) at (0, 0) [bowtie] {$\bowtie$};
        \draw [scalar] (shad) -- (ex0);
        \draw [scalar] (vertR) -- (ex0);
        \draw [finite with arrow] (shad) to (3p);
        \draw [spinning] (vertL)-- (shad);
    \end{tikzpicture},
\end{equation}
which, in symbolic notation reads
\begin{equation}\label{eq:shadowintcrossing}
    \vert  \cO_{\dfb}^\gamma \rangle \bowtie^{\dfb \fb} \langle  \partial_{X_0}{}^{\cA}_{\ \fb}\cO^{\gamma}\vert = - \gamma \ \vert \bX_0^{\cA \dfb} \cO_{\dfb}^\gamma \rangle \bowtie \langle \cO^{\gamma} \vert,
\end{equation}
where we explicitly show the index structure of the gluing operation $\bowtie$ to emphasise that it is an overall singlet of the Levi subgroup. 

We thus arrive at the following expression for the seed block
\begin{equation}
     [\cF^{\gamma, \bullet,\Box}_{\text{seed}}]_{\fa \dfa} =  - \gamma \partial_{\bX_{3}}{}_{\dfa \cA} \bX_0^{\cA \dfb} \langle \cO^{\gamma_1}_{\fa} \cO^{\gamma_2} \cO^{\gamma}_{\dfb} \rangle \bowtie \langle \cO^\gamma \cO^{\gamma_3} \cO^{\gamma_4} \rangle. 
\end{equation}
The last step is to perform the three-point crossing \eqref{eq:crossing} to express the action of $\bX_0$ in the basis of weight-shifting operators acting at a different point. The calculation in the present case is as follows
\begin{equation}\label{eq:internalshiftcrossing}
        \bX_0^{\cA \dfb} \langle \cO^{\gamma_1}_{\fa} \cO^{\gamma_2} \cO^{\gamma}_{\dfb} \rangle = c_1 \partial_{X_1}{}^{\cA}_{\ \fa} \langle \cO^{\gamma_1} \cO^{\gamma_2} \cO^{\gamma} \rangle + c_2 \bX_1^{\cA \dfb} \langle \cO^{\gamma_1}_{\dfb \fa} \cO^{\gamma_2} \cO^{\gamma} \rangle,
\end{equation}
where each three-point structure was constructed in section \ref{sec:Wardids} and $c_1,c_2$ are $6j$ symbols found by explicit calculation to be
\begin{align}
     c_1 \equiv \left\{\begin{array}{ccc}
        \cO^{\gamma_1}_{\fa} & \cO^{\gamma_2} & \cO^{\gamma_1} \\ 
        \ \cO^{\gamma} & \frep & \cO^{\gamma}_{\dfb}
    \end{array}\right\} & = - \frac{1}{\gamma_1}\, , & \ \ c_2 \equiv \left\{\begin{array}{ccc}
        \cO^{\gamma_1}_{\fa} & \cO^{\gamma_2} & \cO^{\gamma_1}_{ \dfb \fa} \\ 
        \ \cO^{\gamma} & \frep & \cO^{\gamma}_{\dfb}
    \end{array}\right\}&= \frac{\gamma_{12,0}}{\gamma_1}\, .
\end{align}
Note that if we had taken a more general exchange with $\ula\neq \bullet$, the above expansion would have more terms.
Then, inserting this into \eqref{eq:seedblockex}, we find the following expression for our seed block
\begin{equation}
    [\cF^{\gamma, \bullet,\Box}_{\text{seed}}]_{\fa \dfa} = \frac{\gamma}{\gamma_1} \partial_{\bX_3}{}_{\dfa \cA} \left( \partial_{X_1}{}^{\cA}_{\ \fa} \langle \cO^{\gamma_1} \cO^{\gamma_2} \cO^\gamma\rangle - \gamma_{12,0} \bX_1^{\cA \dfb} \langle \cO^{\gamma_1}_{\dfb \fa} \cO^{\gamma_2} \cO^\gamma\rangle\right) \bowtie \langle \cO^\gamma \cO^{\gamma_3} \cO^{\gamma_4}\rangle.
\end{equation}
Now, observe that the first term corresponds to a scalar block, whereas the second term corresponds to a block we already considered in section \ref{sec:extblockexample2}. 
But it corresponds to the degenerate case where the exchanged operator is a scalar $\cO^\gamma$, which only contains one tensor structure.
Thus, either weight-shifting operator \eqref{eq:O01WSos1} gives the right answer (with different normalisations). Thus, we choose the first order one and arrive at the following remarkably simple formula
\begin{equation}\label{eq:1seedblock}
    \begin{split}
        [\cF^{\gamma, \bullet,\Box}_{\text{seed}}]_{\fa \dfa} &= \frac{\gamma}{\gamma_1} \partial_{\bX_3}{}_{\dfa \cA} \left( \partial_{X_1}{}^{\cA}_{\ \fa}  +\frac{\gamma_{12,0}}{\gamma_{10,2}} \bX_1^{\cA \dfb}( \hX_{12} X_2 \partial_{X_1})_{\dfb \fa}\right) \cF^{\gamma,\bullet}_{\gamma_1 \gamma_2 \gamma_3 \gamma_4} 
         \\ 
         &= \frac{\gamma}{\gamma_{10,2}}\left((\partial_{\bX_3} \partial_{X_1})_{\dfa \fa} + \gamma_{12,0}  (\partial_{\bX_3}\bX_2 \hX_{21})_{\dfa \fa} \right) \cF^{\gamma,\bullet}_{\gamma_1 \gamma_2 \gamma_3 \gamma_4},
    \end{split}
\end{equation}
where to get the second line we used the identity~\eqref{keyid} and then that $X_1 \partial_{X_1}= - \gamma_1$. 
The above formula for the seed block corresponds to the diagram below
\begin{equation}\label{eq:internalshiftdiagram}
\begin{split}
                \diagramEnvelope{\begin{tikzpicture}[baseline=(align),scale= 0.7]
    \node (vertL) at (-1.5,0) [threept] {};
    \node (vertR) at (1.5,0) [threept] {};
    \node (opO1) at (-2.25,-1.125) [below] {$\cO^{\gamma_2}$};
    \node (opO2) at (-2.25,1.125) [above] {$\cO^{\gamma_1}_{\fa}$};
    \node (opO3) at (2.25,1.125) [above] {$\cO^{\gamma_3}_{\dfa}$};
    \node (opW) at (2.25,-1.125) [below] {$\cO^{\gamma_4}$};
    \node (shad) at (0,0) [bowtie] {$\bowtie$};
    \node at (-0.8,0.1) [above] {$\cO^{\gamma}_{\dfb}$};
    \node at (0.8,0.18) [above] {$\cO^{\gamma}_{\fb}$};
    \draw [scalar] (vertL)-- (opO1);
    \draw [spinning] (vertL)-- (opO2);
    \draw [spinning] (vertL)-- (shad);
    \draw [spinning] (vertR) -- (shad);
    \draw [spinning] (vertR)-- (opO3);
    \draw [scalar] (vertR)-- (opW);
    \node (align) at (0,-.1) {};
\end{tikzpicture}} &= \frac{\gamma}{\gamma_{10,2}}  
\diagramEnvelope{\begin{tikzpicture}[baseline=(align), scale=0.7]
    \node (vertL) at (-1.5,0) [threept] {};
    \node (vertR) at (1.5,0) [threept] {};
    \node (opO2) at (-1.95,0.4) [left] {$\cO^{\gamma_1}$};
    \node (opO3) at (1.95,0.4) [right] {$\cO^{\gamma_3}$};
    \node (opO11) at (-2.25,-1.125) [below] {$\cO^{\gamma_2}$};
    \node (opO22) at (-3.6,1.7) [above] {$\cO^{\gamma_1}_{\fa}$};
    \node (opO33) at (3.6,1.7) [above] {$\cO^{\gamma_3}_{\dfa}$};
    \node (opO44) at (2.25,-1.125) [below] {$\cO^{\gamma_4}$};
    \node (o2) at (2.25,1.05) [threept] {};
    \node (o4) at (-2.25,1.05) [threept] {};
    \node (shad) at (0,0) [bowtie] {$\bowtie$};
    \node at (0,-0.15) [below] {$\cO^{\gamma}$};		
    \draw [scalar] (vertL)-- (opO11);
    \draw [scalar] (vertL)-- (o4);
    \draw [scalar] (vertL)-- (shad);
    \draw [scalar] (vertR) -- (shad);
    \draw [scalar] (vertR)-- (opO44);
    \draw [scalar] (vertR)-- (o2);
    \draw [spinning] (o4)-- (opO22);
    \draw [spinning] (o2)-- (opO33);
    \draw [finite with arrow] (o4)-- (o2);
     \node (align) at (0,-.1) {};
\end{tikzpicture}} \\ 
& \qquad + \frac{\gamma_{12,0} \gamma}{\gamma_{10,2}}  \diagramEnvelope{\begin{tikzpicture}[baseline=(align), scale=0.7]
    \node (vertL) at (-1.2,1) [threept] {};
    \node (vertR) at (1.2,1) [threept] {};
    \node (opO2) at (-2.1,-0.4) [left] {$\cO^{\gamma_2}_{\dfb}$};
    \node (opO3) at (1.9,0.2) [right] {$\cO^{\gamma_3}$};
    \node (opO11) at (-3,3) [above] {$\cO^{\gamma_1}_{\fa}$};
    \node (opO22) at (-3.6,-2.2) [below] {$\cO^{\gamma_2}$};
    \node (opO33) at (3.6,-2.2) [below] {$\cO^{\gamma_3}_{\dfa}$};
    \node (opO44) at (2.1,2.1) [above] {$\cO^{\gamma_4}$};
    \node (o2) at (2.5,-1.2) [threept] {};
    \node (o4) at (-2.5,-1.2) [threept] {};
    \node (D2) at (-1.8,2) [threept] {};
    \node (D1) at (-1.8,0) [threept] {};
    \node (shad) at (0,1) [bowtie] {$\bowtie$};
    \node at (0,1.15) [above] {$\cO^{\gamma}$};	
    \draw [scalar] (vertL)-- (D1);
    \draw [spinning] (D1)-- (o4);
    \draw [scalar] (vertL)-- (shad);
    \draw [scalar] (vertR) -- (shad);
    \draw [scalar] (vertR)-- (opO44);
    \draw [scalar] (vertR)-- (o2);
    \draw [scalar] (o4)-- (opO22);
    \draw [spinning] (o2)-- (opO33);
    \draw [finite with arrow] (o4)-- (o2);
    \draw [scalar] (vertL)-- (D2);
    \draw [spinning] (D2)-- (opO11);
    \draw [finite with arrow] (D1) to[in=-110, out=120](D2);
     \node (align) at (0,-.1) {};
\end{tikzpicture}}.
\end{split}
\end{equation}
As we did for external shifts, we can write the above as a differential operator on the cross-ratio matrix $Z^\fa_{\ \fb}\sim \text{diag}(z_1,\dots, z_m |w_1, \dots, w_n)$ by using \eqref{eq:singletblock} as follows
\begin{align}\label{eq:seedblock11}
{}&(\partial_{\bX_3}\bX_2 \hX_{21})_{\dfa \fb} P F^{\gamma \bullet}(Z)=\\
& P \hX_{31 \dfa \fa} \bigg(
\frac{1}{2} \left(\left(\gamma _{43}-\gamma \right) \delta^\fa_\fb+\left(\gamma -\gamma _3-\gamma _4\right) (Z^{-1})^\fa_{\ \fb}\right)+  (\delta^{\fa_1}_{\fb}-Z^{\fa_1}_{\ \fb})
(\partial_Z)^{\fa}_{\ \fa_1}
\bigg) F^{\gamma \bullet}(Z)\,,
\nonumber
\end{align}
and
\begin{equation}
\begin{aligned}
{}& (\partial_{\bX_3} \partial_{X_1})_{\dfa \fb}P F^{\gamma\bullet}(Z)=\\
& P \hX_{31 \dfa \fa} \bigg(
\frac{1}{4} \Big(\left(\gamma -\gamma _1-\gamma _2\right) \left(\gamma -\gamma _3-\gamma _4\right) (Z^{-1})^\fa_{\ \fb}\\
&-\left(\gamma -\gamma _{43}\right) \left(\gamma -\gamma _1-\gamma _2-\gamma _3-\gamma
   _4+\gamma _{21}+2 (m{-}n)\right) \delta^\fa_\fb\Big)\\
&+\frac{1}{2} \Big(\delta^{\fa_1}_\fb \left(\delta^\fa_{\fa_2} \left(-2 \text{ str} Z+2 \gamma -\gamma _1-\gamma _2-\gamma _3-\gamma _4+2 (m{-}n)\right)+\left(\gamma _{43}-\gamma \right) Z^\fa_{\ \fa_2}\right)\\
&+\left(-\gamma +\gamma _1+\gamma _2+\gamma _3+\gamma _4-\gamma _{21}-2 (m{-}n)\right) \delta^\fa_{\fa_2}
   Z^{\fa_1}_{\ \fb}\Big)
(\partial_Z)^{\fa_2}_{\ \fa_1}\\
&+  (Z^{\fa_3}_{\fa_2}-Z^{\fa_3}_{\fa_5} Z^{\fa_5}_{\fa_2}) 
(\partial_Z)^{\fa_2}_{\ \fb} (\partial_Z)^{\fa}_{\ \fa_3}
\bigg) F^{\gamma \bullet}(Z)\,.
\end{aligned}
\label{seedblock_last}
\end{equation}
Notice that this form explicitly reproduces the solution to the Ward identities shown in section \ref{sec:Wardids}. By substituting in $Z=z$ for $(m,n)=(1,0)$, we reproduce the expected shift in the 1D blocks of \cite{Doobary:2015gia} (see \eqref{1d_c4}).

By following the process outlined above, one can write any $(m,n)$ superspace seed block of the form \eqref{eq:seedblock} as differential operators on the scalar block \eqref{eq:singletblock}. In particular, the first step would be to relate the RHS three-point structure \eqref{eq:seedblock} as follows
\begin{align}
        &\langle \cO_{\ula_2, \ula_1}  \cO_{\umu_2,\bullet} \cO_{ \bullet,\umu_1}\rangle \\
        &=\bPi^{\ula_2}\Pi^{\umu_1} \left(\bD_{0\dfa_0 \cA} \partial_{X_3}{}^{\cA}_{\ \fa_3}\right)^{\otimes|\ula_1/(\ula_1\cap \ula_2)|} \bPi^{\umu_1}\Pi^{\ula_2}\left(\partial_{\bX_{4}}{}_{\dfa_4 \cA} D_0{}^{\cB}_{\ \fa_0}\right)^{\otimes|\ula_2/(\ula_1\cap \ula_2)|}  \langle \cO_{\ula, \ula}  \cO_{\bullet,\bullet} \cO_{\bullet, \bullet}\rangle. \nn
\end{align}
Then, one would have to more all differential operators depending on the exchanged coordinate through the shadow integral as in \eqref{eq:gluecrossing}, to then perform three-point crossing on the LHS three-point structures in order to write the full expression in terms of differential operators on the external insertions.

All in all, the method for internal shifts, combined with that for external shifts shown in section \ref{sec:externalshift}, allows for a systematic derivation of all (super)conformal blocks of theories with $\SL(2m|2n)$ symmetry (and thus any of its real forms) from the simplest set of blocks \eqref{eq:singletblock} of singlets under the rotations of the superspace $\SL(m|n) \times \SL(m|n)$. In the superconformal setting we are interested in, this means that all superblocks for external non-half-BPS supermultiplets can be derived from the half-BPS blocks \cite{Doobary:2015gia, Aprile:2021pwd}.

\subsection{Non-integer dimension}\label{sec:nonintegerdim}
In this section, we give a more extended discussion on how to derive superconformal blocks when either external and/or exchanged representations have non-integer scaling dimension. 
Weight-shifting operators  change the scaling dimension in integer steps only. But in the non-supersymmetric case, since the starting scalar conformal block is known  as a function of arbitrary external ($\Delta_i$) and exchanged ($\Delta$) dimensions the resulting spinning conformal block may also contain such general values. 
In the supersymmetric case however we start from correlators of half-BPS multiplets which have integer scaling dimensions $\Delta = \gamma \in \mathbb{N}$, where $\gamma$ enters the $R$-symmetry representation as $(\gamma;0)$ in $\cN=2$ and as $[0,\gamma,0]$ in $\cN=4$ (see Appendix \ref{sec:repexamples}).

In practice, we build the general weight-shifting operators up from consecutive applications of the fundamental weight-shifting operator with the help of Young diagrams. 
Strictly this process will only ever yield integer shifts, but one can obtain non-integer cases from integer ones by analytically continuing with the help of  quasi-tensors~\cite{Heslop:2001zm,Heslop:2003xu}. This feature occurs for our description of the operators themselves and there is a corresponding statement even in the non-supersymmetric case. Quasi-tensors $\ula[\delta]$ are defined, for integer $\delta$, as follows:
\begin{equation}\label{eq:diagramdelta}
    \begin{aligned}
        (m,n): &\quad    \begin{tikzpicture}[scale=0.65, baseline=(current bounding box.center)]
    \draw[thick] (-0.5,0) rectangle (0,-3);
        \draw[thick] (0,0) rectangle (0.5,-2);
        \draw[thick] (.5,0) rectangle (1.5,-0.5);
        \draw[thick] (.5,-0.5) rectangle (1,-1);
        \draw[<->] (-0.5,0.25) -- (0.5,0.25) node[midway, above] {$n$};
         \draw[<->] (-.75,0) -- (-.75,-1.5) node[midway, left] {$m$};
    \end{tikzpicture}[\delta]\ :=\ \begin{tikzpicture}[scale=0.65, baseline=(current bounding box.center)]
    \draw[thick] (-0.5,0) rectangle (0,-3);
    \draw[thick] (0,0) rectangle (0.5,-2);
    \draw[thick] (0.5,0) -- (4,0);      
    \draw[thick] (4,0) -- (4,-0.5);     
    \draw[thick] (4,-0.5) -- (3,-0.5);     
    \draw[thick] (3.5,-0.5) -- (3.5,-1); 
    \draw[thick] (3.5,-1) -- (3,-1); 
    \draw[thick] (0.5,0) -- (0.5,-1.5);  
    \draw[thick] (3,0) -- (3,-1.5);      
    \draw[thick] (0.5,-1.5) -- (3,-1.5); 
    
    \draw[<->] (0.55,-1.75) -- (3,-1.75) node[midway, below] {$\delta$};
    \draw[<->] (-0.5,0.25) -- (0.5,0.25) node[midway, above] {$n$};
    \draw[<->] (1.25,0) -- (1.25,-1.5) node[midway, right] {$m$};
\end{tikzpicture},
\qquad 
        (m,0): &   \begin{tikzpicture}[scale=0.65, baseline={([yshift=-.6cm]current bounding box.north)}]
        \draw[thick] (.5,0) rectangle (1.5,-0.5);
        \draw[thick] (.5,-0.5) rectangle (1,-1);
         \draw[<->] (.25,0) -- (.25,-1.5) node[midway, left] {$m$};
          \draw[<->,opacity=0] (0.55,-1.75) -- (1.5,-1.75) node[midway, below] {\phantom{$\delta$}};
    \end{tikzpicture}[\delta]\ :=\ \begin{tikzpicture}[scale=0.65, baseline={([yshift=-.6cm]current bounding box.north)}]
    \draw[thick] (0.5,0) -- (4,0);      
    \draw[thick] (4,0) -- (4,-0.5);     
    \draw[thick] (4,-0.5) -- (3,-0.5);     
    \draw[thick] (3.5,-0.5) -- (3.5,-1); 
    \draw[thick] (3.5,-1) -- (3,-1); 
    \draw[thick] (0.5,0) -- (0.5,-1.5);  
    \draw[thick] (3,0) -- (3,-1.5);      
    \draw[thick] (0.5,-1.5) -- (3,-1.5); 
    
    \draw[<->] (0.55,-1.75) -- (3,-1.75) node[midway, below] {$\delta$};
    \draw[<->] (1.25,0) -- (1.25,-1.5) node[midway, right] {$m$};
\end{tikzpicture}.
    \end{aligned}
\end{equation}
The key point is that the dimension (number of components, not to be confused with $\Delta$) of the representation does not depend on $\delta$ and thus $\delta$ itself can be straightforwardly continued to non-integer values. The actual dependence on $\delta$ is very simple; it can be deduced either by thinking through the combinatorics carefully, or alternatively by brute force taking the first few integer cases $\delta=0,1,2$. This is easy to see in the non-supersymmetric case $n=0$ where we see that $\delta$ corresponds to the dimension of $\GL(1)\subset \GL(m)$ which can be non-integer.
As one can see by comparing with Section \ref{sec:cosetspace} and Appendix \ref{sec:repexamples}, long operators are given in analytic superspace by $\cO^\gamma_{\ula}$ with discrete $\gamma$ and `typical' Young diagram $\ula$ 
Then, the parameter $\delta$  contributes \emph{only} to the dimension $\Delta$.

\paragraph{Long to long:}{
 The first and most direct application of our present work is to computing superconformal blocks of correlators involving external quarter-BPS operators where in fact the above subtleties are not really needed. 
  For example, realising the following weight-shifting operation
\begin{equation}
    \langle H H L \rangle \bowtie \langle L H H \rangle \mapsto \langle Q H L'\rangle \bowtie \langle L' Q H \rangle,
\end{equation}
for $H$ half-BPS, $Q$ quarter-BPS and $L\neq L'$ long, with the latter being exchanged. In this case the left-hand-side half-BPS superblock has closed form analytic expressions in $\Delta$. 
Our analytic superspace weight-shifting operators are then able to act on such long representations straightforwardly, since non-trivial dependence on $\Delta$ is only present in the coefficients \eqref{eq:coeffsss}. Therefore, for such operations, our weight-shifting operators work in the same way as the embedding space conformal differential operators of \cite{Karateev:2017jgd}.}

\paragraph{Short to long:}{
More non-trivial cases arise when considering long external operators by performing operations such as 
\begin{equation}
    \langle H H L \rangle \bowtie \langle L H H \rangle \mapsto \langle L' H L\rangle \bowtie \langle L H H \rangle\,.
\end{equation}
Here we map one (or more) half-BPS operator to a long one. 
So here we in principle need weight-shifting operators which shift the dimension by an arbitrary non-integer amount
\begin{equation}\label{eq:hbpstolong}
    \cO^{\gamma \in \mathbb{N}} \mapsto \cO^\gamma_{\ula},
\end{equation}
where $\ula$ is  a long representation with non-integer row lengths.
As discussed such weight-shifting operators can be easily obtained by considering different long Young diagrams $\ula$~\eqref{eq:diagramdelta}, 
with different integer values of the parameter $\delta$ in $\ula$  which encodes the dimension $\Delta$ in the resulting block, and then analytically continuing.

Let us see how this analytic continuation works in more detail.Consider for example a superconformal block for a correlator of four coset scalars (half-BPS) 
\begin{equation}
    \langle \cO^{\gamma_1}_\bullet \cO^{\gamma_2}_{\bullet} \cO^{\gamma}_{\ula} \rangle \bowtie  \langle \cO^{\gamma}_\ula \cO^{\gamma_3}_{\bullet} \cO^{\gamma_4}_{\bullet} \rangle \subset \langle  \cO^{\gamma_1}_\bullet \cO^{\gamma_2}_{\bullet} \cO^{\gamma_3}_{\bullet} \cO^{\gamma_4}_{\bullet}\rangle,
\end{equation}
for integer $\gamma_i$, $i=1,\dots,4.$
Suppose we want to make points 1 and 2 into long operators. The simplest way is to act with successive copies of
\begin{equation}
    (\partial_{\bX_1} \cdot \partial_{X_2})_{\dfa_1 \fa_2} :=(\partial_{\bX_1} )_{\dfa_1 \cA} (\partial_{X_2})^{\cA}_{\ \fa_2}
\end{equation}
and symmetrising into a long $GL(m|n)$ Young diagram \eqref{eq:diagramdelta}. 
The first application yields
\begin{equation}
    \begin{split}
     (\partial_{\bX_1} \cdot \partial_{X_2})_{\dfa_1 \fa_2} \langle  \cO^{\gamma_1}_\bullet \cO^{\gamma_2}_{\bullet} \cO^{\gamma}_{\ula}  \rangle  &=(\partial_{\bX_1} \cdot \partial_{X_2})_{\dfa_1 \fa_2} g_{12}^{\gamma_{12,0}} g_{20}^{\gamma_{20,1}} g_{01}^{\gamma_{01,2}} Y_0^\ula\\ 
        &= \gamma_{12,3} (\gamma_{12,3}-m+n) g_{12}^{\gamma_{12,0}} g_{20}^{\gamma_{20,1}} g_{01}^{\gamma_{01,2}} \hX_{12}{}_{\dfa_1 \fa_2} Y_0^{\ula} \\
        & \propto \langle \cO^{\gamma_1}_{\bullet, \Box} \cO^{\gamma_2}_{\Box,\bullet} \cO^{\gamma}_{\ula}\rangle,
    \end{split}
\end{equation}
for all $m,n$. Similarly, applying this differential operator $|\umu|$ times and symmetrising at both points according to an unspecified Young diagram $\umu$ gives
\begin{equation}
     (\partial_{\bX_1} \cdot \partial_{X_2})^{\umu}  \langle \cO^{\gamma_1}_\bullet \cO^{\gamma_2}_{\bullet} \cO^{\gamma}_{\ula}\rangle  \propto \hX_{12}^{\umu} \langle \cO^{\gamma_1}_\bullet \cO^{\gamma_2}_{\bullet} \cO^{\gamma}_{\ula}\rangle \equiv \langle \cO^{\gamma_1}_{\bullet,\umu} \cO^{\gamma_2}_{\umu,\bullet} \cO_{\ula}^\gamma\rangle\, .
\end{equation}
We then take   $\umu$ to be a long $\GL(m|n)$ representation (i.e. in the form on the right hand side of~\eqref{eq:diagramdelta}).
\footnote{The operators produced here $\cO^{\gamma_1}_\bullet, \cO^{\gamma_2}_{\bullet}$ are not in fact  long in the supersymmetric setting, but only long `on one side' whilst being  half BPS on the other side. In particular  they can not get anomalous dimensions (and in fact don't even exist in $\cN=4$ SYM). To get a more realistic supersymmetric example we would need to do the further step discussed below to obtain symmetric operators. But these are fine in the non supersymmetric setting and serve as the simplest illustrative example of how the continuation to non integer weights works in practice.}

The point is then that one can perform the computation for integer $\delta$ but write the results in a manner which makes the continuation to arbitrary $\delta$ obvious.
To illustrate this in a simple example, take  the 
purely bosonic, non-supersymmetric case, $n=0$ and consider   $\umu=[1^m]$. Then we we have
\begin{equation}
    \umu = \begin{tikzpicture}[scale=0.65, baseline=(current bounding box.center)]
        \draw[thick] (0,0) rectangle (0.5,2.2);
        \draw[<->] (-0.25, 0) -- (-0.25, 2.2) node[midway,left] {$m$};
    \end{tikzpicture} \quad  \Rightarrow  \quad \hX_{12}^{\umu} = (\hX_{12} \dots \hX_{12})_{[\dota_1 \dots \dota_m] [\alpha_1 \dots \alpha_m]} \sim \det \hX_{12} \equiv g_{12}\,.
\end{equation}
Furthermore taking  $\umu=[\delta^m]:=\bullet[\delta]$ (following the notation in~\eqref{eq:diagramdelta}) we get
\begin{equation}\label{eq:mudelta}
    \umu = \begin{tikzpicture}[scale=0.65, baseline=(current bounding box.center)]
        \draw[thick] (0,0) rectangle (1.5,2.2);
        \draw[<->] (-0.25, 0) -- (-0.25, 2.2) node[midway,left] {$m$};
        \draw[<->] (0, -0.25) -- (1.5, -0.25) node[midway,below] {$\delta$};
    \end{tikzpicture} \quad  \Rightarrow  \quad \hX_{12}^{\umu} \sim g_{12}^{\delta}\, .
\end{equation}
Therefore, we can straightforwardly write the following equation involving weight-shifting operators for any (integer) $\delta$:
\begin{equation}
     (\partial_{\bX_1} \cdot \partial_{X_2})^{\umu}  \langle \cO^{\gamma_1} \cO^{\gamma_2} \cO_{\ula}^\gamma\rangle = \langle \cO^{\gamma_1}_{\bullet,\bullet[\delta]} \cO^{\gamma_2}_{\bullet[\delta],\bullet} \cO_{\ula}^\gamma\rangle = \langle \cO^{\gamma_1+\delta}_{\bullet} \cO^{\gamma_2+\delta}_{\bullet} \cO_{\ula}^\gamma\rangle
\end{equation}
Strictly here we have only derived the result for integer  $\delta$, but the analytic continuation is clear for the three-point function.

For the blocks, recall that a conformal block is constructed from the shadow integral
    \begin{align}
        &\langle \cO^{\gamma_1+\delta}_{\bullet} \cO^{\gamma_2+\delta}_{\bullet} \cO_{\ula}^\gamma\rangle \bowtie \langle  \cO_{\ula}^\gamma \cO^{\gamma_3}_{\bullet} \cO^{\gamma_4}_{\bullet}\rangle  \\
        &= \int_{\Gr(m,2m)} dX_0 dX_{0'} \langle \cO^{\gamma_1+\delta}_{\bullet} \cO^{\gamma_2+\delta}_{\bullet} \cO_{\ula}^\gamma (X_0)\rangle \langle \tilde{\cO}^{\gamma,\ula}(X_0)\tilde{\cO}^{\gamma,\ula}(X_{0'})\rangle \langle  \cO_{\ula}^\gamma(X_{0'}) \cO^{\gamma_3}_{\bullet} \cO^{\gamma_4}_{\bullet}\rangle \notag
    \end{align}
in which case we have
    \begin{equation}
        \langle \cO^{\gamma_1+\delta}_{\bullet} \cO^{\gamma_2+\delta}_{\bullet} \cO_{\ula}^\gamma\rangle \bowtie \langle  \cO_{\ula}^\gamma \cO^{\gamma_3}_{\bullet} \cO^{\gamma_4}_{\bullet}\rangle =g_{12}^\delta\langle \cO^{\gamma_1}_{\bullet} \cO^{\gamma_2}_{\bullet} \cO_{\ula}^\gamma\rangle \bowtie \langle  \cO_{\ula}^\gamma \cO^{\gamma_3}_{\bullet} \cO^{\gamma_4}_{\bullet}\rangle,
    \end{equation}
and thus $\delta$ can also be continued in the block. For supersymmetric cases, the process is similar albeit more complicated. In particular the indices in $\umu$ corresponding to the first $n$ columns will remain open (see~\eqref{eq:diagramdelta}, and thus the three-point structures are more complicated to construct.

Note that of course not all cases of analytical continuation are as trivial as the above case. For example, if we instead want to only turn the first operator into a long operator, by successively applying a weight-shifting invariant of the form
\begin{equation}
   \fD_{\dfa_1 \fa_1 } := ( X_2 \cdot D_1)^{\fa_2} _{\ \fa_1}(\partial_{\bX_1}\cdot \partial_{X_2})_{\dfa_1 \fa_2},
\end{equation}
we will instead obtain
\begin{equation}
    \fD^{\umu}  \langle \cO^{\gamma_1}_{\bullet} \cO^{\gamma_2}_{\bullet} \cO_{\ula}^\gamma\rangle \propto Y_1^{\umu} \langle \cO^{\gamma_1}_{\bullet} \cO^{\gamma_2}_{\bullet} \cO_{\ula}^\gamma\rangle \equiv \langle \cO^{\gamma_1}_{\umu} \cO^{\gamma_2}_{\bullet} \cO_{\ula}^\gamma\rangle.
\end{equation}
For $\umu=\bullet[\delta]$ \eqref{eq:mudelta} in $(m,0)$ analytic superspace we then have
\begin{equation}
    Y_{1,20}^{\umu} = \frac{g_{12}^\delta g_{01}^\delta}{g_{20}^{\delta}},
\end{equation}
which has dependence on $X_0$ that will be sensitive to the shadow integral. However, once the three-point function is known for unfixed $\delta$ as described above, the integral may then be performed for unfixed $\delta$. 

But as we have described in earlier sections, the weight-shifting operators can also be used to replace having to perform such integrals.
So how can obtain results with non-integer $\delta$ this way? 
Again the key point is that the methods here perform the difficult job of solving the requirements of superconformal symmetry for any integer $\delta$. Then the step of continuing to arbitrary $\delta$ is a much simpler step. For example
 we also know the form of the bosonic blocks which must  make up the superconformal blocks (in terms of hypergeometric functions whose parameters are linearly dependent on the dimension) and so matching with these one will be able to  derive the case for arbitrary $\delta$ from a few integer $\delta$s.
We leave such calculations in supersymmetric cases for future work.

\subsection{Summary}\label{sec:blocksummary}
For the reader's convenience, we summarise the results of this section and present a clear outline of how to use and generalise them with weight-shifting operators. 

We expressed large classes of superconformal blocks in $(m,n)$ analytic superspace as differential operators on a scalar block
\begin{equation}
    \langle \cO^{\gamma_1} \cO^{\gamma_2} \cO^\gamma_{\ula} \rangle \bowtie \langle \cO^\gamma_{\ula} \cO^{\gamma_3} \cO^{\gamma_4} \rangle.
\end{equation}
\Yboxdim{5pt}
We did so via two different types of operation: external and internal shift. 

The former corresponds to a direct application of the differential basis of three-point functions (developed in section \ref{sec:diffbasis}) to blocks (expressed as shadow integrals \eqref{ij}). We explicitly computed two fundamental cases which underpin any further calculation: chiral and symmetric single index representations at one insertion, with blocks given by
\begin{equation}\label{eq:blockexamples2}
     \langle \cO^{\gamma_1}_{\yng(1),\bullet} \cO^{\gamma_2}_{\bullet,\yng(1)} \cO^\gamma_{\ula} \rangle \bowtie \langle \cO^\gamma_{\ula} \cO^{\gamma_3} \cO^{\gamma_4} \rangle, \qquad \langle \cO^{\gamma_1}_{\yng(1),\yng(1)} \cO^{\gamma_2}\cO^\gamma_{\ula} \rangle \bowtie \langle \cO^\gamma_{\ula} \cO^{\gamma_3} \cO^{\gamma_4} \rangle.
\end{equation}
We expressed the final result as derivatives in the cross-ratio matrix $Z^{\fa}_{\ \fb}$ whose precise form depends on the choice of $(m,n)$. They are given by \eqref{eq:xhatblock}, \eqref{DZ1} in the chiral case and by \eqref{DZ2} in the symmetric case. We also presented the method for performing general external shifts, both chiral \eqref{eq:wsoex11}  and symmetric \eqref{eq:gensymex}, extending \eqref{eq:blockexamples2}.

Internal shifts are more technically challenging since they require commuting differential operators through the shadow integral, as shown in section \ref{sec:internalshift}. We showed the explicit computation of adding a box to one side of the exchanged representation, with the main result \eqref{eq:internalshiftdiagram}  again presented as derivatives in the cross-ratio matrix $Z^{\fa}_{\ \fb}$ \eqref{eq:seedblock11} and \eqref{seedblock_last}. As with the external shift case, we further outlined the general process for the construction of seed blocks in $(m,n)$ analytic superspace.

Note that, by keeping all charges $\gamma_i$ unfixed, all results can apply to a very wide array of representations in different dimensions and amount of supersymmetry (see appendix \ref{sec:repexamples} for translation). In particular, low values of the charge will correspond to short multiplets, and as explained in section \ref{subsec:shortdiffbasis}, our weight-shifting operators map these correctly. This is highly non-trivial, since usually one has to impose an additional Ward identity to constrain tensor structures further. Our analytic superspace differential operators automatically take this into account, even in non-supersymmetric cases. This is not possible in other formalisms (notably, embedding space) where fields are reducible unless the conservation constraint is imposed by hand.

The main application of this work is to have a complete database of all (super) conformal blocks, which can then be used in any (super) conformal bootstrap computation, be it numerical or analytical. 
We have provided all the necessary tools to do so in settings with $\SU(m,m|2n)$ symmetry, via weight-shifting operators that map shortening conditions automatically at the level of correlation functions, even in non-supersymmetric cases.

The most efficient way to use our (and any other set of) weight-shifting operators to this end is by recursion. Indeed, directly performing large changes in quantum numbers, e.g. scalar $\cO^2$ to stress-tensor $T\sim \cO^2_{\yng(2),\yng(2)}$ or half-BPS $\cO^4$ to quarter-BPS $\cO^4_{\yng(1,1),\yng(1,1)}$, is cumbersome. Instead, the process is to recursively add boxes to the Young diagrams by acting on the previous result.
For example, to obtain $$\langle \cO^4_{\yng(1,1),\yng(1,1)} \cO \cO_\ula^\gamma\rangle \bowtie \langle\cO_\ula^\gamma \cO \cO\rangle$$ for some symmetric exchange $\cO^\gamma_\ula$ from the scalar block, one should compute and store the intermediate blocks with the following insertions at point 1:
\begin{equation}\label{eq:recursion}
    \cO^4\mapsto \cO^4_{\yng(1),\bullet} \mapsto \cO^4_{\yng(1),\yng(1)}\mapsto \cO^4_{\yng(1,1),\yng(1)}\mapsto \cO^4_{\yng(1,1),\yng(1,1)}
\end{equation}
where the first two maps were carried out explicitly in section \ref{sec:externalshift} and the shortening condition imposed by $\gamma=4$ restricts to only one three-point tensor structure in the shadow integral and hence one block. As discussed above and in section \ref{subsec:shortdiffbasis}, this is automatically taken into account by our weight-shifting operators, even or conserved currents $\cO^2_{[s],[\bar{s}]}$ in non-supersymmetric CFT. 
The next step is to pick a different type of exchanged representation, provided it exists, and go through the same process. For example,
$$\langle \cO^4_{\yng(1,1),\yng(1,1)} \cO\cO^4_{\yng(1,1),\yng(1,1)} \cO \rangle$$ will admit 1-asymmetric and 2-asymmetric exchanges in the $(12)(34)$ OPE channel, which require a seed block with non-half-BPS multiplets, as discussed in section \ref{sec:seedblockss}. The process is thus to derive the corresponding seed block and repeat the recursion \eqref{eq:recursion}, storing each result.

\Yboxdim{13pt}

\section{Conclusion \& outlook}

We have set up a programme for obtaining superblocks of arbitrary multiplets in theories with $\SU (m,m|2n)$ superspace using $(m,n)$ analytic superspace viewed as a super-Grassmannian. While the primary application is to superconformal field theories, we believe the formalism also holds significant relevance for non-supersymmetric four-dimensional CFTs, often giving simpler expressions in this setting, particularly when dealing with conserved multiplets.
We obtained explicit results for some simple cases in a general $(m,n)$ theory and checked that the $(1,0)$ and $(2,0)$ cases match with the known 1D and 4D spinning conformal blocks.

The focus in this paper has been to apply the techniques developed in~\cite{Karateev:2017jgd} within the analytic superspace formalism, in particular obtaining the corresponding weight-shifting operators.


There are then a number of avenues for future work which we outline below.

A useful approach to obtaining scalar/half-BPS (super)blocks has been via an expansion in Schur or Jack polynomials~\cite{Dolan:2003hv,Doobary:2015gia,Aprile:2021pwd}, which are superconformally invariant completions of the zeroth order term in the OPE expansion of half-BPS superblocks. Initial studies indicate that the analogous building blocks for non-half-BPS superblocks are restricted  Schurs~\cite{Bhattacharyya:2008rb,Collins:2008gc} for which an efficient algorithm has been found recently~\cite{Padellaro:2024rld}. The Casimir has a simple action on this which can then be used to obtain a recursion relation for the coefficients following an analogous path for half-BPS blocks. In particular the coefficients should be universal, i.e.\ $m,n$ independent. This approach would also give results with a more manifest superconformal invariant structure.

More ambitiously, the simplicity of the half-BPS blocks which can be explicitly written in terms of a determinant of hypergeometric functions~\cite{Doobary:2015gia} suggests the more general  blocks could have a similarly simple form.  Computing explicit cases either via weight-shifting operators or by recursion on restricted Schurs will help understand the structure with the aim of writing a general formula in terms of hypergeometric functions.

The previous work on non-supersymmetric spinning blocks has always made use of polarisation vectors to absorb the free indices. As such, the conformal weight-shifting operators of \cite{Karateev:2017jgd} include derivatives in the polarisations. 
This approach can also be used in our superconformal setting.
In general, one would need a different polarisation vector for each index value, to absorb all the superindices $\fa,\dfa$. 
However, certain sufficiently simple classes of supermultiplets, e.g.\ quarter-BPS operators in $\cN=4$ SYM, would only need one type of polarisation vector. 
In fact, the formalism with polarisation vectors for each index would be equivalent to the maximal flag supermanifold,  
where the polarisations are themselves cosets of the two $\SL(m|n)$ subgroups (see~\cite{Howe:1995md}). 
While the form of the superconformal weight-shifting operators would be more complicated, including derivatives in the new (super)polarisations could simplify some calculations at the end and avoid the use of Young symmetrisers. 

The $(m,n)$ analytic superspace we considered in this paper is only relevant to theories with $\SL(2m|2n)$ symmetry, which includes (S)CFTs in 1 and 4 dimensions. 
In fact, this is a particular case ($\theta=1$) of the more general $(m,n,\theta)$ analytic superspace.
Thus, a natural extension of this work would be to generalise to other values of $\theta$, corresponding to theories in 3 and 6 dimensions, including  M2- and M5-brane worldsheet theories. 
Indeed, half-BPS superblocks for arbitrary  $(m,n,\theta)$  are  BC${}_{n|m}$ super-Jacobi functions, natural supersymmetric extensions of Jacobi polynomials of BC${}_{n}$ root systems~\cite{Aprile:2021pwd}. Thus generalising beyond the half-BPS case for arbitrary  $\theta$ theories would  give tensorial generalisations of the BC${}_{n|m}$ super-Jacobi functions.

Another beautiful relation to an area of pure mathematics is the relation between non-spinning blocks and  spherical functions~\cite{Schomerus:2016epl,Schomerus:2017eny,Isachenkov:2017qgn,Buric:2019dfk,Agarwal:2023xwl}. This has  a natural generalisation in this formalism, giving spherical functions of supergroups which would be fascinating to explore.

An aspect of the general formalism we have not emphasised much in the paper is the fact that in the $(0,n)$ theory the unitary operators $\cO^\gamma_{\ula_L,\ula_R}$ correspond to finite-dimensional representations of $\SU (2n)$. The precise relation is given in appendix~\ref{internal}. This makes the various  relations to mathematical results, special functions, spherical functions etc.  very precise. For example, as mentioned above, in the $(0,n,\theta)$ theory the blocks are completely equivalent to BC${}_n$ Jacobi polynomials (rather than the generalisations $BC_{n|m}$ super-Jacobi functions) which are well studied (and are limits of Macdonald polynomials).  
This $(0,n)$ case will correspond to spherical functions of the compact  $\SU (2n)$ group, which is the most studied case and should make contact with existing mathematical work, for example~\cite{Koelink:2021hlp} for $n=2$.
So, there is a nice circle of relations starting with the  $(0,n)$ theory and its connection with finite $\SU (2n)$ results in pure mathematics to give insight into the $(0,n)$ blocks. These then naturally lift to $(m,n)$ superblocks when written in analytic superspace leading supersymmetric uplifts of these results.

The main aim for physicists is of course to make practical use of the general superblocks, either to find better numerical bootstrap bounds or to use analytic bootstrap techniques to make further precision (e.g.\ holographic) computations beyond those obtainable from half-BPS correlators~\cite{Aprile:2017bgs,Aprile:2017qoy,Caron-Huot:2018kta,Alday:2018pdi,Aprile:2019rep,Abl:2020dbx,Aprile:2020mus,Huang:2021xws,Drummond:2022dxw,Alday:2022uxp,Alday:2022xwz,Alday:2023jdk,Alday:2023mvu}.

 Recently, all free theory half-BPS superblock  coefficients were obtained analytically~\cite{Aprile:2025nta}. The free superblock expansion is equivalent to a generalised Cauchy identity whose components are various types of special functions~\cite{Aprile:2021pwd}.
 There should be a huge generalisation of this story for the spinning/non-half-BPS blocks with a tensorial Cauchy identity, which would be fascinating to explore.
We should also point out that the finding of these free theory analytic formulae made use of  generalised non-supersymmetric $\SU (m,m)$ blocks,  $(m,0)$ analytic superspace,  which are much simpler than the full superblocks, but contain  all information about the superblock coefficients and so we would expect this simplification to be crucial in the spinning case too.

Another concrete application requiring the use of the non-half-BPS superblocks arises from the holographic computation of 5-point half-BPS  correlators~\cite{Goncalves:2019znr}. This  in principle makes available a wealth of new data in ${\cal N}=4$ SYM, but it requires the knowledge of more general superblocks in order to extract the data (some initial data has been extracted without the superblocks by making use of the chiral algebra limit in~\cite{Bissi:2021hjk}).

More recently, the program of \cite{Alday:2022uxp,Alday:2022xwz,Alday:2023jdk,Alday:2023mvu} bootstraps string amplitudes on $AdS_5 \times S^5$ by imposing the consistency of the superconformal block expansion with the requirement that string amplitudes are worldsheet integrals over certain single-valued functions. Extending this program to loop amplitudes likely requires studying four-point functions of non-half-BPS operators,
which appear in unitarity cuts of one-loop amplitudes. 
We expect that this paper is an important milestone towards this goal. 
Of particular importance for these applications are  superblocks involving    
long supermultiplet operators and so it would be particularly interesting to investigate these cases in detail including the  additional subtleties  discussed in section~\ref{sec:nonintegerdim} (and briefly in~\ref{longops}).

Finally we wish to point out a very nice by-product of the simple single derivative weight shifting operators 
acting on chiral operators given in~\eqref{wschiral2}, that is in the 
explicit construction of (super)conformal primary fields  in a  scalar field theory (or supersymmetric equivalent like $\cN=4$ SYM). 
Recall that to write down a conformal primary operator in a scalar field theory  directly in coordinates in Minkowski space one would  
 write a linear combination of terms with derivatives acting on different scalar fields and then impose vanishing under special conformal transformations.
 For example spin 1 and higher spin currents are given explicitly as~\cite{Konstein:2000bi}
\begin{equation}
    j^{(1)}=\partial \phi \bar \phi- \phi \partial\bar \phi, \qquad j^{(s)}=\sum_k (-1)^k \begin{pmatrix}
        s\\k
    \end{pmatrix}^2 \partial^k\phi \partial^{s-k}\bar \phi\,.
\end{equation}
A similar  formula in analytic superspace gives the higher spin supercurrents~\cite{Bianchi:2005ze}. More general operators will only look more complicated.
In this Grassmannian formalism however we simply apply the simple weight shifting operator, $(\bar \partial_X)_{\dfa \cA}$ to $\bar \Phi$ (which we 
can without loss of generality assume to be chiral and depend only on $\bX$ not $X$) 
and $(\partial_X)^\cA{}_\fa$ derivatives to $\Phi$ and contract all the $\cA$ indices.\footnote{Note that the equation of motion of the fundamental scalar enforces that when acting with derivatives on a single scalar, all indices are automatically symmetrised.} 
In particular, we can write {\em any} representation with equal numbers of left and right indices in terms of fundamental scalars explicitly as
\begin{align}\label{ops}
    {\mathcal O}^{\gamma_L, \gamma_R}_{\underline \lambda,\underline \mu} = \Pi^{\underline \lambda}_1 \bar \Pi^{\underline \mu}_1\left[\partial^{\lambda_1}_X\Phi_1..\partial^{\lambda_{\gamma_L}}_X\Phi_{\gamma_L} \ \cdot \  \partial^{\mu_1}_{\bX}\bar \Phi_1..\partial^{\mu_{\gamma_R}}_{\bX}\bar \Phi_{\gamma_R}\right]\, .
\end{align}
The Young diagrams $\ula=[\lambda_1,\dots,\lambda_{\gamma_L}]$, $\umu=[\mu_1,\dots,\mu_{\gamma_R}]$ define the representations for the Young projectors $\Pi \bar{\Pi}$, which act on the $\fa$ and $\dfa$ indices. The scalar fields $\Phi_i, \bar \Phi_i$ are considered to be all different for full generality, but in a specific situation they will likely have further structure (e.g.\ adjoint $\SU (N)$ indices over which we take traces). Finally, the dot $\cdot$ indicates contraction of all $\cA$ indices (lower coming from $\partial_{\bX}$ and upper from $\partial_X$).
Note the remarkable simplicity of this result in the Grassmannian formalism compared to position space: here there is no sum.
This application  is of course slightly orthogonal to the rest of the paper, not relating directly to blocks, but in fact it should  be very useful for performing precision explicit (super)conformal computations in concrete theories (for example $\cN=4$ super Yang-Mills). For example strong coupling results in $\cN=4$ SYM rely on the precise identification of free theory operators (see~\cite{Aprile:2019rep,Aprile:2025nta} for example),  a fact that will become increasingly important when going beyond half BPS superblocks.

\acknowledgments

We would like to thank Francesco Aprile, Ilija Buri\'c and  Zhongjie Huang for useful discussions. HPR is supported by an STFC studentship. TH and PH are supported by the STFC Consolidated Grant ST/X000591/1.

\appendix

\section{Notation and useful identities}\label{app:superalgebra}
\paragraph{Index-free notation:}
Throughout the main text, we make extensive use of an index-free notation which follows the order of matrix multiplication. For instance,
\begin{equation}
    (\hX_{ij})_{\dfa \fa} X_{jk}^{\fa \dfb} (\hX_{kl})_{\dfb \fb} X^{\fb \dfc}_{lm} \equiv (\hX_{ij} X_{jk} \hX_{kl} X_{lm})_{\dfa}^{\ \dfc}.
\end{equation}
This leads to neater expressions with less clutter while maintaining all the relevant information. 
One has to be a bit careful however since for expressions involving differential operators, the matrix ordering does not always agree with the ordering of the application of the differential operator. For example, the following two expressions are used frequently throughout the article,
\begin{equation}
    m_R {}_{\dfa}^{\ \dfb} = \bX^{\cA \dfb} \partial_{\bX}{}_{\dfa \cA} \equiv (\partial_{\bX} \bX)_{\dfa}^{\ \dfb}, \qquad \quad X^{\fa}_{\ \cA} (\partial_X)^{\cB}_{\ \fa} \equiv (\partial_X X)^{\cB}_{\ \cA}\,.
\end{equation}

\paragraph{Differentiation:} The action of the two Grassmannian derivatives $\partial_X$ and $\partial_{\bX}$ on all $N$-point structures follows from the following basic identities (recalling that $\hat X_{ij}\coloneq(X_{ji})^{-1}$)
\begin{align}
    (\partial_{X_i}){}^{\cA}_{\ \fa}(\hX_{ji})_{\dfb \fb} &= - (\hX_{ji})_{\dfb \fa}(\bX_j \hX_{ji})^{\cA}_{\ \fb}, & \qquad (\partial_{X_i}){}^{\cA}_{\ \fa} (\hX_{ij})_{\dfb \fb} &= 0, \\
    (\partial_{\bX_i})_{\dfa \cA} (\hX_{ji})_{\dfb \fb} &= 0, & (\partial_{\bX_i})_{\dfa \cA}(\hX_{ij})_{\dfb \fb} &= -(\hX_{ij})_{\dfa \fb} (\hX_{ij}X_j)_{\dfb \cA}.
\end{align}
Relevant corollaries of the above are as follows
\begin{align}\label{a5}
    (\partial_{X_i}){}^{\cA}_{\ \fa} g_{ji} &= - g_{ji} (\bX_j \hX_{ji})^{\cA}_{\ \fa}, & \qquad (\partial_{X_i}){}^{\cA}_{\ \fa} g_{ij} &= 0, \\
    (\partial_{\bX_i})_{\dfa \cA} g_{ji} &= 0, & (\partial_{\bX_i})_{\dfa \cA} g_{ij} &= -g_{ij} (\hX_{ij}X_j)_{\dfa \cA},
\end{align}
and
\begin{equation}
    (\partial_{X_i})^{\cA}_{\ \fa} (Y_{i,jk})_{\dfb \fb} = - (Y_{i,jk})_{\dfb \fa} (\bX_k \hX_{ki})^{\cA}_{\ \fb}, \qquad \quad  (\partial_{\bX_i})_{\dfa \cA} (Y_{i,jk})_{\dfb \fb} = - (Y_{i,jk})_{\dfa \fb} (\hX_{ij} X_j)_{\dfb \cA}.
\end{equation}

\paragraph{Algebra of Isotropy group:}
The commutators can be obtained straightforwardly from \eqref{eq:generators} and are given by
\begin{align}\label{commutators}
[k^{\fa\dfa},p_{\dfb\fb}] = \delta^{\dfa}_{\dfb} m_L{}^{\fa}_{\ \fb} +\delta^{\fa}_{\fb} m_R{}^{\ \dfa}_{\dfb}, \qquad
[k^{\fa\dfa},m_L{}^\fb_{\ \fc}] = -\delta^{\fa}_{\fc} k^{\fb\dfa}, \qquad
[k^{\fa\dfa},m_R{}^{\ \dfb}_{\dfc}] = \delta^{\dfa}_{\dfc} k^{\fa\dfb}.
\end{align}

\paragraph{Grassmannian identities:}

The  key identity~\eqref{g=g} $g_{ij}=g_{ji}$ can be shown as follows. First note that
\begin{align}
    \begin{pmatrix}
        X_i \\ X_j
    \end{pmatrix}(W_i\  \bar X_i) = \begin{pmatrix}
        1&0\\X_j W_i&X_j\bar X_i
    \end{pmatrix}\,.
\end{align}
where we used~\eqref{eq:ginvg}. 
Now taking the (super)determinant of both sides and using that $\det g=1$ (recall~\eqref{inv}) gives
\begin{align}
   \det \begin{pmatrix}
        X_i \\ X_j
    \end{pmatrix} = \det
       X_j\bar X_i = \det X_i \bar X_j\,,
\end{align}
where the second identity  follows simply from the fact that the LHS is clearly invariant on swapping point $i$ and $j$ therefore the RHS must also. A similar computation can be done  involving  $\det (\bar X_i\, \bar X_j)$ and we find
\begin{align}
g_{ij}=g_{ji} =      \det \begin{pmatrix}
        X_i \\ X_j
    \end{pmatrix}^{-1} = \det (\bar X_i\  \bar X_j)^{-1}\,.  
\end{align}

We also have the following key identity
\begin{align}\label{keyid}
    \bar X_i \hat X_{ij} X_j&=- \bar X_j \hat X_{ji} X_i +\mathbb{I}   
\end{align}
which can be proved either by going to the conformal / superspace coordinate frame or more directly by multiplying both sides on the right by the invertible matrix $(\bar X_i, \bar X_j)$ and observing that both sides give the same result $(\bar X_i,0)=-(0,\bar X_j)+(\bar X_i,\bar X_j)$. Notice that this makes use of the key identity $X\bar X=0$.
This identity then implies the identity on 3-point structures:
\begin{align}\label{3ptid}
X_{ki}\hat X_{ij} X_{jk}= - X_{kj}\hat X_{ji} X_{ik}\,. 
\end{align}
Note that if we relaxed the orthogonality condition $X_k\bar X_k=0$ at point $k$ this would instead become
\begin{align}
X_{ki}\hat X_{ij} X_{jk}= - X_{kj}\hat X_{ji} X_{ik}+X_k\bar X_k\,,
\end{align}
and  taking the inverse of both sides recalling the definition $Y_{kij}=\hat X_{ki} X_{ij} \hat X_{jk}$ gives
\begin{align}
    Y_{kji}=-Y_{kij}(1-X_{k}\bar X_k Y_{kij} )^{-1} = -Y_{kij}-Y_{kij}X_{k}\bar X_kY_{kij} +O((X_k\bar X_k))^2\ .
\end{align}
We see that the structure within which $X\bar X$ appears in this identity is consistent with the general ideal constraint we impose~\eqref{idealcon} 
\begin{align}\label{idealcon2}
  {\mathcal Z}_{ \dfa \fa} \coloneq(\hat X_{j1}X_k \bar X_k X_{1i})_{\dfb_j \fb_i}f^{ \fb_i\dfb_j}_{\dfa \fa}=0, \qquad \qquad  f^{ \fb_i\dfb_j}_{ \dfa  \fa}= (\hat X_{ki}X_{ij})_{\dfa}^{\ \dfb_j}  (\hat X_{ij}X_{jk})^{\fb_i}_{\ \fa} \, .
\end{align}
One can also consider relaxing the orthogonality constraint $X\bar X=0$  at the other points $i,j$ in the identity~\eqref{3ptid}. This also gives structures consistent with the ideal constraint~\eqref{idealcon}.

\section{Young projectors}
\label{app:yt}

We will use the same notation~$\ula_{L}$ to denote the Young diagram, and its corresponding $\GL(m|n)$ representation, as well as the multi-index symmetrised according to the corresponding lexicographically ordered Young tableau which we denote $\ula(1)$, thus
\begin{align}
\Yvcentermath1
    \cO_{\ula_L,\ula_R} \equiv \cO_{\underline \fa, \underline \dfa} \qquad \text{where} \qquad \Pi^{\ula_L}_1\cO_{\underline \fa, \underline \dfa}=\bPi^{\ula_R}_1\cO_{\underline \fa, \underline \dfa}=\cO_{\underline \fa, \underline \dfa}\ . 
\end{align}

Throughout the paper we make use of Hermitian Young projectors (HYP)s~\cite{keppeler2014hermitian,alcock2017compact}, $\Pi^{\ula}_i$, and transition operators relating them, $\Pi^{\ula}_i$~\cite{Alcock-Zeilinger:2016cva}. There is a beautiful story involving these which are directly related to Young's semi-normal units which we will make use of (see~\cite{garsia2020lectures} for a nice introduction to Young's semi-normal units and~\cite{Alcock-Zeilinger:2016cva} for their rediscovery as  Hermitian Young projectors and transition operators.) 

The HYPs  project multi-indices into specific irreducible representations (as of course do the more well known Young projectors). Transition operators  can be labelled by two Young tableaux, $\Pi_{T_1,T_2}$ and the HYPs are simply $\Pi_{T}\coloneq\Pi_{T,T}$. We will only consider standard Young tableaux. A standard  Young tableau is a Young diagram with the numbers 1 to n put into each box such that the numbers are strictly ordered both horizontally and vertically.  We often instead directly put the indices in the boxes of the Young diagrams (with an implicitly  understood ordering of the indices e.g.~\eqref{ytindices}).
The standard Young tableaux can be listed by ordering them lexicographically, i.e.\ as you would order them if viewed as digits of numbers (in a base $>n+1$ for $S_n$) written out left to right and top to bottom. Then the $i$th standard tableau in this ordering is denoted by $\ula(i)$. For example, consider the following, which illustrates a simple Young diagram and the corresponding standard Young tableaux:
\begin{equation}\label{eq:YDsimpleexample}
\Yvcentermath1
    \ula= \yng(2,1)\,, \qquad \quad \ula(1)= \young(12,3)\,, \qquad \quad \ula(2)= \young(13,2)\,. 
\end{equation}
We then denote the transition operator from $\ula(j)$ to $\ula(i)$ by
$$\Pi^\ula_{i,j}\coloneq\Pi_{\ula(i),\ula(j)}\,.$$

The transition operators and Young projectors then  represent specific linear combinations of  permutations $\sum_{\sigma \in S_s}a_\sigma \sigma$
for some coefficients $a_\sigma$. The permutations will be acting on indices and we thus  view them very concretely via Kronecker deltas. In other words we identify the action of 
$\sigma \in S_s$ on a tensor with $s$ indices 
$T_{a_1 ..a_s}$ as
\begin{equation}\label{permdel}
    \sigma T_{a_1 ..a_s} \coloneq T_{a_{\sigma^{-1}(1)} ..a_{\sigma^{-1}(s)}}= \delta_{a_1}^{a'_{\sigma(1)}}..\delta_{a_s}^{a'_{\sigma(s)}} T_{a'_1..a'_s}
\end{equation}
which can be viewed pictorially very nicely using the birdtrack formalism.
Similarly whenever we write permutation operators acting on indices we are really writing shorthand for such Kronecker deltas
\begin{align}
\Pi=\sum_{\sigma} a_\sigma \sigma \quad \Rightarrow \quad 
    \Pi T_{a_1..a_s}\coloneq\Pi^{a'_1.a'_s}_{a_1..a_s}T_{a'_1.a'_s}=\sum_\sigma a_\sigma \delta_{a_1}^{a'_{\sigma(1)}}..\delta_{a_s}^{a'_{\sigma(s)}} T_{a'_1..a'_s}\ .
\end{align}
There needs to be an ordering of the indices acted on for this to make sense - this ordering will often not be explicit but is hopefully clear form context. 
This interpretation becomes particularly important when considering contractions of indices amongst permutation operators, which we frequently do. 

For the explicit definition of the HYPs and transition operators win terms of symmetrising and antisymmetrising etc. we refer to the above literature,
\footnote{See for example 2.3.15, 2.3.16 of~\cite{garsia2020lectures} or theorem 6 of~\cite{keppeler2014hermitian}) and theorem 4 of~\cite{Alcock-Zeilinger:2016cva} for the precise definitions.}
instead we point out some of their beautiful properties which we will make use of.

The key equation satisfied by transition operators and HYPs 
(noting that HYPs can be viewed as transition operators with both indices equal: $\Pi^\ula_i=\Pi^\ula_{i,i}$ )
is
\begin{align}\label{units}
\Pi^{\ula}_{i,j}\Pi^{\umu}_{k,l}=\delta_{j,k}\delta_{\ula,\umu}\Pi^{\ula}_{i,l}\ . 
\end{align}
This implies both that HYPs are projectors $
\Pi^{\ula}_i\Pi^{\ula}_i=\Pi^{\ula}_i$ and that  they are orthogonal $\Pi^{\ula}_i\Pi^{\umu}_j=\delta_{ij}\delta_{\ula,\umu}\Pi^{\ula}_i$. 
There are also precisely $n!$ transition operators (including HYPs) and thus they form a special basis in the linear space of permutations, analogous to the natural basis of matrices. 

Another key property  we make use of is 
\begin{align}\label{Tbar}
\Pi_{T_1,T_2}\Pi_{\overline{T}_2}=\Pi_{\overline{T}_1}  \Pi_{T_1,T_2}=\Pi_{T_1,T_2} 
\end{align}
where $\overline T$ represents the Young tableau with the last entry (and its box) deleted, e.g.\ in the example \eqref{eq:YDsimpleexample} we would have
\begin{equation}
    \Yvcentermath1
    \overline{\ula(1)}= \young(12)\,, \qquad \qquad \overline{\ula(2)}= \young(1,2)\,.
\end{equation}

The HYPs also give a complete decomposition of the vector space of multiple-indices into  irreducible subspaces, implementing Schur-Weyl duality. For our purposes, the key equation it satisfies is~\eqref{addbox0} 
\begin{align}\label{addbox2}
\sum_{\ula'_L\in\Box\otimes \ula_L}  \Pi^{\ula'_L}_{j}= \Pi^{\ula_L}_1\,, \qquad \text{where} \qquad \overline{\ula'_L(j)}=\ula_L(1)\ .
\end{align}
Note  $\ula'_L(j)$ is the unique Young tableau of shape $\ula'$ containing $\ula_L(1)$.
This result is theorem 2.8 of~\cite{garsia2020lectures}. On the left hand side the sum is over all HYPs of Young tableaux obtained by adding the last box to the $\ula_L$ Young tableau. On the right hand-side you have the HYP of $\ula_L$ and the last index is not involved in any permutations.

The other key formula the HYPs satisfy is:
\begin{align}\label{HYTbubble}
\Pi^{\ula}_i \left( \sum_k (ks)\right) =\left(\sum_k (k s) \right) \Pi^{\ula}_i = C_{\ula(i)}
\Pi^{\ula}_i\ ,
\end{align}
where $C_{\ula(i)}\coloneq L_{\ula(i)}-H_{\ula(i)}$~\eqref{eq:content} and $s=|\ula|$ is the number of boxes of $\ula$.
This is a special case of theorem 2.13 of~\cite{garsia2020lectures} which originated in the work of Jucys~\cite{jucys1974symmetric} and Murphy~\cite{murphy1981new}. 
The identity~\eqref{eq:mLconstant} and~\eqref{eq:mRconstant} can be straightforwardly derived from this on noting that the action of $m_{L,R}$ in~\eqref{eq:primary} can be succinctly written as a sum of swaps, like the $(\fa_k\fa_s)$  in~\eqref{HYTbubble}
\begin{align}
     (m_L)^{\mathfrak{c}}{}_{\fb} \cO^{\gamma}_{\underline{\fa},\underline{\dfa}}&
        = -\left(\gamma_L-\sum_{k=1}^{s-1}(\fa_k \fb)\right)
         \cO^{\gamma}_{\underline{\fa}, \underline{\dfa}}\delta^{\mathfrak{c}}_{\fb}=-\left(\gamma_L-\sum_{k=1}^{s-1}(ks)\right)^{\underline \fa' \fb'}_{\underline \fa \fb} 
         \cO^{\gamma}_{\underline{\fa}', \underline{\dfa}}\delta^{\mathfrak{c}}_{\fb'}\,.
\end{align}
Then~\eqref{eq:mLconstant} follows directly  from ~\eqref{HYTbubble}
and~\eqref{eq:mRconstant} follows in a similar way.

Next we consider~\eqref{bubble2}. Writing out the indices explicitly this is
\begin{align}
    (\cD_{\cA})^{\ula'_L}_{\ula''_L} (\cD^{\cA})^{\ula_L}_{\ula'_L} \cO_{\ula_L, \ula_R}^{\gamma} \hspace{-1mm} &= c_0 X^{\fb}_{\ \cA} \left(\Pi^{\ula_L'}_{i,1}\right)^{\underline \fa''}_{\underline \fa' \fb'} \hspace{-1mm} \left(\Pi^{\ula'_L}_{1,j}\right)^{\underline \fa \fb}_{\underline \fa''} \hspace{-1mm}\bigg( W {}^{\cA}_{\ \fb} +\bX^{\cA \dfb}\hspace{-2mm}\sum_{\mu_R \in \Box \otimes \ula_R} \hspace{-2mm}c_{\ula_L'\umu_R} \bPi^{\umu_R}p_{\dfb \fb}\bigg) \cO_{\underline \fa, \ula_R}^{\gamma} \nn \\
        & = c_0 \left(\Pi^{\ula_L'}_{i,j}\right)^{\underline \fa \fb}_{\underline \fa' \fb}  \cO_{\underline \fa, \ula_R}^{\gamma}\, .\label{bubble2details}
\end{align}
Then we need to consider a transition operator with the final two indices contracted. This yields a  permutation operator of one fewer elements. Now recalling that  $\overline{\ula'_L(j)}=\ula_L(1),\ \overline{\ula'_L(i)}=\ula''_L(1)$ 
 we have from~\eqref{Tbar} and~\eqref{addbox2} that
 \begin{align}\label{pis}
\Pi^{\umu_L}_k\Pi^{\ula_L'}_{i,j}=\Pi^{\ula'_L}_{i,j} \delta_{\umu_L,\ula''_L}\delta_{k,1}, \qquad  \qquad \Pi^{\ula_L'}_{i,j}\Pi^{\umu_L}_l=\delta_{\umu_L,\ula_L} \delta_{l,1}\ . 
 \end{align} 
Now since the transition operators and HYPs form a basis in the linear space of permutations this implies that
\begin{align}
    (\Pi^{\ula_L'}_{i,j})^{\underline \fa \fb}_{\underline \fa' \fb}  =  \sum_\umu \sum_{k,l} a_{\umu,k,l}(\Pi^\umu_{k,l})^{\underline \fa }_{\underline \fa'} = \delta_{i,j} (\Pi^{\ula_L}_1)^{\underline \fa }_{\underline \fa'}\ .
\end{align}
The second equality holds by applying $\Pi^{\umu_L}_k$ on the left and $\Pi^{\umu_L}_l$ on the right and using~\eqref{pis} to set all but one of the coefficients $a_{\umu,k,l}$ to zero.
 The result~\eqref{bubble2} follows.

In practice, in explicit checks we have not needed the Hermitian Young projector so far and we have used the Mathematica packages xAct~\cite{martin2002xact}, xTensor~\cite{martin2008xtensor} and xTras~\cite{nutma2014xtras} to implement the standard Young symmetrisation.

\Yboxdim{13pt}
\section{From coset coordinates to Grassmannian coordinates}
\label{app:xxbproof}

In section~\ref{sec:fieldsandreps}, we  defined operators on the coset space, with variables $X,\bar W$. But correlators and representations are also naturally given as functions of the orthogonal planes $X$ and $\bX$. In the Grassmannian language, $\bW$ and $W$ are redundancies that should not appear in the final results. This is because elements of the coset space are given by orbits in the equivalence class $g\sim hg$, in which a representative can always be chosen such that $\bW= (0, 1)$.
 It is thus useful to change variables from $g=(X,\bar W)$ to the more natural set of variables $(X, \bX)$. However, these are not independent, since they satisfy  $X \bX=0$ (and all identities derived from $g g^{-1}=1$, see \eqref{eq:ginvg}). Thus we don't have enough degrees of freedom.
Therefore, in order to safely change variables, we must include some of the $\bar W$ variables to account for this. A nice way to define the additional $\bar W$ variables is via an $X$ dependent projector $P^{\cA}{}_{\fa}$ as follows
\begin{align}
	\Big(X,\bar W\Big) &\rightarrow \Big(X,\bar X(X,\bar W), \mathcal{W}(X,\bar W)\Big) 
\end{align}
where
\begin{align} 
	\mathcal{W}_{\dfa \fa} &\coloneq \bW_{\dfa \cA} P^{\cA}{}_{\fa}(X), \qquad X^{\fb}_{\ \cA}P^{\cA}{}_{\fa}=\delta^{\fb}_{\fa}.
\end{align}
This leads to the following new form of the generators acting on functions of $X,\bX$ and $\mathcal{W}$
\begin{equation}\label{eq:gdgnew}
	\left( \begin{array}{cc}
		(m_L)^{\fa}_{\ \fb} &  k^{\fa \dfb} \\ 
		p_{\dfa \fb}  & -(m_R)^{\ \dfa}_{ \dot{\mathfrak{b}}}
	\end{array}\right)= \left(\begin{array}{cc}
		(X \partial_X + \partial_{\mathcal{W}}\mathcal{W} )^{\fa}_{\ \fb} & (\partial_{\mathcal{W}})^{\fa \dfb} \\ 
		(\bW\partial_X- \partial_{\bX} W)_{\dfa \fb} & (-\partial_{\bX} \bX + \mathcal{W} \partial_{\mathcal{W}} )^{\ \dfa}_{ \dot{\mathfrak{b}}}
	\end{array}\right).
\end{equation}
Then, superprimaries are annihilated by $k^{\fa \dfb}=(\partial_{\mathcal{W}}{})^{\fa \dfb}$, that is, they are functions of $X$ and $\bX$ only.

\section{Correspondence between representations}\label{sec:repexamples}
We now show the correspondence between the representations we have constructed in $(m,n)$ analytic superspace and the representation of superconformal representations in terms of Lorentz spin and R-symmetry. We will also relate the various shortening conditions to the notation of~\cite{Cordova:2016emh}.
\subsection{Non-supersymmetric CFT in 1D and 4D}
\paragraph{1 dimension.}
The conformal group of a line is $\SU (1,1)$, whose complexification is $\SL(2)$. Thus, 1D CFT can be described by the $(1,0)$ coset space. $\GL(1)$ Young diagrams consist of one row and thus are only labelled by their length. A field can thus be labelled by three numbers $[\gamma, \lambda_L, \lambda_R]$, which are related to the 1D conformal dimension by
\begin{equation}\label{eq:1Dreps}
  \Delta = \gamma+ \lambda_L + \lambda_R.
\end{equation}
Indeed the Levi subgroup is $S(\GL(1)\times \GL(1)) \cong \mathbb{C}^*$, so adding/removing boxes to $\lambda_L, \lambda_R$ simply corresponds to shifting $\gamma$.
\paragraph{4 dimensions.}
This corresponds to $(2,0)$ analytic superspace. As in the 1D case, both formulations are nearly identical. In particular, $\GL(2)$ Young diagrams have $2$ rows encoding the usual spin. We denote a diagram with two rows by $[\lambda_1, \lambda_2]$, $\lambda_{1}\geq\lambda_2$, where $\lambda_{1},\lambda_2$ are the lengths, i.e.\ number of boxes of each row. Then, the representation with charge $\gamma$ and two Young diagrams $\ula_L=[\lambda_L^1, \lambda_L^2],\ \ula_R=[\lambda_R^1, \lambda_R^2]$ is equivalent to
\begin{equation}\label{eq:4Dreps}
    \ell = \lambda_L^1-\lambda_L^2\,, \qquad \bar{\ell}=\lambda_R^1-\lambda_R^2\,, \qquad \Delta=\gamma+ \frac12 (\lambda_L^1 + \lambda^2_L + \lambda_R^1 + \lambda_R^2)\,,
\end{equation}
where $\ell, \bar{\ell}$ denote the number of Weyl spinor indices.
Note that $\gamma$ can absorb two rows of equal length so that Young diagrams are only given by one row. Alternatively, $\gamma$ can be set to $0$ so that the Young diagrams fully encode the dimension as well as the spin. Indeed note that in 4D, antisymmetrising two Weyl spinor indices gives a scalar i.e.\ only contributes to the scaling dimension $\Delta$.

\subsection{Internal space}
\label{internal}
A very useful special case to understand supersymmetric representations in analytic superspace, is  to  consider representations on the internal space alone. That is, consider a theory whose space-time symmetry group is given by the $R$-symmetry, a compact group $\mathrm{U} (\cN)$. 
This is viewed as the real form of $\GL(|2n)$. Thus, the relevant superspace here is $(0,n)$. Now, the Levi subgroup is $\GL(|n)\times \GL(|n)$. Note that the determinant constraint can be applied to instead consider theories with $\SU (\cN)$ symmetry, such as the $R$-symmetry of 4D $\cN=4$.

In the $(0,n)$ theory the operators $\cO^\gamma_{\ula_L,\ula_R}$ correspond to finite dimensional representations of $\SU (2n)$ and thus the weight-shifting operators, OPEs etc all reduce to tensor products of finite representations of $\SU (2n)$ which have of course been very well studied. This can allow for further tests and checks of our results. We give here the direct translation then between a representation given as a $(0,n)$ operator $\cO^\gamma_{\ula_L,\ula_R}$ and a $\SU (2n)$ irrep.

Following the usual rules for such generalised Young diagrams, $\GL(|n)$ diagrams are simply $\GL(n)$ representations with symmetrisation and antisymmetrisation of indices swapped. This means the Young diagrams are transposed and thus instead of being restricted  to have height less than or equal to $n$, it is instead the width which is cut off. The correspondence between operators on $(0,n)$ analytic superspace operators and $\mathrm{U} (2n)$ representations is given by
\Yboxdim{13pt}
\begin{align}
\cO^{\gamma}_{\ula_L,\ula_R} \quad \leftrightarrow \quad
\begin{tikzpicture}[x=13pt,y=13pt,baseline=(align), scale=0.9]
    \tyng(0,0,16,14,13,12,10,8,6,3);
    \draw [ultra thick, dashed] (12,-3) to (12,-2) to (13,-2) to (13,-1) to (14,-1) to (14,0) to (16,0) to (16,1)to (19,1)to (19,-3) to (12,-3) ;
    \node at (16.5,-1){$\ula_L^T$};
    \draw [ultra thick, dashed] (2,-3)to (10,-3) to (10,-4) to(8,-4) to(8,-5) to (6,-5) to (6,-6) to(3,-6) to(3,-7) to (2,-7) to (2,-3);
     \node at (6.8,-6){$\ula_R^T$};
     \node at (10,-8.1){$\gamma$};
      \draw[thick, stealth-stealth](19,-7.5) to (2.1,-7.5);
      \node at (1,-8.1){$k$};
      \draw[thick, stealth-stealth](0,-7.5) to (1.9,-7.5);
     \node at (-1.1,-1){$n$};
      \draw[thick, stealth-stealth](-0.5,1) to (-.5,-2.9);
     \node at (-1.1,-5){$n$};
      \draw[thick, stealth-stealth](-0.5,-3.1) to (-.5,-7);
    \node (align) at (0,-3) {};
\end{tikzpicture}\ .
\label{map}
\end{align}
Concretely, if $\ula_L$ has column heights $\lambda_{L1},\lambda_{L2},...$ (i.e.\ the transposed Young diagram $\lambda_L^T$ has row lengths $\lambda_{L1},\lambda_{L2},...$)
and similarly $\ula_R$ has column heights $\lambda_{R1},\lambda_{R2},...$ then the corresponding $\mathrm{U} (2n)$ Young diagram $\ula$ has row lengths $\lambda_1, ..., \lambda_{2n}$ given by
\begin{equation}
    \begin{aligned}
        	\lambda_i&=k+\gamma - \lambda_{L(n+1-i)}\,,\qquad &i=1,..,n\,,\\
	\lambda_{i+n}&=k+\lambda_{R(i)}\,,\qquad &i=1,..,n\,,
 \label{rel}
    \end{aligned}
\end{equation}
where $k$ is arbitrary.
Given a charge $\gamma$ and left and right Young diagrams $\ula_L, \ula_R$ whose widths are $\leq n$ and whose combined height is $\leq \gamma$, this map gives  a unique $\mathrm{U} (2n)$ Young diagram (one can add an arbitrary number $k$  of full columns of height  $2n$ to this without altering the $\mathrm{U} (2n)$ rep).

As usual, there is an additional subtlety when taking the unit determinant constraint $\mathrm{U} (2n)\mapsto \SU (2n)$.
Indeed, in this case, the inverse map from the $\SU (2n)$ Young diagram to the operator is not unique. In particular there is freedom to choose the values of $k,\gamma$ within a certain range 
\begin{align}
 \gamma_1+\gamma_2 &\geq \gamma \geq \lambda_1-k\,, \qquad 0\leq k\leq \lambda_{2n}\ .
\end{align}
After they are fixed, the left and right Young diagrams, $\lambda_{Li}$,  $\lambda_{Ri}$,  are then fixed by~\eqref{rel}.
This non uniqueness simply reflects  the fact that the  operator itself is not uniquely labelled. Rather the following operators all correspond to the same representation
$\cO^{\gamma}_{\lambda_L,\lambda_R} \sim \cO^{\gamma+1}_{[n,\lambda_L],[\lambda_R]}\sim \cO^{\gamma+1}_{[\lambda_L],[n,\lambda_R]}$ (where by $[n,\lambda_L]$ we mean the Young diagram with a row of length $n$ appended above it). 

\subsection{4D $\cN=2$}\label{sec:N2repsexamples}
This corresponds to $(2,1)$ analytic superspace associated with the symmetry group $\SL(4|2)$.
The transformations under superspace rotations are now given by two $\GL(2|1)$ Young diagrams. The key difference from the non-supersymmetric case is that these diagrams can now include an arbitrarily long first column.

Four-dimensional $\cN=2$ superconformal representations obey symmetric unitarity bounds and shortening conditions for both $Q$ and $\bar{Q}$. This symmetry, often referred to as \emph{two-sidedness} is encoded by the two distinct $\GL(2|1)$ Young diagrams on the supercoset space. In particular, shortening conditions for $Q$ are described by the left diagram, while those for $\bar{Q}$ correspond to the right one.
Indeed the shortening conditions of the representation as a whole are entirely inherited from the corresponding shortening conditions of the finite-dimensional representations of $\SL(2|1)$. 
The $Q$ and $\bar{Q}$ shortening conditions in \cite{Cordova:2016emh} are denoted as $B, A_2, A_1$ and $L$, where $B$ denotes isolated reps (i.e.\ there are no nearby unitary reps as one varies the dimension), $L$ long reps, and $A$ short reps at the threshold of the unitarity bound (which occur with two types of shortening conditions). 
They map to our notation is then as follows
\begin{equation}\label{eq:ABLshorteningN2}
    B \leftrightarrow [0], \qquad
    A_2\leftrightarrow\Box, \qquad
    A_1\leftrightarrow [\lambda>1], \qquad 
    L\leftrightarrow [\lambda_1,\lambda_2,1^\mu], \qquad
\end{equation}
where $1^\mu$ denotes $\mu$ rows of length $1$, so the first column of $L$ has height $\mu+2$. 
As in the non-supersymmetric cases, the length of each row is directly related to the spin $\ell$ and $\bar{\ell}$. 
The new arbitrarily long column is linked with the $\SU (2)$ part of the $R$-symmetry ($\mathrm{U} (2)$) representation, whereas 
the remaining $\mathbb{C}^*$ charge is related to the difference in the number of boxes between left and right Young diagrams. 
Table \ref{tab:4DN4reps2} shows particular examples or multiplets with physically relevant fields and table \ref{tab:4DN4reps1} shows the full correspondence between $(2,1)$ supercoset space representations and four-dimensional $\cN=2$ superconformal representations in the notation of \cite{Cordova:2016emh} which we used in the main text \eqref{eq:multiplet2}. 
\begin{table}[ht]
    \centering
    \renewcommand{\arraystretch}{1.5}
    \scalebox{0.9}{\begin{tabular}{|c||c|c|}
    \hline
        Multiplet & $[\gamma, \ula_L, \ula_R]$& Notation\\ \hhline{|=::==|}
        Hypermultiplet & $[1,\bullet, \bullet]$& $\cO^1$ \\ \hline
        Flavour current multiplet& $[2,\bullet, \bullet]$& $\cO^2$ \\ \hline
         \ Yang-Mills multiplet \  & $[1,\Box, \bullet]$ &$ \cO^1_{\fa}$\\ \hline
         \ Extra-SUSY current \  & $[2,\Box, \bullet]$ &$ \cO^2_{\fa}$\\ \hline
         Higher-spin currents & \ $[2,[\ell+1],\bullet]$ \ \ & \  $ \cO^2_{(\fa_1 \dots \fa_{\ell+1})}$ \ \\ \hline
         Stress-tensor multiplet & $[2,\Box,\Box]$ & $\cO^2_{\dfa \fa}$ \\ \hline
\end{tabular}}
    \caption{Examples of physically relevant 4D $\cN=2$ multiplets in the $(2,1)$ coset space.}
    \label{tab:4DN2notablereps}
\end{table}
\newpage
\begin{landscape}
\begin{table}[h]
    \centering
    \resizebox{1.28\textwidth}{!}{\renewcommand{\arraystretch}{1.8}
\setlength{\tabcolsep}{6pt}
\begin{tabular}{|c||c|c|c|c|}
\hline
& $\mathbf{\bar{L}}$&  $\mathbf{\bar{A}_1}$ & $\mathbf{\bar{A}_2}$ & $\mathbf{\bar{B}}$ \\ 
\hhline{|=::====|} 
\multirow{3}{*}{$\mathbf{L}$} & $\left[\gamma, [\lambda_L^1, \lambda_L^2, 1^{\mu_L}], [\lambda_R^1, \lambda_R^2, 1^{\mu_R}]\right]$ &  $\left[\gamma, [\lambda_L^1, \lambda_L^2, 1^\mu], [\lambda_R>1]\right]$ & $[\gamma, [\lambda^1, \lambda^2, 1^\mu], \Box]$ & $\left[\gamma, [\lambda^1, \lambda^2, 1^\mu],\bullet \right]$ \\
& $[\lambda_L^1-\lambda_L^2,\lambda_R^1-\lambda_R^2]_{\Delta}^{(\gamma-\mu_{L,R}-4;\delta\mu+\delta|\ula|)}$ &  $[\lambda^1_L-\lambda_L^2\lambda_R-1]_{\Delta}^{(\gamma-\mu-3;\mu+1+\delta|\ula|)}$ & $[\lambda^1-\lambda^2,0]_{\Delta}^{(\gamma-\mu-3;|\ula_L|+\mu+1)}$ & $[\lambda^1-\lambda^2,0]^{(\gamma-\mu-2;|\ula_L|+\mu+2)}_{\Delta}$ \\ 
 & $\Delta=\gamma+\frac12(\lambda_{L,R}^1+\lambda_{L,R}^2)-2$ & $\Delta=\gamma+\frac12(\lambda_{L}^1+\lambda_{L}^2 + \lambda_R-3)$& $\Delta=\gamma+\frac12(\lambda^1+\lambda^2)-1$ & $\Delta=\gamma+\frac12(\lambda^1+\lambda^2)-1$ \\
\hline
\multirow{3}{*}{$\mathbf{A}_1$} & $\left[\gamma, [\lambda_L>1], [\lambda_R^1, \lambda_R^2, 1^{\mu}]\right]$& $\left[\gamma, [\lambda_L>1], [\lambda_R>1]\right]$ & $[\gamma, [\lambda>1], \Box]$ & $\left[\gamma, [\lambda>1],\bullet \right]$ \\
& $[\lambda_L-1,\lambda_R^1-\lambda_R^2]_{\Delta}^{(\gamma-\mu-3;\delta|\ula|-\delta\mu-1)}$ &$[\lambda_L-1,\lambda_R-1]_{\Delta}^{(\gamma-2;\lambda_L-\lambda_R)}$ & $[\lambda-1,0]_{\Delta}^{(\gamma-2;\lambda-1)}$ & $[\lambda-1,0]^{(\gamma-1;1+\lambda)}_{\Delta}$  \\
& $\Delta=\gamma+\frac12(\lambda_{R}^1+\lambda_{R}^2 + \lambda_L-3)$ & $\Delta=\gamma+\frac12(\lambda_L+\lambda_R-2)$ & $\Delta= \gamma+\frac12(\lambda-1)$ & $\Delta=\gamma+\frac12(\lambda-1)$ \\
\hline
\multirow{3}{*}{$\mathbf{A}_2$} & $\left[\gamma, \Box, [\lambda^1, \lambda^2, 1^{\mu}]\right]$ &$[\gamma, \Box , [\lambda>1]]$ & $\left[\gamma, \Box,\Box \right]$ & $\left[\gamma, \Box,\bullet \right]$ \\
&$[0,\lambda^1-\lambda^2]_{\Delta}^{(\gamma-\mu-3;-|\ula_R|-\mu-1)}$ &$[0,\lambda-1]_{\Delta}^{(\gamma-2;1-\lambda)}$ & $[0,0]^{(\gamma-2;0)}_{\Delta}$ & $[0,0]^{(\gamma-1;2)}_{\Delta}$ \\
& $\Delta=\gamma+\frac12(\lambda^1+\lambda^2)-1$ & $\Delta= \gamma+\frac12(\lambda-1)$ & $\Delta= \gamma$ & $\Delta= \gamma$ \\
\hline
\multirow{3}{*}{$\mathbf{B}$} &  $\left[\gamma, \bullet, [\lambda^1, \lambda^2, 1^{\mu}]\right]$ & $\left[\gamma,\bullet ,[\lambda>1] \right]$ & $\left[\gamma, \bullet,\Box \right]$ & $[\gamma, \bullet,\bullet]$ \\
&$[0,\lambda^1-\lambda^2]_{\Delta}^{(\gamma-\mu-2;-|\ula_R|-\mu-2)}$ &$[0,\lambda-1]^{(\gamma-1;-1-\lambda)}_{\Delta}$ & $[0,0]^{(\gamma-1;-2)}_{\Delta}$ & $[0,0]^{(\gamma;0)}_{\Delta}$ \\
& $\Delta=\gamma+\frac12(\lambda^1+\lambda^2)-1$ & $\Delta= \gamma+\frac12(\lambda-1)$ & $\Delta=\gamma$ & $\Delta= \gamma$ \\
\hline
 \end{tabular}}
 \caption{$(2,1)$ analytic superspace description of all 4D $\cN=2$ representations. The notation $\bullet_{L,R}\coloneq\bullet_L+\bullet_R$ and $\delta \bullet = \bullet_L - \bullet_R$ is used to avoid clutter.}
 \label{tab:4DN2reps1}
\end{table}
\end{landscape}
\newpage

\subsection{4D $\cN=4$}
This corresponds to $(2,2)$ analytic superspace and the symmetry group $\SL(4|4)$.
As in the $\cN=2$ case, we have two-sided multiplets with $Q$ and $\bar{Q}$ shortening conditions, albeit they are now encoded by two $\GL(2|2)$ Young diagrams. These can now have two arbitrarily long columns describing the three Dynkin labels of the $\SU (4)$ R-symmetry representation (together with $\gamma$). 
Note that $\SL(2m|2n)$ is not semi-simple when $m=n$, which means that representations are actually of $\mathrm{PSL}(2m|2n)\coloneq \SL(2m|2n)/\mathcal{Z}$ where $\mathcal{Z}$ is the centre. 
This induces an additional degeneracy in the $\GL(2|2)\times \GL(2|2)$ representations, which can be fixed by requiring that left and right Young diagrams have the same number of boxes.
As one may expect, this is analogous to the R-symmetry being $\SU (4)$ instead of $\mathrm{U}(4)$. Indeed, recall from the previous section that in 4D $\cN=2$, the $\mathbb{C}^*$ charge $r$ depended on the difference in the number of boxes between left and right diagrams.

The correspondence between $Q$ and $\bar{Q}$ shortening conditions and $\GL(2|2)$ Young diagrams is as follows
\begin{equation}
    B  \leftrightarrow [1^\mu], \quad A_2 \leftrightarrow [2,1^\mu], \quad A_1  \leftrightarrow [\lambda>2, 1^\mu], \quad L \leftrightarrow [\lambda_1,\lambda_2>1,2^{\mu_1},1^{\mu_2}].
\end{equation}
Table \ref{tab:4DN4reps2} shows particular examples or multiplets with physically relevant fields and table \ref{tab:4DN4reps1} shows the full correspondence between $(2,2)$ supercoset space representations and four-dimensional $\cN=4$ superconformal representations in the notation of \cite{Cordova:2016emh} given in \eqref{eq:N4multiplet}. 
\vspace{-0.3cm}
\Yboxdim{8pt}
\begin{table}[h]
    \centering
\renewcommand{\arraystretch}{1.5} 
\setlength{\tabcolsep}{4pt} 
    \scalebox{0.9}{\begin{tabular}{|c||c|c|}
    \hline
        Multiplet & $[\gamma, \ula_L, \ula_R]$& Notation\\ \hhline{|=::==|}
         Stress-tensor& $[2,\bullet, \bullet]$& $\cO^2$ \\ \hline
         Higher KK half modes& $[p>2,\bullet, \bullet]$& $\cO^p$ \\ \hline
         Free higher-spin currents $\cO_{01}$ & $[2,\Box,\Box]$ & $\cO^2_{\dfa \fa}$ \\ \hline
         First interacting $1/4$ BPS $\cO_{02}$ & $\Yvcentermath1 [2,\,\yng(1,1)\,,\,\yng(1,1)\,]$ & $\cO^2_{[\dfa \dfb] [\fa \fb]}$ \\ \hline
\end{tabular}}
    \caption{Examples of four-dimensional $\cN=4$ supermultiplets. }
    \label{tab:4DN4reps2}
\end{table}

\newpage

\begin{landscape}
\begin{table}[ht]
\centering
\resizebox{1.28\textwidth}{!}{\renewcommand{\arraystretch}{1.8}
\setlength{\tabcolsep}{6pt}
\begin{tabular}{|c||c|c|c|c|}
\hline
& $\mathbf{\bar{L}}$&  $\mathbf{\bar{A}_1}$ & $\mathbf{\bar{A}_2}$ & $\mathbf{\bar{B}}$ \\ 
\hhline{|=::====|} 
\multirow{3}{*}{$\mathbf{L}$} & $\left[\gamma, [\lambda_L^1, \lambda_L^2, 2^{\mu_L^1}, 1^{\mu_L^2}], [\lambda_R^1, \lambda_R^2, 2^{\mu_R^1},1^{\mu_R^2}]\right]$ &  $\left[\gamma, [\lambda_L^1, \lambda_L^2, 2^{\mu_L^1}, 1^{\mu_L^2}], [\lambda_R>2,1^{\mu_R}]\right]$ & $[\gamma, [\lambda^1, \lambda^2, 2^{\mu^1_L}, 1^{\mu^2_L}], [2,1^{\mu_R}]]$ & $\left[\gamma, [\lambda^1, \lambda^2, 2^{\mu_L^1}, 1^{\mu_L^2}],[1^{\mu_R}] \right]$ \\
& $[\lambda_L^1-\lambda_L^2,\lambda_R^1-\lambda_R^2]_{\Delta}^{(\mu^2_L,\gamma-\mu^{1,2}_{L,R}-4,\mu^2_R)}$ &  $[\lambda_L^1-\lambda_L^2,\lambda_R-2]_{\Delta}^{(\mu^2_L,\gamma-\mu^{1,2}_{L}-\mu_R-3,\mu_R)}$ & $[\lambda^1-\lambda^2,0]_{\Delta}^{(\mu^2_L,\gamma-\mu_L^{1,2}-\mu_R-3,\mu_R)}$ & $[\lambda^1-\lambda^2,0]^{(\mu^2_L,\gamma-\mu_L^{1,2}-\mu_R-2,\mu_R)}_{\Delta}$ \\ 
 & $\Delta=\gamma+\frac12(\lambda_{L,R}^1+\lambda_{L,R}^2)-4$ & $\Delta=\gamma+\frac12(\lambda_{L}^1+\lambda_{L}^2 +\lambda_R)-3$& $\Delta=\gamma+\frac12 (\lambda^1 +\lambda^2) -2$ & $\Delta=\gamma+\frac12 (\lambda^1 +\lambda^2) -2$ \\
\hline
\multirow{3}{*}{$\mathbf{A}_1$} & $\left[\gamma, [\lambda_L>2, 1^{\mu_L}], [\lambda_R^1, \lambda_R^2, 2^{\mu_R^1},1^{\mu_R^2}]\right]$ & $\left[\gamma, [\lambda>2,1^\mu], [\lambda \pm \nu, 1^{\mu\mp\nu}]\right]$ & $\left[\gamma, [\lambda>2, 1^\mu], [2,1^{\mu+\lambda-2}]\right]$ & $\left[\gamma, [\lambda>2, 1^\mu],[1^{\mu+\lambda}] \right]$ \\
& $[\lambda_L-2,\lambda_R^1-\lambda_R^2]_{\Delta}^{(\mu_L,\gamma-\mu^{1,2}_{R}-\mu_L-3,\mu^2_R)}$ &$[\lambda-2,\lambda\pm\nu-2]_{\Delta}^{(\mu,\gamma-2\mu \pm \nu-2,\mu \mp \nu)}$ & $[\lambda-2,0]_{\Delta}^{(\mu,\gamma-2\mu-\lambda,\mu+\lambda-2)}$ & $[\lambda-2,0]^{(\mu,\gamma-\lambda-2\mu-1,\mu+\lambda)}_{\Delta}$  \\
& $\Delta=\gamma+\frac12(\lambda_{L}+\lambda_{R}^1+\lambda_{R}^2)-3$ & $\Delta=\gamma+\lambda-2\pm\frac{\nu}{2}$ & $\Delta= \gamma+\frac12(\lambda-2)$ & $\Delta=\gamma+\frac12(\lambda-2)$ \\
\hline
\multirow{3}{*}{$\mathbf{A}_2$} & $\left[\gamma, [2,1^{\mu_L}], [\lambda^1, \lambda^2, 2^{\mu^1_R},1^{\mu^2_R}]\right]$ & $\left[\gamma, [2,1^{\mu+\lambda-2}],[\lambda>2, 1^\mu]\right]$ & $\left[\gamma, [2,1^\mu]\right]$ & $\left[\gamma, [2,1^\mu], [1^{\mu+2}] \right]$ \\
& $[0,\lambda^1-\lambda^2]_{\Delta}^{(\mu_L,\gamma-\mu^{1,2}_{R}-\mu_L-3,\mu_R^2)}$ & $\left[0, \lambda-2\right]^{(\mu+\lambda-2, \gamma-2\mu - \lambda, \mu)}_\Delta$ & $[0,0]^{(\mu,\gamma-2\mu-2,\mu)}_{\gamma}$ & $[0,0]^{(\mu,\gamma-2\mu-3,\mu+2)}_{\gamma}$ \\
& $\Delta=\gamma+ \frac12 (\lambda^1 +\lambda^2)-2$ & $\Delta= \gamma+\frac12(\lambda-2)$ & $\Delta= \gamma$ & $\Delta= \gamma$ \\
\hline
\multirow{3}{*}{$\mathbf{B}$} & $\left[\gamma, [1^{\mu_L}],[\lambda^1, \lambda^2, 2^{\mu_R^1},1^{\mu_R^2}]\right]$ & $\left[\gamma, [1^{\mu+\lambda}],[\lambda>2, 1^\mu]\right]$ & $\left[\gamma, [1^{\mu+2}],[2,1^\mu]\right]$ & $\left[\gamma, [1^\mu]\right]$ \\
& $[0,\lambda^1-\lambda^2]_{\Delta}^{(\mu_L,\gamma-\mu^{1,2}_{L}-\mu_R-2,\mu_R^2)}$ &$[0,\lambda-2]_{\Delta}^{(\mu+\lambda,\gamma-\lambda-2\mu-1,\mu)}$ & $[0,0]_\gamma^{(\mu+2, \gamma-2\mu-3, \mu)}$ & $[0,0]^{(\mu,\gamma-2\mu,\mu)}_{\gamma}$  \\
& $\Delta=\gamma+ \frac12 (\lambda^1 +\lambda^2)-2$ & $\Delta=\gamma+ \frac{1}{2}(\lambda-2)$ & $\Delta=\gamma$ & $\Delta= \gamma$ \\
\hline
 \end{tabular}}
 \caption{$(2,2)$ analytic superspace description of all 4D $\cN=4$ representations. The restriction $|\ula_L|=|\ula_R|$ has not been imposed for multiplets with either $L$ or $\bar{L}$ shortening conditions to avoid clutter, but should be implied. We used the notation $\bullet_{L,R}\coloneq\bullet_L+\bullet_R$ and $\bullet^{1,2}=\bullet^1+\bullet^2$.}
 \label{tab:4DN4reps1}
\end{table}
\end{landscape}
\Yboxdim{13pt}

\section{Removing $W$,$\bW$ dependence in weight-shifting operators}\label{app:cosettoGR}
The weight-shifting operator~\eqref{wsgen}, and hence the resulting operator $\cO'$ \eqref{oprime}, appear to depend on all the coset variables \eqref{eq:g} and \eqref{inv}. 
But we know that since it is primary it must be rewritable in 
 a way independent of $W,\bW$ by using  identities \eqref{eq:ginvg} (see section~\ref{sec:Grassmannianfields} and appendix~\ref{app:xxbproof}).
So we  wish to derive weight-shifting operators depending explicitly on $X_i, \bX_i$ only i.e.\ directly on the Grassmannian. In the main text (section \ref{sec:grassmannianWSO}) we presented a more direct and general derivation of such differential operators. Here, we show the brute force method from the analytic superspace form of the weight-shifting operator \eqref{oprime} with a few simple examples.

Consider acting on a scalar $\cO^\gamma$ by acting a dotted index. By $\eqref{oprime}$, we have
\begin{equation}\label{eq:ansatzscalar}
\Ylinethick{.2pt}
\Yboxdim{6pt}
    {\cO}^{\prime\gamma}_{\dfa} = c_0\left(\bar W_{\dfa \cA }+ \frac1\gamma X^{\fa}_{\ \cA} p_{\dfa \fa} \right) \cO^\gamma\,.
\end{equation}
 Then, using the expression for $p=\bW \partial_X -\partial_{\bX} W$ in~\eqref{kinXXb}, the above becomes
\begin{align}{\cO}^{\prime\gamma}_{\dfa} &= c_0\Phi^{\cA} ( \bar W +\frac1\gamma\bar W \partial_X X -  \frac1\gamma\partial_{\bar X}W X)_{\dfa \cA} \cO^\gamma\notag \\
&=-\frac{c_0}{\gamma} \frep^\cA \left(\partial_{\bar X} - \gamma\bar W - \partial_{\bar X}\bar X\bar W  -\bar W \partial_{ X}X \right)_{\dfa \cA} \cO^\gamma\,,
\label{honest}
\end{align}
where we used $WX=\mathbb{I}-\bar X \bar W$~\eqref{eq:ginvg} to get the second line. Firstly, note that $\partial_{\bX} \bX \equiv m_R$ and $m_R \cO^\gamma = -\gamma_R \cO^\gamma$ \eqref{eq:primary}. Then, the above can be written in a manifestly Grassmannian ($\bW, W$ independent) form by using the following identity\footnote{The simplest way to derive this is by acting on an insertion of generic scalar correlator $\prod_ig_{1i}^{\gamma_L^i} g_{i1}^{\gamma_R^i}$, where $\gamma_{L,R}= \sum_i \gamma_{L,R}^i$.}
\begin{equation}
    (\bW \partial_X X)_{\dfa \cA} \cO^\gamma = - \frac{\gamma_L}{\gamma_R} \partial_{\bX}{}_{\dfa \cA} \cO^\gamma- \frac{1}{\gamma_R} (\partial_{\bX} \partial_X X)_{\dfa \cA}\cO^\gamma - \gamma_L \bW_{\dfa \cA} \cO^\gamma.
\end{equation}
Then, by noting that $\gamma_L+\gamma_R=\gamma$, equation \eqref{eq:ansatzscalar} simplifies to
\begin{align}\label{scalarexample}
{\cO}^{\prime\gamma}_{\dfa} = -\frac{c_0}{\gamma_R}\frep^\cA (\partial_{\bar X}+\frac1\gamma \partial_{\bX} \partial_X X)_{\dfa \cA} \cO^{\gamma}\, .
\end{align}
By comparing with \eqref{eq:weightshiftingdef}, the above defines a weight-shifting operator $\bar{\cD}_{\cA}$ transforming in the antifundamental $\overline{\frep}$, which acts on a coset scalar field $\cO^{\gamma}$ by adding a dotted index (so maps $\ula_R=\bullet\mapsto \ula_R'=\Box$) as follows (fixing $c_0=-\gamma_R$)
\begin{equation}\label{scalarwsop}
    (\bar{\cD}_{\cA})_{\dfa} \cO^{\gamma} =  (\partial_{\bar X}+\frac1\gamma \partial_{\bX} \partial_X X)_{\dfa \cA} \cO^{\gamma},
\end{equation}
as expected from the main text.

\section{Differential bubble property}\label{app:hardbubble}
In this appendix we provide details on the general derivation of the bubble property \eqref{eq:bubble} in the case where both weight-shifting operators are differential \eqref{eq:collectedWSO}.
In particular, we want to show that the following vanishes
\begin{align}\label{eq:hardbubbleEX}
     &(\bar{\cD}_{\cA})^{\ula_R}_{\ula'_R} (\cD^{\cA})^{\ula_L}_{\ula'_L}
        \\
    &= c_0^2 \bar{\Pi}^{\ula_R'}_{1,i}\bigg( \bW_{\dfa \cA} +X^{\fb}_{\ \cA}\!\!\sum_{\umu_L\in\ula'_L\otimes\Box} c_{\umu_L\!(j)\ula_R'(i)} \Pi^{\umu_L}_jp_{\dfa \fb}\bigg)\Pi^{\ula'_L}_{1,k} \bigg( W {}^{\cA}_{\ \fa} -\bX^{\cA \dfb}\!\!\sum_{\umu_R\in\ula_R\otimes\Box} c_{\ula_L'\!(k)\umu_R(l)} \bar{\Pi}^{\umu_R}_{l}p_{\dfb \fa}\bigg). \nn
\end{align}
We defined the indices $i,j,k,l$ specifying the Young tableau by
$$\overline{\ula'_R(i)}=\ula_R(1),  \quad \overline{\ula'_L(k)}=\ula_L(1), \quad \overline{\umu_L(j)}=\ula'_L(1), \quad \overline{\umu_R(l)}=\ula_R(1),$$
where $\overline T$ is the Young tableau obtained by deleting the last box from $T$.
Clearly the term $\bW W$ vanishes but no other terms do so. Let us divide through by the overall normalisation $c_0^2$ and expand the brackets to obtain
\begin{equation}\nn
    \begin{split}
       &- \bPi^{\ula_R'}_{1,i} \delta^{\dfb}_{\dfa} \Pi^{\ula_L'}_{1,k} \sum_{\umu_R \in \ula_R \otimes \Box} c_{\ula_L', \umu_R} \bPi^{\umu_R}_{l} p_{\dfb \fa} 
        + \bPi^{\ula_R'}_{1,i}  X^{\fb}_{\ \cA} \sum_{\umu_L \in \ula_L' \otimes \Box} c_{\umu_L, \ula'_R} \Pi^{\umu_L}_{j} p_{\dfa \fb} \Pi^{\ula_L'}_{1,k} W^{\cA}_{\ \fa} \\
        &\quad - \bPi^{\ula_R'}_{1,i} X^{\fb}_{\ \cA} \sum_{\umu_L \in \ula_L' \otimes \Box} c_{\umu_L, \ula'_R} \Pi^{\umu_L}_{j} p_{\dfa \fb} \Pi^{\ula_L'}_{1,k} \bX^{\cA \dfb} \sum_{\umu_R \in \ula_R \otimes \Box} c_{\ula_L', \umu_R} \bPi^{\umu_R}_l p_{\dfb \fa}.
    \end{split}
\end{equation}
Then, the first term simplifies by orthogonality of Young projectors $\bPi$ \eqref{units}, and the third term does so by $[p,\bX]=- W$, $[p,W]=0$ and $X \bX=0,\ XW=1$, giving
\begin{equation}\label{eq:hardbubble1}
    \begin{split}
         & - \bPi^{\ula_R'}_{1,i} \Pi^{\ula_L'}_{1,k}  c_{\ula_L', \ula'_R}  p_{\dfa \fa} 
         + \bPi^{\ula_R'}_{1,i} \sum_{\umu_L \in \ula_L' \otimes \Box} c_{\umu_L, \ula'_R} \delta^{\fb}_{\fc} \Pi^{\umu_L}_{j} p_{\dfa \fb} \Pi^{\ula_L'}_{1,k} \delta^{\fc}_{\fa} 
        \\
        & \quad + c_{\ula_L', \ula'_R} \sum_{\umu_L \in \ula_L' \otimes \Box} c_{\umu_L, \ula'_R} \delta^{\fb}_{\fc} \Pi^{\umu_L}_{j} \delta^{\fc}_{\fb} \Pi^{\ula_L'}_{1,k} \bPi^{\ula_R'}_{1,i} p_{\dfa \fa}\,.
    \end{split}
\end{equation}
\Yboxdim{8pt}
To proceed we need to understand and simplify various contractions of Young operators. We claim the following: 
\begin{equation}\label{eq:Pi_ids}
    \begin{aligned}
          \delta^{\fb}_{\fc} \Pi^{\umu_L}_j \delta^{\fc}_{\fb}\Pi^{\ula'_L}_1  &=  \frac{h_{\ula'_L}}{h_{\umu_L}}\left(m-n +C_{\umu_L\!(j)}\right)\Pi^{\ula'_L}_1 \,,\\
  \delta^{\fb}_{\fc}\Pi^{\umu_L}_{j} p_{\dfa \fb} \Pi^{\ula_L'}_{1,k} \delta^{\fc}_{\fa} &=\frac{h_{\ula'_L}(m-n +C_{\umu_L\!(j)})}{h_{\umu_L}(C_{\umu_L\!(j)}-C_{\ula'_L\!(k)})} \Pi^{\ula_L'}_{1,k}   p_{\dfa \fa} \,,
    \end{aligned}
\end{equation}
where $h_{\umu_L}$ is the product of hook lengths of $\umu_L$. The RHS of both equations can be fixed up to normalisation by the orthogonality properties of Young projectors \eqref{units} and \eqref{Tbar} . 
The normalisation can be determined from the definition of the Hermitian Young projectors in terms of standard Young projectors (see 2.3.16 of~\cite{garsia2020lectures} and theorem 6 of~\cite{keppeler2014hermitian}). 
The first follows straightforwardly, and the second one likely admits a similar proof, but we have instead verified it for sufficiently generic Young tableaux $\ula_L'$, $\umu_L$.
Now, noting the  identity
\begin{equation}\label{eq:coeffdecomp}
    c_{\ula'_L,\ula'_R}c_{\umu_L, \ula'_R}= \frac{c_{\ula'_L,\ula'_R}-c_{\umu_L, \ula'_R}}{C_{\umu_L\!(j)}-C_{\ula'_L\!(k)}},
\end{equation}
together with \eqref{eq:Pi_ids}, expression~\eqref{eq:hardbubble1} simplifies to
\begin{equation}\label{eq:hardbubble3}
    \begin{split}
         &  c_{\ula_L', \ula'_R}\left( -1 + \sum_{\umu_L \in \ula_L' \otimes \Box} \frac{h_{\ula'_L}}{h_{\umu_L}}+ \sum_{\umu_L \in \ula_L' \otimes \Box} \frac{h_{\ula'_L}(m-n+C_{\ula'_L\!(k)})}{h_{\umu_L}(C_{\umu_L\!(j)}-C_{\ula'_L\!(k)})}\right)\bPi^{\ula_R'}_{1,i} \Pi^{\ula_L'}_{1,k}   p_{\dfa \fa} \,,
    \end{split}
\end{equation}
which vanishes
due to the identities
\begin{align}
    \sum_{\umu_L \in \ula_L' \otimes \Box} \frac{1}{h_{\umu_L}}= \frac1{h_{\ula_L'}}, \qquad \qquad \sum_{\umu_L \in \ula_L' \otimes \Box} \frac{1}{h_{\umu_L}(C_{\umu_L\!(j)}-C_{\ula'_L\!(k)})}=0\,.
\end{align}
The first of these is a well known identity (see 2.3.35 of~\cite{garsia2020lectures}). The second we have checked in many cases but have not proven in general. E.g.\ for $\ula'_L=[\lambda]$, then the second equation above reads
\begin{equation}
    \frac{1}{(\lambda+1)!}- \frac{1}{\lambda} \frac{1}{(\lambda+1)\times (\lambda-1)!}=0.
\end{equation}

\section{Consistency checks in 1D and 4D}\label{app:checks}

In this appendix we provide some explicit checks of our results from section \ref{sec:blocks}  against known conformal blocks in 1 and 4 dimensions.

\subsection{1 dimension}

We can describe 1D CFTs by choosing $(m,n)=(1,0)$. In this case we have for the prefactor \eqref{eq:K4} and the conformal block \cite{Doobary:2015gia}
\bea
    P^\gamma_{\Delta_1, \Delta_2, \Delta_3, \Delta_4} &= x_{12}^{-\frac{\Delta_1+\Delta_2}{2}} x_{34}^{-\frac{\Delta_3+\Delta_4}{2}} \left(\frac{x_{24}}{x_{14}}\right)^{-\frac{1}{2}\Delta_{21}}\left(\frac{x_{14}}{x_{13}}\right)^{-\frac{1}{2}\Delta_{43}}\left(\frac{x_{13}x_{24}}{x_{12}x_{34}}\right)^{-\frac{1}{2}\gamma}\,,\\
F^{\gamma, \lambda}_{\Delta_{21},\Delta_{43}} (z) &= z^\lambda \, {}_2 F_1\left(\lambda+ \frac12 (\gamma+\Delta_{21}), \lambda+ \frac12 (\gamma-\Delta_{43}), 2 \lambda + \gamma ; z\right)\,.
\eea{eq:K4_1D}
There is some degeneracy in this description, as operators in 1D depend only on their scaling dimension
\beq
\Delta = \gamma + \lambda\,,
\eeq
but we will keep both labels $\gamma$ and $\lambda$ here to connect with our more general results. Consequently, all that the weight-shifting operators do in 1D is to shift scaling dimensions.

Let us begin by testing the two weight-shifting operators defined in \eqref{wso_ex1} that turn a scalar block into a block with indices as in $\langle \cO_{\fa} \cO_{\dfa} \cO \cO \rangle$. For the first one we simply have
\beq
\fD^{[1]}_{\fa \dfa} P^\gamma_{\Delta_1 \Delta_2 \Delta_3 \Delta_4}
F^{\gamma, \lambda}_{\Delta_{21},\Delta_{43}} (z) 
= \frac{1}{x_{21}} P^\gamma_{\Delta_1 \Delta_2 \Delta_3 \Delta_4}
F^{\gamma, \lambda}_{\Delta_{21},\Delta_{43}} (z) 
= - P^\gamma_{\Delta_1+1, \Delta_2+1, \Delta_3, \Delta_4}
F^{\gamma, \lambda}_{\Delta_{21},\Delta_{43}} (z)\,,
\label{1d_c1}
\eeq
so it shifts up $\Delta_1$ and $\Delta_2$ as expected.
For the second operator we can use \eqref{DZ10} and \eqref{DZ1} to check that similarly
\begin{align}
{}&\fD^{[2]}_{\fa \dfa} P^\gamma_{\Delta_1 \Delta_2 \Delta_3 \Delta_4}
F^{\gamma, \lambda}_{\Delta_{21},\Delta_{43}} (z) \nn \\
&= P^\gamma_{\Delta_1 \Delta_2 \Delta_3 \Delta_4}
\frac{1}{x_{21}}
\bigg(
\frac{1}{4} \left(\left(\gamma -\Delta_1-\Delta_2\right) \left(2 -\gamma -\Delta_1-\Delta_2\right) +\left(\gamma +\Delta_{21}\right) \left(\gamma -\Delta_{43}\right) z \right) \nn \\
&\quad+\frac{1}{2} \Big( \left(2 \gamma +\Delta_{21}-\Delta_{43}+2\right) z^2 -2 \gamma   z \Big) \partial_z 
+z^2 (z-1) \partial_z^2 
\bigg)
F^{\gamma, \lambda}_{\Delta_{21},\Delta_{43}} (z) \label{1d_c2} \\
&=  -\frac{1}{4} (\Delta_1+\Delta_2-2 \lambda-\gamma) (\Delta_1+\Delta_2+2 \lambda+\gamma-2)
P^\gamma_{\Delta_1+1, \Delta_2+1, \Delta_3, \Delta_4}
F^{\gamma, \lambda}_{\Delta_{21},\Delta_{43}} (z)\,.\nn
\end{align}
We performed the analogous checks for both operators in \eqref{eq:O01WSos1} that create a block of the form $\langle \cO_{\dfa\fa} \cO \cO \cO \rangle$,
finding that as expected
\bea
\fD^{[1]}_{\dfa \fa} P^\gamma_{\Delta_1, \Delta_2, \Delta_3, \Delta_4}
F^{\gamma, \lambda}_{\Delta_{21},\Delta_{43}} (z) 
\propto \fD^{[2]}_{\dfa \fa} P^\gamma_{\Delta_1, \Delta_2, \Delta_3, \Delta_4}
F^{\gamma, \lambda}_{\Delta_{21},\Delta_{43}} (z) 
\propto P^\gamma_{\Delta_1+2, \Delta_2, \Delta_3, \Delta_4}
F^{\gamma, \lambda}_{\Delta_{21}-2,\Delta_{43}}\,.
\eea{check3}
Finally, we tested the internal shift operator in equations \eqref{eq:1seedblock} to \eqref{seedblock_last} by acting on $P^\gamma_{\Delta_1, \Delta_2, \Delta_3, \Delta_4} F^{\gamma, 0}_{\Delta_{21},\Delta_{43}} (z)$ and confirming that the result is proportional to
\beq
P^{\gamma+1}_{\Delta_1+1, \Delta_2, \Delta_3+1, \Delta_4} F^{\gamma+1, 0}_{\Delta_{21}-1,\Delta_{43}-1} (z)\,.
\label{1d_c4}
\eeq

\subsection{4 dimensions}

In four dimensions we can compare our weight-shifting operators to the comprehensive work and Mathematica package \cite{Cuomo:2017wme}.
We use the kinematic factor
\bea
    P_{\gamma_1, \gamma_2, \gamma_3, \gamma_4} &= (x^2_{12})^{-\frac{\gamma_1+\gamma_2}{2}} (x^2_{34})^{-\frac{\gamma_3+\gamma_4}{2}} \left(\frac{x^2_{24}}{x^2_{14}}\right)^{-\frac{1}{2}\gamma_{21}}\left(\frac{x^2_{14}}{x^2_{13}}\right)^{-\frac{1}{2}\gamma_{43}}\,,
\eea{eq:K4_4D}
and $F^{\Delta,\ell,\bar{\ell}}_{\gamma_{21},\gamma_{43}} (z,\bar{z})$ is the usual 4D conformal block, but for our checks it is enough to know that it is a function of $\gamma_{21}$, $\gamma_{43}$, $z$ and $\bar{z}$. The $\gamma_i$'s are related to scaling dimensions and spins by
\beq
\gamma_i = \Delta_i - \frac{\ell_i+\bar{\ell}_i}{2}\,.
\eeq

We begin again with the case $\langle \cO_{\fa} \cO_{\dfa} \cO \cO \rangle$.
By using the Mathematica package \cite{Cuomo:2017wme} we find the following two spinning conformal blocks
\bea
\Xi[I_{12}] P_{\Delta_1, \Delta_2, \Delta_3, \Delta_4} F^{\Delta,\ell,\bar{\ell}}_{\Delta_{21},\Delta_{43}} (z,\bar{z})
={}& P_{\Delta_1-\frac12, \Delta_2-\frac12, \Delta_3, \Delta_4} \left(\frac{\eta _1 \overline{\eta }_2}{z}-\frac{\xi _1 \overline{\xi }_2}{\overline{z}}\right)
 F^{\Delta,\ell,\bar{\ell}}_{\Delta_{21},\Delta_{43}} (z,\bar{z})\,,
\eea{CFTs4D_1}
and
\bea
{}&\Xi[\nabla_{21}] \Xi[D_{21}] \Xi[D_{12}] P_{\Delta_1, \Delta_2, \Delta_3, \Delta_4} F^{\Delta,\ell,\bar{\ell}}_{\Delta_{21},\Delta_{43}} (z,\bar{z})\\
&= P_{\Delta_1-\frac12, \Delta_2-\frac12, \Delta_3, \Delta_4}
 \bigg(4 \eta_1 \overline{\eta }_2 (z-1) z \partial_z^2 -4 \xi _1 \overline{\xi }_2 \left(\overline{z}-1\right) \overline{z} \partial_{\overline{z}}^2\\
& + \frac{2  \eta _1 \overline{\eta }_2 \left(2 \left(z^2-\overline{z}\right)+\left(\Delta _{21}-\Delta _{43}\right) z \left(z-\overline{z}\right)\right)-4 \xi_1 \overline{\xi }_2 (z-1) z }{z-\overline{z}} \partial_z\\
   &+\frac{2 \xi _1 \overline{\xi }_2 \left(2 (\overline{z}^2-z)+\left(\Delta _{21}-\Delta _{43}\right)
  \overline{z} \left(\overline{z}-z\right) \right)-4 \eta _1 \overline{\eta }_2 \left(\overline{z}-1\right) \overline{z}}{z-\overline{z}} \partial_{\overline{z}} \\
   &+\frac{\Delta _{21} \left(\eta _1 \overline{\eta}_2 \left(\Delta _{21}-\Delta _{43} z-4\right) \overline{z}-\xi _1  \overline{\xi }_2  \left(\Delta _{21}-\Delta _{43} \overline{z}-4\right)z\right)}{z \overline{z}}\bigg)
    F^{\Delta,\ell,\bar{\ell}}_{\Delta_{21},\Delta_{43}} (z,\bar{z})\,.
\eea{CFTs4D_2}
Note that $\eta _1 \overline{\eta }_2$ and $\xi _1 \overline{\xi }_2$ are the two tensor structures in the conformal frame defined for this case in \cite{Cuomo:2017wme}. In order to write our tensor structures in the conformal frame, we have to set
\bea
x_1^\mu &= (0,0,0,0)\,,\\
x_2^\mu &= ((\bar{z}-z)/2,0,0,(z+\bar{z})/2)\,,\\
x_3^\mu &= (0,0,0,1)\,,\\
x_4^\mu &= (0,0,0,\infty)\,,
\eea{conformal_frame}
and compute
\beq
(\overline{\xi }_2, \overline{\eta }_2) \cdot \hX_{21} \cdot Z^n \cdot (\xi_1, \eta_1)^T
= (\overline{\xi }_2, \overline{\eta }_2) \cdot 
\begin{pmatrix}
-\frac{1}{\bar{z}} & 0\\
0 & \frac{1}{z}
\end{pmatrix} \cdot 
\begin{pmatrix}
\bar{z}^n & 0\\
0 & z^n
\end{pmatrix} \cdot 
\begin{pmatrix}
\xi_1\\
\eta_1
\end{pmatrix}
=\frac{z^n \eta _1 \overline{\eta }_2}{z}-\frac{\bar{z}^n \xi _1 \overline{\xi }_2}{\overline{z}}.
\label{cf_1}
\eeq
With this, we find that the first weight-shifting operator of \eqref{wso_ex1} gives precisely the same result as \eqref{CFTs4D_1}
\beq
(\xi_1, \eta_1)^\fa (\overline{\xi }_2, \overline{\eta }_2)^\dfa \fD^{[1]}_{\fa \dfa} P_{\gamma_1 \gamma_2 \gamma_3 \gamma_4}
F^{\Delta,\ell,\bar{\ell}}_{\Delta_{21},\Delta_{43}} (z, \bar{z}) 
= P_{\gamma_1 \gamma_2 \gamma_3 \gamma_4} \left(\frac{\eta _1 \overline{\eta }_2}{z}-\frac{\xi _1 \overline{\xi }_2}{\overline{z}}\right)
 F^{\Delta,\ell,\bar{\ell}}_{\Delta_{21},\Delta_{43}} (z,\bar{z})\,.
 \label{4d_c1}
\eeq
For the second structure we insert \eqref{cf_1} into the operator in \eqref{op_tk} and find agreement with a linear combination of \eqref{CFTs4D_1} and \eqref{CFTs4D_2}
\bea
{}& (\xi_1, \eta_1)^\fa (\overline{\xi }_2, \overline{\eta }_2)^\dfa  \fD^{[2]}_{\fa \dfa} P_{\gamma_1 \gamma_2 \gamma_3 \gamma_4}
F^{\Delta,\ell,\bar{\ell}}_{\Delta_{21},\Delta_{43}} (z, \bar{z}) \\
&=\left( \frac14 \Xi[\nabla_{21}] \Xi[D_{21}] \Xi[D_{12}]
+ \Delta_1(\Delta_2-2) \Xi[I_{12}]\right)
 P_{\Delta_1 \Delta_2 \Delta_3 \Delta_4} F^{\Delta,\ell,\bar{\ell}}_{\Delta_{21},\Delta_{43}} (z,\bar{z})\,.
\eea{check5}

For the case $\langle \cO_{\dfa \fa} \cO \cO \cO\rangle$ we compute the tensor structures in the conformal frame analogously to \eqref{cf_1}
\beq
(\overline{\xi }_1, \overline{\eta }_1) \cdot Y_{1,24} \cdot Z^n \cdot (\xi_1, \eta_1)^T
= -\frac{z^n \eta _1 \overline{\eta }_1}{z}+\frac{\bar{z}^n \xi _1 \overline{\xi }_1}{\overline{z}}\,,
\eeq
and find that the first operator in \eqref{eq:O01WSos1} does match with the operator $\Xi[D_{12}]$ from \cite{Cuomo:2017wme}
\beq
(\overline{\xi }_1, \overline{\eta }_1)^\dfa (\xi_1, \eta_1)^\fa  \fD^{[1]}_{\dfa \fa} P_{\gamma_1 \gamma_2 \gamma_3 \gamma_4}
F^{\Delta,\ell,\bar{\ell}}_{\Delta_{21},\Delta_{43}} (z, \bar{z})= \frac12 \Xi[D_{12}]
 P_{\Delta_1 \Delta_2 \Delta_3 \Delta_4} F^{\Delta,\ell,\bar{\ell}}_{\Delta_{21},\Delta_{43}} (z,\bar{z})\,.
 \label{4d_c3}
\eeq

\bibliographystyle{JHEP}

\bibliography{refe}

\providecommand{\href}[2]{#2}\begingroup\raggedright\begin{thebibliography}{100}

\bibitem{Mack:1969rr}
G.~Mack and A.~Salam, \emph{{Finite-Component Field Representations of the Conformal Group}}, \href{https://doi.org/10.1016/0003-4916(69)90278-4}{\emph{Annals Phys.} {\bfseries 53} (1969) 174}.

\bibitem{Polyakov:1974gs}
A.M.~Polyakov, \emph{{Non-Hamiltonian approach to quantum field theory}}, {\emph{Zh. Eksp. Teor. Fiz.} {\bfseries 66} (1974) 23}.

\bibitem{FERRARA1973161}
S.~Ferrara, A.F.~Grillo and R.~Gatto, \emph{Tensor representations of conformal algebra and conformally covariant operator product expansion}, \href{https://doi.org/10.1016/0003-4916(73)90446-6}{\emph{Annals Phys.} {\bfseries 76} (1973) 161}.

\bibitem{Ferrara:1974nf}
S.~Ferrara, A.F.~Grillo, G.~Parisi and R.~Gatto, \emph{Covariant expansion of the conformal four-point function}, \href{https://doi.org/10.1016/0550-3213(72)90587-1}{\emph{Nucl. Phys. B} {\bfseries 49} (1972) 77}.

\bibitem{Rattazzi:2008pe}
R.~Rattazzi, V.S.~Rychkov, E.~Tonni and A.~Vichi, \emph{{Bounding scalar operator dimensions in 4D CFT}}, \href{https://doi.org/10.1088/1126-6708/2008/12/031}{\emph{JHEP} {\bfseries 12} (2008) 031} [\href{https://arxiv.org/abs/0807.0004}{{\ttfamily 0807.0004}}].

\bibitem{Caracciolo:2009bx}
F.~Caracciolo and V.S.~Rychkov, \emph{{Rigorous Limits on the Interaction Strength in Quantum Field Theory}}, \href{https://doi.org/10.1103/PhysRevD.81.085037}{\emph{Phys. Rev. D} {\bfseries 81} (2010) 085037} [\href{https://arxiv.org/abs/0912.2726}{{\ttfamily 0912.2726}}].

\bibitem{Poland:2011ey}
D.~Poland, D.~Simmons-Duffin and A.~Vichi, \emph{{Carving Out the Space of 4D CFTs}}, \href{https://doi.org/10.1007/JHEP05(2012)110}{\emph{JHEP} {\bfseries 05} (2012) 110} [\href{https://arxiv.org/abs/1109.5176}{{\ttfamily 1109.5176}}].

\bibitem{El-Showk:2012cjh}
S.~El-Showk, M.F.~Paulos, D.~Poland, S.~Rychkov, D.~Simmons-Duffin and A.~Vichi, \emph{{Solving the 3D Ising Model with the Conformal Bootstrap}}, \href{https://doi.org/10.1103/PhysRevD.86.025022}{\emph{Phys. Rev. D} {\bfseries 86} (2012) 025022} [\href{https://arxiv.org/abs/1203.6064}{{\ttfamily 1203.6064}}].

\bibitem{Poland:2018epd}
D.~Poland, S.~Rychkov and A.~Vichi, \emph{{The Conformal Bootstrap: Theory, Numerical Techniques, and Applications}}, \href{https://doi.org/10.1103/RevModPhys.91.015002}{\emph{Rev. Mod. Phys.} {\bfseries 91} (2019) 015002} [\href{https://arxiv.org/abs/1805.04405}{{\ttfamily 1805.04405}}].

\bibitem{Dolan:2000ut}
F.A.~Dolan and H.~Osborn, \emph{{Conformal four point functions and the operator product expansion}}, \href{https://doi.org/10.1016/S0550-3213(01)00013-X}{\emph{Nucl. Phys. B} {\bfseries 599} (2001) 459} [\href{https://arxiv.org/abs/hep-th/0011040}{{\ttfamily hep-th/0011040}}].

\bibitem{Dolan:2003hv}
F.A.~Dolan and H.~Osborn, \emph{{Conformal partial waves and the operator product expansion}}, \href{https://doi.org/10.1016/j.nuclphysb.2003.11.016}{\emph{Nucl. Phys. B} {\bfseries 678} (2004) 491} [\href{https://arxiv.org/abs/hep-th/0309180}{{\ttfamily hep-th/0309180}}].

\bibitem{Dolan:2011dv}
F.A.~Dolan and H.~Osborn, \emph{{Conformal Partial Waves: Further Mathematical Results}},  \href{https://arxiv.org/abs/1108.6194}{{\ttfamily 1108.6194}}.

\bibitem{Costa:2011mg}
M.S.~Costa, J.~Penedones, D.~Poland and S.~Rychkov, \emph{{Spinning Conformal Correlators}}, \href{https://doi.org/10.1007/JHEP11(2011)071}{\emph{JHEP} {\bfseries 11} (2011) 071} [\href{https://arxiv.org/abs/1107.3554}{{\ttfamily 1107.3554}}].

\bibitem{Costa:2011dw}
M.S.~Costa, J.~Penedones, D.~Poland and S.~Rychkov, \emph{{Spinning Conformal Blocks}}, \href{https://doi.org/10.1007/JHEP11(2011)154}{\emph{JHEP} {\bfseries 11} (2011) 154} [\href{https://arxiv.org/abs/1109.6321}{{\ttfamily 1109.6321}}].

\bibitem{Penedones:2015aga}
J.~Penedones, E.~Trevisani and M.~Yamazaki, \emph{{Recursion Relations for Conformal Blocks}}, \href{https://doi.org/10.1007/JHEP09(2016)070}{\emph{JHEP} {\bfseries 09} (2016) 070} [\href{https://arxiv.org/abs/1509.00428}{{\ttfamily 1509.00428}}].

\bibitem{Costa:2016hju}
M.S.~Costa, T.~Hansen, J.a.~Penedones and E.~Trevisani, \emph{{Projectors and seed conformal blocks for traceless mixed-symmetry tensors}}, \href{https://doi.org/10.1007/JHEP07(2016)018}{\emph{JHEP} {\bfseries 07} (2016) 018} [\href{https://arxiv.org/abs/1603.05551}{{\ttfamily 1603.05551}}].

\bibitem{Costa:2016xah}
M.S.~Costa, T.~Hansen, J.a.~Penedones and E.~Trevisani, \emph{{Radial expansion for spinning conformal blocks}}, \href{https://doi.org/10.1007/JHEP07(2016)057}{\emph{JHEP} {\bfseries 07} (2016) 057} [\href{https://arxiv.org/abs/1603.05552}{{\ttfamily 1603.05552}}].

\bibitem{Cuomo:2017wme}
G.F.~Cuomo, D.~Karateev and P.~Kravchuk, \emph{{General Bootstrap Equations in 4D CFTs}}, \href{https://doi.org/10.1007/JHEP01(2018)130}{\emph{JHEP} {\bfseries 01} (2018) 130} [\href{https://arxiv.org/abs/1705.05401}{{\ttfamily 1705.05401}}].

\bibitem{Karateev:2017jgd}
D.~Karateev, P.~Kravchuk and D.~Simmons-Duffin, \emph{{Weight Shifting Operators and Conformal Blocks}}, \href{https://doi.org/10.1007/JHEP02(2018)081}{\emph{JHEP} {\bfseries 02} (2018) 081} [\href{https://arxiv.org/abs/1706.07813}{{\ttfamily 1706.07813}}].

\bibitem{Erramilli:2019njx}
R.S.~Erramilli, L.V.~Iliesiu and P.~Kravchuk, \emph{{Recursion relation for general 3d blocks}}, \href{https://doi.org/10.1007/JHEP12(2019)116}{\emph{JHEP} {\bfseries 12} (2019) 116} [\href{https://arxiv.org/abs/1907.11247}{{\ttfamily 1907.11247}}].

\bibitem{Dymarsky:2017xzb}
A.~Dymarsky, J.~Penedones, E.~Trevisani and A.~Vichi, \emph{{Charting the space of 3D CFTs with a continuous global symmetry}}, \href{https://doi.org/10.1007/JHEP05(2019)098}{\emph{JHEP} {\bfseries 05} (2019) 098} [\href{https://arxiv.org/abs/1705.04278}{{\ttfamily 1705.04278}}].

\bibitem{Reehorst:2019pzi}
M.~Reehorst, E.~Trevisani and A.~Vichi, \emph{{Mixed Scalar-Current bootstrap in three dimensions}}, \href{https://doi.org/10.1007/JHEP12(2020)156}{\emph{JHEP} {\bfseries 12} (2020) 156} [\href{https://arxiv.org/abs/1911.05747}{{\ttfamily 1911.05747}}].

\bibitem{He:2023ewx}
Y.-C.~He, J.~Rong, N.~Su and A.~Vichi, \emph{{Non-Abelian currents bootstrap}}, \href{https://doi.org/10.1007/JHEP03(2024)175}{\emph{JHEP} {\bfseries 03} (2024) 175} [\href{https://arxiv.org/abs/2302.11585}{{\ttfamily 2302.11585}}].

\bibitem{Dymarsky:2017yzx}
A.~Dymarsky, F.~Kos, P.~Kravchuk, D.~Poland and D.~Simmons-Duffin, \emph{{The 3d Stress-Tensor Bootstrap}}, \href{https://doi.org/10.1007/JHEP02(2018)164}{\emph{JHEP} {\bfseries 02} (2018) 164} [\href{https://arxiv.org/abs/1708.05718}{{\ttfamily 1708.05718}}].

\bibitem{Chang:2024whx}
C.-H.~Chang, V.~Dommes, R.S.~Erramilli, A.~Homrich, P.~Kravchuk, A.~Liu, M.S.~Mitchell, D.~Poland and D.~Simmons-Duffin, \emph{{Bootstrapping the 3d Ising stress tensor}}, \href{https://doi.org/10.1007/JHEP03(2025)136}{\emph{JHEP} {\bfseries 03} (2025) 136} [\href{https://arxiv.org/abs/2411.15300}{{\ttfamily 2411.15300}}].

\bibitem{Maldacena:1997re}
J.M.~Maldacena, \emph{{The Large N limit of superconformal field theories and supergravity}}, \href{https://doi.org/10.4310/ATMP.1998.v2.n2.a1}{\emph{Adv. Theor. Math. Phys.} {\bfseries 2} (1998) 231} [\href{https://arxiv.org/abs/hep-th/9711200}{{\ttfamily hep-th/9711200}}].

\bibitem{Dolan:2001tt}
F.A.~Dolan and H.~Osborn, \emph{{Superconformal symmetry, correlation functions and the operator product expansion}}, \href{https://doi.org/10.1016/S0550-3213(02)00096-2}{\emph{Nucl. Phys. B} {\bfseries 629} (2002) 3} [\href{https://arxiv.org/abs/hep-th/0112251}{{\ttfamily hep-th/0112251}}].

\bibitem{Dolan:2004mu}
F.A.~Dolan, L.~Gallot and E.~Sokatchev, \emph{{On four-point functions of 1/2-BPS operators in general dimensions}}, \href{https://doi.org/10.1088/1126-6708/2004/09/056}{\emph{JHEP} {\bfseries 09} (2004) 056} [\href{https://arxiv.org/abs/hep-th/0405180}{{\ttfamily hep-th/0405180}}].

\bibitem{Nirschl:2004pa}
M.~Nirschl and H.~Osborn, \emph{{Superconformal Ward identities and their solution}}, \href{https://doi.org/10.1016/j.nuclphysb.2005.01.013}{\emph{Nucl. Phys. B} {\bfseries 711} (2005) 409} [\href{https://arxiv.org/abs/hep-th/0407060}{{\ttfamily hep-th/0407060}}].

\bibitem{Doobary:2015gia}
R.~Doobary and P.~Heslop, \emph{{Superconformal partial waves in Grassmannian field theories}}, \href{https://doi.org/10.1007/JHEP12(2015)159}{\emph{JHEP} {\bfseries 12} (2015) 159} [\href{https://arxiv.org/abs/1508.03611}{{\ttfamily 1508.03611}}].

\bibitem{Beem:2014zpa}
C.~Beem, M.~Lemos, P.~Liendo, L.~Rastelli and B.C.~van Rees, \emph{{The $ \mathcal{N}=2 $ superconformal bootstrap}}, \href{https://doi.org/10.1007/JHEP03(2016)183}{\emph{JHEP} {\bfseries 03} (2016) 183} [\href{https://arxiv.org/abs/1412.7541}{{\ttfamily 1412.7541}}].

\bibitem{Bobev:2015jxa}
N.~Bobev, S.~El-Showk, D.~Mazac and M.F.~Paulos, \emph{{Bootstrapping SCFTs with Four Supercharges}}, \href{https://doi.org/10.1007/JHEP08(2015)142}{\emph{JHEP} {\bfseries 08} (2015) 142} [\href{https://arxiv.org/abs/1503.02081}{{\ttfamily 1503.02081}}].

\bibitem{Bissi:2015qoa}
A.~Bissi and T.~\L{}ukowski, \emph{{Revisiting $ \mathcal{N}=4 $ superconformal blocks}}, \href{https://doi.org/10.1007/JHEP02(2016)115}{\emph{JHEP} {\bfseries 02} (2016) 115} [\href{https://arxiv.org/abs/1508.02391}{{\ttfamily 1508.02391}}].

\bibitem{Li:2016chh}
Z.~Li and N.~Su, \emph{{The Most General $4\mathcal{D}$ $\mathcal{N}=1$ Superconformal Blocks for Scalar Operators}}, \href{https://doi.org/10.1007/JHEP05(2016)163}{\emph{JHEP} {\bfseries 05} (2016) 163} [\href{https://arxiv.org/abs/1602.07097}{{\ttfamily 1602.07097}}].

\bibitem{Bobev:2017jhk}
N.~Bobev, E.~Lauria and D.~Mazac, \emph{{Superconformal Blocks for SCFTs with Eight Supercharges}}, \href{https://doi.org/10.1007/JHEP07(2017)061}{\emph{JHEP} {\bfseries 07} (2017) 061} [\href{https://arxiv.org/abs/1705.08594}{{\ttfamily 1705.08594}}].

\bibitem{Bissi:2022mrs}
A.~Bissi, A.~Sinha and X.~Zhou, \emph{{Selected topics in analytic conformal bootstrap: A guided journey}}, \href{https://doi.org/10.1016/j.physrep.2022.09.004}{\emph{Phys. Rept.} {\bfseries 991} (2022) 1} [\href{https://arxiv.org/abs/2202.08475}{{\ttfamily 2202.08475}}].

\bibitem{Heslop:2022xgp}
P.~Heslop, \emph{{The SAGEX Review on Scattering Amplitudes, Chapter 8: Half BPS correlators}}, \href{https://doi.org/10.1088/1751-8121/ac8c71}{\emph{J. Phys. A} {\bfseries 55} (2022) 443009} [\href{https://arxiv.org/abs/2203.13019}{{\ttfamily 2203.13019}}].

\bibitem{Beem:2013qxa}
C.~Beem, L.~Rastelli and B.C.~van Rees, \emph{{The $\mathcal N=4$ Superconformal Bootstrap}}, \href{https://doi.org/10.1103/PhysRevLett.111.071601}{\emph{Phys. Rev. Lett.} {\bfseries 111} (2013) 071601} [\href{https://arxiv.org/abs/1304.1803}{{\ttfamily 1304.1803}}].

\bibitem{Beem:2016wfs}
C.~Beem, L.~Rastelli and B.C.~van Rees, \emph{{More ${\mathcal N}=4$ superconformal bootstrap}}, \href{https://doi.org/10.1103/PhysRevD.96.046014}{\emph{Phys. Rev. D} {\bfseries 96} (2017) 046014} [\href{https://arxiv.org/abs/1612.02363}{{\ttfamily 1612.02363}}].

\bibitem{Chester:2021aun}
S.M.~Chester, R.~Dempsey and S.S.~Pufu, \emph{{Bootstrapping $ \mathcal{N} $ = 4 super-Yang-Mills on the conformal manifold}}, \href{https://doi.org/10.1007/JHEP01(2023)038}{\emph{JHEP} {\bfseries 01} (2023) 038} [\href{https://arxiv.org/abs/2111.07989}{{\ttfamily 2111.07989}}].

\bibitem{Chester:2024bij}
S.M.~Chester, R.~Dempsey and S.S.~Pufu, \emph{{Higher-derivative corrections in M-theory from precision numerical bootstrap}},  \href{https://arxiv.org/abs/2412.14094}{{\ttfamily 2412.14094}}.

\bibitem{Aprile:2017bgs}
F.~Aprile, J.M.~Drummond, P.~Heslop and H.~Paul, \emph{{Quantum Gravity from Conformal Field Theory}}, \href{https://doi.org/10.1007/JHEP01(2018)035}{\emph{JHEP} {\bfseries 01} (2018) 035} [\href{https://arxiv.org/abs/1706.02822}{{\ttfamily 1706.02822}}].

\bibitem{Aprile:2017qoy}
F.~Aprile, J.M.~Drummond, P.~Heslop and H.~Paul, \emph{{Loop corrections for Kaluza-Klein AdS amplitudes}}, \href{https://doi.org/10.1007/JHEP05(2018)056}{\emph{JHEP} {\bfseries 05} (2018) 056} [\href{https://arxiv.org/abs/1711.03903}{{\ttfamily 1711.03903}}].

\bibitem{Caron-Huot:2018kta}
S.~Caron-Huot and A.-K.~Trinh, \emph{{All tree-level correlators in AdS$_{5}$\texttimes{}S$_{5}$ supergravity: hidden ten-dimensional conformal symmetry}}, \href{https://doi.org/10.1007/JHEP01(2019)196}{\emph{JHEP} {\bfseries 01} (2019) 196} [\href{https://arxiv.org/abs/1809.09173}{{\ttfamily 1809.09173}}].

\bibitem{Alday:2018pdi}
L.F.~Alday, A.~Bissi and E.~Perlmutter, \emph{{Genus-One String Amplitudes from Conformal Field Theory}}, \href{https://doi.org/10.1007/JHEP06(2019)010}{\emph{JHEP} {\bfseries 06} (2019) 010} [\href{https://arxiv.org/abs/1809.10670}{{\ttfamily 1809.10670}}].

\bibitem{Aprile:2019rep}
F.~Aprile, J.~Drummond, P.~Heslop and H.~Paul, \emph{{One-loop amplitudes in AdS$_{5}$\texttimes{}S$^{5}$ supergravity from $ \mathcal{N} $ = 4 SYM at strong coupling}}, \href{https://doi.org/10.1007/JHEP03(2020)190}{\emph{JHEP} {\bfseries 03} (2020) 190} [\href{https://arxiv.org/abs/1912.01047}{{\ttfamily 1912.01047}}].

\bibitem{Abl:2020dbx}
T.~Abl, P.~Heslop and A.E.~Lipstein, \emph{{Towards the Virasoro-Shapiro amplitude in AdS$_{5} \times S^{5}$}}, \href{https://doi.org/10.1007/JHEP04(2021)237}{\emph{JHEP} {\bfseries 04} (2021) 237} [\href{https://arxiv.org/abs/2012.12091}{{\ttfamily 2012.12091}}].

\bibitem{Aprile:2020mus}
F.~Aprile, J.M.~Drummond, H.~Paul and M.~Santagata, \emph{{The Virasoro-Shapiro amplitude in AdS$_{5}$ \texttimes{} S$^{5}$ and level splitting of 10d conformal symmetry}}, \href{https://doi.org/10.1007/JHEP11(2021)109}{\emph{JHEP} {\bfseries 11} (2021) 109} [\href{https://arxiv.org/abs/2012.12092}{{\ttfamily 2012.12092}}].

\bibitem{Huang:2021xws}
Z.~Huang and E.Y.~Yuan, \emph{{Graviton scattering in AdS$_{5}$\texttimes{} S$^{5}$ at two loops}}, \href{https://doi.org/10.1007/JHEP04(2023)064}{\emph{JHEP} {\bfseries 04} (2023) 064} [\href{https://arxiv.org/abs/2112.15174}{{\ttfamily 2112.15174}}].

\bibitem{Drummond:2022dxw}
J.M.~Drummond and H.~Paul, \emph{{Two-loop supergravity on AdS$_{5}$\texttimes{}S$^{5}$ from CFT}}, \href{https://doi.org/10.1007/JHEP08(2022)275}{\emph{JHEP} {\bfseries 08} (2022) 275} [\href{https://arxiv.org/abs/2204.01829}{{\ttfamily 2204.01829}}].

\bibitem{Alday:2022uxp}
L.F.~Alday, T.~Hansen and J.A.~Silva, \emph{{AdS Virasoro-Shapiro from dispersive sum rules}}, \href{https://doi.org/10.1007/JHEP10(2022)036}{\emph{JHEP} {\bfseries 10} (2022) 036} [\href{https://arxiv.org/abs/2204.07542}{{\ttfamily 2204.07542}}].

\bibitem{Alday:2022xwz}
L.F.~Alday, T.~Hansen and J.A.~Silva, \emph{{AdS Virasoro-Shapiro from single-valued periods}}, \href{https://doi.org/10.1007/JHEP12(2022)010}{\emph{JHEP} {\bfseries 12} (2022) 010} [\href{https://arxiv.org/abs/2209.06223}{{\ttfamily 2209.06223}}].

\bibitem{Alday:2023jdk}
L.F.~Alday, T.~Hansen and J.A.~Silva, \emph{{Emergent Worldsheet for the AdS Virasoro-Shapiro Amplitude}}, \href{https://doi.org/10.1103/PhysRevLett.131.161603}{\emph{Phys. Rev. Lett.} {\bfseries 131} (2023) 161603} [\href{https://arxiv.org/abs/2305.03593}{{\ttfamily 2305.03593}}].

\bibitem{Alday:2023mvu}
L.F.~Alday and T.~Hansen, \emph{{The AdS Virasoro-Shapiro amplitude}}, \href{https://doi.org/10.1007/JHEP10(2023)023}{\emph{JHEP} {\bfseries 10} (2023) 023} [\href{https://arxiv.org/abs/2306.12786}{{\ttfamily 2306.12786}}].

\bibitem{Fitzpatrick:2014oza}
A.L.~Fitzpatrick, J.~Kaplan, Z.U.~Khandker, D.~Li, D.~Poland and D.~Simmons-Duffin, \emph{{Covariant Approaches to Superconformal Blocks}}, \href{https://doi.org/10.1007/JHEP08(2014)129}{\emph{JHEP} {\bfseries 08} (2014) 129} [\href{https://arxiv.org/abs/1402.1167}{{\ttfamily 1402.1167}}].

\bibitem{Khandker:2014mpa}
Z.U.~Khandker, D.~Li, D.~Poland and D.~Simmons-Duffin, \emph{{$ \mathcal{N} $ = 1 superconformal blocks for general scalar operators}}, \href{https://doi.org/10.1007/JHEP08(2014)049}{\emph{JHEP} {\bfseries 08} (2014) 049} [\href{https://arxiv.org/abs/1404.5300}{{\ttfamily 1404.5300}}].

\bibitem{Lemos:2016xke}
M.~Lemos, P.~Liendo, C.~Meneghelli and V.~Mitev, \emph{{Bootstrapping $\mathcal{N}=3$ superconformal theories}}, \href{https://doi.org/10.1007/JHEP04(2017)032}{\emph{JHEP} {\bfseries 04} (2017) 032} [\href{https://arxiv.org/abs/1612.01536}{{\ttfamily 1612.01536}}].

\bibitem{Li:2017ddj}
D.~Li, D.~Meltzer and A.~Stergiou, \emph{{Bootstrapping mixed correlators in 4D $ \mathcal{N} $ = 1 SCFTs}}, \href{https://doi.org/10.1007/JHEP07(2017)029}{\emph{JHEP} {\bfseries 07} (2017) 029} [\href{https://arxiv.org/abs/1702.00404}{{\ttfamily 1702.00404}}].

\bibitem{Cornagliotto:2017dup}
M.~Cornagliotto, M.~Lemos and V.~Schomerus, \emph{{Long Multiplet Bootstrap}}, \href{https://doi.org/10.1007/JHEP10(2017)119}{\emph{JHEP} {\bfseries 10} (2017) 119} [\href{https://arxiv.org/abs/1702.05101}{{\ttfamily 1702.05101}}].

\bibitem{Ramirez:2018lpd}
I.A.~Ram\'\i{}rez, \emph{{Towards general super Casimir equations for $4D$ ${\mathcal N}=1$ SCFTs}}, \href{https://doi.org/10.1007/JHEP03(2019)047}{\emph{JHEP} {\bfseries 03} (2019) 047} [\href{https://arxiv.org/abs/1808.05455}{{\ttfamily 1808.05455}}].

\bibitem{Kos:2018glc}
F.~Kos and J.~Oh, \emph{{2d small N=4 Long-multiplet superconformal block}}, \href{https://doi.org/10.1007/JHEP02(2019)001}{\emph{JHEP} {\bfseries 02} (2019) 001} [\href{https://arxiv.org/abs/1810.10029}{{\ttfamily 1810.10029}}].

\bibitem{Buric:2019rms}
I.~Buric, V.~Schomerus and E.~Sobko, \emph{{Superconformal Blocks: General Theory}}, \href{https://doi.org/10.1007/JHEP01(2020)159}{\emph{JHEP} {\bfseries 01} (2020) 159} [\href{https://arxiv.org/abs/1904.04852}{{\ttfamily 1904.04852}}].

\bibitem{Buric:2020buk}
I.~Buri\'c, V.~Schomerus and E.~Sobko, \emph{{The superconformal $X$-ing equation}}, \href{https://doi.org/10.1007/JHEP10(2020)147}{\emph{JHEP} {\bfseries 10} (2020) 147} [\href{https://arxiv.org/abs/2005.13547}{{\ttfamily 2005.13547}}].

\bibitem{Buric:2020qzp}
I.~Buri\'c, V.~Schomerus and E.~Sobko, \emph{{Crossing symmetry for long multiplets in 4D $ \mathcal{N} $ = 1 SCFTs}}, \href{https://doi.org/10.1007/JHEP04(2021)130}{\emph{JHEP} {\bfseries 04} (2021) 130} [\href{https://arxiv.org/abs/2011.14116}{{\ttfamily 2011.14116}}].

\bibitem{Li:2018mdl}
Z.~Li, \emph{{Superconformal partial waves for stress-tensor multiplet correlator in 4D$ \mathcal{N} $ = 2 SCFTs}}, \href{https://doi.org/10.1007/JHEP05(2020)101}{\emph{JHEP} {\bfseries 05} (2020) 101} [\href{https://arxiv.org/abs/1806.11550}{{\ttfamily 1806.11550}}].

\bibitem{Rakshit:2023baq}
S.~Rakshit and S.~Mukhopadhyay, \emph{{Superconformal blocks for stress-tensor and chiral operators for 4D ${\mathcal {N}}=2$ superconformal field theories}}, \href{https://doi.org/10.1140/epjc/s10052-024-12903-6}{\emph{Eur. Phys. J. C} {\bfseries 84} (2024) 567} [\href{https://arxiv.org/abs/2301.02782}{{\ttfamily 2301.02782}}].

\bibitem{Galperin:1984av}
A.~Galperin, E.~Ivanov, S.~Kalitsyn, V.~Ogievetsky and E.~Sokatchev, \emph{{Unconstrained N=2 Matter, Yang-Mills and Supergravity Theories in Harmonic Superspace}}, \href{https://doi.org/10.1088/0264-9381/1/5/004}{\emph{Class. Quant. Grav.} {\bfseries 1} (1984) 469}.

\bibitem{Howe:1995md}
P.S.~Howe and G.G.~Hartwell, \emph{{A Superspace survey}}, \href{https://doi.org/10.1088/0264-9381/12/8/005}{\emph{Class. Quant. Grav.} {\bfseries 12} (1995) 1823}.

\bibitem{Hartwell:1994rp}
G.G.~Hartwell and P.S.~Howe, \emph{{(N, p, q) harmonic superspace}}, \href{https://doi.org/10.1142/S0217751X95001820}{\emph{Int. J. Mod. Phys. A} {\bfseries 10} (1995) 3901} [\href{https://arxiv.org/abs/hep-th/9412147}{{\ttfamily hep-th/9412147}}].

\bibitem{Howe:2001je}
P.S.~Howe and P.C.~West, \emph{{AdS / SCFT in superspace}}, \href{https://doi.org/10.1088/0264-9381/18/16/305}{\emph{Class. Quant. Grav.} {\bfseries 18} (2001) 3143} [\href{https://arxiv.org/abs/hep-th/0105218}{{\ttfamily hep-th/0105218}}].

\bibitem{Heslop:2001dr}
P.J.~Heslop and P.S.~Howe, \emph{{A Note on composite operators in N=4 SYM}}, \href{https://doi.org/10.1016/S0370-2693(01)00961-3}{\emph{Phys. Lett. B} {\bfseries 516} (2001) 367} [\href{https://arxiv.org/abs/hep-th/0106238}{{\ttfamily hep-th/0106238}}].

\bibitem{Heslop:2001gp}
P.J.~Heslop and P.S.~Howe, \emph{{OPEs and three-point correlators of protected operators in N=4 SYM}}, \href{https://doi.org/10.1016/S0550-3213(02)00023-8}{\emph{Nucl. Phys. B} {\bfseries 626} (2002) 265} [\href{https://arxiv.org/abs/hep-th/0107212}{{\ttfamily hep-th/0107212}}].

\bibitem{Heslop:2001zm}
P.J.~Heslop, \emph{{Superfield representations of superconformal groups}}, \href{https://doi.org/10.1088/0264-9381/19/2/309}{\emph{Class. Quant. Grav.} {\bfseries 19} (2002) 303} [\href{https://arxiv.org/abs/hep-th/0108235}{{\ttfamily hep-th/0108235}}].

\bibitem{Heslop:2002hp}
P.J.~Heslop and P.S.~Howe, \emph{{Four point functions in N=4 SYM}}, \href{https://doi.org/10.1088/1126-6708/2003/01/043}{\emph{JHEP} {\bfseries 01} (2003) 043} [\href{https://arxiv.org/abs/hep-th/0211252}{{\ttfamily hep-th/0211252}}].

\bibitem{Heslop:2003xu}
P.J.~Heslop and P.S.~Howe, \emph{{Aspects of N=4 SYM}}, \href{https://doi.org/10.1088/1126-6708/2004/01/058}{\emph{JHEP} {\bfseries 01} (2004) 058} [\href{https://arxiv.org/abs/hep-th/0307210}{{\ttfamily hep-th/0307210}}].

\bibitem{alcock2017compact}
J.~Alcock-Zeilinger and H.~Weigert, \emph{Compact hermitian young projection operators}, \href{https://doi.org/10.1063/1.4983478}{\emph{Journal of Mathematical Physics} {\bfseries 58} (2017) }.

\bibitem{Alcock-Zeilinger:2016cva}
J.~Alcock-Zeilinger and H.~Weigert, \emph{{Transition Operators}}, \href{https://doi.org/10.1063/1.4983479}{\emph{J. Math. Phys.} {\bfseries 58} (2017) 051703} [\href{https://arxiv.org/abs/1610.08802}{{\ttfamily 1610.08802}}].

\bibitem{garsia2020lectures}
A.M.~Garsia and O.~Egecioglu, \emph{Lectures in algebraic combinatorics}, {\emph{Lecture Notes in Math} {\bfseries 2277} (2020) }.

\bibitem{Aprile:2017xsp}
F.~Aprile, J.M.~Drummond, P.~Heslop and H.~Paul, \emph{{Unmixing Supergravity}}, \href{https://doi.org/10.1007/JHEP02(2018)133}{\emph{JHEP} {\bfseries 02} (2018) 133} [\href{https://arxiv.org/abs/1706.08456}{{\ttfamily 1706.08456}}].

\bibitem{Aprile:2025nta}
F.~Aprile, J.M.~Drummond, P.J.~Heslop and M.~Santagata, \emph{{A formula for the block expansion in free CFTs and applications to ${\cal N}=4$ SYM at strong coupling}},  \href{https://arxiv.org/abs/2502.14077}{{\ttfamily 2502.14077}}.

\bibitem{Howe:1994ms}
P.S.~Howe and M.I.~Leeming, \emph{{Harmonic superspaces in low dimensions}}, \href{https://doi.org/10.1088/0264-9381/11/12/004}{\emph{Class. Quant. Grav.} {\bfseries 11} (1994) 2843} [\href{https://arxiv.org/abs/hep-th/9408062}{{\ttfamily hep-th/9408062}}].

\bibitem{Howe:2000nq}
P.S.~Howe, \emph{{Aspects of the D = 6, (2,0) tensor multiplet}}, \href{https://doi.org/10.1016/S0370-2693(00)01304-6}{\emph{Phys. Lett. B} {\bfseries 503} (2001) 197} [\href{https://arxiv.org/abs/hep-th/0008048}{{\ttfamily hep-th/0008048}}].

\bibitem{Heslop:2004du}
P.J.~Heslop, \emph{{Aspects of superconformal field theories in six dimensions}}, \href{https://doi.org/10.1088/1126-6708/2004/07/056}{\emph{JHEP} {\bfseries 07} (2004) 056} [\href{https://arxiv.org/abs/hep-th/0405245}{{\ttfamily hep-th/0405245}}].

\bibitem{Aprile:2021pwd}
F.~Aprile and P.~Heslop, \emph{{Superconformal Blocks in Diverse Dimensions and BC Symmetric Functions}}, \href{https://doi.org/10.1007/s00220-023-04740-7}{\emph{Commun. Math. Phys.} {\bfseries 402} (2023) 995} [\href{https://arxiv.org/abs/2112.12169}{{\ttfamily 2112.12169}}].

\bibitem{Dirac:1936fq}
P.A.M.~Dirac, \emph{{Wave equations in conformal space}}, \href{https://doi.org/10.2307/1968455}{\emph{Annals Math.} {\bfseries 37} (1936) 429}.

\bibitem{Simmons-Duffin:2012juh}
D.~Simmons-Duffin, \emph{{Projectors, Shadows, and Conformal Blocks}}, \href{https://doi.org/10.1007/JHEP04(2014)146}{\emph{JHEP} {\bfseries 04} (2014) 146} [\href{https://arxiv.org/abs/1204.3894}{{\ttfamily 1204.3894}}].

\bibitem{stembridge1985characterization}
J.R.~Stembridge, \emph{A characterization of supersymmetric polynomials}, \href{https://doi.org/10.1016/0021-8693(85)90115-2}{\emph{Journal of algebra} {\bfseries 95} (1985) 439}.

\bibitem{ZuckermanFunctors}
G.~Zuckerman, \emph{Tensor products of finite and infinite dimensional representations of semisimple lie groups}, \href{https://doi.org/10.2307/1971097}{\emph{Annals of Mathematics} {\bfseries 106} (1977) 295}.

\bibitem{jantzen_highest_weight_modules}
J.C.~Jantzen, \emph{Moduln mit einem höchsten Gewicht}, vol.~750 of \emph{Lecture Notes in Mathematics}, Springer (1979), \href{https://doi.org/10.1007/BFb0069521}{10.1007/BFb0069521}.

\bibitem{knapp_vogan_cohomological_induction}
A.W.~Knapp and D.A.~Vogan, \emph{Cohomological Induction and Unitary Representations}, vol.~45 of \emph{Princeton Mathematical Series}, Princeton University Press, Princeton, NJ (1995).

\bibitem{Witten:2003nn}
E.~Witten, \emph{{Perturbative gauge theory as a string theory in twistor space}}, \href{https://doi.org/10.1007/s00220-004-1187-3}{\emph{Commun. Math. Phys.} {\bfseries 252} (2004) 189} [\href{https://arxiv.org/abs/hep-th/0312171}{{\ttfamily hep-th/0312171}}].

\bibitem{Cordova:2016emh}
C.~Cordova, T.T.~Dumitrescu and K.~Intriligator, \emph{{Multiplets of Superconformal Symmetry in Diverse Dimensions}}, \href{https://doi.org/10.1007/JHEP03(2019)163}{\emph{JHEP} {\bfseries 03} (2019) 163} [\href{https://arxiv.org/abs/1612.00809}{{\ttfamily 1612.00809}}].

\bibitem{CastedoEcheverri:2015mkz}
A.~Castedo~Echeverri, E.~Elkhidir, D.~Karateev and M.~Serone, \emph{{Deconstructing Conformal Blocks in 4D CFT}}, \href{https://doi.org/10.1007/JHEP08(2015)101}{\emph{JHEP} {\bfseries 08} (2015) 101} [\href{https://arxiv.org/abs/1505.03750}{{\ttfamily 1505.03750}}].

\bibitem{Schreier:1971um}
E.J.~Schreier, \emph{{Conformal symmetry and three-point functions}}, \href{https://doi.org/10.1103/PhysRevD.3.980}{\emph{Phys. Rev. D} {\bfseries 3} (1971) 980}.

\bibitem{Osborn:1993cr}
H.~Osborn and A.C.~Petkou, \emph{{Implications of conformal invariance in field theories for general dimensions}}, \href{https://doi.org/10.1006/aphy.1994.1045}{\emph{Annals Phys.} {\bfseries 231} (1994) 311} [\href{https://arxiv.org/abs/hep-th/9307010}{{\ttfamily hep-th/9307010}}].

\bibitem{CastedoEcheverri:2016dfa}
A.~Castedo~Echeverri, E.~Elkhidir, D.~Karateev and M.~Serone, \emph{{Seed Conformal Blocks in 4D CFT}}, \href{https://doi.org/10.1007/JHEP02(2016)183}{\emph{JHEP} {\bfseries 02} (2016) 183} [\href{https://arxiv.org/abs/1601.05325}{{\ttfamily 1601.05325}}].

\bibitem{Bissi:2021hjk}
A.~Bissi, G.~Fardelli and A.~Manenti, \emph{{Rebooting quarter-BPS operators in $ \mathcal{N} $ = 4 super Yang-Mills}}, \href{https://doi.org/10.1007/JHEP04(2022)016}{\emph{JHEP} {\bfseries 04} (2022) 016} [\href{https://arxiv.org/abs/2111.06857}{{\ttfamily 2111.06857}}].

\bibitem{Bhattacharyya:2008rb}
R.~Bhattacharyya, S.~Collins and R.~de~Mello~Koch, \emph{{Exact Multi-Matrix Correlators}}, \href{https://doi.org/10.1088/1126-6708/2008/03/044}{\emph{JHEP} {\bfseries 03} (2008) 044} [\href{https://arxiv.org/abs/0801.2061}{{\ttfamily 0801.2061}}].

\bibitem{Collins:2008gc}
S.~Collins, \emph{{Restricted Schur Polynomials and Finite N Counting}}, \href{https://doi.org/10.1103/PhysRevD.79.026002}{\emph{Phys. Rev. D} {\bfseries 79} (2009) 026002} [\href{https://arxiv.org/abs/0810.4217}{{\ttfamily 0810.4217}}].

\bibitem{Padellaro:2024rld}
A.~Padellaro, S.~Ramgoolam and R.~Suzuki, \emph{{Eigenvalue systems for integer orthogonal bases of multi-matrix invariants at finite N}}, \href{https://doi.org/10.1007/JHEP02(2025)111}{\emph{JHEP} {\bfseries 02} (2025) 111} [\href{https://arxiv.org/abs/2410.13631}{{\ttfamily 2410.13631}}].

\bibitem{Schomerus:2016epl}
V.~Schomerus, E.~Sobko and M.~Isachenkov, \emph{{Harmony of Spinning Conformal Blocks}}, \href{https://doi.org/10.1007/JHEP03(2017)085}{\emph{JHEP} {\bfseries 03} (2017) 085} [\href{https://arxiv.org/abs/1612.02479}{{\ttfamily 1612.02479}}].

\bibitem{Schomerus:2017eny}
V.~Schomerus and E.~Sobko, \emph{{From Spinning Conformal Blocks to Matrix Calogero-Sutherland Models}}, \href{https://doi.org/10.1007/JHEP04(2018)052}{\emph{JHEP} {\bfseries 04} (2018) 052} [\href{https://arxiv.org/abs/1711.02022}{{\ttfamily 1711.02022}}].

\bibitem{Isachenkov:2017qgn}
M.~Isachenkov and V.~Schomerus, \emph{{Integrability of conformal blocks. Part I. Calogero-Sutherland scattering theory}}, \href{https://doi.org/10.1007/JHEP07(2018)180}{\emph{JHEP} {\bfseries 07} (2018) 180} [\href{https://arxiv.org/abs/1711.06609}{{\ttfamily 1711.06609}}].

\bibitem{Buric:2019dfk}
I.~Buri\'c, M.~Isachenkov and V.~Schomerus, \emph{{Conformal Group Theory of Tensor Structures}}, \href{https://doi.org/10.1007/JHEP10(2020)004}{\emph{JHEP} {\bfseries 10} (2020) 004} [\href{https://arxiv.org/abs/1910.08099}{{\ttfamily 1910.08099}}].

\bibitem{Agarwal:2023xwl}
P.~Agarwal, R.C.~Brower, T.G.~Raben and C.-I.~Tan, \emph{{Embedding space approach to Lorentzian CFT amplitudes and causal spherical functions}}, \href{https://doi.org/10.1103/PhysRevD.110.086019}{\emph{Phys. Rev. D} {\bfseries 110} (2024) 086019} [\href{https://arxiv.org/abs/2302.06469}{{\ttfamily 2302.06469}}].

\bibitem{Koelink:2021hlp}
E.~Koelink and J.~Liu, \emph{{$BC_2$ type multivariable matrix functions and matrix spherical functions}},  \href{https://arxiv.org/abs/2110.02287}{{\ttfamily 2110.02287}}.

\bibitem{Goncalves:2019znr}
V.~Gon\c{c}alves, R.~Pereira and X.~Zhou, \emph{{$20'$ Five-Point Function from $AdS_5\times S^5$ Supergravity}}, \href{https://doi.org/10.1007/JHEP10(2019)247}{\emph{JHEP} {\bfseries 10} (2019) 247} [\href{https://arxiv.org/abs/1906.05305}{{\ttfamily 1906.05305}}].

\bibitem{Konstein:2000bi}
S.E.~Konstein, M.A.~Vasiliev and V.N.~Zaikin, \emph{{Conformal higher spin currents in any dimension and AdS / CFT correspondence}}, \href{https://doi.org/10.1088/1126-6708/2000/12/018}{\emph{JHEP} {\bfseries 12} (2000) 018} [\href{https://arxiv.org/abs/hep-th/0010239}{{\ttfamily hep-th/0010239}}].

\bibitem{Bianchi:2005ze}
M.~Bianchi, P.J.~Heslop and F.~Riccioni, \emph{{More on La Grande Bouffe}}, \href{https://doi.org/10.1088/1126-6708/2005/08/088}{\emph{JHEP} {\bfseries 08} (2005) 088} [\href{https://arxiv.org/abs/hep-th/0504156}{{\ttfamily hep-th/0504156}}].

\bibitem{keppeler2014hermitian}
S.~Keppeler and M.~Sj\"odahl, \emph{{Hermitian Young Operators}}, \href{https://doi.org/10.1063/1.4865177}{\emph{J. Math. Phys.} {\bfseries 55} (2014) 021702} [\href{https://arxiv.org/abs/1307.6147}{{\ttfamily 1307.6147}}].

\bibitem{jucys1974symmetric}
A.-A.~Jucys, \emph{Symmetric polynomials and the center of the symmetric group ring}, \href{https://doi.org/10.1016/0034-4877(74)90019-6}{\emph{Reports on Mathematical Physics} {\bfseries 5} (1974) 107}.

\bibitem{murphy1981new}
G.E.~Murphy, \emph{A new construction of young's seminormal representation of the symmetric groups}, \href{https://doi.org/10.1016/0021-8693(81)90205-2}{\emph{Journal of Algebra} {\bfseries 69} (1981) 287}.

\bibitem{martin2002xact}
J.M.~Mart{\'\i}n-Garc{\'\i}a, A.~Garc{\'\i}a-Parrado, A.~Stecchina, B.~Wardell, C.~Pitrou, D.~Brizuela et~al., \emph{xact: Efficient tensor computer algebra for mathematica}, {\emph{URL: http://xact.es} (2002) }.

\bibitem{martin2008xtensor}
J.M.~Mart{\'\i}n-Garc{\'\i}a, \emph{xtensor: A free fast abstract tensor manipulator},  in \emph{The Eleventh Marcel Grossmann Meeting}, pp.~1552--1554, World Scientific, 2008, \href{https://doi.org/10.1142/9789812834300_0192}{DOI}.

\bibitem{nutma2014xtras}
T.~Nutma, \emph{xtras: A field-theory inspired xact package for mathematica}, \href{https://doi.org/10.1016/j.cpc.2014.02.006}{\emph{Computer Physics Communications} {\bfseries 185} (2014) 1719}.

\end{thebibliography}\endgroup

\end{document}